\documentclass[11pt,expanded,copyright]{fsuthesis}

\usepackage{amsmath}
\usepackage{bm}
\usepackage{braket}
\usepackage{tensor}
\usepackage{mathtools}
\usepackage{amssymb}
\usepackage{color}
\usepackage{array}
\usepackage{chngcntr}
\usepackage{cite}

\numberwithin{equation}{section}
\allowdisplaybreaks
\DeclareMathOperator{\Tr}{\text{Tr}}

\author{Hiram G. Menendez Santiago}
\title{Angular Distributions, Polarization Observables, Spin Density Matrices and Statistical Tensors in Photoproduction of Two Pseudoscalar Mesons off a Nucleon}
\college{College of Arts and Sciences}
\degree{Doctor of Philosophy}
\manuscripttype{Dissertation}
\degreeyear{2020}
\defensedate{June 25, 2020}

\committeeperson{Wei Yang}{University Representative}
\committeeperson{Simon Capstick}{Committee Member}
\committeeperson{Volker Crede}{Committee Member}
\committeeperson{Efstratios Manousakis}{Committee Member}

\newcommand{\minus}{\scalebox{0.75}{-}}
\newcommand{\mt}{& \hphantom{=}}

\counterwithout{equation}{section}

\begin{document}




\frontmatter 

\maketitle            
\makecommitteepage 

\begin{acknowledgments}
As is the case with any endeavor requiring hard work and dedication, this work would not have been possible without the help of some wonderful individuals that have earned my gratitude. No one has helped me more in the development of this research than my adviser, Dr. Winston Roberts. Without his guidance, completion of this work would have been significantly harder. I would not have been able to arrive at some of the results in this work if not for his insight and experience. The fact that we would often both arrive at the same conclusions by thinking in completely different ways was of immense value. But even more so were the times when his insight would cast doubt into some of the results I would obtain. Having someone always available that could challenge my work and ideas made this research project much better than it would have been. I am also grateful for our many discussions on topics in physics that were, at best, tangentially related to this work, and to those not related in any way to earning my degree. 

I would like to thank the many wonderful friends that I have met after arriving in Tallahassee, including many who left town long ago to pursue greater things. First, thanks to the members of the Friday night board games group: Jonathan Baron, Nellie Speirs Baron, Andrue Christopher Henderson, Sierra Henderson, Chaille Kent, Stephen Kolar and Taylor Van Winkle. Those nights were consistently some of the best highlights of my weeks. The games were great, but all of you were phenomenal. Thanks also to the awesome friends I met at a trip to New Orleans, where we bonded in our misery at not being able to sit down and rest our feet unless we bought alcohol we did not intend to drink: Archishman Ghosh, Vishav Pandhi, Dhevathi Rajagopalan, and Garret Dan Vo. We could always on Archis to come up with new ideas for fun things to do. I'm happy that Archis and Dhev had the chance to attend Friday night board games, if only for a brief time. I am also grateful to have had the company of Dhev while working long hours at All Saints Caf\'{e}, which made the work experience so much more enjoyable. I wish to thank my office mate and friend Zulkaida Akbar, with whom I also spent many times at All Saints Caf\'{e}. I cherish all the good times I've had with him and I am proud to have helped him obtain his driving license. I am also happy to have helped Dhev obtain his driving license as well. I was lucky to have another group of friends from a different game group: Ryan Baird, Dan Mock, and Alex Parker. I was introduced to many great games thanks to them, and I also enjoyed our many conversations. Many thanks to Anish Bhardwaj, one of the first people I met at the physics department, and all the times we spent at Sweet Pea Caf\'{e}. Thanks also to Nabin Rijal. I will miss the free beer we got twice a week at Poor Paul's Poorhouse. Last but not least, I am extremely grateful to have met Weng Ramirez. I never would have guessed I would ever visit The Philippines and have such an amazing time. Thank you so much for all you've done for me.

But I also have to thank those friends who I have met many years before the beginning of my graduate studies. While I have not been able to physically spend time with them in a long time, the days I get to contact them are always great. These include Ricardo Arzola, Janet Col\'{o}n Castellano, Walter Morales, Kendrick Ng and Juan Carlos S\'{a}nchez. 

I wish to give a huge thanks to the Florida State University Department of Physics, for giving me the chance to pursue this degree despite my very limited experience in the field of physics. It has been a wonderful experience to finally get work in this field, which is something I had wanted to do for a long time. I would like to thank the excellent professors that make up this department. Many thanks to Dr. Jorge Piekarewicz, for his excellent lectures on classical electrodynamics and for his office's open door policy that led to us having some interesting conversations. Thanks also to Dr. Efstratios Manousakis, for his also excellent lectures on quantum mechanics, his carefully handwritten lecture notes that I still keep, and for the discussions we've had, especially on the foundations of quantum mechanics. Many thanks to Dr. Simon Capstick, for always being available in his office to answer any kinds of questions that came to my mind. I also wish to thank Dr. Volker Crede, for our helpful discussions regarding my research and the experiments being done at JLab. From outside the Department of Physics, I would like to thank Dr. Wei Yang for taking some time out of his work to serve on my graduate committee. And finally, many thanks to Dr. Don Robson, whose comments regarding my research during my presentation at a nuclear seminar led me to significantly expand the scope of this research.

I am eternally grateful to the members of my family, among them my sisters Lorena Men\'{e}ndez and Hiradith Men\'{e}ndez, and my favorite niece Lorena Gabriela Men\'{e}ndez, . While we are each living our own separate lives far away from each other, it makes me happy to know that I have a group of people in my life that I know I can always count on, no matter the time, place, or circumstances.

And finally, but certainly not least, I give my heartfelt thanks to my mother, Judith Ivette Santiago Rivera. I can say, without a shadow of a doubt, that there has never been anyone in the world who has helped, taught, worried about and cared about me to the extent that she has. Everything I have accomplished in life, including this work, would not have been possible without the immense influence she has had in shaping my life. 

I would like to thank the FSU College of Arts and Sciences and the FSU Office of Research for making this work possible. This work was partially supported by the US Department of Energy through award DE-SC0002615.
\end{acknowledgments}

\tableofcontents
\listoftables
\listoffigures






\begin{abstract}
In two meson photoproduction off a nucleon, for the case in which two of the three final state hadrons are products of the decay of an intermediate resonance, general expressions for its decay distribution and for polarization observables are derived in a model independent way. These are functions of either the spin density matrix elements (SDME's) or statistical tensors of the resonance, and the angles of its decay products. The expressions are general enough that it also describes cases where more than one resonance of arbitrary quantum numbers contributes, including interference effects. They can therefore be used to extract the SDME's or statistical tensors of the resonances that contribute to the reaction. 
\end{abstract}

\mainmatter



\chapter{Introduction} \label{Introduction}
The field of hadronic nuclear physics aims to understand phenomena arising due to the strong nuclear force, and the properties of particles that 
interact via said force, called hadrons. Its existence was deduced from the fact that the electromagnetic interaction could not be  responsible for binding protons and neutrons in an atomic nucleus, since they have positive and neutral charge, respectively. In the electromagnetic
interaction, positively charged particles would experience a repulsive force, and neutral particles would not participate in said
interaction. It was therefore clear that protons and neutrons had to also participate in another
type of interaction, stronger and shorter in range than the electromagnetic one. This interaction came to be known alternatively as the
strong nuclear force, the strong force, or the strong interactions. This section highlights some of the important discoveries that have 
increased our understanding of it throughout the years after its discovery.

In 1932, Werner Heisenberg made the observation that, if the electromagnetic interaction could somehow be ``turned off'', protons
and neutrons would interact via the strong force identically. In fact, if additionally it were the case that the proton and the neutron had
identical masses, there would be no way to distinguish between them. Since the electromagnetic interaction is orders of magnitude weaker than
the strong one, and since the proton and neutron mass is almost identical, if all protons in a nucleus were to be replaced
with neutrons and vice versa, the mass of the resulting nucleus would be almost identical to the previous one. 

Heisenberg therefore introduced the concept of an approximate symmetry which came to be known as isospin  \cite{heisenberg_uber_1989}. The reason for this name is due to
its analogy with the concept of spin. All the fundamental particles known at the time have spin 1/2 (in units of $\hslash$), and came in two states with
different spin projection along a coordinate axis:$+$1/2 and $-$1/2. In the same way, the proton and the neutron could be considered to be
different isospin states of a single isospin 1/2 particle, the nucleon. In one convention, the proton and neutron are considered to
have isospin projection of $+$1/2 and $-$1/2, respectively.

Just as with spin symmetry, isospin symmetry is described by the group $SU(2)$. The nucleon belongs to its fundamental representation, and is 
therefore an isospin doublet. Since this is an approximate symmetry, the members of this doublet (the proton and the neutron) are almost degenerate in mass. And just as angular momentum is conserved in interactions due to rotational symmetry, total isospin is conserved in the strong interactions, but not in the  electromagnetic and weak interactions. One example of the usefulness of isospin symmetry is that it can be used to relate the reaction rates of different strong processes that are related by rotations in isospin space.

The first hadrons other than the nucleon were discovered in cosmic ray experiments. In these,
detectors such as bubble chambers were constructed in order to detect the particles produced when cosmic rays collide with Earth's 
atmosphere. It was in these experiments that the charged pions were discovered in 1947 \cite{lattes_processes_1947}.The neutral
pion was discovered three years later \cite{bjorklund_high_1950, steinberger_evidence_1950, carlson_lxiii_1950, panofsky_gamma-ray_1950, panofsky_gamma-ray_1951}. The pions were actually predicted to exist earlier in 1935 by Hideki Yukawa \cite{yukawa_interaction_1935}, who described it as the 
mediator of the strong interaction among nucleons (in the same way that the photon mediates the electromagnetic interactions). 
He also correctly predicted its mass based on the range of the interaction.

Unlike the nucleon, which has half-integer
spin and is therefore a fermion, the pions have integer spin 0 and are therefore bosons. Hadrons that are fermions are called
baryons, while those that are bosons are called mesons. There are three pions with different electric charges that are almost degenerate in mass:
$\pi^{+}$, $\pi^{0}$, and $\pi^{-}$. This same year, the kaons were also discovered \cite{rochesterdr_evidence_1947}, also having different electric charges and almost
degenerate masses: $K^{0}$, $K^{+}$, $K^{-}$, and $\overline{K}^{0}$.

Particle accelerators with energies in the hundreds of MeV were eventually built. This led to the discovery of many more ``fundamental''
particles during the 1950's and 1960's, such as the $\Lambda^0$ baryon \cite{hopper_evidence_1950}, the $\eta$ and $\eta'$ mesons \cite{pevsner_evidence_1961, kalbfleisch_observation_1964, goldberg_existence_1964}, the $\Delta$ baryons \cite{hahn_neutrons_1952, anderson_total_1952}, the $\Sigma$ baryons \cite{plano_demostration_1957}, etc. Finding patterns
in the properties of these particles (sometimes informally called the ``particle zoo'') as a way to categorize them into different classes
became important. 

It was found that, in the same way that the states of the nucleon formed an isospin doublet, so too could these newly
discovered hadrons be assigned to different representations of $SU(2)$ isospin, with each of their members having nearly degenerate masses. The three
pion states form an isospin triplet,  $K^{0}$ and $K^{+}$ form a doublet, $K^{-}$ and $\overline{K}^{0}$ form another doublet, $\Lambda^0$ forms a singlet, $\eta$ and $\eta'$ each forms a singlet, the $\Sigma$ baryons form a triplet ($\Sigma^{+}$, 
$\Sigma^{0}$, and $\Sigma^{-}$), the delta baryons form a quadruplet ($\Delta^{++}$, $\Delta^{+}$, $\Delta^{0}$ and $\Delta^{-}$), etc.

Another property that was used to categorize them came from the observation that some of these hadrons had mean lifetimes that were orders of
magnitude larger than others. Because of this ``strange'' behavior, they were given the name strange hadrons. Their long lifetimes were 
explained by the introduction of a new quantum number called strangeness \cite{gell-mann_isotopic_1953, nakano_charge_1953, gell-mann_interpretation_1956, nishijima_theory_1956}, which is conserved in the strong and electromagnetic interactions 
but not in the weak interaction. Therefore, the former interactions can produce particle pairs with net strangeness of zero in relatively
large amounts. But since each particle in the pair carries strangeness, their ground states can only decay via the weak interaction, which leads them to have longer mean lifetimes.

The isospin multiplets that grouped the different hadrons into groups were still too numerous. A more general organization scheme called the
eightfold way was developed in 1961 by Murray Gell-Mann and Yuval Ne'eman \cite{gell-mann_eightfold_1961, neeman_derivation_1961}. In this scheme, the hadrons were organized into larger $SU(3)$ supermultiplets, with its members labeled by their isospin
projection ($SU(2)$ is a subgroup of $SU(3)$) and strangeness. The name eightfold way is a reference to the fact that $SU(3)$'s associated
lie algebra is eight-dimensional, i.e., it has eight generators (unlike $SU(2)$, which has 3). The degeneracy in the mass of the members of
these multiplets was not as exact as in $SU(2)$ isospin, making $SU(3)$ a less exact symmetry than the former, but still good enough to be
useful. Experimental validity for the eightfold way came with the discovery of the omega baryon
($\Omega^{-}$) in 1964 \cite{barnes_observation_1964}.

The mesons were organized into a singlet and an octet representation, while the baryons were organized into an octet and decuplet representation. But no collection of particles seemed to belong to the fundamental representation, the triplet. It was also known that tensor representations of groups can be decomposed into irreducible representations. For example, the representation $3 \otimes \bar{3}$
(the number refers to the dimensions of the representation, while the bar refers to the conjugate representation) can be decomposed into
$8 \oplus 1$, the octet and singlet representations found in the mesons. The representation $3 \otimes 3 \otimes 3$ decomposes into $10 \oplus 8 \oplus 8
\oplus 1$, which contains the decuplet and octet representations found in the baryons. It was therefore proposed, independently by Murray Gell-Mann and 
George Zweig in 1964 \cite{gell-mann_schematic_1964, zweig_su_1964}, that the hadrons were actually bound states of point-like spin-1/2
constituents belonging to the triplet representation. These were called quarks, and this proposal gave rise to the first quark model.

Since the quarks belong to the three-dimensional representation, they come
in three types, or flavors, called up, down, and strange ($u$, $d$ and $s$). Their associated antiquarks ($\bar{u}$, $\bar{d}$ and $\bar{s}$) 
belong to the conjugate of this representation. As their names imply, the up and down quark have zero strangeness, while the strange quark
and strange antiquark have strangeness with opposite sign. The baryons could therefore be organized into an
octet and a decuplet because they are bound states of three quarks (for the antibaryons, three antiquarks). The mesons could be organized
into a singlet and an octet because they are bound states of a quark and an antiquark. Since the hadrons have integer electric charge, the
quarks must therefore have fractional charges (+2/3 for the up quark, -1/3 for the down and strange quark. The antiquarks, have
the opposite charge). This approximate $SU(3)$ symmetry came to be known
as flavor symmetry, since it describes the symmetry between different flavored hadrons. 

The reason that this symmetry is not exact is in part due to the fact that the quarks have different masses. $SU(2)$ isospin is a better symmetry than $SU(3)$ flavor because the up and down quark masses are
very nearly identical. The latter is more badly broken because the strange quark has a significantly higher mass than the other
two.

While this quark model was very successful in describing the properties of the known hadrons at the time, its biggest perceived failure was
that quarks were never found in isolation, leading to discussions about whether quarks were ``real'' or the quark model was just a
mathematical bookkeeping tool, hiding as of yet unknown physics that would explain the properties of the hadrons. It would not be until the 
late 1960's and 1970's that deep inelastic scattering experiments done at the Stanford Linear Accelerator Center (SLAC) confirmed that the
nucleon was made up of fractionally-charged point-like constituents \cite{bloom_high-energy_1969, breidenbach_observed_1969}. 

Eventually, three other flavors of quarks were discovered: the charm \cite{augustin_discovery_1974, aubert_experimental_1974}, bottom \cite{herb_observation_1977}, and top quarks \cite{abe_f_et_al_cdf_collaboration_observation_1995, abachi_s_et_al_d0_collaboration_observation_1995}, with electric charges +2/3, -1/3, and +2/3, 
respectively. This means that the hadrons can, in principle, be organized into $SU(N)$ multiplets, with $N$ being the number of quark flavors
(It should be noted, however, that the top quark does not form bound states because its lifetime is too short).
However, these three quarks are much higher in mass than the three light quarks, so the symmetry ends up being much more badly broken. 

Another problem came from the observation that the $\Delta^{++}$ baryon required the spins of its three up quarks to be parallel and have zero total orbital angular momenta. It therefore seemed that this baryon could not have an antisymmetric wave function as is required of fermions by the Pauli exclusion principle. A proposed solution was to postulate the existence of a new quantum number, called color \cite{greenberg_spin_1964}. The idea
was that the quarks came in three different color states: red, blue, and green. This also offered an explanation for the lack of
observation of quarks in isolation by postulating the color hypothesis, which states that the color quantum number in all observed hadrons is described by a totally antisymmetric, color-singlet wave function. Therefore, a red, blue, and green quark would form bound states that have net color ``white'', i.e., it would be a color neutral state, while a colored quark and antiquark of the corresponding anti-color could also form a color-neutral bound state. Since individual quarks have net color,
they can never be observed in isolation. They can only be confined as color-neutral bound states that manifest as the hadrons that are observed
in experiments. This idea is often called color confinement. In mathematical language, a new $SU(3)$ color symmetry was postulated in which, in color space, quarks and antiquarks belonged to the fundamental representation and its conjugate, respectively, and that only bound states belonging to 
the singlet representation in color space (color neutral) are realized in nature.

Color confinement also has the implication that the energy of the bound quarks would increase if they get farther away from each other. 
Therefore, it was expected that if one of the quarks gets struck in a scattering event and gets pulled apart from the other quarks, energy 
gained by the system due to their strong interactions may be enough for new quark-antiquark pairs to be created from the vacuum, which could
then form new hadrons with the outgoing quark and the other remaining quarks.

While more experiments were providing clues that would hopefully lead to a general theory of the strong interactions, the electromagnetic
interaction at the quantum level was much better understood. It's theory was called quantum electrodynamics (QED), which is a field theory
of fermions and the photon. In addition to having the required global Lorentz symmetry, the U(1) global symmetry already present in the Dirac
equation that described relativistic spin-1/2 fermions was promoted to a local U(1) gauge symmetry. Local symmetries require the presence of
massless vector gauge bosons
belonging to the adjoint representation of the symmetry group. The adjoint representation of U(1) is one dimensional,
so only one vector gauge boson is needed, the photon. Gauge bosons are said to be the force carriers that mediate the interaction. Therefore,
interactions among particles can emerge from the requirement of local gauge invariance.

Since QED was extremely successful in correctly calculating known experimental quantities to very high precision (such as the fine structure 
constant), the idea to formulate the strong interactions as a local gauge theory eventually emerged and the $SU(3)$ color symmetry was promoted to a local
gauge symmetry \cite{nambu_systematics_1966, han_three-triplet_1965}. The group $SU(3)$ is non-abelian, unlike U(1) which is abelian. This
made the field theory of the strong interactions more complicated than QED, and had properties not found in the latter. For example, since
the adjoint representation of $SU(3)$ is eight dimensional, the gauge bosons of the theory belong to a color octet. The theory therefore
had 8 gauge bosons with different color charges, called gluons, in contrast to the single one in QED. Since the gluons themselves have a
color charge, they can interact directly with each other. By contrast, photons cannot
directly interact with each other since they lack an electric charge. This theory of the 
strong interactions was given the name quantum chromodynamics (QCD).

QCD has a property called asymptotic freedom \cite{gross_ultraviolet_1973, politzer_reliable_1973}, by which the strength of the strong 
interaction is less at higher momentum transfers (or shorter distance scales). This is due to the fact that the coupling strength between quarks and gluons receives quantum corrections that depend on the interaction energy, giving rise to a renormalized coupling. Unlike in QED, in QCD the renormalized coupling becomes smaller at larger energies. This means that in high energy collisions, the quarks and gluons barely 
interact with each other. Therefore,
calculations using the techniques of perturbation theory can be applied in this energy regime in QCD to make precise predictions. These
calculations have been put to the test in many experiments, and were very successful in describing experimental data. 

Due to the immense success of QCD, it is nowadays widely considered the correct theory of the strong 
interactions. However, perturbation theory cannot be applied to low-energy phenomena in the non-perturbative energy regime, where
the interaction strength becomes large. Therefore, perturbative QCD is incapable of describing a lot of phenomena of interest, such as the
hadron spectrum, the nature of color confinement, the hadronization process (in which individual quarks scattered during reactions form
hadrons), and the momentum, position, or spin distributions of the quarks that form the hadrons. The phenomena of color confinement has also not been able to 
be proven analytically from QCD. 

The only known way to extract information about low-energy phenomena from first-principles QCD calculations is from a formalism known as lattice QCD (LQCD) \cite{wilson_confinement_1974, polyakov_compact_1975, wegner_duality_1971}, in which spacetime is discretized to form a lattice in order to perform numerical computations. For example,
evidence for color confinement has been gathered from LQCD \cite{creutz_monte_1980, creutz_asymptotic-freedom_1980}. The drawback of this
method is the immense amount of time and computational resources needed. Many techniques have been used to cut down the computation time to 
manageable levels, such as using pion masses that are heavier than their real values as input, but at the cost of having less accurate results. 
But advances in in computer technology and computational techniques have allowed more realistic computations to be performed. Despite these
drawbacks, LQCD has been very successful in describing the hadron spectrum, correctly predicting, with some uncertainty, the masses of the known resonances \cite{fucito_hadron_1982, creutz_monte_1983, durr_ab_2008, langguth_monte_1984, bernard_qcd_2001, liu_recent_2014}.

We can therefore distinguish between the short-distance perturbative regime, where strong interaction physics is best described in
terms of nearly free, point-like quarks and gluons, and the long-distance non-perturbative regime, where it is best described in
terms of hadrons. For many of the long-distance phenomena of interest, there is often no known way
to obtain general analytic expressions for quantities of interest from QCD. For example, while the momentum distribution of the quarks and gluons in the hadrons can be 
measured from experiments \cite{gluck_dynamical_2008, nadolsky_implications_2008, martin_parton_2009, ball_first_2010}, there is no known way to find general expressions for them from QCD.

Physicists have therefore relied on models to describe the long-distance physics of hadrons, each with varying degrees of
success and range of applicability. For example, starting in the 1950's, a variety of quark models have been used that have given good
predictions to the hadron spectra and other properties such as couplings and decay rates \cite{hey_baryon_1983, capstick_quark_2000}. Many fall under the class known as potential quark
models, in which the interaction between constituent quarks is described with a non-local potential. Many of the potentials used in these
models were in fact inspired by QCD, i.e., known properties of QCD were used as guidance in developing them. Early quark models were 
non-relativistic, but eventually ones that did incorporate special relativity were developed. 

There are also low-energy effective descriptions of QCD. These come about when the energy scale of the interactions being studied is much smaller than some large energy scale that describes some short distance phenomena. From QCD, perturbative expansions can be done with respect to the ratio of these energy scales, and the techniques of perturbation theory can be
used to make calculations. These are called
effective field theories. One of them is heavy quark effective theory (HQET) \cite{eichten_effective_1990, georgi_effective_1990}, which gives an exact description in the limit where the mass
of one
of the quarks in the hadron goes to infinity. It has been used to give  good descriptions of hadrons which contain at least one charm
or bottom quark. Another such effective field theory is chiral perturbation theory (\raisebox{2pt}{$\chi$}PT) \cite{weinberg_phenomenological_1979, weinberg_nuclear_1990} which gives an exact description
in the limit where the quark masses go to zero.

In summary, since the discovery of the nuclear strong interactions responsible for the binding of protons and neutrons into nuclei at the 
center of atoms, experimental and theoretical efforts throughout the years have gradually increased our understanding of it, leading to the 
development of QCD, the theory of the strong interactions. Despite this, many aspects of the strong interaction are not well understood, 
especially in the low energy non-perturbative regime. This thesis presents my theoretical efforts in furthering our understanding of the strong interactions in this regime.

\chapter{Motivation} \label{Motivation}
One area of experimental research that was fundamental in furthering our understanding of the strong interaction is hadron spectroscopy,
which is concerned with searches of resonances arising in scattering experiments, and establishing their mass and decay widths. A compilation
of the known resonances and their respective masses and widths have been compiled in the Review of Particle Physics publication by the
Particle Data Group (PDG) collaboration \cite{pa_zyla_et_al_partilce_data_group_review_2020}. Predictions for the hadron spectrum have been extracted from constituent quark models since their
development in the 1960's. Many of the resonances predicted by these models have been found in experiments, and their calculated properties
(masses, widths, couplings, etc.) have matched the measured values with a good degree of success. Years later, LQCD has also been able to
make predictions about the hadron spectrum and, again, many of the predicted resonances were found in experiments, with the calculated
masses roughly matching those measured in experiments.

In the case of baryon spectroscopy, much of the information that has been gathered on the low-lying non-strange and strange baryon resonances
have been obtained from $\pi N$ and $\bar{K} N$ scattering \cite{klempt_baryon_2010}. This energy region is complicated, with 
numerous overlapping and broad resonances, making their identification difficult. Thus, measurements of the total and 
differential scattering cross sections are not enough to disentangle the contributions of the many resonances to the scattering amplitude.

To remedy this, Partial Wave Analysis (PWA) has been employed as a tool in analyzing the experimental data \cite{peters_primer_2006, cutkosky_pion-nucleon_1979, arndt_nucleon-nucleon_1983}, where the scattering
amplitude is decomposed into a sum of partial waves of definite orbital angular momenta, total isospin, and total angular momenta. This helps in the search for resonances because
each one of a particular spin, parity, and total isospin receives contributions from a single $\pi N$ partial wave. The goal of these analyses is to extract the partial
wave amplitudes from experimental data, since their behavior as a function of the kinematic variables (center-of-mass energy, momentum
transfer, etc.) can be interpreted in terms of the resonances that contribute to the scattering process, thus facilitating the identification
of resonant contributions. Partial wave amplitudes can be found
if the helicity amplitudes of the scattering process are known, which are the scattering amplitudes for processes in which the spin state of the particles involved is known. PWA's therefore require polarization experiments to be carried out, which make use of polarized beams
and/or targets, and may be able to measure the polarization of the reaction products. The observables of interest that are used to extract
the amplitudes are called polarization observables \cite{pichowsky_polarization_1996, roberts_polarization_2005}, which is one of the main topics of this work. 

As more resonances were discovered, it became clear that, while many of the resonances predicted by
constituent quark models were found, there were also a significant number that were not \cite{isgur_hyperfine_1977, isgur_p-wave_1978, isgur_positive-parity_1979, koniuk_where_1980, isgur_erratum_1981, burkert_nucleon_2016, capstick_quark_2000, saghai_search_2007}. LQCD also predicts an overabundance
of resonances \cite{edwards_excited_2011}. This is not necessarily a failure of quark models. One possible proposed solution is that some of the degrees of
freedom within the quark model are ``frozen'', leading to a reduction in the number of resonances. This can happen if two of the constituent
quarks in the baryon behaves as a collective unit. The baryon will then
effectively be a quark-diquark pair \cite{anselmino_diquarks_1993, ida_baryon_1966, lichtenberg_baryon_1967, lichtenberg_quark-diquark_1968}, and the reduction in the number of degrees of freedom would reduce the number of expected baryon
resonances. While early formulation of ``static'' quark-diquark models have recently been excluded by experiments, ``dynamic'' quark-diquark models have also been proposed as a possible solution. However, even with this reduction, the quark-diquark model still predicts more resonances that have been observed.

Another possible solution is that the missing/undiscovered resonances couple weakly to the production channel ($\pi N$). This hypothesis has
been verified in quark model calculations \cite{koniuk_where_1980, koniuk_baryon_1980, capstick_n_1993, capstick_quasi-two-body_1994, capstick_strange_1998}. These same
models suggest that many of these missing resonances couple strongly to channels such as $\pi \Delta$ and $\rho N$ (as of yet, LQCD calculations for strong decay couplings of exited baryons have not been carried out). 

Due to this, experiments using different production mechanisms, such as photoproduction and electroproduction, have been proposed. These have
been carried out in a number of facilities \cite{ireland_photoproduction_2020}, such as CEBAF, ESRF, MAMI, ELSA, SPring-8 and ELPH. At CEBAF, for example, polarized electron and photon beams are produced \cite{leemann_continuous_2001}, and its CLAS detector \cite{brooks_clas_2000} contains a polarized target (FroST) \cite{keith_jefferson_2012}. It is, however, not able to 
directly measure the polarization of the reaction products (but the polarization of weakly decaying particles, such as the $\Lambda^0$
baryon, can be deduced from the angular distribution of their decay products).

Some reactions of recent interest have been single and double pion photoproduction, $\gamma N \rightarrow \pi N$ and
$\gamma N \rightarrow \pi \pi N$, where the photon and/or the target nucleon are polarized. Since it is well known that the latter reaction has contributions from quasi-two-body states, reactions such as $\gamma N \rightarrow \rho N \rightarrow \pi \pi N$ and $\gamma N \rightarrow \pi \Delta \rightarrow \pi \pi N$ have also been of interest \cite{ballam_bubble-chamber_1972}. In these reactions, one of the hadrons in the quasi-two-body state undergoes a two-body decay (in the previous examples, $\rho \rightarrow \pi \pi$ and $\Delta \rightarrow \pi N$). It has often been of interest to measure the spin properties of this decaying hadron. 

In scattering experiments, the beam and target are made up of a large number of particles with spin. The spin state of every single particle
that is a member of the beam or target is not known. Rather, only partial information is ever known: the expectation value of the spin projection 
among all the particles, also known as the polarization of the system. Therefore, the beam and target are not pure quantum states, but rather statistical ensembles of single particle spin states, with each member of the ensemble having a probability of being realized. This is called a mixed quantum state, and it cannot be described by a state vector, but rather by a density matrix \cite{d_a_varshalovich_quantum_1988, blum_density_2012}. The density matrix formalism will be discussed in section \ref{SpinDensityMatrix}. All spin information about the beam, target, and recoiling particles is therefore contained in their respective spin density matrix. Experiments have control over the spin state of the beam and target, which means the spin density matrix elements (SDME's) of the spin and target will be known. For an initial spin configuration, the dynamics,
i.e., the helicity amplitudes of the process will determine the SDME's of the recoil baryon.

Spin density matrices are not the only ways to represent the spin of a mixed state. Alternatively, the expectation values of the polarization operators, called statistical tensors, can also be used \cite{d_a_varshalovich_quantum_1988}. While recently in the literature the topic of polarization observables has almost always been discussed in terms of the SDME's, this thesis will utilize both representations. This is because using statistical tensors instead of SDME's offer many advantages that will be discussed throughout this thesis. Their basic properties are discussed in section \ref{StatisticalTensors}.

Recently there has been interest in so-called complete experiments. These are scattering experiments in which enough polarization measurements are made to be able to extract all of the helicity amplitudes of the process (except for an overall phase). These types of measurements are called polarization observables, and will be discussed in section \ref{PolarizationObservables}.

Achieving complete experiments involves performing various scattering experiments where the beam and/or target is polarized in some chosen direction, and may also involve measuring the recoil baryon's polarization \cite{barker_complete_1975, keaton_amplitude_1996, chiang_completeness_1997}. These allow for the measurement of polarization observables at different kinematic points (scattering angle, momentum transfer, etc.), which can then be used to extract the SDME's of the recoil baryon for the chosen configuration of the beam and target polarizations. Since the values of these SDME's are determined by the helicity amplitudes, they can in turn be used to extract them. 

One method that has been used is to express the angular distribution of the decay products of the resonance forming part of the quasi-two-body
state as functions of its SDME's \cite{gottfried_connection_1964, schilling_analysis_1970, thews_high-energy_1968, kloet_spin_1998, titov_selected_2008}. For example, the following expression for the decay distribution of a photoproduced vector meson has
appeared in the literature, often in slightly different form,
\begin{equation} \label{WDistributionRho} 
\begin{split}
W(\theta^{*},\phi^{*};V) & = \rho_{00}(V) \cos^{2}\theta^{*} + \frac{1}{2}(\rho_{11}(V)+\rho_{-1-1}(V)) \sin^{2}\theta^{*} \\
     \mt -\frac{1}{\sqrt{2}}\Re[\rho_{10}(V)-\rho_{0-1}(V)]\sin2\theta^{*}\cos\phi^{*} \\
     \mt +\frac{1}{\sqrt{2}}\Im[\rho_{10}(V)-\rho_{0-1}(V)]\sin2\theta^{*}\sin\phi^{*} \\
     \mt - \Re[\rho_{1-1}(V)]\sin^{2}\theta^{*} \cos2\phi^{*}
                    + \Im[\rho_{1-1}(V)]\sin^{2}\theta^{*} \sin2\phi^{*},                     
 \end{split}
\end{equation}
while the following expression for the angular distribution of a spin-3/2 baryon has also appeared, also often in slightly different form,
\begin{equation} \label{WDistributionDelta1}
\begin{split}
W(\theta^{*},\phi^{*};3/2) & = \frac{5}{8}(\rho_{11}(3/2) +\rho_{-1-1}(3/2))(1 + \frac{3}{5} \cos 2 \theta^{*}) \\
    \mt +\frac{3}{4}(\rho_{33}(3/2)+\rho_{-3-3}(3/2))\sin^{2}\theta^{*} \\
    \mt -\frac{\sqrt{3}}{2}\Re[\rho_{31}(3/2)-\rho_{-1-3}(3/2)]\sin 2 \theta^{*} \cos \phi^{*} \\
    \mt +\frac{\sqrt{3}}{2}\Im[\rho_{31}(3/2)-\rho_{-1-3}(3/2)]\sin 2 \theta^{*} \sin \phi^{*} \\
    \mt -\frac{\sqrt{3}}{2}\Re[\rho_{3-1}(3/2)+\rho_{1-3}(3/2)]\sin^{2} \theta^{*} \cos 2 \phi^{*} \\
    \mt +\frac{\sqrt{3}}{2}\Im[\rho_{3-1}(3/2)+\rho_{1-3}(3/2)] \sin^{2} \theta^{*} \sin 2 \phi^{*}.
\end{split}
\end{equation}
The angles $\theta^{*}$ and $\phi^{*}$ are the polar and azimuthal angles, respectively, of the three-momentum vector of one of the decay
products, defined in the rest frame of the decaying resonance (in this frame, the other decay product has opposite three-momentum). The $\rho(V)$'s and the $\rho(3/2)$'s are the SDME's of a decaying vector meson and spin-3/2 baryon, respectively. It should be noted that, while not 
explicitly labeled in these expressions, these SDME's are also functions of the polarization state of the beam and target. We will later show that this dependence can be factored out of the SDME's. 
Both of these expressions can help in extracting the SDME's of the photoproduced meson or baryon by, for example, measuring the 
angular distributions of the mesons and the nucleon in the final state and performing fits with the SDME's as parameters.

The purpose of this research was to generalize this formalism in two ways. First, by considering the decay of resonances of arbitrary spin and second, by considering the case where more than one quasi-two-body state contributes to the reaction, taking into account interference effects. The relevance of this is that, as stated previously in section \ref{Introduction}, the energy region in which the missing resonances are being sought has many broad overlapping resonances. It would therefore be
useful to consider double-meson photoproduction under the assumption that the (possibly many) quasi-two-body state reaction channels contribute and interfere with one another.

This work makes no attempt at calculating the SDME's or statistical tensors of the decaying resonance from any particular model. Rather, our purpose is to find a general relationship between them and the angular distributions of the decaying resonance, and also between them and the polarization observables in a model independent way. 

Our entire analysis is completely independent of quantum numbers
other than spin. Therefore, it also applies to any other double-meson photoproduction reaction such as
\begin{equation} \label{Reactions}
    \begin{split}
        \gamma N & \rightarrow \pi \pi N, \\
        \gamma N & \rightarrow \pi \eta N, \\
        \gamma N & \rightarrow \eta \eta N, \\
        \gamma N & \rightarrow K \bar{K} N, \\
        \gamma N & \rightarrow \pi K Y, \\
        \gamma N & \rightarrow K K \Xi,
    \end{split}
\end{equation}

\chapter{Kinematics} \label{Kinematics}
Our reaction of interest is of the general form
\begin{equation} \label{Reaction}
  \gamma N \rightarrow M_{1} M_{2} B,
\end{equation}
where $N$ is the target nucleon, $M_{1}$ and $M_{2}$ are spin-0 mesons, and $B$ is a spin-1/2 baryon. An example of a process of this form 
that is being studied is double-pion photoproduction,
\begin{equation} \label{ReactionDoublePionPhotoproduction}
  \gamma N \rightarrow \pi \pi N.
\end{equation}
We will consider three different types of channels, which we will call pathways A, B, and C. The general form of pathway A is
\begin{equation} \label{PathwayA}
  \gamma N \rightarrow M^{*} B \rightarrow M_{1} M_{2} B, 
\end{equation}
in which a meson resonance or arbitrary spin $M^{*}$ decays into $M_{1}$ and $M_{2}$. The form of pathway B is
\begin{equation} \label{PathwayB}
  \gamma N \rightarrow M_{2} B^{*} \rightarrow M_{1} M_{2} B,
\end{equation}
in which a baryon resonance of arbitrary spin $B^{*}$ decays into $M_{1}$ and $B$. Pathway C has the form
\begin{equation} \label{PathwayC}
  \gamma N \rightarrow M_{1} B^{*} \rightarrow M_{1} M_{2} B,
\end{equation}
in which a baryon resonance of arbitrary spin $B^{*}$ decays into $M_{2}$ and $B$. 

An example of a contribution to double-pion photoproduction of the form of pathway A is vector meson photoproduction, such as the
photoproduction of a $\rho$ meson,
\begin{equation} \label{RhoPhotoproduction}
  \gamma N \rightarrow \rho N \rightarrow \pi_{1} \pi_{2} N.
\end{equation}
The subscripts on the pions are used to indicate that each is described by different scattering angles. An example of a contribution to the
same process of the form of path B and path C is the photoproduction of a spin-3/2 baryon, such as the
photoproduction of a $\Delta$ baryon,
\begin{equation} \label{DeltaPhotoproduction}
  \begin{split}
    & \gamma N \rightarrow \pi_{1} \Delta \rightarrow \pi_{1} \pi_{2} N, \\
    & \gamma N \rightarrow \pi_{2} \Delta \rightarrow \pi_{1} \pi_{2} N.
  \end{split}
\end{equation}
Note how for the case of baryon photoproduction where $M_1$ and $M_2$ are the same meson, both pathways B and C will contribute, since the baryon can decay into either of the two mesons. 

Energy-momentum conservation requires that
\begin{equation} \label{eqn 1}
k+p=q_{1}+q_{2}+p',
\end{equation}
where $k$, $p$ and $p'$ are the four-vectors of the photon, target nucleon and recoil baryon $B$, respectively, while $q_{1}$ and $q_{2}$ are
the four-vectors of the final state mesons $M_{1}$ and $M_{2}$.

\begin{figure}
  \begin{center}
    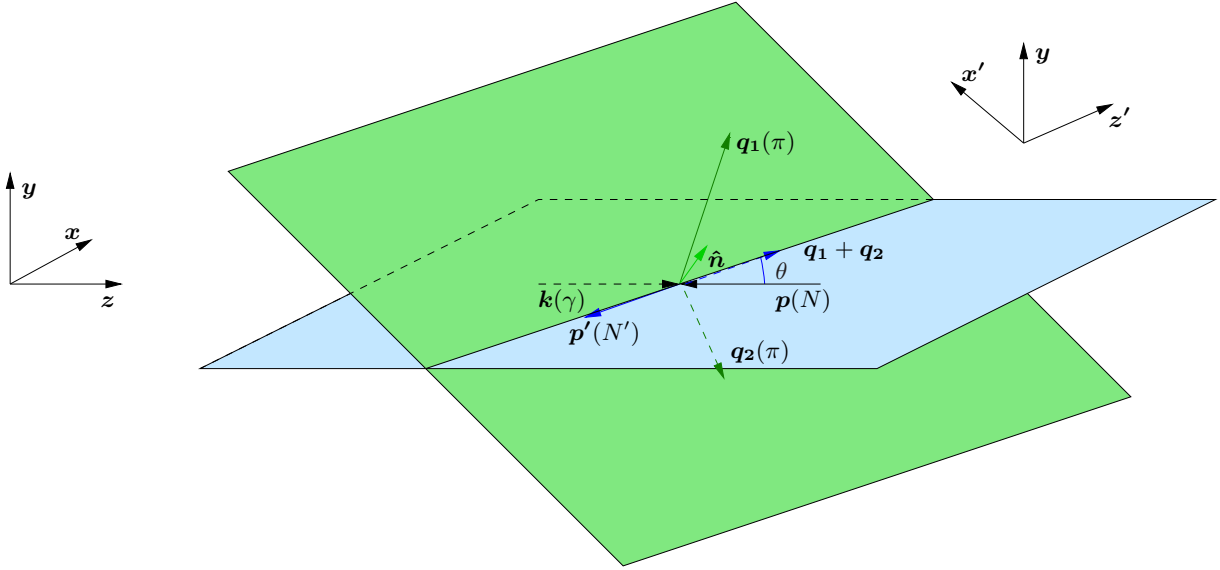
    \caption{Kinematics for the reaction $\gamma N \rightarrow \pi \pi N$ in the overall center-of-mass frame. The coordinate system is
      defined such that $\bm{k}$, $\bm{p}$ and $\bm{p'}$ lie on the $z$-$x$, plane (blue). Conservation of momentum constrains the three
      momenta of the final states to lie on a plane (green), shown in the figure with the vector normal to it, $\bm{\hat{n}}$. 
      Another coordinate system with the same $y$-axis but with its $z'$-axis pointing in the direction of $\bm{q_{1}}+\bm{q_{2}}$ is also 
      shown.}
    \label{Fig: Kinematics1}
  \end{center}
\end{figure}

The coordinate system that will be used in the overall center of mass frame will depend on the pathway of interest. For pathway A it is
(see fig. \ref{Fig: Kinematics1}):
\begin{equation} \label{CoordinateSystemA}
\bm{\hat{z}}=\frac{\bm{k}}{|\bm{k}|},\ \bm{\hat{y}}=\frac{\bm{k} \times \bm{q_{12}}}{|\bm{k} \times \bm{q_{12}}|},\ 
\bm{\hat{x}}=\bm{\hat{y}} \times \bm{\hat{z}}, 
\end{equation}
where $\bm{k}$ is the spatial part of $k$ and $\bm{q_{12}}$ is the spatial part of the four-momentum
\begin{equation} \label{q12}
 q_{12}=q_{1}+q_{2}.
\end{equation}
Note that $q_{12}$ is the four-vector of the meson resonance $M^{*}$. The $z$-$x$-plane is called the 
reaction plane, since the hadrons involved in the initial two-body reaction $\gamma N \rightarrow M^{*} B$ will lie on this plane. This is shown as the blue plane in fig. \ref{Fig: Kinematics1}. 

The coordinate system in the center-of-mass frame for pathways B and C are, respectively,
\begin{equation} \label{CoordinateSystemB}
  \begin{split}
    & \bm{\hat{z}}=\frac{\bm{k}}{|\bm{k}|},\ \bm{\hat{y}}=\frac{\bm{k} \times \bm{q_{23}}}{|\bm{k} \times \bm{q_{23}}|},\ 
      \bm{\hat{x}}=\bm{\hat{y}} \times \bm{\hat{z}}, \\
    & \bm{q}_{23} \equiv \bm{q}_{2} + \bm{p}', 
  \end{split}
\end{equation}
and
\begin{equation} \label{CoordinateSystemC}
  \begin{split}
    & \bm{\hat{z}}=\frac{\bm{k}}{|\bm{k}|},\ \bm{\hat{y}}=\frac{\bm{k} \times \bm{q_{31}}}{|\bm{k} \times \bm{q_{31}}|},\ 
	\bm{\hat{x}}=\bm{\hat{y}} \times \bm{\hat{z}}, \\
    & \bm{q}_{31} \equiv \bm{q}_{1} + \bm{p}'.
  \end{split}
\end{equation}
Therefore, in these last two coordinate systems, it is the the baryon $B$ and one of the mesons, $M_{1}$ or $M_{2}$, that define the reaction
plane. 

\begin{figure}
  \begin{center}
    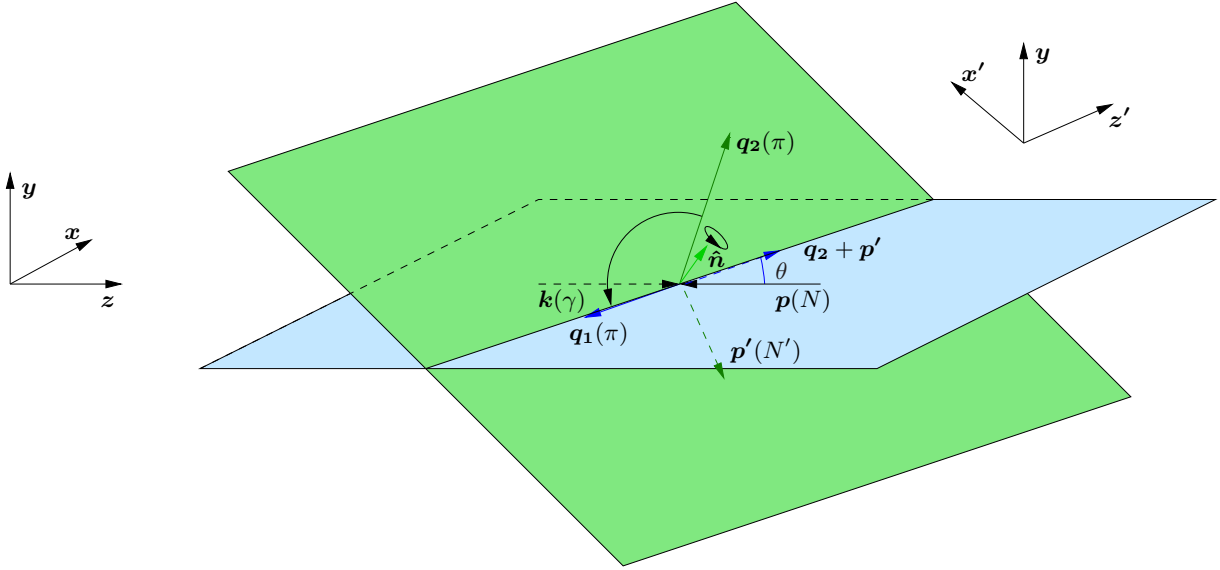
    \caption{Kinematics for the reaction $\gamma N \rightarrow \pi \pi N$ in the overall center-of-mass frame with the coordinate system
      defined such that $\bm{q_{1}}$ now lies on the $z$-$x$ plane (blue). We can arrive at this coordinate system from the one in fig. 
      \ref{Fig: Kinematics1} by applying a rotation along the normal to the green plane $\bm{\hat{n}}$. From here we can reach the center-of-mass frame of $\bm{q_{2}}$
      and $\bm{p}$ by boosting in the negative $z'$ direction.}
    \label{Fig: Kinematics2}
  \end{center}
\end{figure}

Conservation of momentum restricts the momenta $\bm{q_{1}}$, $\bm{q_{2}}$ and $\bm{p}'$ to lie on a plane. For pathway A, this is shown as the green plane 
in fig. \ref{Fig: Kinematics1}. To change from each of the three coordinate systems in eqns. (\ref{CoordinateSystemA}), (\ref{CoordinateSystemB}) and (\ref{CoordinateSystemC}), a rotation around the vector normal to that plane, $\hat{\bm{n}}$, can be applied, as
shown in fig. \ref{Fig: Kinematics2}.

For simplicity, we will use the coordinate system described in eqn. \eqref{CoordinateSystemA} (shown in 
fig. \ref{Fig: Kinematics1}) for the rest of this section. In it, the momentum of the baryon $B$, $\bm{p}'$, is on the reaction plane. The four-vectors in the center of mass frame are
\begin{equation} \label{eqn 5}
\begin{split}
k=\begin{pmatrix} |\bm{k}|, & 0, & 0, & |\bm{k}| \end{pmatrix},
  & \quad q_{12}=\begin{pmatrix} E_{q_{12}}, & |\bm{q_{12}}|\sin \theta, & 0, & |\bm{q_{12}}|\cos \theta \end{pmatrix}, \\
p=\begin{pmatrix} \sqrt{s}-|\bm{k}|, & 0, & 0, & -|\bm{k}| \end{pmatrix},
  & \quad p'=\begin{pmatrix} E_{p'}, & -|\bm{q_{12}}|\sin \theta, & 0, & -|\bm{q_{12}}|\cos \theta \end{pmatrix},
\end{split}
\end{equation}
where $\sqrt{s}$ is the center of mass energy, while $E_{q_{12}}$, $E_{p'}$ and $|\bm{q_{12}}|$ are
\begin{equation} \label{eqn 6}
\begin{split}
E_{q_{12}}=\frac{s+s_{M_{1}M_{2}}-m^{2}_{B}}{2\sqrt{s}}, \\
E_{p'}=\frac{s-s_{M_{1}M_{2}}+m^{2}_{B}}{2\sqrt{s}}, \\
|\bm{q_{12}}|=\frac{\lambda^{1/2}(s,s_{M_{1}M_{2}},m^{2}_{B})}{2\sqrt{s}}, \\
s_{M_{1},M_{2}}=(q_{1}+q_{2})^2,
\end{split}
\end{equation}
where $m_{B}$ is the mass of the recoil baryon, $s_{M_{1}M_{2}}$ is the square of the invariant mass of the two-pseudoscalar-meson system in the final state, and
\begin{equation} \label{eqn 7}
\lambda(x,y,z)=x^{2}+y^{2}+z^{2}-2xy-2yz-2xz,
\end{equation}
is the K\"{a}ll\'{e}n function. The angle $\theta$ between $\bm{k}$ and $\bm{q_{12}}$ is called the scattering angle,
\begin{equation} \label{eqn 8}
\cos(\theta)=\frac{\bm{k}}{|\bm{k}|} \cdot \frac{\bm{q_{12}}}{|\bm{q_{12}}|}.
\end{equation}

\begin{figure}
  \begin{center}
    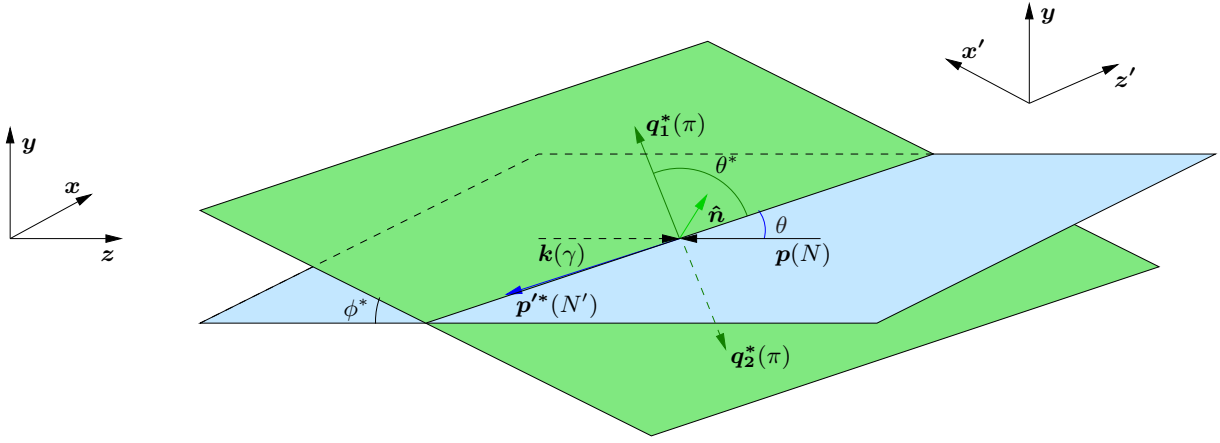
    \caption{Kinematics for the reaction $\gamma N \rightarrow \pi \pi N$ in the pion-pion center-of-mass frame. It is reached from the 
      overall center-of-mass frame shown in fig. \ref{Fig: Kinematics1} by boosting along the $z'$ direction. The three-momenta of
      the pions lies on the green plane. Their angular distribution is given by $\theta^{*}$ and $\phi^{*}$, which are the polar and azimuthal
      angles of $\bm{q^{*}_{1}}$ in the $x'$-$y$-$z'$ coordinate system.}
    \label{Fig: Kinematics3}
  \end{center}
\end{figure}

Since two of the final state particles in the reaction will come from a decay, it is convenient to describe another coordinate system with its
$z$-axis in the direction of the decaying particle's three-momentum, and its $y$-axis perpendicular to this three-vector and $\bm{\hat{k}}$.
For example, if $M_{1}$ and $M_{2}$ come from the decay of $M^{*}$, we define the this coordinate system,
\begin{equation} \label{eqn 9}
\bm{\hat{z}'}=\frac{\bm{q_{12}}}{|\bm{q_{12}}|},\ \bm{\hat{y}'}=\frac{\bm{k} \times \bm{q_{12}}}{\bm{k} \times \bm{q_{12}}}=\bm{\hat{y}},
     \ \bm{\hat{x}'}=\bm{\hat{y}'} \times \bm{\hat{z}'},
\end{equation}
which we will call the primed coordinate system, and it will be used in the rest frame of the decaying resonance.
In this frame, the four-momenta of the decay products $q^{*}_{1}$ and $q^{*}_{2}$ are
\begin{equation} \label{eqn 10}
\begin{split}
q^{*}_{1} & = \begin{pmatrix} E^{*}_{q_{1}}, & |\bm{q^{*}_{1}}|\sin \theta^{*} \cos \phi^{*},  & |\bm{q^{*}_{1}}|\sin \theta^{*} \sin \phi^{*},
  & |\bm{q^{*}_{1}}|\cos \theta^{*} \end{pmatrix}, \\
q^{*}_{2} & = \begin{pmatrix} E^{*}_{q_{2}}, & -|\bm{q^{*}_{1}}|\sin \theta^{*} \cos \phi^{*},  & -|\bm{q^{*}_{1}}|\sin \theta^{*} \sin \phi^{*},
  & -|\bm{q^{*}_{1}}|\cos \theta^{*} \end{pmatrix}, \\
p'^{*} & = \begin{pmatrix} E_{p'}^{*}, & -|\bm{p}'^{*}| \sin \theta, & 0, & -|\bm{p}'^{*}| \cos \theta \end{pmatrix},
\end{split}
\end{equation}
where
\begin{equation} \label{eqn 11}
\begin{split}
E^{*}_{q_{1}} & = \frac{s_{M_{1}M_{2}}+m^{2}_{M_{1}}-m^{2}_{M_{2}}}{2\sqrt{s_{M_{1}M_{2}}}}, \\
E^{*}_{q_{2}} & = \frac{s_{M_{1}M_{2}}-m^{2}_{M_{1}}+m^{2}_{M_{2}}}{2\sqrt{s_{M_{1}M_{2}}}}, \\
E_{p'}^{*} & = \frac{s-s_{M_{1}M_{2}}-m_{B}^{2}}{2\sqrt{s_{M_{1}M_{2}}}}, \\
|\bm{q^{*}_{1}}|& = \frac{\lambda^{1/2}(s_{M_{1}M_{2}},m^{2}_{M_{1}},m^{2}_{M_{2}})}{2\sqrt{s_{M_{1}M_{2}}}}, \\
|\bm{p}'^{*}| & = \frac{\lambda^{1/2}(s, s_{M_{1}M_{2}}, m_{B}^2)}{2\sqrt{s_{M_{1}M_{2}}}}, 
\end{split}
\end{equation}
$m_{M_{1}}$($m_{M_{2}}$) is the mass of the $M_{1}$($M_{2}$) final state meson and the decay angles $\theta^{*}$ and $\phi^{*}$ are the
polar and azimuthal angles of the unit vector $\bm{\hat{\pi}}=\frac{\bm{q_{1}^{*}}}{|\bm{q_{1}^{*}}|}$ with respect to the primed coordinate
system,
\begin{equation} \label{eqn 12}
\cos(\theta^{*})=\bm{\hat{\pi}} \cdot \bm{\hat{z}'}, \ \cos(\phi^{*})=\frac{\bm{\hat{y}'} \cdot (\bm{\hat{z}'} \times 
\bm{\hat{\pi}})}{|\bm{\hat{z}'} \times \bm{\hat{\pi}}|}, \ \sin(\phi^{*})=-\frac{\bm{\hat{x}'} \cdot (\bm{\hat{z}'} \times 
\bm{\hat{\pi}})}{|\bm{\hat{z}'} \times \bm{\hat{\pi}}|}.
\end{equation}
This coordinate system is shown in fig. \ref{Fig: Kinematics3}. Note how the angle $\phi^{*}$ is also the angle between the plane containing the vectors $q_{1}^{*}$, $q_{2}^{*}$ and $p'^{*}$ (shown as the green plane), and the reaction plane (shown as the blue plane).
\chapter{The Spin Density Matrix} \label{SpinDensityMatrix}
The spin state of a quantum system is represented by a state vector belonging to a Hilbert space, and is in general a coherent superposition of states from a chosen basis. For example, the state vector of a massive one particle system with spin $S$ can be represented as a linear superposition
of $2 S + 1$ orthogonal states. These can be chosen to be the eigenstates of the spin projection operator along the $z$-axis $\hat{S}_{z}$,
each with eigenvalue $s_{m}$,
\begin{equation} \label{StateVector}
 \ket{\Psi(S)} = \sum_{m = -S}^S \psi^{z}_{m} \ket{s_{m}, z}, 
\end{equation}
where $m$ labels the eigenvalues and $\psi^{z}_{m}$ are complex amplitudes. These are also called pure states. 

The state vector contains all of the information that can be known about the system: if the values of all $\psi^{z}_{m}$ are known (up to an overall phase), complete information about the system is known. In particular, the expectation values of any spin observable, with 
associated hermitian operator $\hat{S}_{O}$, can be found from
\begin{equation} \label{ExpectationValueStateVector}
 \langle S_{O} \rangle = \frac{\braket{\Psi|\hat{S}_{O}|\Psi}}{\braket{\Psi|\Psi}},
\end{equation}
where usually the state vector is normalized so that $\braket{\Psi|\Psi} = 1$.

However, one could have a more complicated system, such as a beam or target
used in a scattering experiment made up of a large number of particles with spin. The state of this system would have to include the amplitudes $\psi^{z}_{m}$ of all of the particles in the beam, making it too complicated, and this level of fine detail is usually not known or of interest. Instead, having only partial information on the system is of interest, such as the expectation value of an observable, e.g., a spin projection, of all the particles as a whole. The system is therefore considered a statistical ensemble of many possible pure states $\ket{\Psi_{i}}$, where the $i$-th state has a probability $\omega_{i}$ of being realized during a measurement (such that $\sum_{i} \omega_{i} = 1$). These types of systems are called mixed states, and cannot be represented by a state vector.

The expectation values of mixed states can be expressed in terms of a weighted average of the expectation values of every member of the ensemble,
\begin{align} \label{WeightedSum}
  \langle S_{O} \rangle =  \frac{\sum_{i} \omega_{i} \braket{\Psi_{i}|\hat{S}_{O}|\Psi_{i}}}
    {\sum_{i} \omega_{i} \braket{\Psi_{i}|\Psi_{i}}},
\end{align}
where $S_{O}$ is some spin observable, such as a spin projection. This can be expressed as
\begin{equation} \label{TraceOfExpectationValue}
  \begin{split}
    \langle S_{O} \rangle & =  \frac{\Tr\Big[\Big(\sum_{i} \omega_{i} \ket{\Psi_{i}}\bra{\Psi_{i}}\Big) \hat{S}_{O}\Big]}
	{\Tr\Big[\sum_{i} \omega_{i} \ket{\Psi_{i}}\bra{\Psi_{i}}\Big]}.
  \end{split}
\end{equation}
If we define the density matrix of the system as
\begin{equation} \label{DensityMatrix}
 \hat{\rho} = \sum_{i} \omega_{i} \ket{\Psi_{i}} \bra{\Psi_{i}},
\end{equation}
we can rewrite the expectation value, 
\begin{equation} \label{ExpectationValuesDensityMatrix}
  \langle S_{O} \rangle = \frac{\Tr\Big[\hat{\rho} \hat{S}_{O}\Big]}{\Tr\big[\hat{\rho}\big]}.
\end{equation}

If the states in the ensemble are normalized,
\begin{equation} \label{NormalizedDensityMatrix}
  \Tr\big[\hat{\rho}\big] = 1 \qquad \textrm{when} \ \braket{\Psi_{i}|\Psi_{i}} = 1 \  \textrm{for all} \ i.
\end{equation}

A mixed state is therefore described by a density matrix, since it contains all the information of interest that can be known about the system.
It can be used to calculate the expectation values of any observable using eqn. (\ref{ExpectationValuesDensityMatrix}).
If the observables of interest are related to spin, as in this thesis, it is customarily called the spin density matrix. The spin state of 
these mixed states is completely known if all spin density matrix elements (SDME's) are known (up to an overall phase). 

The dimensions of the spin density matrix depend on the dimensions of the Hilbert space of the states in the ensemble. If the states
in the ensemble are made up of massive particles of spin $S$, the spin density matrix will have dimensions of $(2S+1) \times (2S+1)$. For example, if the ensemble is made up of spin-1/2 particles, whose Hilbert space is spanned by a basis of two orthogonal states
(states with spin projections of $1/2$ and $-1/2$ along the $z$-axis, for example), the spin density matrix of the system will be
$2 \times 2$. For massless vector bosons, such as the photon, it will always be a $2 \times 2$ matrix. 

From the definition of the density matrix matrix in eqn. (\ref{DensityMatrix}), it can be shown that density matrices are hermitian,
\begin{equation} \label{Hermiticity}
 \hat{\rho}^{\dag} = \hat{\rho},
\end{equation}
as is required to guarantee that expectation values of experimental observables are real.

As an example, the spin density matrix of a spin-1/2 particle (such as a nucleon), or a massless particle with spin (such as a photon),
will have the general form
\begin{equation} \label{TwoByTwoDensityMatrix}
 \hat{\rho}_{\textrm{2-dim}} = \begin{pmatrix}
                                \rho_{11} & \rho_{12} \\
                                \rho_{21} & \rho_{22}
                               \end{pmatrix}
			     = \begin{pmatrix}
                                \rho_{11} & \rho_{12} \\
                                \rho^{*}_{12} & \rho_{22}
                               \end{pmatrix}.
\end{equation}
Note how the hermiticity condition of eqn. (\ref{Hermiticity}) reduces the number of independent parameters needed to fully specify a complex $2\times2$ matrix from 8 real numbers to to 4. From eqn. (\ref{NormalizedDensityMatrix}), when the matrix is normalized to 1, only 3 parameters are independent. We can therefore describe the polarization state of these systems by specifying the three components of the vector
\begin{equation} \label{PolarizationVector}
  \vec{\Lambda} = \begin{pmatrix}
                   \Lambda^{x} \\
                   \Lambda^{y} \\
                   \Lambda^{z}
                  \end{pmatrix}
                = \frac{1}{S} \begin{pmatrix}
                   \langle S^{x} \rangle \\
                   \langle S^{y} \rangle \\
                   \langle S^{z} \rangle
                  \end{pmatrix},
\end{equation}
where $S$ is total spin and the $S^{i}$'s are the expectation values of the spin projections along the $i$-axes calculated from eqn. (\ref{ExpectationValuesDensityMatrix}). The factor of $\frac{1}{S}$ is a convention used to 
guarantee that the magnitude of this vector is at most 1. This vector is called the polarization vector, or 
the degree of polarization. Each of the components could be interpreted as
\begin{equation} \label{PolarizationVectorInterpretation}
	\Lambda^{i} = \frac{n_{+i} - n_{-i}}{n},
\end{equation}
where $n_{+i}$ is the number of particles in the ensemble with positive spin projection along the $i$-axis, vice versa for $n_{-i}$, and $n$ is the total number of particles in the ensemble.

\begin{figure}
  \begin{center}
    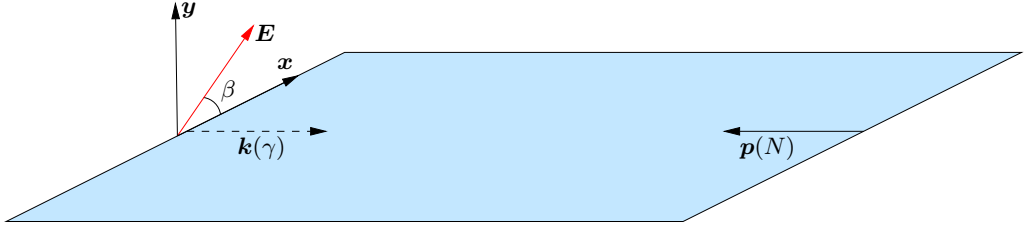
    \caption{The direction of the electric field vector of linearly polarized photons is defined as the angle, $\beta$, it makes with respect
      to the $x$-axis of the $x$-$y$ plane.}
    \label{Fig: ElectricFieldVector}
  \end{center}
\end{figure}
In the case of the photon, the polarization vector can alternatively be expressed as
\begin{equation} \label{Beta}
  \begin{split}
    \Lambda^{x (\gamma)} & = \delta_{l} \cos2\beta, \\
    \Lambda^{y (\gamma)} & = \delta_{l} \sin2\beta, \\
    \Lambda^{z (\gamma)} & = \delta_{\odot},
  \end{split}
\end{equation}
where $\delta_{l}$ and $\delta_{\odot}$ are the degrees of linear and circular polarization, respectively, and $\beta$ is the angle that the
photon's electric field vector makes with respect to the $x$-axis on the $x$-$y$ plane, shown in fig. \ref{Fig: ElectricFieldVector}.
For linearly polarized photons, $\delta_{l} = 1$ and $\delta_{\odot}=0$, while for circularly polarized photons, $\delta_{l}=0$ and $\delta_{\odot} = 1$ (If both are non-zero, the photon
is said to be elliptically polarized, but such cases will not be considered in this thesis). 

We say that a system is fully polarized if all the states in the ensemble are the same. A fully polarized state is therefore a pure state.
Its spin density matrix is given by
\begin{equation} \label{DensityMatrixFullyPolarized}
  \hat{\rho}_{\textrm{pol}} = \ket{\Psi}\bra{\Psi},
\end{equation}
were every member of the ensemble is in state $\ket{\Psi}$. As an example, we will write the density matrix for fully polarized spin-1/2
particles. We will choose the eigenstates of the operator for the spin projection along the $z$-axis,
\begin{equation} \label{SpinZOperator}
 \hat{S}_{z} = \frac{1}{2} \begin{pmatrix}
                1 & \hphantom{-} 0 \\
                0 & - 1
               \end{pmatrix},
\end{equation}
with basis states,
\begin{equation} \label{BasisStatesSpinZ}
  \ket{+, z} = \begin{pmatrix}
                \ 1 \ \\
                \ 0 \
               \end{pmatrix}, \qquad
  \ket{-, z} = \begin{pmatrix}
                \ 0 \ \\
                \ 1 \
               \end{pmatrix}.
\end{equation}
The spin density matrices for a state where all particles have spin projection of $+1/2$ along the $z$-axis, and one where all particles
have spin projection of $-1/2$ along said axis are
\begin{equation} \label{PolarizedZDensityMatrices}
  \hat{\rho}_{\textrm{pol}}(+, z) = \begin{pmatrix}
				      1 \quad 0 \\
				      0 \quad 0
                     \end{pmatrix}, \qquad
  \hat{\rho}_{\textrm{pol}}(-, z) = \begin{pmatrix}
				      0 \quad 0 \\
				      0 \quad 1
                     \end{pmatrix}.
\end{equation}
The polarization vectors of these states are
\begin{equation} \label{PolarizationVectorsZ}
  \vec{\Lambda}(+, z) = \begin{pmatrix}
			 \ 0 \ \\
			 \ 0 \ \\
			 \ 1 \
			\end{pmatrix}, \qquad
  \vec{\Lambda}(-, z) = \begin{pmatrix}
			  \ 0 \ \\
			  \ 0 \ \\
			   -1 \
			\end{pmatrix}.
\end{equation}
We therefore say that are states are fully polarized along the $z$-axis.

As another example, we could have every state in the ensemble be an eigenstates of the operator for spin projection along the $y$-axis,
\begin{equation} \label{SpinYOperator}
 \hat{S}_{y} = \frac{1}{2} \begin{pmatrix}
                0 & - i \\
                i & 0
               \end{pmatrix},
\end{equation}
with corresponding eigenstates
\begin{equation} \label{BasisStatesSpinY}
  \ket{+, y} = \frac{1}{\sqrt{2}} \begin{pmatrix}
                \ 1 \  \\
                \ i \ 
               \end{pmatrix}, \qquad
  \ket{-, y} = \frac{1}{\sqrt{2}} \begin{pmatrix}
                \ 1 \ \\
                -i \ 
               \end{pmatrix}.
\end{equation}
The density matrix for a polarized system with spin projection along the $y$-axis of $1/2$ and $-1/2$ are, respectively,
\begin{equation} \label{PolarizedYDensityMatrices}
  \hat{\rho}_{\textrm{pol}}(+, y) = \frac{1}{2} \begin{pmatrix}
				      \ 1 \quad -i \\
				      \ i \quad \ 1
                     \end{pmatrix}, \qquad
  \hat{\rho}_{\textrm{pol}}(-, y) = \frac{1}{2} \begin{pmatrix}
				      \ 1 \quad \ i \\
				      -i \quad \ 1
                     \end{pmatrix}.
\end{equation}
The polarization vectors for these states are
\begin{equation} \label{PolarizationVectorsY}
  \vec{\Lambda}(+, y) = \begin{pmatrix}
			 \ 0 \ \\
			 \ 1 \ \\
			 \ 0 \
			\end{pmatrix}, \qquad
  \vec{\Lambda}(-, y) = \begin{pmatrix}
			  \ 0 \ \\
			   -1 \ \\
			  \ 0 \
			\end{pmatrix}.
\end{equation}

A state can in general be fully polarized along any axis. The state vector of a spin-1/2 particle with spin projection $+1/2$ along the axis
\begin{equation} \label{QuantizationAxis}
	\hat{\bm{\omega}} = \begin{pmatrix}
					\sin\theta \cos\phi \\
					\sin\theta \sin\phi \\
					\cos\theta
	               \end{pmatrix},
\end{equation}
described by polar and azimuthal angles $\theta$ and $\phi$, is given by
\begin{equation} \label{StateVectorArbitraryAxis}
	\ket{\psi(\hat{\bm{\omega}}),z} = \begin{pmatrix}
										\cos(\theta/2) \\
										\sin(\theta/2)e^{i\phi}
	                                \end{pmatrix}
\end{equation}
in the $z$ basis. Therefore, the general expression (up to a phase) for the normalized spin density matrix of a spin-1/2 system fully polarized along the $\hat{\bm{\omega}}(\theta,\phi)$ direction is
\begin{equation} \label{DensityMatrixFullyPolarizedArbitraryAxis}
	\hat{\rho}_{\text{pol}}(\hat{\bm{\omega}}) 
	= \begin{pmatrix}
			\cos^{2}(\theta/2) & \cos(\theta/2)\sin(\theta/2)e^{-i\phi} \\
		\cos(\theta/2)\sin(\theta/2)e^{i\phi}	& \sin^{2}(\theta/2)
		\end{pmatrix}.
\end{equation}
In the case of a fully polarized system, the polarization vector is equal to $\hat{\bm{\omega}}$,
\begin{equation} \label{PolarizationVectorFullyPolarizedArbitraryAxis}
	\vec{\Lambda} = \begin{pmatrix}
						\sin\theta \cos\phi \\
						\sin\theta \sin\phi \\
						\cos\theta
	               \end{pmatrix}.
\end{equation}
Note how only two parameters are needed to describe the previous polarization vector, not three. This is because for systems that are fully polarized, the magnitude of its polarization vector equals 1,
\begin{equation} \label{PolarizationVectorPolarized}
  |\vec{\Lambda}_{\textrm{pol}}|^{2} = \Lambda_{x}^{2} + \Lambda_{y}^{2} + \Lambda_{z}^{2} = 1
\end{equation}
(assuming the spin density matrix's trace is normalized to 1).

For example, the state vector of a spin-1/2 particle with spin projection $+1/2$ along the direction described by polar angles $\theta = 60^{\circ}$ and $\phi = 90^{\circ}$ is 
\begin{equation} \label{PureStateExample}
  \ket{\psi} = \begin{pmatrix}
		  \frac{\sqrt{3}}{2} \\
		  \frac{i}{2}
               \end{pmatrix},
\end{equation}
has density matrix
\begin{equation} \label{DensityMatrixExample}
  \hat{\rho}(\psi) = \begin{pmatrix}
		  \frac{3}{4} & - \frac{i \sqrt{3}}{4} \\
		  \frac{i \sqrt{3}}{4} & \frac{1}{4}
               \end{pmatrix},
\end{equation}
and polarization vector
\begin{equation} \label{PolarizationVectorExample}
  \vec{\Lambda}(\psi) = \begin{pmatrix}
			  0 \\
			  \frac{\sqrt{3}}{2} \\
			  \frac{1}{2}
                        \end{pmatrix}.
\end{equation}

A necessary and sufficient condition for a density matrix to describe a fully polarized state is
\begin{equation} \label{NecessaryAndSufficientCondition}
  \hat{\rho}^{2}_{\textrm{pol}} = \hat{\rho}_{\textrm{pol}},
\end{equation}
as can be verified for the density matrices in the previous examples. The density matrix of a fully polarized system is therefore a projection operator. 

A mixed spin state is said to be fully unpolarized if the members of the ensemble have equal probability to be in any one of the states that forms
a complete basis. Using an arbitrary set of basis states, ${\ket{m}}$, where each has the same probability $N$ of being realized in the 
ensemble, the density matrix of an unpolarized state is
\begin{equation} \label{UnpolarizedDensityMatrix}
  \hat{\rho} = \sum_{m} \omega_{m} \ket{m} \bra{m} = N \Big(\sum_{m} \ket{m} \bra{m}\Big) = N \hat{I},
\end{equation}
where the last equality follows from the completeness property of basis states, and $\hat{I}$ is the identity matrix. Therefore, the spin
density matrix of fully unpolarized states is always proportional to the identity matrix, regardless of the chosen basis. When the trace 
of a spin-$S$ spin density matrix is normalized to 1, $N = 1/(2S+1)$.

For example, in a beam of spin-1/2 particles in which the spin projection along the $z$-axis of half of them is $1/2$ and 
for the other half is $-1/2$, its density matrix is
\begin{equation} \label{Unpolarized}
  \begin{split}
  \rho_{\textrm{unpol}} & =  \frac{1}{2} \Big( \ket{+, z}\bra{+, z}\ +\ \ket{-, z}\bra{-, z} \Big),  \\
  \rho_{\textrm{unpol}} & = \frac{1}{2} \begin{pmatrix}
					  1 & 0 \\
					  0 & 1
					\end{pmatrix}.
  \end{split}
\end{equation}
It's polarization vector is
\begin{equation} \label{PolarizationVectorUnpolarized}
  \vec{\Lambda}_{\textrm{unpol}} = \begin{pmatrix}
					\ 0 \ \ \\
				    \ 0 \ \ \\
				    \ 0 \ \
                                   \end{pmatrix}.
\end{equation}
A partially polarized state will have a polarization vector with magnitude between $0$ and $1$,
\begin{equation} \label{PolarizationVectorPariallyPolarized}
  0 < |\vec{\Lambda}|^2 < 1 \qquad \textrm{for partially polarized system}. 
\end{equation}

A pure spin $S$ state, $\ket{\psi^{S}}$, transforms under rotations as
\begin{equation} \label{RotationPureState}
	\begin{split}
	\ket{\psi'^{S}_{m}} & = \hat{D}(\alpha,\beta,\gamma)\ket{\psi^{S}_{m}}, \\
	\psi'^{S}_{m} & = \sum_{m' = -S}^{S} D^{S}_{m m'}(\alpha,\beta,\gamma) \psi^{S}_{m'}, \\
	\psi^{S}_{m} & \equiv \braket{S, m|\psi^{S}},
	\end{split}
\end{equation}
where the $D^{S}_{m m'}$'s are the well known Wigner $\mathcal{D}$-functions and $\alpha$, $\beta$, and $\gamma$ are the Euler angles describing the rotation. Therefore, it can be shown from eqn. (\ref{DensityMatrix}) that the spin density matrix of a spin-$S$ system transforms under rotations as 
\begin{equation} \label{RotationDensityMatrix}
  \begin{split}
    \hat{\rho}'(S) & = \hat{D}^{S}(\alpha,\beta,\gamma) \hat{\rho}(S) \hat{D}^{S \dagger}(\alpha,\beta,\gamma), \\
    \rho'_{m m'}(S) & = \sum_{n, n' = -S}^{S} D^{S}_{m n}(\alpha,\beta,\gamma) \rho_{n n'}(S) D^{S *}_{m' n'}(\alpha,\beta,\gamma).
  \end{split}
\end{equation}

This previous equation and the definition of the spin density matrix in eqn. (\ref{DensityMatrix}) can be used to show, for example, that any spin density matrix for a polarized state (like the one in eqn. 
(\ref{DensityMatrixExample})) can be transformed by rotations into the ones in eqn. (\ref{PolarizedZDensityMatrices}). It can also be used to
show that a fully unpolarized state will take the form in eqn. (\ref{Unpolarized}) in any basis, i.e., the identity matrix is invariant under rotations of the coordinate system.

Since complex numbers are described by two real numbers, a spin density matrix is described by has $2\times(2S+1)\times(2S+1)$ real numbers.
But they are also hermitian matrices, so only half of them, $(2S+1)^{2}$, will be independent (one less if the matrix is normalized to have 
a trace of one). For example, the spin density matrix of a spin-1/2 particle will have 4 independent parameters, while a massive spin-1
particle will have 9 (massless vector bosons will have 4 independent parameters, since their spin density matrix will always be $2\times2$). This is why
the normalized spin density matrix of a spin-1/2 and spin-1 particle can be fully described by the three components of the polarization vector $\vec{\Lambda}$. 

\chapter{Statistical Tensors} \label{StatisticalTensors}
Since only some of the elements of a given spin density matrix are independent, it is only necessary to specify the real and imaginary parts
of these independent elements to completely define the matrix. However, it is possible to specify instead a different
set of quantities, each one being a linear combination of the independent matrix elements, such that they transform under different
representations of the rotation group. For example, the spin density matrix for a massive spin-1 particle, described by $9$ independent
quantities, can be written as
\begin{equation} \label{DensityMatrixDecomposition}
\begin{split}
 & \hat{\rho}(S=1) = \frac{1}{3} \mathcal{N} (\hat{I} + \sum_{i}v^{i}\hat{S}_{i} + \sum_{i,j}q^{ij}\hat{Q}_{ij}), \\
 & i,j \in \{x,y,z\},
\end{split}
\end{equation}
where $\hat{I}$ is the $3\times3$ identity matrix, the $\hat{S}_{i}$'s are the Cartesian components of the spin operator, and the
$\hat{Q}_{ij}$'s are the Cartesian components of the quadrupole operator,
\begin{equation} \label{QuadrapoleOperator}
\begin{split}
& \hat{Q}_{ij} \equiv \frac{1}{2}(\hat{S}_{i}\hat{S}_{j} + \hat{S}_{j}\hat{S}_{i} - \frac{4}{3}\delta_{ij}\hat{I}), \\
& \hat{Q}_{ij} = \hat{Q}_{ji}, \ \sum_{i}\hat{Q}_{ii}=0. 
\end{split}
\end{equation}
From the form of the spin density matrix in eqn. (\ref{DensityMatrixDecomposition}), along with eqns. (\ref{ExpectationValuesDensityMatrix}), 
(\ref{PolarizationVector}), and (\ref{QuadrapoleOperator}), it can be shown that $v^{i}$ is equal to the polarization vector,
\begin{equation} \label{vEqualsPolarizationVector}
  v^{i} = \Lambda^{i}.
\end{equation}

$\mathcal{N}$ is a rotational scalar, while the $v^{i}$'s and the $Q^{ij}$ transform under tensor representations of the rotation group,
\begin{equation} \label{VectorQuadrapoleRotation}
\begin{split}
& v^{i'} = \sum_{j}\mathcal{R}\indices{^i_j}(\vec{\omega})v^{j}, \\
& q^{ij'} = \sum_{kl}\mathcal{R}\indices{^i_k}(\vec{\omega})\mathcal{R}\indices{^j_l}(\vec{\omega})q^{kl},
\end{split}
\end{equation}
where the $\mathcal{R}\indices{^i_j}$'s are the elements of the three dimensional representation of the rotation group, for a rotation 
along the axis pointing in the direction of $\vec{\omega}$ by an angle $|\vec{\omega}|$.\ Note that since the $q^{ij}$'s
are elements of a traceless, symmetric matrix, only 5 of its elements are independent. Therefore, $\mathcal{N}$, the $v^{i}$'s and the
$q^{ij}$'s together form 9 independent quantities which completely specify the
spin density matrix. These quantities are the eigenvalues of hermitian operators (the $\hat{S}_{i}$'s and $\hat{Q}_{ij}$'s) and as such, are
purely real, i.e., they are experimental observables.

As seen in eqn. (\ref{VectorQuadrapoleRotation}), $q^{ij}$ does not transform under an irreducible representation of the rotation group. It is
therefore convenient to decompose the spin density matrix as in eqn. (\ref{DensityMatrixDecomposition}) but into matrices that 
transform under irreducible representation,
\begin{equation} \label{TExpansion}
  \hat{\rho}(S) = \sum_{L=0}^{2S}\sum_{M=-L}^{L} t_{LM}(S)\hat{T}^{\dagger}_{LM}(S),
\end{equation}
where the matrices $\hat{T}_{LM}(S)$'s are called the polarization operators and the $t_{LM}(S)$'s are called the statistical tensors
of rank $L$. The matrix elements of the $\hat{T}_{LM}(S)$'s are defined as
\begin{equation} \label{TDefinition1}
 [T_{LM}(S)]_{m m'} = (-1)^{S-m'} C^{LM}_{S m; S - m'},
\end{equation}
where the $C^{LM}_{Sm;S-m'}$'s are the well known Clebsch-Gordan coefficients (the matrix elements are therefore real). The $t_{LM}$'s are the expectation values of the polarization operators,
\begin{equation} \label{STDefinition1}
 t_{LM}(S)=\langle \hat{T}_{LM}(S) \rangle = \Tr[\hat{\rho}(S)\hat{T}_{LM}(S)] = \sum_{m,m'=-S}^{S}(-1)^{S-m}\rho_{mm'}(S)C^{LM}_{sm';s-m}.
\end{equation}
The inverse relation of eqn. (\ref{STDefinition1}) is
\begin{equation} \label{InverseRelation}
  \rho_{mm'}(S)=\sum_{L=0}^{2S} \sum_{M=-L}^{L} (-1)^{S-m} t_{LM}(S) C^{LM}_{Sm';S-m}.
\end{equation}
Under rotations, the $t_{LM}(S)$'s transform as 
\begin{equation} \label{STRotations}
 t_{LM}'(S) = \sum_{M=-L}^{L}D^{L}_{MM'}(\vec{\omega})t_{LM'}(S),
\end{equation}
where the $D^{L}_{MM'}(\vec{\omega})$ are the elements of the Wigner 
$\mathcal{D}$-matrix, called the Wigner $\mathcal{D}$-functions. The Wigner-$\mathcal{D}$ matrices are the $2L+1$ dimensional irreducible representations of the rotation group.

Note that, unlike the Cartesian tensors ($v^{i}$, $q^{ij}$, and other higher rank tensors), the $t_{LM}$'s
are complex quantities. However, with the standard Clebsch-Gordan phase convention, we find
\begin{equation} \label{STHermiticity}
 t_{LM}^{*}(S)=(-1)^{M}t_{L-M}(S),
\end{equation}
so that the total number of independent parameters remains the same. Written in terms of the statistical tensors, the total number of 
independent parameters $N_{p}$ required to describe the spin state of a spin-$S$ particle is
\begin{equation} \label{IndependentParameters}
  N_{p} = \sum_{L=0}^{2S} (2L+1).
\end{equation}
While in the Cartesian rotation operators $R\indices{^i_j}$ the $i$'s and $j$'s label Cartesian components ($x$, $y$ and $z$), the $M$'s of the tensor operators label spherical components, i.e., the polarization operators are spherical tensors operators. The Wigner-Eckart theorem therefore applies to them,
\begin{equation} \label{Wigner-EckartTheorem}
  \braket{jm|\hat{T}_{LM}|j'm'} = C^{jm}_{j'm';LM}\braket{j||\hat{T}_{L}||j'},
\end{equation}
where $\bra{jm}$ and $\ket{j'm'}$ are angular momentum eigenstates, and the factor $\braket{j||\hat{T}^{L}||j'}$ is a quantity that does not
depend on $M$, $m$, or $m'$, called the reduced matrix element. By writing the spin density matrix in terms of tensor operators, rotational
symmetry can be exploited to simplify the calculation of matrix elements by using the Wigner-Eckart theorem.

In summary, there are two equivalent ways to describe the spin state of a statistical ensemble of states: either by specifying the real and 
imaginary parts of all independent elements of the spin density matrix, or by specifying the values of the statistical tensors. The two 
are related by the linear relations in eqns. (\ref{STDefinition1}) and (\ref{InverseRelation}), so it is easy to rewrite any equation 
involving the matrix elements in terms of the the tensors, and vice versa. The advantage of the tensors is that they have good
transformation properties under rotations, and can be used to take advantage of the rotational symmetry of the system of interest via the
Wigner-Eckart theorem.

\chapter{SDME'\MakeLowercase{s} and Helicity Amplitudes in $\gamma N \rightarrow M_{1} M_{2} B$}
The quantum states describing the system before and after the scattering process are related via the transition matrix $\hat{M}$, 
\begin{equation} \label{InitialToFinalState}
 \ket{\Psi^{i}(M_{1} M_{2} B)} = \hat{M} \ket{\Psi^{i}(\gamma N)},
\end{equation}
where $\ket{\Psi^{i}(\gamma N)}$ and $\ket{\Psi^{i}(M_{1} M_{2} B)}$ are the state vectors of the initial and final systems, $\gamma N$ and $M_{1} M_{2} B$, and $i$ labels the different members of the statistical ensemble of the mixed state.

The initial states in the ensemble consist of a beam ($\gamma$) and target (N). Each will be a pure state 
consisting of a photon and a nucleon. Since they are uncorrelated, the state vector of this system is a tensor product
of the state vectors of the individual subsystems,
\begin{equation} \label{TensorProductStateTemp}
  \ket{\Psi^{i}(\gamma N)} = \ket{\Psi^{i}(\gamma)}\otimes\ket{\Psi^{i}(N)}.
\end{equation}
In this thesis, we will always describe spin states using the helicity basis, in which the quantization axis is chosen to be in the same direction as the momentum of the particle. Since both the beam and target have two possible spin projections ($\pm1$ for the photon and $\pm1/2$
for the nucleon), the spin wave function of the system will have four orthogonal states, and can be represented as,
\begin{equation} \label{InitialState}
 \ket{\Psi^{i}(\gamma N)} = \begin{pmatrix*}[l]
			  \psi^{i}_{1 \frac{1}{2}}(\gamma N) \\
			  \psi^{i}_{1 \minus \frac{1}{2}}(\gamma N) \\
			  \psi^{i}_{\minus 1 \frac{1}{2}}(\gamma N) \\
			  \psi^{i}_{\minus 1 \minus \frac{1}{2}}(\gamma N)
			\end{pmatrix*},
\end{equation}
where $\psi^{i}_{jk}(\gamma N)$ is the probability amplitudes for the $i$-th member of the ensemble to have helicities $j$ and $k$ for the bean and nucleon, respectively. Similarly, the final state is
\begin{equation} \label{FinalState}
 \ket{\Psi^{i}(M_{1} M_{2} B)} = \begin{pmatrix*}[l]
				    \psi^{i}_{\frac{1}{2}}(M_{1} M_{2} B; \theta) \\
				    \psi^{i}_{\minus \frac{1}{2}}(M_{1} M_{2} B; \theta)
				 \end{pmatrix*}.
\end{equation}
The subindices refer only to the spin projection of the baryon $B$, since $M_{1}$ and $M_{2}$ are spinless. These two amplitudes are also functions of the scattering angle $\theta$. Since the initial state is given by a $4\times1$ column matrix while the final state is $2\times1$, the transition matrix is $2\times4$,
\begin{equation} \label{TransitionMatrix}
 \hat{M}(\theta) = \begin{pmatrix} 
            M_{\frac{1}{2}; 1 \frac{1}{2}}(\theta) & M_{\frac{1}{2}; 1 \minus \frac{1}{2}}(\theta)
	      & M_{\frac{1}{2}; \minus 1 \frac{1}{2}}(\theta) & M_{\frac{1}{2}; \minus 1 \minus \frac{1}{2}}(\theta) \\
	    M_{\minus \frac{1}{2}; 1 \frac{1}{2}}(\theta) & M_{\minus \frac{1}{2}; 1 \minus \frac{1}{2}}(\theta)
	      & M_{\minus \frac{1}{2}; \minus 1 \frac{1}{2}}(\theta) & M_{\minus \frac{1}{2}; \minus 1 \minus \frac{1}{2}}(\theta) \\
           \end{pmatrix}.
\end{equation}
The elements of the transition matrix are known as the helicity amplitudes. They are the probability amplitudes for an initial state to transition into a particular final state. For example, $M_{\frac{1}{2}; 1 \frac{1}{2}}(\theta)$ is the probability amplitude for an initial state with helicities $1$ and $1/2$ to transition into a $M_{1} M_{2}B$ state with recoil baryon helicity $1/2$ and scattering angle $\theta$ (defined in section \ref{Kinematics}).

As mentioned in section \ref{Motivation}, a complete experiment is one in which enough measurements can be made to be able to extract the values of all helicity amplitudes (except an overall phase) at different kinematic points (different values of $\theta$ and other kinematic variables). The dependence on the kinematic variables of these amplitudes can be interpreted in terms of the resonances that contribute to the scattering process. 

Since the beam, target, and recoil baryon are mixed states represented by density matrices, it is necessary to find how the SDME's are related to the helicity amplitudes. The states in the ensemble are product states, which means the spin density matrix of the ensemble is also a tensor product of two spin density 
matrices,
\begin{equation} \label{TensorProductDensityMatrix}
  \hat{\rho}(\gamma N) = \hat{\rho}(\gamma)\otimes\hat{\rho}(N).
\end{equation}
Since the pure states in the ensemble shown in eqn. (\ref{InitialState}) are $4\times1$ column matrices, its spin density matrix is $4\times4$,
\begin{equation} \label{InitialDensityMatrix}
 \hat{\rho}(\gamma N) = \begin{pmatrix*}[l]
                         \rho_{1 \frac{1}{2}; 1 \frac{1}{2}}(\gamma N) 
                              & \rho_{1 \frac{1}{2}; 1 \minus\frac{1}{2}}(\gamma N) 
                              & \rho_{1 \frac{1}{2}; \minus1 \frac{1}{2}}(\gamma N)
                              & \rho_{1 \frac{1}{2}; \minus1 \minus\frac{1}{2}}(\gamma N) \\
                         \rho_{1 \minus\frac{1}{2}; 1 \frac{1}{2}}(\gamma N) 
                              & \rho_{1 \minus\frac{1}{2}; 1 \minus\frac{1}{2}}(\gamma N) 
                              & \rho_{1 \minus\frac{1}{2}; \minus1 \frac{1}{2}}(\gamma N)
                              & \rho_{1 \minus\frac{1}{2}; \minus1 \minus\frac{1}{2}}(\gamma N) \\     
                         \rho_{\minus1 \frac{1}{2}; 1 \frac{1}{2}}(\gamma N) 
                              & \rho_{\minus1 \frac{1}{2}; 1 \minus\frac{1}{2}}(\gamma N) 
                              & \rho_{\minus1 \frac{1}{2}; \minus1 \frac{1}{2}}(\gamma N) 
                              & \rho_{\minus1 \frac{1}{2}; \minus1 \minus\frac{1}{2}}(\gamma N) \\
                         \rho_{\minus1 \minus\frac{1}{2}; 1 \frac{1}{2}}(\gamma N) 
                              & \rho_{\minus1 \minus\frac{1}{2}; 1 \minus\frac{1}{2}}(\gamma N) 
                              & \rho_{\minus1 \minus\frac{1}{2}; \minus1 \frac{1}{2}}(\gamma N) 
                              & \rho_{\minus1 \minus\frac{1}{2}; \minus1 \minus\frac{1}{2}}(\gamma N) \\
                        \end{pmatrix*}.
\end{equation}
Note that the ``row'' and ``column'' indices are actually made up of two indices: one for the helicity of the beam, and one for the 
helicity of the target. Combining the definition of spin density matrices shown in eqn. (\ref{DensityMatrix}) with eqn. (\ref{InitialToFinalState}), we can see that the relation between the spin density matrices of the initial and final
state is
\begin{equation} \label{InitialToFinalDensityMatrix}
 \hat{\rho}(M_{1}M_{2}B) = \hat{M}\hat{\rho}(\gamma N)\hat{M}^{\dagger}.
\end{equation}
The spin density matrix that describes the ensemble of the final three hadrons will therefore be a $2 \times 2$
matrix,
\begin{equation} \label{FinalDensityMatrix}
 \hat{\rho}(M_{1} M_{1} B) = \begin{pmatrix*}[l]
                         \rho_{\frac{1}{2}; \frac{1}{2}}(M_{1} M_{1} B;\theta) 
                              & \rho_{\frac{1}{2}; \minus\frac{1}{2}}(M_{1} M_{1} B;\theta) \\
                         \rho_{\minus \frac{1}{2}; \frac{1}{2}}(M_{1} M_{1} B;\theta) 
                              & \rho_{\minus \frac{1}{2};\minus \frac{1}{2}}(M_{1} M_{1} B;\theta)
                        \end{pmatrix*}.
\end{equation}

As we saw in eqn. (\ref{DensityMatrixDecomposition}), every spin density matrix can be written as linear combination of a particular set of
matrices. Since the spin density matrix of both the beam and the target are $2 \times 2$, they can each be decomposed into four matrices:
the identity matrix, and the operators for the three components of spin. In Cartesian coordinates, the spin operator of the beam and target 
are, respectively,
\begin{equation} \label{SpinOperators}
    \begin{split}
        \hat{S}_{i}(\gamma) & = \hat{\sigma}_{i}, \\
        \hat{S}_{i}(N) & = \frac{1}{2} \hat{\sigma}_{i},
    \end{split}
\end{equation}
where $i$ is a Cartesian
component and the $\hat{\sigma}_{i}$'s are the well known Pauli matrices,
\begin{equation} \label{PauliMatrices}
  \hat{\sigma}_{x} = \begin{pmatrix}
               0 & \hphantom{-} 1 \\
               1 & \hphantom{-} 0
               \end{pmatrix}, \quad       
  \hat{\sigma}_{y} = \begin{pmatrix}
                0 & -i \\
                i & \hphantom{-} 0
               \end{pmatrix}, \quad
  \hat{\sigma}_{z} = \begin{pmatrix}
               1 & \hphantom{-} 0 \\
               0 & -1
               \end{pmatrix}.
\end{equation}
Therefore, the spin density matrix can be expressed as
\begin{equation} \label{InitialDensityMatrixDecomposition}
 \begin{split}
\hat{\rho}(\gamma N) & = \hat{\rho}(\gamma)\otimes\hat{\rho}(N) \\
  & = \frac{1}{2}(\hat{I}
      +\sum_{i}\Lambda^{i (\gamma)}\hat{\sigma}_{i})\otimes
      \frac{1}{2} (\hat{I}
      +\sum_{i}\Lambda^{i (N)} \hat{\sigma}_{i}) \\
  & = \frac{1}{4}(\hat{I} \otimes \hat{I} + \sum_{i} \Lambda^{i (\gamma)} \hat{\sigma}_{i} \otimes \hat{I} 
     + \sum_{i} \Lambda^{i (N)} I \otimes \hat{\sigma}_{i} 
     + \sum_{i,j} \Lambda^{i (\gamma)} \Lambda^{j (N)} \hat{\sigma}_{i} \otimes \hat{\sigma}_{j}), \\
i, j & \in \{x, y, z\}.
\end{split}
\end{equation}

The $\Lambda^{i (N)}$'s and $\Lambda^{i (\gamma)}$'s are the components of the polarization vector of the photon and nucleon, 
respectively. The prefactor of $1/2$ in the expression arises from
the fact that both spin density matrices are normalized such that $\Tr[\hat{\rho}] = 1$. Since the indices $i$ and $j$ run over three values,
the density matrix has been decomposed into a linear combination of 16 matrices. Each of these matrices is a tensor product of two $2\times2$
matrices, so each of these matrices will be $4\times4$. An example of one of these matrices is
\begin{align} \label{16MatricesExample1}
    & \hat{I}=\begin{pmatrix}
		1 & 0 \\
		0 & 1
	      \end{pmatrix}, \quad
      \hat{\sigma}_{z}=\begin{pmatrix}
			  1 & 0 \\
			  0 & -1
			\end{pmatrix}, \nonumber \\
    & \hat{I} \otimes \hat{\sigma}_{z}=\begin{pmatrix}
					1 & 0 \\
					0 & 1
					\end{pmatrix}
      \otimes \hat{\sigma}_{z} = \begin{pmatrix}
				  \hat{\sigma}_{z} & 0 \\
				  0 & \hat{\sigma}_{z}
				  \end{pmatrix}
      = \begin{pmatrix}
	1 & 0 & 0 & 0 \\
	0 & -1 & 0 & 0 \\
	0 & 0 & 1 & 0 \\
	0 & 0 & 0 & -1
	\end{pmatrix}.
\end{align}
Another example is
\begin{align} \label{16MatricesExample2}
    & \hat{\sigma}_{x}=\begin{pmatrix}
			  0 & 1 \\
			  1 & 0
	      \end{pmatrix}, \quad
      \hat{\sigma}_{y}=\begin{pmatrix}
			  0 & -i \\
			  i & 0
			\end{pmatrix}, \nonumber \\
    & \hat{\sigma}_{x} \otimes \hat{\sigma}_{y}=\begin{pmatrix}
						  0 & 1 \\
						  1 & 0
						\end{pmatrix}
      \otimes \hat{\sigma}_{y} = \begin{pmatrix}
				    0 & \hat{\sigma}_{y} \\
				    \hat{\sigma}_{y} & 0
				  \end{pmatrix}
      = \begin{pmatrix}
	0 & 0 & 0 & -i \\
	0 & 0 & i & 0 \\
	0 & -i & 0 & 0 \\
	i & 0 & 0 & 0
	\end{pmatrix}.
\end{align}

Combining eqns. (\ref{InitialToFinalDensityMatrix}) and (\ref{InitialDensityMatrixDecomposition}) gives us the spin
density matrix of the recoiling baryon, as a sum of 16 $2 \times 2$ matrices,
\begin{equation} \label{16Matrices}
\hat{\rho}(B) =  \hat{\rho}^{0}+\sum_{i}\Lambda^{i (\gamma)} \hat{\rho}^{i (\gamma)} 
                            + \sum_{i} \Lambda^{i (N)} \hat{\rho}^{i (N)}
                            + \sum_{i,j} \Lambda^{i (\gamma)} \Lambda^{j (N)} \hat{\rho}^{ij (\gamma N)},
\end{equation}
where the 16 matrices are defined as
\begin{equation} \label{DensityMatrixSigmaTensorProduct}
  \begin{split}
    \hat{\rho}^{0}(B) & = \hat{M} (\hat{I} \otimes \hat{I}) \hat{M}^{\dagger}, \\
    \hat{\rho}^{i (\gamma)}(B) & = \hat{M} (\hat{\sigma}^{i} \otimes \hat{I}) \hat{M}^{\dagger}, \\
    \hat{\rho}^{i (N)}(B) & = \hat{M} (\hat{I} \otimes \hat{\sigma}^{i}) \hat{M}^{\dagger}, \\
    \hat{\rho}^{ij (\gamma N)}(B) & = \hat{M} (\hat{\sigma}^{i} \otimes \hat{\sigma}^{j}) \hat{M}^{\dagger}.
  \end{split}
\end{equation}
In terms of the matrix elements,the expression is 
\begin{align} \label{DensityMatrixSigmaMatrixElements}
    \rho^{0}_{\lambda_{B} \lambda'_{B}}(B)
      & =  \frac{1}{4}\sum_{\lambda_{\gamma}\lambda_{N}} 
                        M_{\lambda_{B}; \lambda_{\gamma}\lambda_{N}}
                        M^{*}_{\lambda'_{B}; \lambda_{\gamma}\lambda_{N} }, \nonumber \\
	\rho^{i (\gamma)}_{\lambda_{B} \lambda'_{B}}(B)
      & =\frac{1}{4} \sum_{\substack{\lambda_{\gamma} \lambda'_{\gamma} \\
                        \lambda_{N}}} 
                        M_{\lambda_{B}; \lambda_{\gamma}\lambda_{N}}
                        \sigma^{i}_{\lambda_{\gamma} \lambda'_{\gamma}}
                        M^{*}_{\lambda'_{B}; \lambda'_{\gamma}\lambda_{N} }, \\
    \rho^{i (N)}_{\lambda_{B} \lambda'_{B}}(B)
      & =  \frac{1}{4}\sum_{\substack{\lambda_{\gamma} \nonumber \\
                        \lambda_{N} \lambda'_{N}}} 
                        M_{\lambda_{B}; \lambda_{\gamma}\lambda_{N}}
                        \sigma^{i}_{\lambda_{N} \lambda'_{N}}
                        M^{*}_{\lambda'_{B}; \lambda_{\gamma}\lambda'_{N} }, \\
    \rho^{ij (\gamma N)}_{\lambda_{B} \lambda'_{B}}(B)
      & =\frac{1}{4} \sum_{\substack{\lambda_{\gamma} \lambda'_{\gamma} \\
                        \lambda_{N} \lambda'_{N}}} 
                        M_{\lambda_{B}; \lambda_{\gamma}\lambda_{N}}
                        \sigma^{i}_{\lambda_{\gamma}\lambda'_{\gamma}}\sigma^{j}_{\lambda_{N} \lambda'_{N}}
                        M^{*}_{\lambda'_{B}; \lambda'_{\gamma}\lambda'_{N}}.\nonumber
\end{align}
These can be further simplified by using the explicit forms of the Pauli matrices in the $z$ basis found in eqn. (\ref{PauliMatrices}). For example,
\begin{align} \label{DensityMatrixSigmaMatrixElementsExample}
 & \rho^{x (\gamma)}_{\lambda_{B} \lambda'_{B}}(B)
     =  \frac{1}{4}\sum_{\lambda_{\gamma} \lambda_{N}} 
                        M_{\lambda_{B}; \lambda_{N} \lambda_{\gamma}}
                        M^{*}_{\lambda'_{B}; \lambda_{N} -\lambda_{\gamma}}, \nonumber \\
 & \rho^{y (\gamma)}_{\lambda_{B} \lambda'_{B}}(B)
     =-\frac{i}{4} \sum_{\lambda_{N} \lambda_{\gamma}} 
                        \lambda_{\gamma} M_{\lambda_{B}; \lambda_{N} \lambda_{\gamma}}
                        M^{*}_{\lambda'_{B}; \lambda_{N} -\lambda_{\gamma}}, \\
 & \rho^{z (\gamma)}_{\lambda_{B} \lambda'_{B}}(B)
     = \frac{1}{4} \sum_{\lambda_{N} \lambda_{\gamma}} 
                        \lambda_{\gamma} M_{\lambda_{B}; \lambda_{N} \lambda_{\gamma}}
                        M^{*}_{\lambda'_{B}; \lambda_{N} \lambda_{\gamma}}. \nonumber
\end{align}
These expression are general for the spin density matrix of any spin-1/2 final state whose initial state has two particles, each with two possible spin projections. 

Note that each of the elements of these 16 matrices do not depend
on the initial polarization of the beam and target. They are, however, functions of the scattering angle $\theta$. Since they do not depend on the 
initial polarizations, they only contain information about the dynamics of the
reaction. This can be seen from eqn. (\ref{DensityMatrixSigmaMatrixElements}), 
which shows that the elements of these 16 matrices depend only on the
helicity amplitudes.

Since each of the 16 matrices in the expansion of $\hat{\rho}(B)$ are $4\times4$ and hermitian, each is described by four independent parameters (eqn. (\ref{IndependentParameters})). There are therefore a total of 64 parameters. Note from eqn. (\ref{DensityMatrixSigmaMatrixElements}) that each of these parameters are bilinear combinations of the helicity amplitudes. Therefore, if the values of these parameters are determined from experiments at each kinematic point, they can be used to extract the values of the amplitudes (up to an overall phase), entailing a complete experiment.

\chapter{Polarization Observables} \label{PolarizationObservables}
A question that naturally arises is: How can the polarization state of the reaction products in a scattering experiment be determined for a 
given polarization of the beam and target?, i.e., which experimental quantities should be measured in order to determine the 
independent parameters discussed in the previous section? These quantities are
the polarization observables, which will be defined later in this section. 

The way to extract them experimentally is to define 64 quantities known as the polarization observables, which are linear combinations of the 64 independent parameters. We can therefore determine the SMDE's of the 16 matrices by measuring the observables. Note that while the density matrix $\hat{\rho}(B)$ has only four independent parameters, these are functions of the initial polarizations $\vec{\Lambda}^{(\gamma)}$ and $\vec{\Lambda}^{(N)}$. This dependence on the initial polarization was factored out in eqn. (\ref{16Matrices}) so that the SDME's of the 16 expansion matrices only depend on the helicity amplitudes. The polarization observables can only be used to extract the SDME's of these expansion matrices.

The observables are categorized as either single, double, or triple polarization observables, based on how many of the particles with spin (beam, target, and recoil baryon) are involved in the measurement. We will first show their definitions in terms of the SDME's and we will later show how they relate to polarized cross sections. The unpolarized observable is
\begin{equation} \label{Observables1}
 I_{0}=\Tr[\hat{\rho}^{0}].
\end{equation}
The three single polarization beam observables are
\begin{equation} \label{Observables2}
    \begin{split}
        I_{0} I^{c} & =\Tr[\hat{\rho}^{x (\gamma)}], \\
        I_{0} I^{s} & =\Tr[\hat{\rho}^{y (\gamma)}], \\
        I_{0} I^{\odot} & =\Tr[\hat{\rho}^{z (\gamma)}].
    \end{split}
\end{equation}
The three single polarization target observables are
\begin{equation} \label{Observables3}
 I_{0} P_{i}=\Tr[\hat{\rho}^{i (N)}].
\end{equation}
The three single polarization recoil observables are
\begin{equation} \label{Observables4}
    I_{0} P_{i'}=\Tr[\hat{\rho}^{0}\hat{\sigma}^{i}].
\end{equation}
More generally, the spin operator is used in place of $\hat{\sigma}^{i}$'s for higher spin particles. The 9 beam-target double polarization
observables are
\begin{equation} \label{Observables5}
 \begin{split}
        I_{0} P^{c}_{i} & =\Tr[\hat{\rho}^{x i (\gamma N)}], \\
        I_{0} P^{s}_{i} & =\Tr[\hat{\rho}^{y i (\gamma N)}], \\
        I_{0} P^{\odot}_{i} & =\Tr[\hat{\rho}^{z i (\gamma N)}].
    \end{split}
\end{equation}
The 9 beam-recoil double polarization observables are
\begin{equation} \label{Observables6}
    \begin{split}
        I_{0} P^{c}_{i'} & =\Tr[ 
            \hat{\rho}^{x (\gamma)}\hat{\sigma}^{i}], \\
        I_{0} P^{s}_{i'} & =\Tr[
            \hat{\rho}^{y (\gamma)}\hat{\sigma}^{i}], \\
        I_{0} P^{\odot}_{i'} & =\Tr[
            \hat{\rho}^{z (\gamma)}\hat{\sigma}^{i}].
    \end{split}
\end{equation}
The 9 target-recoil double polarization observables are
\begin{equation} \label{Observables7}
    I_{0} \mathcal{O}_{\alpha \beta'} = \Tr[
        \hat{\rho}^{\alpha (N)}\hat{\sigma}^{\beta'}].
\end{equation}

The 27 triple polarization observables are
\begin{equation} \label{Observables8}
    \begin{split}
        I_{0} \mathcal{O}^{c}_{\alpha \beta'} & =
            \Tr[\hat{\rho}^{x \alpha (\gamma N)}\hat{\sigma}^{\beta'}], \\
        I_{0} \mathcal{O}^{s}_{\alpha \beta'} & =
            \Tr[\hat{\rho}^{y \alpha (\gamma N)}\hat{\sigma}^{\beta'}], \\
        I_{0} \mathcal{O}^{\odot}_{\alpha \beta'} & =
            \Tr[\hat{\rho}^{z \alpha (\gamma N)}\hat{\sigma}^{\beta'}].
    \end{split}
\end{equation}

To measure the observables, we need to be able to relate them to differential cross sections. These are in turn related to the scattering amplitude, which is given in terms of the scattered state $\ket{\Psi}$ and the state found after a measurement $\ket{\Phi}$ by,
\begin{equation} \label{CrossSectionPureState}
  \sigma \propto |\braket{\Phi|\Psi}|^{2}.
\end{equation}
While the symbol $\sigma$ is usually used for total cross sections, in order to simplify the
notation we will take it to refer to the differential cross section (Remember that the SDME's are not only functions of the spin projections, but also of the scattering angle $\theta$).
If the scattered state is instead a mixed state, it is not represented by the state vector $\ket{\Psi}$ but rather by a density matrix $\hat{\rho}(\Psi)$. It can be shown from the definition of the 
density matrix in eqn. (\ref{DensityMatrix}) that the probability of finding a member of the statistical ensemble in state $\Phi$ after 
a measurement is given by
\begin{equation} \label{CrossSectionMixedState}
  \sigma \propto \braket{\Phi|\hat{\rho}(\Psi)|\Phi}.
\end{equation}
We will soon show that the polarization observables are related to ratios of sums of cross sections, which means that the proportionality constant that turns eqns. (\ref{CrossSectionPureState}) and (\ref{CrossSectionMixedState}) into an equality is irrelevant. Therefore, for the rest of this thesis we will instead use
\begin{equation} \label{CrossSectionMixedState2}
  \sigma \propto \braket{\Phi|\hat{\rho}(\Psi)|\Phi} \Rightarrow \sigma = \braket{\Phi|\hat{\rho}(\Psi)|\Phi}.
\end{equation} 

Since we can now relate matrix elements of density matrices to cross sections, we can relate the observables defined in eqns.
(\ref{Observables1}) to (\ref{Observables8}) to cross sections. The expression for the spin density matrix of the recoil baryon shown in eqn. (\ref{16Matrices}) depends on the state of the beam and target through the quantities $\vec{\Lambda}^{(\gamma)}$ and
$\vec{\Lambda}^{(N)}$. We will therefore write it as
\begin{equation} \label{DensityMatrixAndPolarizationVectors}
 \hat{\rho}(B) \equiv \hat{\rho}(\vec{\Lambda}^{(\gamma)}, \vec{\Lambda}^{(N)}).
\end{equation}
Each of the polarization vectors in the previous equation have three components, so the spin density matrix of the recoil baryon requires six quantities to specify the spin state of the beam and target that gave rise to it. However, in the rest of this section we will assume that the experiments will be set up such that the beam and target are either polarized along one of
the three Cartesian coordinate axes ($x$, $y$ or $z$), or unpolarized. Therefore, it will depend on two quantities: the two degrees of polarization of the beam and target along an axis. From eqn. (\ref{CrossSectionMixedState2}) we see that the cross section for measuring the recoil nucleon with spin projection $\lambda_{i}$ along the
$i$-axis from a reaction in which the photon beam and target nucleon are fully polarized along the $j$- and $k$-axis respectively, is given by
\begin{equation} \label{CrossSectionAndDensityMatrix}
  \sigma_{i;jk}(\pm;\pm,\pm) 
    = \braket{\lambda_{i} = \pm \scalebox{.8}{$\frac{1}{2}$}|\hat{\rho}(\Lambda^{j (\gamma)} = \pm 1,\Lambda^{k (N)} = \pm 1)|
    \lambda_{i} = \pm \scalebox{.8}{$\frac{1}{2}$}}.
\end{equation}
This gives us eight possible polarized cross sections.

For example, the cross section for the recoil nucleon to be measured with spin projection $-1/2$ along the $y$-axis when the beam is polarized with spin projection $-1$ along the 
$z$-axis (circularly polarized beam) and the 
target is polarized with spin projection $1/2$ along the $x$-axis is
\begin{equation} \label{CrossSectionAndDensityMatrixExample1}
  \sigma_{y;zx}(-;-,+) 
    = \braket{\lambda_{y} = - \scalebox{.8}{$\frac{1}{2}$}|\hat{\rho}(\Lambda^{z (\gamma)} = - 1,\Lambda^{x (N)} = + 1)|
    \lambda_{y} = - \scalebox{.8}{$\frac{1}{2}$}}.
\end{equation}
If the beam or target are unpolarized, $\Lambda^{i} = 0$ for the corresponding hadron. By using this identity that can be shown from eqn. (\ref{16Matrices}),
\begin{equation} \label{Finding16Matrices1}
	\hat{\rho}(\vec{\Lambda}^{(\gamma)},\Lambda^{z (N)} = 0) = \frac{1}{2}\Big(\hat{\rho}(\vec{\Lambda}^{(\gamma)},\Lambda^{z (N)} = 1) + \hat{\rho}(\vec{\Lambda}^{(\gamma)},\Lambda^{z (N)} = -1)\Big),
\end{equation}
we can show that, if in the expression shown in eqn. (\ref{CrossSectionAndDensityMatrixExample1}) the target baryon was instead unpolarized, the cross section would be
\begin{equation} \label{CrossSectionAndDensityMatrixExample2}
  \begin{split}
    \sigma_{y;zz}(-;-,0) 
      & = \braket{\lambda_{y} = - \scalebox{.8}{$\frac{1}{2}$}|\hat{\rho}(\Lambda^{z (\gamma)} = - 1,\Lambda^{z (N)} = 0)|
		\lambda_{y} = - \scalebox{.8}{$\frac{1}{2}$}} \\
	& = \bra{\lambda_{y} = - \scalebox{.8}{$\frac{1}{2}$}}\frac{1}{2}\Big(\hat{\rho}(\Lambda^{z (\gamma)} = - 1,\Lambda^{z (N)} = 1) \\
	\mt \qquad + \hat{\rho}(\Lambda^{z (\gamma)} = - 1,\Lambda^{z (N)} = -1)\Big)\ket{\lambda_{y} = - \scalebox{.8}{$\frac{1}{2}$}} \\
	& = \frac{1}{2}\Big[\braket{\lambda_{y} = - \scalebox{.8}{$\frac{1}{2}$}|\hat{\rho}(\Lambda^{z (\gamma)} = - 1,\Lambda^{z (N)} = 1)|
	\lambda_{y} = - \scalebox{.8}{$\frac{1}{2}$}} \\
	& \qquad + \braket{\lambda_{y} = - \scalebox{.8}{$\frac{1}{2}$}|\hat{\rho}(\Lambda^{z (\gamma)} = - 1,\Lambda^{z (N)} = -1)|
	\lambda_{y} = - \scalebox{.8}{$\frac{1}{2}$}}\Big] \\
	& = \frac{1}{2}\big[\sigma_{y;zz}(-;-,+) + \sigma_{y;zz}(-;-,-)\big].
  \end{split}
\end{equation}
The last line of the previous expression shows that a cross section with an unpolarized beam or target is equal to the average of the polarized cross sections with the two possible orthogonal spin states in some basis. Similarly, if the polarization of the recoil nucleon is not measured, its cross section is equal
to the sum of the cross sections with all possible polarizations in some basis. For example, if the photon has spin projection $-1$ along the $z$-axis, the target has spin projection $+1/2$ along the $x$-axis and the recoil polarization is not measured, the cross section is 
\begin{equation} \label{CrossSectionAndDensityMatrixExample3}
  \begin{split}
    \sigma_{z;zx}(0;-,+) 
      & = \sigma_{z;zx}(+;-,+) + \sigma_{z;zx}(-;-,+) \\
		& = \braket{\lambda_{z} = \scalebox{.8}{$\frac{1}{2}$}|\hat{\rho}(\Lambda^{z (\gamma)} = - 1,\Lambda^{x (N)} = 1)|
	    \lambda_{z} = \scalebox{.8}{$\frac{1}{2}$}} \\
      \mt + \braket{\lambda_{z} = - \scalebox{.8}{$\frac{1}{2}$}|\hat{\rho}(\Lambda^{z (\gamma)} = - 1,\Lambda^{x (N)} = 1)|
	    \lambda_{z} = - \scalebox{.8}{$\frac{1}{2}$}} \\
      & = \Tr[\hat{\rho}(\Lambda^{z (\gamma)} = - 1,\Lambda^{x (N)} = 1)].
  \end{split}
\end{equation}
Therefore, the trace of the spin density matrix is taken when the recoiling particle's polarization is not measured. 

We can now write the observables in terms of polarized differential cross sections. For the observable $I_{0}$, we get
\begin{equation} \label{Observable1CrossSection}
  \begin{split}
    I_{0} & \equiv \Tr[\rho^{0}] \\
      & = \Tr[\hat{\rho}(\Lambda^{z (\gamma)} = 0, \Lambda^{z (N)} = 0)] \\
      & = \sigma_{z;zz}(0;0,0).
  \end{split}
\end{equation}
In other words, this observable is simply the unpolarized cross section (up to a spin-independent constant). Written in terms of polarized cross sections,
\begin{equation} \label{Observable1CrossSection2}
  \begin{split}
    I_{0} & = \frac{1}{4}[\sigma_{z;zz}(+;+,+) + \sigma_{z;zz}(+;+,-) + \sigma_{z;zz}(-;+,+) + \sigma_{z;zz}(-;+,-) \\
      \mt+ \sigma_{z;zz}(+;-,+) + \sigma_{z;zz}(+;-,-) + \sigma_{z;zz}(-;-,+) + \sigma_{z;zz}(-;-,-)] \\
      & = \frac{1}{4} \sigma_{0},
  \end{split}
\end{equation}
where we have defined the quantity inside the brackets on the first line as $\sigma_{0}$ (the factor of $\frac{1}{4}$ comes from averaging over the initial polarizations for the photon and the target nucleon). 

To find expressions for the beam or target polarization observables, we need to be able to find the 16 matrices $\hat{\rho}^{i}$ from $\hat{\rho}(\vec{\Lambda}^{(\gamma)}, \vec{\Lambda}^{(N)})$. To find $\hat{\rho}^{x(N)}$, for example, we use eqn. (\ref{16Matrices}) to find,
\begin{equation} \label{Finding16Matrices2}
	\hat{\rho}^{x} = \frac{1}{2}\Big(\hat{\rho}(\Lambda^{x(\gamma)} = 0, \Lambda^{x(N)} = 1) - \hat{\rho}(\Lambda^{x(\gamma)} = 0, \Lambda^{x(N)} = -1)\Big).
\end{equation}
Therefore, the observable $P^{x}$ is
given by
\begin{equation} \label{Observable2CrossSection}
  \begin{split}
    P^{x} & \equiv \frac{\Tr[\hat{\rho}^{x (N)}]}{\Tr[\hat{\rho}^{0}]} \\
      & = \frac{\frac{1}{2} \big\{ \Tr[\hat{\rho}(\Lambda^{z (\gamma)} = 0,\Lambda^{x (N)} = 1)] 
	- \Tr[\hat{\rho}(\Lambda^{z (\gamma)} = 0,\Lambda^{x (N)]} = - 1)] \big\}}
	{\frac{1}{2} \big\{ \Tr[\hat{\rho}(\Lambda^{z (\gamma)} = 0,\Lambda^{x (N)} = 1)]
	+\Tr[\hat{\rho}(\Lambda^{z (\gamma)} = 0,\Lambda^{x (N)} = -1)]  \big\} } \\
      & = \frac{\sigma_{z;zx}(0;0,+) - \sigma_{z;zx}(0;0,-)}{\sigma_{z;zx}(0;0,+) + \sigma_{z;zx}(0;0,-)}.
  \end{split}
\end{equation}
This shows that the target polarization observables are simply the asymmetry in the polarized cross sections along some axis, normalized over the unpolarized cross section, at each value of the scattering angle. Therefore, in order to measure this observable in the lab,
the experimenter would have to run the experiment with the target polarized along the positive $x$-axis and measure the scattering 
cross section without measuring the spin of the recoil nucleon, and then repeat the measurement with the experiment set with the target polarized along
the negative $x$-axis. The difference between these two quantities divided by their sum gives you the observable. If written in terms of all the polarized
cross sections, we get,
\begin{equation} \label{Observable2CrossSection2}
    \begin{split}
    P^{x} & = \big[ \sigma_{z;zx}(+;+,+) + \sigma_{z;zx}(+;-,+) +                               \sigma_{z;zx}(-;+,+) + \sigma_{z;zx}(-;-,+) \\
	  \mt - \sigma_{z;zx}(+;+,-) - \sigma_{z;zx}(+;-,-) - \sigma_{z;zx}(-;+,-) - \sigma_{z;zx}(-;-,-) \big] \Bigg/ \sigma_{0}.
    \end{split}
\end{equation}
In general, every single observable will be a linear combination of all of the eight polarized cross sections normalized over the unpolarized cross section. For each observable, the relative minus signs among the eight terms will be different.

To find the expressions for the recoil observables, we need to use the identity
\begin{equation} \label{SigmaMatrixRelation}
  \Tr[\hat{A}\hat{\sigma}^{z}] = \braket{+,z|\hat{A}|+,z} - \braket{-,z|\hat{A}|-,z},
\end{equation}
where $\hat{A}$ is any $2\times2$ matrix. Therefore, the observable $P_{z'}$ can be expressed as
\begin{equation} \label{Observable3CrossSection}
  \begin{split}
    P_{z'} & \equiv \frac{\Tr[\hat{\rho}^{0}\hat{\sigma}^{z}]}{\Tr[\hat{\rho}^{0}]} \\
      & = \frac{\braket{+,z|\hat{\rho}(\Lambda^{z (\gamma)} = 0,\Lambda^{z (N)} = 0)|+,z}
	- \braket{-,z|\hat{\rho}(\Lambda^{z (\gamma)} = 0,\Lambda^{z (N)} = 0)|-,z}}
	{\Tr[\hat{\rho}(\Lambda^{z (\gamma)} = 0,\Lambda^{z (N)} = 0]} \\
      & = \frac{\sigma_{z;zz}(+;0,0) - \sigma_{z;zz}(-;0,0)}{\sigma_{z;zz}(+;0,0) + \sigma_{z;zz}(-;0,0)}.
  \end{split}
\end{equation}
Once again, we get an asymmetry in the cross sections of measuring the recoil nucleon along some axis, normalized over the 
unpolarized cross section.  

In order to find the expression for $P_{x'}$ or $P_{y'}$, we need these relations, which are correct up to an irrelevant phase factor,
\begin{equation} \label{RotationRelations}
  \begin{split}
    & \hat{\sigma}^{x} = \hat{\mathcal{R}}^{y}\big(\scalebox{1}{$\frac{\pi}{2}$}\big)\hat{\sigma}^{z}
	\hat{\mathcal{R}}^{y}\big(\scalebox{1}{$\minus \frac{\pi}{2}$}\big), \quad
	\hat{\sigma}^{y} = \hat{\mathcal{R}}^{x}\big(\scalebox{1}{$\minus \frac{\pi}{2}$}\big)\hat{\sigma}^{z}
	\hat{\mathcal{R}}^{x}\big(\scalebox{1}{$\frac{\pi}{2}$}\big), \\
    & \hat{\mathcal{R}}^{y}\big(\scalebox{1}{$\frac{\pi}{2}$}\big) \ket{\pm, z} = \ket{\pm, x}, \quad
      \hat{\mathcal{R}}^{x}\big(\scalebox{1}{$\minus \frac{\pi}{2}$}\big) \ket{\pm, z} = \ket{\pm, x},
  \end{split}
\end{equation}
where $\hat{\mathcal{R}}^{i}\big(\scalebox{1}{$\pm \frac{\pi}{2}$}\big)$ is the rotation matrix around the $i$-axis by $\pm \pi/2$
radians. The expression for the observable $P_{x}$ is therefore
\begin{align} \label{Observable4CrossSection}
    P_{x'} & \equiv \frac{\Tr[\hat{\rho}^{0}\hat{\sigma}^{x}]}{\Tr[\hat{\rho}^{0}]} \nonumber \\
      & = \frac{\Tr[\hat{\rho}^{0}\hat{\mathcal{R}}^{y}\big(\scalebox{1}{$\frac{\pi}{2}$}\big)\hat{\sigma}^{z}
	\hat{\mathcal{R}}^{y}\big(\scalebox{1}{$\minus \frac{\pi}{2}$}\big)]}{\Tr[\hat{\rho}^{0}]} \nonumber \\
      & = \frac{\Tr[\hat{\mathcal{R}}^{y}\big(\scalebox{1}{$\minus \frac{\pi}{2}$}\big)
	\hat{\rho}^{0}\hat{\mathcal{R}}^{y}\big(\scalebox{1}{$\frac{\pi}{2}$}\big)\hat{\sigma}^{z} ]}{\Tr[\hat{\rho}^{0}]} \\
      & = \frac{\braket{+, z|\hat{\mathcal{R}}^{y}\big(\scalebox{1}{$\minus \frac{\pi}{2}$}\big)
	\hat{\rho}^{0}\hat{\mathcal{R}}^{y}\big(\scalebox{1}{$\frac{\pi}{2}$}\big)|+, z} - \braket{-, z|\hat{\mathcal{R}}^{y}
	\big(\scalebox{1}{$\minus \frac{\pi}{2}$}\big) \hat{\rho}^{0}\hat{\mathcal{R}}^{y}\big(\scalebox{1}{$\frac{\pi}{2}$}\big)|-, z)}}
	{\Tr[\hat{\rho}^{0}]} \nonumber \\
      & = \frac{\braket{+, x|\hat{\rho}^{0}|+, x} - \braket{-, x|\hat{\rho}^{0}|-, x)}}
	{\Tr[\hat{\rho}^{0}]} \nonumber \\
      & = \frac{\sigma_{x;zz}(+;0,0) - \sigma_{x;zz}(-;0,0)}{\sigma_{x;zz}(+;0,0) + \sigma_{x;zz}(-;0,0)}, \nonumber
\end{align}
where in the third line of the previous equation we used the cyclic property of the trace.
Once again, we get an asymmetry, this time for measuring the recoil baryon's spin along the $x$-axis. 

For a double polarization observable, we find (using the beam-target observable $P^{s}_{x}$ as
an example),
\begin{align} \label{Observable5CrossSection}
    P^{s}_{x} & \equiv \frac{\Tr[\hat{\rho}^{yx}]}{\Tr[\hat{\rho}^{0}]}, \nonumber \\
      & = \frac{1}{2} \bigg\{ \frac{1}{2} \Big( \Tr[\hat{\rho}(\Lambda^{y (\gamma)} = 1, \Lambda^{x (N)} =1)]
	- \Tr[\hat{\rho}(\Lambda^{y (\gamma)} = 1, \Lambda^{x (N)} = -1)] \Big) \nonumber \\ 
      \mt - \frac{1}{2} \Big( \Tr[\hat{\rho}(\Lambda^{y (\gamma)} = -1, \Lambda^{x (N)} = 1)]
	- \Tr[\hat{\rho}(\Lambda^{y (\gamma)} = -1, \Lambda^{x (N)} = -1)] \Big) \bigg\} \Bigg/ \Tr[\hat{\rho}^{0}] \\
      & = \frac{\frac{1}{2} \bigg\{ \frac{1}{2}[\sigma_{z;yx}(0;+,+) - \sigma_{z;yx}(0;+,-)] - \frac{1}{2}[\sigma_{z;yx}(0;-,+) -
	\sigma_{z;yx}(0;-,-)] \bigg\}}{\frac{1}{4}[\sigma_{z;yx}(0;+,+) + \sigma_{z;yx}(0;+,-) + \sigma_{z;yx}(0;-,+) 
	+ \sigma_{z;yx}(0;-,-)]} \nonumber \\
      & = \frac{\sigma_{z;yx}(0;+,+) - \sigma_{z;yx}(0;+,-) - \sigma_{z;yx}(0;-,+) +
	\sigma_{z;yx}(0;-,-)}{\sigma_{z;yx}(0;+,+) + \sigma_{z;yx}(0;+,-) + \sigma_{z;yx}(0;-,+) 
	+ \sigma_{z;yx}(0;-,-)}. \nonumber
\end{align}
Note from the third line of the last equation that a double polarization observable is an asymmetry of an asymmetry, i.e., an 
asymmetry in the target is measured for each of the two values of the beam polarization, and then the asymmetry between these two values is
taken.

Recall that eqn. (\ref{DensityMatrixSigmaMatrixElements}) shows that the 64 SDME's are equal to bilinear combinations of the helicity amplitudes in eqn. (\ref{TransitionMatrix}). The observables can therefore also be expressed in terms of said amplitudes. As such, measurements of observables can be used to extract the helicity amplitudes of the process at different kinematic points. As previously mentioned, the behaviour of these amplitudes as a function of the kinematic variables is needed to carry out the PWA's, which could establish the presence of the resonances contributing to the reactions. Eqn. (\ref{TransitionMatrix}) shows there are 8 complex amplitudes in double pion photoproduction, for a total 16 parameters. Since a state vector describes a quantum state up to an overall phase factor, it should come as no surprise that 15 parameters can be extracted from the measurement of the observables. 

Substituting the expressions in eqn. (\ref{DensityMatrixSigmaMatrixElements}) into the definitions of the observables in eqns. (\ref{Observables1})-(\ref{Observables8}) allows us to write the observables in terms of the helicity amplitudes. The unpolarized cross section is
\begin{align} \label{ObservableTransitionMatrix1} 
    I_{0} \equiv \Tr[\rho^{0}] & = \frac{1}{4}\Big(|M_{+,++}|^{2} + |M_{+,-+}|^{2} + |M_{+,+-}|^{2} + |M_{+,--}|^{2} \nonumber \\
      \mt + |M_{-,++}|^{2} + |M_{-,-+}|^{2} + |M_{-,+-}|^{2} + |M_{-,--}|^{2}\Big), \nonumber \\
    \sigma_{0} & =  |M_{+,++}|^{2} + |M_{+,-+}|^{2} + |M_{+,+-}|^{2} 
        + |M_{+,--}|^{2} \\
        \mt + |M_{-,++}|^{2} + |M_{-,-+}|^{2}
            + |M_{-,+-}|^{2} + |M_{-,--}|^{2},\nonumber
\end{align}
which is what is expected for the unpolarized cross section. To show a few more examples, the single polarization observables for the
beam are given by
\begin{equation} \label{ObservableTransitionMatrix2}
  \begin{split} 
    I^{\odot} \equiv \frac{\Tr[\rho^{z (\gamma)}]}{\Tr[\rho^{0}]}
      & = \bigg(\Big[|M_{+,++}|^{2} + |M_{+,-+}|^{2} + |M_{-,++}|^{2} + |M_{-,-+}|^{2}\Big] \\
      \mt - \Big[|M_{+,+-}|^{2} + |M_{+,--}|^{2} + |M_{-,+-}|^{2} + |M_{-,--}|^{2}\Big]\bigg) \Bigg/ \sigma_{0},
  \end{split}
\end{equation}
\begin{equation} \label{ObservableTransitionMatrix3}
  \begin{split}
    I^{c} \equiv \frac{\Tr[\hat{\rho}^{x (\gamma)}]}{\Tr[\hat{\rho}^{0}]}, 
     & = 2 \Re\Big[M_{+;++}M^{*}_{+;+-} + M_{+;-+}M^{*}_{+;--} \\
     \mt + M_{-;++}M^{*}_{-;+-} + M_{-;-+}M^{*}_{-;--}\Big] \Bigg/ \sigma_{0},
  \end{split}
\end{equation}
\begin{equation} \label{ObservableTransitionMatrix4}
  \begin{split}
    I^{s} \equiv \frac{\Tr[\hat{\rho}^{y (\gamma)}]}{\Tr[\hat{\rho}^{0}]}, 
     & = 2 \Im\Big[M_{+;++}M^{*}_{+;+-} + M_{+;-+}M^{*}_{+;--} \\
     \mt + M_{-;++}M^{*}_{-;+-} + M_{-;-+}M^{*}_{-;--}\Big] \Bigg/ \sigma_{0}.
  \end{split}
\end{equation}
Note that all observables are therefore the ratio of a bilinear sum of the transition amplitudes and the unpolarized cross section.

Eqn. (\ref{16Matrices}) shows the spin density matrix of the nucleon $\hat{\rho}(B)$ is a function of 16 expansion matrices. But since the observables are functions of the SDME's of these matrices, it can alternatively be expressed in terms of the observables. To show this, we will need these identities,
\begin{equation} \label{PauliMatricesIdentities}
\Tr[\hat{\sigma}^{i}] = 0, \quad \Tr[\hat{\sigma}^{i}
        \hat{\sigma}^{j}] = \delta_{ij}.
\end{equation}
Since any $2\times2$ matrix can be expressed as a linear combination of the $2\times2$ identity and three Pauli matrices, we can do so for the 16 matrices in the expansion in eqn. (\ref{16Matrices}). As an example, we will write the matrix $\hat{\rho}^{x (N)}$ as
\begin{equation} \label{ExpansionMatrixExpansion}
    \hat{\rho}^{x (N)} = A + \sum_{i}B_{xj}\hat{\sigma}^{j},
\end{equation}
where $A$ and the $B_{xj}$'s are as-of-yet-unknown expansion coefficients (the first term has an implied factor of a $2\times2$ identity matrix). To find $A$, we take the trace on both sides of the equation,
\begin{equation} \label{FindingA}
    \begin{split}
        & I_{0}\Tr[\hat{\rho}^{x (N)}] = 2 I_{0} A = P_{x}, \\
        & \rightarrow A = \frac{1}{2} \frac{P_{x}}{I_0}
    \end{split}
\end{equation}
(The factor of $2$ comes from taking the trace of the identity matrix). To find the $B_{xj}$'s, we multiply both sides by a Pauli matrix and take the trace,
\begin{equation} \label{FindingB}
\begin{split}
& I_{0} \Tr[\hat{\rho}^{x (N)}\hat{\sigma}^{i}] = 2 I_{0}B_{xi}
    = \mathcal{O}_{xi'}, \\
& \rightarrow B_{xi}
    = \frac{1}{2} \frac{\mathcal{O}_{xi'}}{I_{0}}.
\end{split}
\end{equation}
Therefore,
\begin{equation} \label{ExpansionMatrixObservables}
    \hat{\rho}^{x (N)} = P_{x} + \sum_{i}\mathcal{O}_{xi'}\hat{\sigma}^{i}.
\end{equation}
Applying this derivation to all 16 matrices, we find
\begin{equation} \label{16MatricesObservables}
  \begin{split}   
    \hat{\rho}^{0} & = \frac{1}{2}\Big(I_{0} + \sum_{i'} P_{i'}\hat{\sigma}^{i'}\Big), \\
    \hat{\rho}^{i (N)} & = \frac{1}{2}I_{0}\Big(P_{i} + \sum_{\beta'} \mathcal{O}_{i\beta'} \hat{\sigma}^{\beta'}\Big), \\
    \hat{\rho}^{x (\gamma)} & = \frac{1}{2}I_{0}\Big(I^{c} + \sum_{i'} P^{c}_{i'} \hat{\sigma}^{i'}\Big), \\
    \hat{\rho}^{y (\gamma)} & = \frac{1}{2}I_{0}\Big(I^{s} + \sum_{i'} P^{s}_{i'} \hat{\sigma}^{i'}\Big), \\
    \hat{\rho}^{z (\gamma)} & = \frac{1}{2}I_{0}\Big(I^{\odot} + \sum_{i'} P^{\odot}_{i'} \hat{\sigma}^{i'}\Big), \\
    \hat{\rho}^{xi (\gamma N)} & = \frac{1}{2}I_{0}\Big(P^{c}_{i} + \sum_{\beta'} \mathcal{O}^{c}_{i\beta'} \hat{\sigma}^{\beta'}\Big), \\
    \hat{\rho}^{yi (\gamma N)} & = \frac{1}{2}I_{0}\Big(P^{s}_{i} + \sum_{\beta'} \mathcal{O}^{s}_{i\beta'} \hat{\sigma}^{\beta'}\Big), \\
    \hat{\rho}^{zi (\gamma N)} & = \frac{1}{2}I_{0}\Big(P^{\odot}_{i} + \sum_{\beta'} \mathcal{O}^{\odot}_{i\beta'} \hat{\sigma}^{\beta'}\Big). \\
  \end{split}
\end{equation} 
By substituting these expressions into eqn. (\ref{16Matrices}) we get
\begin{align} \label{DensityMatrixObservables}
    \hat{\rho}(B)I\big(\vec{\Lambda}^{(\gamma)},\vec{\Lambda}^{(N)}\big) & = \frac{1}{2}I_{0}\Bigg\{ \Big(1 
      + \sum_{i}\Lambda^{i (N)}P_{i} + \sum_{i'}\hat{\sigma}^{i'}P_{i'}
      + \sum_{\alpha, \beta'} \Lambda^{\alpha (N)} \hat{\sigma}^{\beta'} \mathcal{O}_{\alpha \beta'}\Big) \nonumber \\
      \mt + \delta_{\odot}\Big(I^{\odot} 
      + \sum_{i}\Lambda^{i (N)}P^{\odot}_{i} + \sum_{i'}\hat{\sigma}^{i'}P^{\odot}_{i'} 
      + \sum_{\alpha, \beta'}\Lambda^{\alpha (N)}\hat{\sigma}^{\beta'}\mathcal{O}^{\odot}_{\alpha \beta'}\Big) \\
      \mt + \delta_{l}\bigg[\sin2\beta \Big(I^{s} 
      + \sum_{i}\Lambda^{i (N)}P^{s}_{i} + \sum_{i'}\hat{\sigma}^{i'}P^{s}_{i'} 
      + \sum_{\alpha, \beta'}\Lambda^{\alpha (N)}\hat{\sigma}^{\beta'}\mathcal{O}^{s}_{\alpha \beta'} \Big)  \nonumber \\
      \mt + \cos2\beta \Big(I^{c} 
      + \sum_{i}\Lambda^{i (N)}P^{c}_{i} + \sum_{i'}\hat{\sigma}^{i'}P^{c}_{i'} 
      + \sum_{\alpha, \beta'}\Lambda^{\alpha (N)}\hat{\sigma}^{\beta'}\mathcal{O}^{c}_{\alpha \beta'} \Big)\bigg] \Bigg\}, \nonumber
\end{align}
where in this expression the density matrix $\hat{\rho}(B)$ is normalized so that its trace is equal to $1$, 
$I
$ is the reaction rate when the spin of the recoil nucleon is not measured and is
a function of the initial polarizations $\vec{\Lambda}^{(\gamma)}$ and $\vec{\Lambda}^{(N)}$.

As established previously, the spin density matrix $\hat{\rho}(B)$, can also be expressed in terms of the statistical tensors instead of the SDME's of the 16 matrices in eqn. (\ref{16Matrices}). By combining this equation with eqn. (\ref{TExpansion}), we get
\begin{equation} \label{Matrix16Tensors}
\hat{\rho}(B)=\sum_{L=0}^{1} \sum_{M=-L}^{L} (t_{LM}^{0} + \sum_{i} \Lambda^{i (\gamma)} t_{LM}^{i (\gamma)} + 
      \sum_{i} \Lambda^{i (N)} t_{LM}^{i (N)} + \sum_{i,j} \Lambda^{i (\gamma)} \Lambda^{j (N)} t_{LM}^{ij (\gamma N)}) \hat{T}^{\dagger}_{LM},
\end{equation}
where, for example,
\begin{equation} \label{MatrixToTensor}
 t_{LM}^{i (\gamma)} = \sum_{m,m'=-1/2}^{1/2} (-1)^{\frac{1}{2}-m} \rho^{i (\gamma)}_{mm'} C^{LM}_{\frac{1}{2}m'\frac{1}{2}-m}
\end{equation}
(The sum goes over $-1/2$ and $1/2$ because the recoil nucleon $B$ is spin-1/2). The inverse relation of eqn. (\ref{MatrixToTensor}) is
\begin{equation} \label{16InverseRelation}
\begin{split}
\rho^{i (\gamma)}_{mm'} & = \sum_{L=0}^{1} \sum_{M=-L}^{L} t_{LM}^{i (\gamma)} [\hat{T}_{LM}]_{m'm}, \\ 
& = \sum_{L=0}^{1} \sum_{M=-L}^{L} (-1)^{\frac{1}{2}-m'} t_{LM}^{i (\gamma)} C^{LM}_{\frac{1}{2}m';\frac{1}{2}-m}.
\end{split}
\end{equation}
By comparing eqns. (\ref{Matrix16Tensors}) and (\ref{TExpansion}), we see that just as the spin density matrix can be expanded into a linear combination
of 16 other matrices, the statistical tensors can also be expanded into a linear combination of 16 other tensors,
\begin{equation} \label{Tensor16Tensors}
 t_{LM} = t_{LM}^{0} + \sum_{i} \Lambda^{i (\gamma)} t_{LM}^{i (\gamma)} + \sum_{i} \Lambda^{i (N)} t_{LM}^{i (N)} 
      + \sum_{ij} \Lambda^{i (\gamma)} \Lambda^{j (N)} t_{LM}^{ij (\gamma N)}.
\end{equation} 

To find the expression relating the polarization observables to these statistical tensors, all that is needed is to combine equation 
(\ref{16InverseRelation}) with eqns. (\ref{Observables1}) through (\ref{Observables8}) and use the trace property of the polarization
operators,
\begin{equation} \label{TracePolarizationOperator1}
 \Tr[\hat{T}_{LM}(S)] = \sqrt{2S+1} \delta_{L0} \delta_{M0}.
\end{equation}
The unpolarized observable is 
\begin{equation} \label{ObservableTensor1}
  I_{0} = \sqrt{2} t^{0}_{00}.
\end{equation}
The three beam single polarization observables are given by
\begin{equation} \label{ObservableTensor2}
  \begin{split}
    I_{0} I^{c} & = \sqrt{2} t^{x (\gamma)}_{00}, \\
    I_{0} I^{s} & = \sqrt{2} t^{y (\gamma)}_{00}, \\
    I_{0} I^{\odot} & = \sqrt{2} t^{z (\gamma)}_{00}.
  \end{split}
\end{equation}
All of the observables that do not involve the recoil baryon will have the same form as eqns. (\ref{ObservableTensor1}) and (\ref{ObservableTensor2}), with the only difference among them being the superscript of $t_{00}$ ($x(N)$, $z (\gamma)$, $y (\gamma N)$, etc).
To find the expressions for the recoil polarizations, we need to express the Pauli matrices in terms of the spherical
components of the recoil baryon's spin operator,
\begin{equation} \label{CartesianToSpherical}
 \begin{split}
  \hat{\sigma}_{z} = & 2 \hat{S}_{0} = \sqrt{2} \hat{T}_{10}, \\
  \hat{\sigma}_{x} = & 2 (\hat{S}_{-1} - \hat{S}_{+1}) = \sqrt{2} (\hat{T}_{1-1} - \hat{T}_{11}), \\
  \hat{\sigma}_{y} = & i 2 (\hat{S}_{-1} + \hat{S}_{+1}) = i \sqrt{2} (\hat{T}_{1-1} + \hat{T}_{11}),
  \end{split}
\end{equation}
so that we can we can exploit the property
\begin{equation} \label{TracePolarizationOperator2}
 \Tr[\hat{T}_{L_{1}M_{1}}\hat{T}_{L_{2}M_{2}}] = (-1)^{M_{1}} \delta_{L_{1}L_{2}} \delta_{M_{1} -M_{2}}. 
\end{equation}
Using eqns. (\ref{CartesianToSpherical}), (\ref{TracePolarizationOperator2}) and (\ref{STHermiticity}) along with the equations for the observables that involve the recoil baryon from eqns. (\ref{Observables1}) to (\ref{Observables8}), we get for example
\begin{equation} \label{ObservableTensor3}
 \begin{split}
  I_{0} P_{z'} = & \sqrt{2} t^{0}_{10}, \\
  I_{0} P_{x'} = & \sqrt{2} (t^{0}_{11} - t^{0}_{1-1}) = 2 \sqrt{2} \Re[t^{0}_{11}], \\ 
  I_{0} P_{y'} = & -i \sqrt{2} (t^{0}_{11} + t^{0}_{1-1}) = 2 \sqrt{2} \Im[t^{0}_{11}],
 \end{split}
\end{equation}
where in the last two lines the hermiticity property of the tensors shown in eqn. (\ref{STHermiticity}) was used. All observables that involve the recoil baryon will have this form. The ones involving the $z$-projection of the recoil baryon will be proportional to a $t_{10}$, while those involving the $x$- and $y$-projections of the recoil baryon are proportional to the real and imaginary parts of a $t_{11}$, respectively. The only difference among them will be the superscript of the tensor. For example, the observable
\begin{equation} \label{ObservableTensor4}
  \begin{split}
    I_{0} \mathcal{O}_{zx'} & = 2 \sqrt{2} \Re[t^{z (N)}_{11}]
  \end{split}
\end{equation}
has a $z (N)$ superscript because it involves the $z$-projection of the target, and is proportional to the real part of the tensor because it involves the $x$-projection of the recoil baryon.

For comparison, the same observables in eqns. (\ref{ObservableTensor2}) and (\ref{ObservableTensor3}) in terms of the SDME's are,
\begin{equation} 
  \begin{split}
    I_{0} I^{c} & = \rho^{x (\gamma)}_{\frac{1}{2} \frac{1}{2}} + \rho^{x (\gamma)}_{\minus \frac{1}{2} \minus \frac{1}{2}}, \\
    I_{0} I^{s} & = \rho^{y (\gamma)}_{\frac{1}{2} \frac{1}{2}} + \rho^{y (\gamma)}_{\minus \frac{1}{2} \minus \frac{1}{2}}, \\
    I_{0} I^{\odot} & = \rho^{z (\gamma)}_{\frac{1}{2} \frac{1}{2}} + \rho^{z (\gamma)}_{\minus \frac{1}{2} \minus \frac{1}{2}}, \\
  \end{split}
\end{equation}
and
\begin{equation}
  \begin{split}
    I_{0} P_{z'} & = \rho^{0}_{\frac{1}{2} \frac{1}{2}} - \rho^{0}_{\minus \frac{1}{2} \minus \frac{1}{2}}, \\
    I_{0} P_{x'} & = \rho^{0}_{\frac{1}{2} \minus \frac{1}{2}} + \rho^{0}_{\minus \frac{1}{2} \frac{1}{2}} 
       = 2 \Re \Big[ \rho^{0}_{\frac{1}{2} \minus \frac{1}{2}} \Big], \\
    I_{0} P_{y'} & = i \bigg( \rho^{0}_{\frac{1}{2} \minus \frac{1}{2}} - \rho^{0}_{\minus \frac{1}{2} \frac{1}{2}} \bigg)
       = -2 \Im \Big[ \rho^{0}_{\frac{1}{2} \minus \frac{1}{2}} \Big].
  \end{split}
\end{equation}
While the SDME's and the statistical tensors contain the same information, it can be seen from this comparison that expressions for the observables in terms of the tensors are simpler than the ones in terms of the SDME's. This, in addition to the fact that the tensors transform under irreducible representations of the rotation group (the well known Wigner $\mathcal{D}$-matrices) and the fact that it allows the use of the Wigner-Eckart theorem shown in eqn. (\ref{Wigner-EckartTheorem}) in theoretic calculations, makes them a more natural choice in describing the observables. Nevertheless, since the use of SDME's is widespread in the literature, we will express all results in this thesis in terms of both the SDME's and the statistical tensors.

\chapter{Angular Distributions in $\gamma N \rightarrow M^{*} B \rightarrow M_{1} M_{2} B$, $\gamma N \rightarrow M_{1} B^{*} \rightarrow M_{1} M_{2} B$, and $\gamma N \rightarrow M_{2} B^{*} \rightarrow M_{1} M_{2} B$} \label{AngularDistribution}
The previous section dealt with polarization observables in photoproduction reactions without considering the reaction mechanism. Since it is well known that many of the contributions to this reaction are quasi-two-body states, we will consider them in the rest of this thesis. As mentioned in section \ref{Kinematics}, the three possible channels are $\gamma N \rightarrow M^{*} B \rightarrow M_{1} M_{2} B$, $\gamma N \rightarrow M_{1} B^{*} \rightarrow M_{1} M_{2} B$, and $\gamma N \rightarrow M_{2} B^{*} \rightarrow M_{1} M_{2} B$, where $M_{1}$, $M_{2}$ and $M^{*}$ are mesons while $B$ and $B^{*}$ are baryons. The goal of this section is to find an expressions that relates the decay distribution and the polarization observables of the photoproduction reaction to the SDME's or statistical tensors of the decaying hadron ($M'$ or $B'$). For now, we will make the 
assumption that only one such quasi-two-body state contributes to the reaction. The examples we'll consider are the reactions $\gamma N \rightarrow  V N \rightarrow M_{1} M_{2} N$, where $V$ is a vector meson, $\gamma N \rightarrow  M_{1} B_{3/2} \rightarrow M_{1} M_{2} N$, where $B_{3/2}$  is a spin-3/2 baryon, and $\gamma N \rightarrow M_{2} B_{3/2} \rightarrow M_{1} M_{2} N$. This procedure is also laid out in a somewhat different mathematical language in refs. \cite{schilling_analysis_1970, gottfried_connection_1964}.

The $VB$ system has a massive spin-1 hadron (3 orthogonal spin states) and a spin-1/2 hadron (2 orthogonal states). Therefore, the state vector in the ensemble can be represented by a $6\times1$ column matrix,
\begin{align} \label{StateVectorVectorMeson}
\ket{\Psi^{i}(VB)}=\begin{pmatrix}
               \psi^{i}_{1\frac{1}{2}}(VB) \\ \psi^{i}_{1 \minus \frac{1}{2}}(VB) \\ \psi^{i}_{0 \frac{1}{2}}(VB) \\ \psi^{i}_{0 \minus \frac{1}{2}}(VB) \\ \psi^{i}_{\minus 1 \frac{1}{2}}(VB) \\ \psi^{i}_{\minus 1 \minus \frac{1}{2}}(VB)
            \end{pmatrix},
\end{align}
where $i$ labels a members of the statistical ensemble in the mixed state. Since the initial state shown in eqn. (\ref{InitialState}) is represented by a $4\times1$ column matrix, the transition matrix will be $6\times4$,
\begin{align} \label{TransitionMatrix2}
 \hat{M}(\theta; \gamma N \rightarrow VB) = \begin{pmatrix} 
            M_{1 \frac{1}{2}; 1 \frac{1}{2}}(\theta) & M_{1 \frac{1}{2}; 1 \minus \frac{1}{2}}(\theta)
	      & M_{1 \frac{1}{2}; \minus 1 \frac{1}{2}}(\theta) & M_{1 \frac{1}{2}; \minus 1 \minus \frac{1}{2}}(\theta) \\
	    M_{1 \minus \frac{1}{2}; 1 \frac{1}{2}}(\theta) & M_{1 \minus \frac{1}{2}; 1 \minus \frac{1}{2}}(\theta)
	      & M_{1 \minus \frac{1}{2}; \minus 1 \frac{1}{2}}(\theta) & M_{1 \minus \frac{1}{2}; \minus 1 \minus \frac{1}{2}}(\theta) \\
	      M_{0 \frac{1}{2}; 1 \frac{1}{2}}(\theta) & M_{0 \frac{1}{2}; 1 \minus \frac{1}{2}}(\theta)
	      & M_{0 \frac{1}{2}; \minus 1 \frac{1}{2}}(\theta) & M_{0 \frac{1}{2}; \minus 1 \minus \frac{1}{2}}(\theta) \\
	      M_{0 \minus \frac{1}{2}; 1 \frac{1}{2}}(\theta) & M_{0 \minus \frac{1}{2}; 1 \minus \frac{1}{2}}(\theta)
	      & M_{0 \minus \frac{1}{2}; \minus 1 \frac{1}{2}}(\theta) & M_{0 \minus \frac{1}{2}; \minus 1 \minus \frac{1}{2}}(\theta) \\ M_{\minus 1 \frac{1}{2}; 1 \frac{1}{2}}(\theta) & M_{\minus 1 \frac{1}{2}; 1 \minus \frac{1}{2}}(\theta)
	      & M_{\minus 1 \frac{1}{2}; \minus 1 \frac{1}{2}}(\theta) & M_{\minus 1 \frac{1}{2}; \minus 1 \minus \frac{1}{2}}(\theta) \\
	      M_{\minus 1 \minus \frac{1}{2}; 1 \frac{1}{2}}(\theta) & M_{\minus 1 \minus \frac{1}{2}; 1 \minus \frac{1}{2}}(\theta)
	      & M_{\minus 1 \minus \frac{1}{2}; \minus 1 \frac{1}{2}}(\theta) & M_{\minus 1 \minus \frac{1}{2}; \minus 1 \minus \frac{1}{2}}(\theta)
           \end{pmatrix}.
\end{align}
From the definition of the density matrix in eqn. (\ref{DensityMatrix}), the density matrix of the $VB$ system is $6\times6$. 

Just like we have done in eqn. (\ref{InitialToFinalDensityMatrix}), we can obtain the density matrix of the $VB$ system from that of the initial state by,
\begin{equation} \label{InitialToFinalDensityMatrixVB}
\hat{\rho}(VB)=\hat{M}\hat{\rho}(\gamma N)\hat{M}^{\dag},
\end{equation}
or, in terms of the individual matrix elements,
\begin{equation} \label{InitialToFinalDensityMatrixVBIndices}
\rho(VB)_{\lambda_{V} \lambda_{B}; \lambda_{V}' \lambda_{B}'}=\sum_{\substack{\lambda_{\gamma} \lambda_{N} \\ 
     \lambda_{\gamma}' \lambda_{N}'}}M_{\lambda_{V} \lambda_{B},  \lambda_{\gamma} \lambda_{N}}\rho_{\lambda_
{\gamma} \lambda_{N}; \lambda_{\gamma}' \lambda_{N}'}(\gamma N)M^{*}_{\lambda_{V}' \lambda_{B}'; \lambda_{\gamma}' \lambda_{N}'},
\end{equation}
where the $\lambda_{\gamma}$, $\lambda_{N}$, $\lambda_{V}$, and $\lambda_{B}$ are the helicities of the beam, target, vector meson, and recoil nucleon respectively. Next, the state will transition once again, this time from the $VB$ state to the $M_{1} M_{2} B$ state through the decay of the vector meson into $M_{1}$ and $M_{2}$. The transition matrix for the two body decay of an arbitrary particle at rest of spin $S$, $A \rightarrow B C$, is proportional to a Wigner $\mathcal{D}$-matrix,
\begin{equation} \label{WignerDecayMatrix}
   M_{\lambda_{B} \lambda_{C}; \lambda_{A}} \propto D^{S*}_{\lambda_{A} (\lambda_{B}-\lambda_{C})}. 
\end{equation}
Since the $B$ subsystem is a spectator during the decay of the vector meson, the transition matrix is therefore the tensor product of two transition matrices, one acting on the $V$ 
subsystem and the other on the $B$ subsystem,
\begin{equation} \label{TransitionMatrixTensorProduct}
\hat{M}(\theta^{*}, \phi^{*}; VB \rightarrow M_{1} M_{2} B) = \hat{D}^{1 \dagger}(\theta^{*}, \phi^{*}) \otimes \hat{I},
\end{equation} 
where $\hat{I}$ is the identity matrix, and $\hat{D}^{1\dagger}$ is the $1 \times 3$ matrix
\begin{equation} \label{WignerDMatrixVectorMesonDecay}
\begin{split}
\hat{D}^{1 \dagger} & = c \begin{pmatrix}
D^{1*}_{10}(\theta^{*},\phi^{*}) & D^{1*}_{00}(\theta^{*},\phi^{*}) & D^{1*}_{-1 0}(\theta^{*},\phi^{*})
\end{pmatrix} \\
& = c \begin{pmatrix}
-\frac{e^{-i\phi^{*}}\sin(\theta^{*})}{\sqrt{2}} & \cos(\theta^{*}) & 
\frac{e^{i\phi^{*}}\sin(\theta^{*})}{\sqrt{2}}
\end{pmatrix}.
\end{split}
\end{equation}
The ${D^{1*}_{\lambda_{V} 0}}$ are complex conjugates of the Wigner $\mathcal{D}$-functions, which are functions of the decay angles $\theta^{*}$ and $\phi^{*}$
defined in section \ref{Kinematics}, and $c$ is a proportionality constant that won't be relevant in these discussions, and so will be dropped for the rest of this thesis. The state of the $B$ subsystem is 
being multiplied by the identity matrix because the final nucleon does not transition into new particles 
(it does not decay). We can use the transition matrix shown in eqn. (\ref{TransitionMatrixTensorProduct}) to find the spin density matrix of the $M_{1}M_{2}B$ system,
\begin{equation} \label{InitialToFinalDensityMatrixVectorMeson}
\hat{\rho}(M_{1} M_{2} B)=(\hat{D}^{1\dagger} \otimes \hat{I})\hat{\rho}(VB)(\hat{D}^{1} \otimes \hat{I}),
\end{equation}
or, using index notation,
\begin{equation} \label{InitialToFinalDensityMatrixVectorMesonIndex}
    \rho_{\lambda_{B}\lambda_{B}'}(M_{1}M_{2}B)=\sum_{\lambda_{V}\lambda_{V}'}D^{1*}_{\lambda_V 0}(\theta^{*},\phi^{*})\rho_{\lambda_{V} \lambda_{B};\lambda_{V}'\lambda_{B}'}(BV) D^{1}_{\lambda_{V}' 0}(\theta^{*},\phi^{*}).
\end{equation}
We showed in eqn. (\ref{CrossSectionMixedState2}) that the trace of the spin density matrix is proportional to the cross section. We therefore take the trace of the previous equation,
\begin{equation} \label{TraceDensityMatrixFinalStateVectorMeson}
    \begin{split}
        \Tr[\hat{\rho}(M_{1}M_{2}B)] = & \sum_{\lambda_{B}}\rho_{\lambda_{B},\lambda_{B}}(M_{1}    M_{2}B) \\
        = & \sum_{\lambda_{V}\lambda_{V}'}D^{1*}_{\lambda_V 0}(\theta^{*},\phi^{*})
        \Big[\sum_{\lambda_{B}}\rho_{\lambda_{V}\lambda_{B};\lambda_{V}'\lambda_{B}}(VB)\Big] 
        D^{1}_{\lambda_{V}' 0}(\theta^{*},\phi^{*}).
    \end{split}
\end{equation}
The expression in square brackets in the second line of the previous equation is called a partial trace over the spin density matrix,
\begin{equation} \label{PartialTraceVectorMeson}
    \begin{split}
        \Big[\sum_{\lambda_{B}}\rho(VB)_{\lambda_{V}\lambda_{B};\lambda_{V}'\lambda_{B}}\Big] \equiv & \rho_{\lambda_{V}\lambda_{V}'}(V), \\
        \Tr_{B}[\hat{\rho}(VB)] \equiv & \hat{\rho}(V).
     \end{split}
\end{equation}
The subscript $B$ on the second line of the previous equation means that we are taking the trace only over the indices of the recoil baryon $B$. In terms of experiments, taking the trace over recoil baryon indices means that its spin is not being measured. This removes the spin information of the recoil baryon that the density matrix contains. What we end up with is therefore considered the spin density matrix of the vector meson $V$ subsystem. We can then consider only the process of the decay of the $V$ into $M_{1}$ and $M_{2}$. Equation (\ref{TraceDensityMatrixFinalStateVectorMeson}) therefore simplifies to
\begin{equation} \label{TraceDensityMatrixFinalStateVectorMesonSimplified}
    \begin{split}
        \Tr[\hat{\rho}(M_{1}M_{2})] = & \sum_{\lambda_{V}\lambda_{V}'}D^{1*}_{\lambda_V 0}(\theta^{*},\phi^{*})
        \rho_{\lambda_{V},\lambda_{V}'}(V)D^{1}_{\lambda_{V}' 0}(\theta^{*},\phi^{*}) \\
        \equiv & W(\theta^{*}, \phi^{*}; V),
    \end{split}
\end{equation}
(note that since $M_{1}$ and $M_{2}$ are spinless, its density matrix $\hat{\rho}(M_{1}M_{2})$ is just a real number. While in general a trace must be taken, in this example there is technically no need to take a trace). Since the trace is proportional to the cross section of a decay process, and since its angular dependence is known, we can also call it the angular distribution of the decay products $M_{1}$ and $M_{2}$, which we designate as $W(\theta^{*}, \phi^{*}; V)$. Using the explicit forms of the Wigner $\mathcal{D}$-functions, the decay rate is
\begin{equation} \label{WDistributionV} 
\begin{split}
W(\theta^{*},\phi^{*};V) & = \rho_{00}(V) \cos^{2}\theta^{*} + \frac{1}{2}(\rho_{11}(V)+\rho_{-1-1}(V)) \sin^{2}\theta^{*} \\
     \mt -\frac{1}{\sqrt{2}}\Re[\rho_{10}(V)-\rho_{0-1}(V)]\sin2\theta^{*}\cos\phi^{*} \\
     \mt +\frac{1}{\sqrt{2}}\Im[\rho_{10}(V)-\rho_{0-1}(V)]\sin2\theta^{*}\sin\phi^{*} \\
     \mt - \Re[\rho_{1-1}(V)]\sin^{2}\theta^{*} \cos2\phi^{*} + \Im[\rho_{1-1}(V)]\sin^{2}\theta^{*} \sin2\phi^{*},                      
 \end{split}
\end{equation}
where we have used the hermiticity condition of the density matrix in eqn. (\ref{Hermiticity}) to express $W$ only in terms of the independent elements. Note that, since the diagonal elements are purely real, the distribution is real, as expected. 

In the case of the $\gamma N \rightarrow M_{1} B^{*} \rightarrow M_{1} M_{2} B_{3/2}$, we can also find the distribution of the decay products using the same procedure. This time the $\gamma N$ systems in the ensemble transition into a $M_{1} B'$ state with spin state vector
\begin{equation} \label{StateVectorB'}
\ket{\Psi^{i}(M_{1} B_{3/2})}=\begin{pmatrix}
                 \psi^{i}_{\frac{3}{2}}(M_{1} B_{3/2}) \\ \psi^{i}_{\frac{1}{2}}(M_{1} B_{3/2}) \\ \psi^{i}_{-\frac{1}{2}}(M_{1} B_{3/2}) \\ \psi^{i}_{-\frac{3}{2}}(M_{1} B_{3/2})
                 \end{pmatrix}.
\end{equation}
Having four possible spin states, the spin transition matrix is therefore $4 \times 4$,
\begin{equation} \label{TransitionMatrix3}
 \hat{M}(\theta; \gamma N \rightarrow M_{1} B_{3/2}) = \begin{pmatrix} 
            M_{\frac{3}{2}; 1 \frac{1}{2}}(\theta) & M_{\frac{3}{2}; 1 \minus \frac{1}{2}}(\theta)
	      & M_{\frac{3}{2}; \minus 1 \frac{1}{2}}(\theta) & M_{\frac{3}{2}; \minus 1 \minus \frac{1}{2}}(\theta) \\
	    M_{\frac{1}{2}; 1 \frac{1}{2}}(\theta) & M_{\frac{1}{2}; 1 \minus \frac{1}{2}}(\theta)
	      & M_{\frac{1}{2}; \minus 1 \frac{1}{2}}(\theta) & M_{\frac{1}{2}; \minus 1 \minus \frac{1}{2}}(\theta) \\
	      M_{\minus \frac{1}{2}; 1 \frac{1}{2}}(\theta) & M_{\minus \frac{1}{2}; 1 \minus \frac{1}{2}}(\theta)
	      & M_{\minus \frac{1}{2}; \minus 1 \frac{1}{2}}(\theta) & M_{\minus \frac{1}{2}; \minus 1 \minus \frac{1}{2}}(\theta) \\
	      M_{\minus \frac{3}{2}; 1 \frac{1}{2}}(\theta) & M_{\minus \frac{3}{2}; 1 \minus \frac{1}{2}}(\theta)
	      & M_{\minus \frac{3}{2}; \minus 1 \frac{1}{2}}(\theta) & M_{\minus \frac{3}{2}; \minus 1 \minus \frac{1}{2}}(\theta)
           \end{pmatrix}.
\end{equation}
Just as in the example of a decaying vector meson, we can write the spin density matrix of the $M_{1} M_{2} B$ system in terms of the spin density matrix of the $M_{1} B_{3/2}$ system, 
\begin{equation} \label{InitialToFinalDensityMatrixThreeHalfsBaryon}
\hat{\rho}(M_{1} M_{2} B)=\hat{D}^{3/2\dagger} \hat{\rho}(M_{1} B_{3/2}) \hat{D}^{3/2}.
\end{equation}
Since $M_{1}$ is spinless and only a spectator in the decay of $B_{3/2}$, we can instead write it as
\begin{equation} \label{InitialToFinalDensityMatrixThreeHalfsBaryon2}
    \hat{\rho}(M_{2} B)=\hat{D}^{3/2\dagger} \hat{\rho}(B_{3/2}) \hat{D}^{3/2},
\end{equation}
where we call $\hat{\rho}(B_{3/2})$ the spin density matrix of the baryon subsystem $B_{3/2}$. Using index notation, it can be expressed as
\begin{equation} \label{InitialToFinalDensityMatrixThreeHalfsBaryonIndex}
    \rho_{\lambda_{B}\lambda_{B}'}(M_{2}B)=\sum_{\lambda_{B_{3/2}}\lambda_{B_{3/2}}'}D^{3/2*}_{\lambda_{B_{3/2}} \lambda_{B}}(\theta^{*},\phi^{*})\rho_{\lambda_{B_{3/2}} \lambda_{B_{3/2}}'}(B_{3/2}) D^{3/2}_{\lambda_{B_{3/2}}' \lambda_{B}'}(\theta^{*},\phi^{*}),
\end{equation}
where the transition matrix for the decay is now
\begin{equation} \label{eqn 34}
\hat{\cal{D}}^{3/2\dagger}=  c\begin{pmatrix}
                 D^{3/2*}_{\frac{3}{2}\frac{1}{2}}(\theta^{*},\phi^{*}) &
                 D^{3/2*}_{\frac{1}{2}\frac{1}{2}}(\theta^{*},\phi^{*}) &
                 D^{3/2*}_{-\frac{1}{2}\frac{1}{2}}(\theta^{*},\phi^{*}) &
                 D^{3/2*}_{-\frac{3}{2}\frac{1}{2}}(\theta^{*},\phi^{*}) \\
                 D^{3/2*}_{-\frac{3}{2}-\frac{1}{2}}(\theta^{*},\phi^{*}) &
                 D^{3/2*}_{-\frac{1}{2}-\frac{1}{2}}(\theta^{*},\phi^{*}) &
                 D^{3/2*}_{-\frac{1}{2}-\frac{1}{2}}(\theta^{*},\phi^{*}) &
                 D^{3/2*}_{-\frac{3}{2}-\frac{1}{2}}(\theta^{*},\phi^{*})
                 \end{pmatrix}, 
\end{equation}
where $c$ is a proportionality constant. This matrix in terms of the explicit expressions for the Wigner $\mathcal{D}$-functions
is too large to be displayed in this article. Just as in the example of the decay of the vector meson $V$, taking the trace of the matrix in eqn. (\ref{InitialToFinalDensityMatrixThreeHalfsBaryonIndex}) and using the explicit expressions for the Wigner $\mathcal{D}$-functions gives us the angular distribution of the decay products, $M_{2}$ and $B'$,
\begin{equation} \label{WDistributionBThreeHalfs}
\begin{split}
W(\theta^{*},\phi^{*};B_{3/2}) & = \frac{5}{8}(\rho_{11}(B_{3/2}) +\rho_{-1-1}(B_{3/2}))(1 + \frac{3}{5} \cos 2 \theta^{*}) \\
    \mt +\frac{3}{4}(\rho_{33}(B_{3/2})+\rho_{-3-3}(B_{3/2}))\sin^{2}\theta^{*} \\
    \mt -\frac{\sqrt{3}}{2}\Re[\rho_{31}(B_{3/2})-\rho_{-1-3}(B_{3/2})]\sin 2 \theta^{*} \cos \phi^{*} \\
    \mt +\frac{\sqrt{3}}{2}\Im[\rho_{31}(B_{3/2})-\rho_{-1-3}(B_{3/2})]\sin 2 \theta^{*} \sin \phi^{*} \\
    \mt -\frac{\sqrt{3}}{2}\Re[\rho_{3-1}(B_{3/2})+\rho_{1-3}(B_{3/2})]\sin^{2} \theta^{*} \cos 2 \phi^{*} \\
    \mt +\frac{\sqrt{3}}{2}\Im[\rho_{3-1}(B_{3/2})+\rho_{1-3}(B_{3/2})] \sin^{2} \theta^{*} \sin 2 \phi^{*}. \\
\end{split}
\end{equation}
For the case of the reaction $\gamma N \rightarrow M_{2} B_{3/2} \rightarrow M_{1}M_{2}B$, the same procedure just described applies. The only difference is that in this case the variables $\theta^{*}$ and $\phi^{*}$ describe different angles, because they describe the decay distribution of $M_{1}$ and $B$ instead of $M_{2}$ and $B$. 

As mentioned in section \ref{Kinematics}, the coordinate system used to define the angles $\theta^{*}$ and $\phi^{*}$ is different for each of the three pathways shown in eqns. (\ref{PathwayA}), (\ref{PathwayB}) and (\ref{PathwayC}). The $z'$-axis is always defined as pointing in the opposite direction as the three-momentum of the spectator hadron, and the $\theta^{*}$ and $\phi^{*}$ are the polar and azimuthal angle of one of the decaying hadrons with respect to this axis, as shown in fig. \ref{Fig: Kinematics3}. 

As we have seen in eqn. (\ref{TExpansion}) and (\ref{TDefinition1}), we can express spin density matrix elements in terms of statistical
tensors. In terms of these, the angular distribution for the decay of a $V$ meson is given by
\begin{equation} \label{WTDistributionRho}
\begin{split}
 W(\theta^{*},\phi^{*};V) & = \frac{1}{\sqrt{3}}t_{00}(V)
                      -\frac{1}{2\sqrt{6}}t_{20}(V)(1 + 3 \cos 2 \theta^{*}) \\
       \mt +\Re[t_{21}(V)] \sin 2 \theta^{*} \cos \phi^{*}+\Im[t_{21}(V)] \sin 2 \theta^{*} \sin \phi^{*} \\
       \mt -\Re[t_{22}(V)] \sin^{2}\theta^{*} \cos 2 \phi^{*} - \Im[t_{22}(V)] \sin^{2} \theta \sin 2\phi^{*},
\end{split}
\end{equation}
and the angular distribution for the decay of a $B_{3/2}$ baryon is given by
\begin{equation} \label{WTDistributionDelta}
\begin{split}
 W(\theta^{*},\phi^{*};B_{3/2}) & = \frac{1}{2} t_{00}(3/2)
                      -\frac{1}{8}t_{20}(3/2)(1 + 3 \cos 2 \theta^{*}) \\
       \mt + \frac{1}{2}\sqrt{\frac{3}{2}}\Re[t_{21}(3/2)] \sin 2 \theta^{*} \cos \phi^{*}
                      + \frac{1}{2} \sqrt{\frac{3}{2}}\Im[t_{21}(3/2)] \sin 2\theta^{*} \sin \phi^{*} \\
       \mt -\frac{1}{2}\sqrt{\frac{3}{2}}\Re[t_{22}(3/2)] \sin^{2}\theta^{*} \cos 2 \phi^{*}
                      -\frac{1}{2}\sqrt{\frac{3}{2}}\Im[t_{22}(3/2)] \sin^{2}\theta^{*} \sin 2 \phi^{*}.
\end{split}
\end{equation}
As can be seen from these last two equations, only statistical tensors of even rank appear in the expressions for the distribution (rank 1
tensors don't appear in the expression for the distribution of a $V$ meson, while rank 1 and 3 tensors don't appear in the expression
for the distribution of the $B_{3/2}$ baryon). As long as the decay process conserves parity, this property  holds true even when interference effects of different resonances are taken into account, as we will show in section \ref{GeneralExpression}.

Note how the functions of the angles multiplied by each of the tensors in the sum of the last two equations is a spherical harmonic,
\begin{equation} \label{SphericalHarmonics1}
    \begin{split}
        W(\theta^{*},\phi^{*};V) & = 2 \sqrt{\frac{\pi}{3}} t_{00}(V) Y_{00}(\theta^{*}, \phi^{*})
                      -2\sqrt{\frac{2\pi}{15}} t_{20}(V)Y_{20}(\theta^{*}, \phi^{*}) \\
       \mt -4\sqrt{\frac{2\pi}{15}}\Re[t_{21}(V)] \Re[Y_{21}(\theta^{*}, \phi^{*})]+4\sqrt{\frac{2\pi}{15}}\Im[t_{21}(V)]\Im[Y_{21}(\theta^{*}, \phi^{*})] \\
       \mt -4\sqrt{\frac{2\pi}{15}}\Re[t_{22}(V)] \Re[Y_{22}(\theta^{*}, \phi^{*})]+4\sqrt{\frac{2\pi}{15}}\Im[t_{22}(V)]\Im[Y_{22}(\theta^{*}, \phi^{*})],
    \end{split}
\end{equation}
and
\begin{align} \label{SphericalHarmonics2}
        W(\theta^{*},\phi^{*};B_{3/2}) & = \sqrt{\pi} t_{00}(3/2) Y_{00}(\theta^{*}, \phi^{*})
                      -\sqrt{\frac{\pi}{5}} t_{20}(3/2)Y_{20}(\theta^{*}, \phi^{*}) \nonumber \\
       \mt -2\sqrt{\frac{\pi}{5}}\Re[t_{21}(3/2)] \Re[Y_{21}(\theta^{*}, \phi^{*})]+2\sqrt{\frac{\pi}{5}}\Im[t_{21}(3/2)]\Im[Y_{21}(\theta^{*}, \phi^{*})] \\
       \mt -2\sqrt{\frac{\pi}{5}}\Re[t_{22}(3/2)] \Re[Y_{22}(\theta^{*}, \phi^{*})]+2\sqrt{\frac{\pi}{5}}\Im[t_{22}(3/2)]\Im[Y_{22}(\theta^{*}, \phi^{*})].\nonumber
\end{align}
These last two equations show another advantage of using the statistical tensors instead of the SDME's: the angular functions multiplied by the tensors of given $L$ and $M$ are proportional to the well known spherical harmonics with same $L$ and $M$. By contrast, in the expressions in terms of the SDME's in eqns. (\ref{WDistributionV}) and (\ref{WDistributionBThreeHalfs}), the quantities multiplied by the angular functions are linear combinations of SDME's, with no obvious connection between them and the angular function that multiply them. In section \ref{GeneralExpression} we will show that this is in general true for any quasi-two-body state in which the decay is parity conserving.

\chapter{Polarization Observables in Terms of SDME's and Statistical Tensors of the Quasi-Two-Body States} \label{ObservablesQuasiTwoBodyState}
Given the assumption that the photoproduction process goes through a quasi-two-body state, the polarization observables can be expressed
in terms of the SDME's or statistical tensors of the resonances that decay. We will first examine this for for pathway A, using the example of a decaying vector meson. Afterwards, we will examine this for pathways B and C for the example of a decaying spin-3/2 baryon.

We have shown in equation (\ref{16Matrices}) how the spin density matrix of the final $M_{1} M_{2} B$ system can be decomposed into a sum of 16 matrices. But this can be done for any spin density matrix, including the one for the vector meson V,
\begin{equation} \label{VMatrixDecomposition}
 \hat{\rho}(V)=\rho^{0} + \sum_{i} \Lambda^{i (\gamma)}
	\hat{\rho}^{i (\gamma)}(V) +\sum_{i}\Lambda^{i (N)} \hat{\rho}^{i (N)}(V) + \sum_{i,j} \Lambda^{i (\gamma)} \Lambda^{j (N)} \hat{\rho}^{ij (\gamma N)}(V),
\end{equation}
where the SDME's of the 16 matrices are
\begin{equation} \label{DensityMatrixSigmaMatrixElementsV}
  \begin{split}
    \rho^{0}_{\lambda_{V} \lambda'_{V} }(V)
      & =  \frac{1}{4}\sum_{\substack{\lambda_{\gamma}\lambda_{N}\\ \lambda_{B}}} M_{\lambda_{V} \lambda_{B}; \lambda_{\gamma} \lambda_{N}} M^{*}_{\lambda'_{V} \lambda_{B}; \lambda_{\gamma} \lambda_{N}}, \\ 
    \rho^{i (\gamma)}_{\lambda_{V} \lambda'_{V} }(V)
      & =\frac{1}{4} \sum_{\substack{\lambda_{\gamma} \lambda'_{\gamma}\\ \lambda_{N} \lambda_{B}}} 
                        M_{\lambda_{V} \lambda_{B}; \lambda_{\gamma} \lambda_{N}}
                        \sigma^{i}_{\lambda_{\gamma} \lambda'_{\gamma}}
                        M^{*}_{\lambda'_{V} \lambda_{B}; \lambda'_{\gamma} \lambda_{N}}, \\
	\rho^{i (N)}_{\lambda_{V} \lambda'_{V} }(V)
      & =  \frac{1}{4}\sum_{\substack{\lambda_{N} \lambda'_{N} \\
                        \lambda_{\gamma}\lambda_{B}}} 
                        M_{\lambda_{V} \lambda_{B}; \lambda_{\gamma} \lambda_{N}}
                        \sigma^{i}_{\lambda_{N} \lambda'_{N}}
                        M^{*}_{\lambda'_{V} \lambda_{B}; \lambda_{\gamma} \lambda'_{N}}, \\
    \rho^{ij (\gamma N)}_{\lambda_{V}\lambda'_{V} }(V)
      & =\frac{1}{4} \sum_{\substack{\lambda_{\gamma} \lambda'_{\gamma}  				\\ \lambda_{N} \lambda'_{N} \\ \lambda_{B}}} 
                        M_{\lambda_{V} \lambda_{B}; \lambda_{\gamma} \lambda_{N}}\sigma^{i}_{\lambda_{\gamma}\lambda'_{\gamma}}
                        \sigma^{j}_{\lambda_{N} \lambda'_{N}}
                        M^{*}_{\lambda'_{V} \lambda_{B}; \lambda'_{\gamma} \lambda'_{N}}.
  \end{split}
\end{equation}
Note how the $\lambda_{B}$ indices on the right-hand sides of the equations are summed over, since the partial trace was taken.
We can use the explicit forms of the Pauli matrices to simplify the expression further. If we label their elements with the integers $-1$ and $1$ for the first and second row/column index, respectively, we can write them in terms of the Kronecker delta,
\begin{align} \label{PauliMatrixKroneckerDelta}
        \sigma^{x}_{m, m'} & = \delta_{m, -m'}, \nonumber \\
        \sigma^{y}_{m, m'} & = -i m  \delta_{m, -m'}, \\
        \sigma^{z}_{m, m'} & = -m \delta_{m, m'}, \nonumber \\
        & \nonumber \\
        m = \lambda_{\gamma} \quad & \text{or} \quad m = 2 \lambda_{N}, \nonumber
\end{align}
where $i$ is the imaginary number. Examples of some of these matrices when the previous equation is applied are
\begin{equation} \label{DensityMatrixSigmaMatrixElementsExampleV}
\begin{split}
 & \rho^{x (\gamma)}_{\lambda_{V} \lambda'_{V}}(V)
     =  \frac{1}{4}\sum_{\substack{\lambda_{\gamma} \lambda_{N}\\ \lambda_{B}}} 
                        M_{\lambda_{V} \lambda_{B}; \lambda_{\gamma} \lambda_{N}}
                        M^{*}_{\lambda'_{V} \lambda_{B}; -\lambda_{\gamma} \lambda_{N}}, \\
 & \rho^{y (\gamma)}_{\lambda_{V} \lambda'_{V}}(V)
     =-\frac{i}{4} \sum_{\substack{\lambda_{\gamma} \lambda_{N}\\ \lambda_{B}}} 
                        \lambda_{\gamma} M_{\lambda_{V} \lambda_{B}; \lambda_{N} \lambda_{\gamma}}
                        M^{*}_{\lambda'_{V} \lambda_{B}; -\lambda_{\gamma} \lambda_{N}}, \\
 & \rho^{z (\gamma)}_{\lambda_{V} \lambda'_{V}}(V)
     = \frac{1}{4} \sum_{\substack{\lambda_{\gamma} \lambda_{N}\\ \lambda_{B}}} 
                        \lambda_{\gamma} M_{\lambda_{V} \lambda_{B}; \lambda_{\gamma} \lambda_{N}}
                        M^{*}_{\lambda'_{V} \lambda_{B}; \lambda_{\gamma} \lambda_{N}}.
\end{split}
\end{equation}

By substituting eqn. (\ref{VMatrixDecomposition}) into eqn. (\ref{TraceDensityMatrixFinalStateVectorMesonSimplified}), we see that the distribution $W$ can also be decomposed into 16
terms,
\begin{equation} \label{WDecomposition}
 \begin{split}
     W(\theta^{*},\phi^{*};V) & = W^{0}(\theta^{*},\phi^{*};V) + \sum_{i} \Lambda^{i (\gamma)} W^{i (\gamma)}(\theta^{*},\phi^{*};V) 
          + \sum_{i} \Lambda^{i (N)} W^{i (N)}(\theta^{*},\phi^{*};V) \\ 
          \mt+ \sum_{i,j} \Lambda^{i (\gamma)} \Lambda^{j (N)} W^{ij (\gamma N)}(\theta^{*},\phi^{*};V).
 \end{split}
\end{equation}
This last equation is valid not just for vector mesons, but for any kind of decaying particle. The $W^{i}$'s are given by
\begin{equation} \label{WisRho}
 \begin{split}
 W^{0}(\theta^{*},\phi^{*};V) & = \sum_{\lambda_{V}\lambda_{V}'}
                      D^{1 *}_{0\lambda_{V}}(\theta^{*},\phi^{*})
                      \rho^{0}_{\lambda_{V},\lambda_{V}'}(V)
                      D^{1}_{0\lambda_{V}'}(\theta^{*},\phi^{*}), \\
 W^{i (\gamma)}(\theta^{*},\phi^{*};V) & = \sum_{\lambda_{V}\lambda_{V}'}
                      D^{1 *}_{0\lambda_{V}}(\theta^{*},\phi^{*})
                      \rho^{i (\gamma)}_{\lambda_{V},\lambda_{V}'}(V)
                      D^{1}_{0\lambda_{V}'}(\theta^{*},\phi^{*}), \\
 W^{i (N)}(\theta^{*},\phi^{*};V) & = \sum_{\lambda_{V}\lambda_{V}'}
                      D^{1 *}_{0\lambda_{V}}(\theta^{*},\phi^{*})
                      \rho^{i (N)}_{\lambda_{V},\lambda_{V}'}(V)
                      D^{1}_{0\lambda_{V}'}(\theta^{*},\phi^{*}), \\
W^{ij (\gamma N)}(\theta^{*},\phi^{*};V) & = \sum_{\lambda_{V}\lambda_{V}'}
                      D^{1 *}_{0\lambda_{V}}(\theta^{*},\phi^{*})
                      \rho^{ij (\gamma N)}_{\lambda_{V},\lambda_{V}'}(V)
                      D^{1}_{0\lambda_{V}'}(\theta^{*},\phi^{*}). \\                      
 \end{split}
\end{equation}
Each of these $W^{i}$'s will have exactly the same mathematical form as $W$ in eqn. (\ref{WDistributionV}) after the substitution 
$\rho_{\lambda_{V} \lambda'_{V}} \rightarrow \rho^{i}_{\lambda_{V} \lambda'_{V}}$. For example, $W^{y(\gamma)}$ takes the form
\begin{equation} \label{W2DistributionV}
\begin{split} 
 W^{y(\gamma)}(\theta^{*},\phi^{*};V) & = \rho^{y(\gamma)}_{00}(V) \cos^{2}\theta^{*} + \frac{1}{2}(\rho^{y(\gamma)}_{11}(V)+\rho^{y(\gamma)}_{-1-1}(V)) \sin^{2}\theta^{*} \\
     \mt -\frac{1}{\sqrt{2}}\Re[\rho^{y(\gamma)}_{10}(V)-\rho^{y(\gamma)}_{0-1}(V)]\sin2\theta^{*}\cos\phi^{*}\\
     \mt + \frac{1}{\sqrt{2}}\Im[\rho^{y(\gamma)}_{10}(V)-\rho^{y(\gamma)}_{0-1}(V)]\sin2\theta^{*}\sin\phi^{*} \\
     \mt - \Re[\rho^{y(\gamma)}_{1-1}(V)]\sin^{2}\theta^{*} \cos2\phi^{*} + \Im[\rho^{y(\gamma)(V)}_{1-1}(V)]\sin^{2}\theta^{*} \sin2\phi^{*}.                     
 \end{split}
\end{equation}
In eqn. (\ref{Tensor16Tensors}) we showed how statistical tensors can also be decomposed into other 16 statistical tensors. The equations that relates the 16 SDME's to the 16 tensors were shown in eqns. (\ref{MatrixToTensor}) and (\ref{16InverseRelation}). To find the $W^{i}$'s in terms of these tensors, we simply substitute eqn. (\ref{16InverseRelation}) into the definition of the $W^{i}$'s in eqn. (\ref{WisRho}) to get, for example,
\begin{equation} \label{W2TDistributionV}
\begin{split}
 W^{y(\gamma)}(\theta^{*},\phi^{*};V) & = \frac{1}{\sqrt{3}}t^{y(\gamma)}_{00}(V) -\frac{1}{2\sqrt{6}}t^{y(\gamma)}_{20}(V)(1 + 3 \cos 2 \theta^{*}) \\
       \mt +\Re[t^{y(\gamma)}_{21}(V)] \sin 2 \theta^{*} \cos \phi^{*}-\Im[t^{y(\gamma)}_{21}(V)] \sin 2 \theta^{*} \sin \phi^{*} \\
       \mt -\Re[t^{y(\gamma)}_{22}(V)] \sin^{2}\theta^{*} \cos 2 \phi^{*} + \Im[t^{y(\gamma)}_{22}(V)] \sin^{2} \theta \sin 2\phi^{*}.
\end{split}
\end{equation}
Note how it has the same mathematical form as the the expression for $W$ in terms of the tensors shown in eqn. (\ref{WTDistributionRho}) except that now the tensors have the appropriate superscript. 

To find how these 16 $W^{i}$'s are related to the polarization observables, we use eqn. (\ref{DensityMatrixObservables}) and take the trace on both sides. The terms proportional to the Pauli matrices, $\sigma^{i}$, on the right-hand side will vanish. The left-hand side is the definition of the distribution $W$,
\begin{equation} \label{WDefinition}
    W \equiv \Tr[\hat{\rho}(M_{1}M_{2}B)].
\end{equation}
After taking said trace and comparing it with eqn. (\ref{WDecomposition}), we get
\begin{align} \label{WObservables}
  W^{0}(\theta^{*}, \phi^{*};V) & = I_{0}(\theta^{*}, \phi^{*}),\nonumber \\
  W^{x (\gamma)}(\theta^{*}, \phi^{*};V) & = I_{0} \cos2\beta I^{c}(\theta^{*}, \phi^{*}),\nonumber \\
  W^{y (\gamma)}(\theta^{*}, \phi^{*};V) & = I_{0} \sin2\beta I^{s}(\theta^{*}, \phi^{*}), \\
  W^{z (\gamma)}(\theta^{*}, \phi^{*};V) & = I_{0} I^{\odot}(\theta^{*}, \phi^{*}),\nonumber \\
  W^{i (N)}(\theta^{*}, \phi^{*};V) & = I_{0} P_{i}(\theta^{*}, \phi^{*}),\nonumber \\
  W^{ix (\gamma N)}(\theta^{*}, \phi^{*};V) & = I_{0} \cos2\beta P_{i}^{c}(\theta^{*}, \phi^{*}),\nonumber \\
  W^{iy (\gamma N)}(\theta^{*}, \phi^{*};V) & = I_{0} \sin2\beta P_{i}^{s}(\theta^{*}, \phi^{*}),\nonumber \\
  W^{iz (\gamma N)}(\theta^{*}, \phi^{*};V) & = I_{0} P_{i}^{\odot}(\theta^{*}, \phi^{*}). \nonumber
\end{align}
This last equation is not only valid for the decay of vector mesons, but also for any type of decaying particle. In deriving these expressions, a sum over the spin indices of the recoil baryon baryon was done. This represents the fact that the polarization of the recoiling particle is not measured. As such, the $W^{i}$'s are related to the observables that do not involve the recoil baryon. 

For the cases of pathways B and C with a decaying spin-3/2 baryon, $\gamma N \rightarrow M_{1} B_{3/2} \rightarrow M_{1}M_{1}B$, and $\gamma N \rightarrow M_{2} B_{3/2} \rightarrow M_{1}M_{1}B$, we follow the same procedure and decompose the decaying baryon's spin density matrix into 16 matrices,
\begin{equation} \label{BMatrixDecomposition}
\begin{split}
 \hat{\rho}(B_{3/2}) & =\rho^{0} + \sum_{i} \Lambda^{i (\gamma)}
	\hat{\rho}^{i (\gamma)}(B_{3/2}) \\
	\mt + \sum_{i}\Lambda^{i (N)} \hat{\rho}^{i (N)}(B_{3/2}) + \sum_{i,j} \Lambda^{i (\gamma)} \Lambda^{j (N)} \hat{\rho}^{ij (\gamma N)}(B_{3/2}),
\end{split}
\end{equation}
where
\begin{align} \label{DensityMatrixSigmaMatrixElementsExampleB'}
 & \rho^{0}_{\lambda_{B_{3/2}} \lambda'_{B_{3/2}}}(B_{3/2})
     = \frac{1}{4}\sum_{\lambda_{\gamma}\lambda_{N}} 
                        M_{0\lambda_{B_{3/2}};\lambda_{\gamma} \lambda_{N}}
                        M^{*}_{0 \lambda'_{B_{3/2}};\lambda_{\gamma} \lambda_{N}}, \nonumber \\
 & \rho^{i (\gamma)}_{\lambda_{B_{3/2}} \lambda'_{B_{3/2}}}(B_{3/2})
     =\frac{1}{4} \sum_{\substack{\lambda_{\gamma} \lambda'_{\gamma}
                         \\ \lambda_{N}}} 
                        M_{0\lambda_{B_{3/2}}; \lambda_{\gamma} \lambda_{N}}
                        \sigma^{i}_{\lambda_{\gamma} \lambda'_{\gamma}}
                        M^{*}_{0\lambda'_{B_{3/2}}; \lambda'_{\gamma} \lambda_{N}}, \\
 & \rho^{i (N)}_{\lambda_{B_{3/2}} \lambda'_{B_{3/2}}}(B_{3/2})
     =  \frac{1}{4}\sum_{\substack{
                        \lambda_{\gamma} \\
                        \lambda_{N} \lambda'_{N}}} 
                        M_{0\lambda_{B_{3/2}}; \lambda_{\gamma}\lambda_{N}}
                        \sigma^{i}_{\lambda_{N} \lambda_{N}'}
                        M^{*}_{0 \lambda'_{B_{3/2}}; \lambda_{\gamma} \lambda'_{N}}, \nonumber \\
 & \rho^{ij (\gamma N)}_{\lambda_{B_{3/2}} \lambda'_{B_{3/2}}}(B_{3/2})
     =\frac{1}{4} \sum_{\substack{\lambda_{\gamma} \lambda'_{\gamma} \\
                        \lambda_{N} \lambda'_{N}}} 
                        M_{0 \lambda_{B_{3/2}};  \lambda_{\gamma}\lambda_{N}}\sigma^{i}_{\lambda_{\gamma}\lambda'_{\gamma}}
                        \sigma^{j}_{\lambda_{N} \lambda'_{N}}
                        M^{*}_{0 \lambda'_{B_{3/2}}; \lambda'_{\gamma}  \lambda'_{N}}, \nonumber
\end{align}
(compare with equation (\ref{DensityMatrixSigmaMatrixElementsV}) for the vector meson V). The angular distribution of its decay products can again be decomposed into 16 terms,
\begin{align} \label{DeltaWDecomposition}
     & W(\theta^{*},\phi^{*};B_{3/2}) = W^{0}(\theta^{*},\phi^{*};B_{3/2}) + \sum_{i} \Lambda^{i (\gamma)} W^{i (\gamma)}(\theta^{*},\phi^{*};B_{3/2}) \\ 
         & \qquad  + \sum_{i} \Lambda^{i (N)} W^{i (N)}(\theta^{*},\phi^{*};B_{3/2}) + \sum_{i,j} \Lambda^{i (\gamma)} \Lambda^{j (N)}W^{ij (N \gamma)}(\theta^{*},\phi^{*};B_{3/2}). \nonumber
\end{align}
where
\begin{align} \label{WisDelta}
 W^{0}(\theta^{*},\phi^{*};B_{3/2}) & = \sum_{\substack{\lambda_{B_{3/2}}\lambda_{B_{3/2}}' \\ \lambda_{B}}}
                      D^{3/2 *}_{\lambda_{B_{3/2}}\lambda_{B}}(\theta^{*},\phi^{*})
                      \rho^{0}_{\lambda_{B_{3/2}},\lambda_{B_{3/2}}'}(B_{3/2})
                      D^{3/2}_{\lambda_{B_{3/2}}'\lambda_{B}}(\theta^{*},\phi^{*}), \nonumber \\
 W^{i (\gamma)}(\theta^{*},\phi^{*};B_{3/2}) & = \sum_{\substack{\lambda_{B_{3/2}}\lambda_{B_{3/2}}' \\ \lambda_{B}}}
                      D^{3/2 *}_{\lambda_{B_{3/2}}\lambda_{B}}(\theta^{*},\phi^{*})
                      \rho^{i (\gamma)}_{\lambda_{B_{3/2}},\lambda_{B_{3/2}}'}(B_{3/2})
                      D^{3/2}_{\lambda_{B_{3/2}}'\lambda_{B}}(\theta^{*},\phi^{*}), \\
 W^{i (N)}(\theta^{*},\phi^{*};B_{3/2}) & = \sum_{\substack{\lambda_{B_{3/2}}\lambda_{B_{3/2}}' \\ \lambda_{B}}}
                      D^{3/2 *}_{\lambda_{B_{3/2}}\lambda_{B}}(\theta^{*},\phi^{*})
                      \rho^{i (N)}_{\lambda_{B_{3/2}},\lambda_{B_{3/2}}'}(B_{3/2})
                      D^{3/2}_{\lambda_{B_{3/2}}'\lambda_{B}}(\theta^{*},\phi^{*}), \nonumber \\
W^{ij (\gamma N)}(\theta^{*},\phi^{*};B_{3/2}) & = \sum_{\substack{\lambda_{B_{3/2}}\lambda_{B_{3/2}}' \\ \lambda_{B}}}
                      D^{3/2 *}_{\lambda_{B_{3/2}}\lambda_{B}}(\theta^{*},\phi^{*})
                      \rho^{ij (\gamma N)}_{\lambda_{B_{3/2}},\lambda_{B_{3/2}}'}(B_{3/2})
                      D^{3/2}_{\lambda_{B_{3/2}}'\lambda_{B}}(\theta^{*},\phi^{*}) \nonumber
\end{align}
(note the sum over the index $\lambda_{B}$, since we are taking a trace). Each of these $W^{i}$'s will have exactly the same mathematical form as $W$ in eqn. (\ref{WDistributionBThreeHalfs}) after the substitution $\rho_{\lambda_{B_{3/2}}, \lambda_{B_{3/2}}'}(B_{3/2}) \rightarrow \rho^{i}_{\lambda_{B_{3/2}}, \lambda'_{B_{3/2}}}(B_{3/2})$. For example, $W^{y (\gamma)}$ is given by
\begin{align} \label{W2DistributionB'}
W^{y (\gamma)}(\theta^{*},\phi^{*};B_{3/2}) & = \frac{5}{8}(\rho^{y (\gamma)}_{11}(B_{3/2}) 
                   +\rho^{y (\gamma)}_{-1-1}(B_{3/2}))(1 + \frac{3}{5} \cos 2 \theta^{*})\nonumber \\
    \mt +\frac{3}{4}(\rho^{y (\gamma)}_{33}(B_{3/2})+\rho^{y (\gamma)}_{-3-3}(B_{3/2}))\sin^{2}\theta^{*} \\
    \mt -\frac{\sqrt{3}}{2}\Re[\rho^{y (\gamma)}_{31}(B_{3/2})
                        -\rho^{y (\gamma)}_{-1-3}(B_{3/2})]\sin 2 \theta^{*} \cos \phi^{*}\nonumber \\
    \mt +\frac{\sqrt{3}}{2}\Im[\rho^{y (\gamma)}_{31}(B_{3/2})
                        -\rho^{y (\gamma)}_{-1-3}(B_{3/2})]\sin 2 \theta^{*} \sin \phi^{*}\nonumber \\
    \mt -\frac{\sqrt{3}}{2}\Re[\rho^{y (\gamma)}_{3-1}(B_{3/2})
                        +\rho^{y (\gamma)}_{1-3}(B_{3/2})]\sin^{2} \theta^{*} \cos 2 \phi^{*}\nonumber \\
    \mt +\frac{\sqrt{3}}{2}\Im[\rho^{y (\gamma)}_{3-1}(B_{3/2})
                        +\rho^{y (\gamma)}_{1-3}(B_{3/2})] \sin^{2} \theta^{*} \sin 2 \phi^{*}.\nonumber
\end{align}
In terms of statistical tensors, this distribution is
\begin{align} \label{W2TDistributionDelta}
 W^{y (\gamma)}(\theta^{*},\phi^{*};B_{3/2}) & = t^{y (\gamma)}_{00}(B_{3/2})
                      -\frac{1}{4}t^{y (\gamma)}_{20}(B_{3/2})(1 + 3 \cos 2 \theta^{*})\nonumber \\
       \mt +\sqrt{\frac{3}{2}}\Re[t^{y (\gamma)}_{21}(B_{3/2})] \sin 2 \theta^{*} \cos \phi^{*}\nonumber \\
       \mt -\sqrt{\frac{3}{2}}\Im[t^{y (\gamma)}_{21}(B_{3/2})] \sin 2\theta^{*} \sin \phi^{*} \\
       \mt -\frac{1}{2}\sqrt{\frac{3}{2}}\Re[t^{y (\gamma)}_{22}(B_{3/2})] \sin^{2}\theta^{*} \cos 2 \phi^{*}\nonumber \\
		\mt +\frac{1}{2}\sqrt{\frac{3}{2}}\Im[t^{y (\gamma)}_{22}(B_{3/2})] \sin^{2}\theta^{*} \sin 2 \phi^{*}. \nonumber
\end{align}
The polarization observables for this case are again given by eqn. (\ref{WObservables}),
\begin{align} \label{B'WObservables}
  W^{0}(\theta^{*}, \phi^{*};B_{3/2}) & = I_{0}(\theta^{*}, \phi^{*}), \nonumber \\
  W^{i (N)}(\theta^{*}, \phi^{*}; B_{3/2}) & = I_{0} P^{i}(\theta^{*}, \phi^{*}),\nonumber \\
  W^{x (\gamma)}(\theta^{*}, \phi^{*}; B_{3/2}) & = I_{0} \cos2\beta I^{c}(\theta^{*}, \phi^{*}),\nonumber \\
  W^{y (\gamma)}(\theta^{*}, \phi^{*}; B_{3/2}) & = I_{0} \sin2\beta I^{s}(\theta^{*}, \phi^{*}), \\
  W^{z (\gamma)}(\theta^{*}, \phi^{*}; B_{3/2}) & = I_{0} I^{\odot}(\theta^{*}, \phi^{*}),\nonumber \\
  W^{ix (\gamma N)}(\theta^{*}, \phi^{*}; B_{3/2}) & = I_{0} \cos2\beta P^{ic}(\theta^{*}, \phi^{*}),\nonumber \\
  W^{iy (\gamma N)}(\theta^{*}, \phi^{*}; B_{3/2}) & = I_{0} \sin2\beta P^{is}(\theta^{*}, \phi^{*}),\nonumber \\
  W^{iz (\gamma N)}(\theta^{*}, \phi^{*}; B_{3/2}) & = I_{0} P^{i\odot}(\theta^{*}, \phi^{*}). \nonumber
\end{align}

Note that while the dependence on the scattering angle $\theta$ is contained inside of the SDME's or statistical tensors of the decaying hadron, their $\theta^{*}$ and $\phi^{*}$ dependence is not. Rather, they are fully contained in the trigonometric functions in
eqns. (\ref{W2DistributionV}), (\ref{W2DistributionB'}), (\ref{WObservables}) and (\ref{B'WObservables}). For the expressions involving the SDME's, these trigonometric functions are made up of two multiplied elements of Wigner $\mathcal{D}$-functions. For those involving the statistical tensors, the trigonometric functions are spherical harmonics. Since the dependence of the distribution and observables on these angles is known, the expressions derived here can be used to apply fits to the data with the SDME's or statistical tensors as fit parameters. Note, however, that so far we have made the assumption that only one quasi-two-body state process contributes to the process. We will later consider the case of multiple channels contributing.

\chapter{Parity Invariance Considerations} \label{ParityInvariance}
Parity is conserved in electromagnetic and strong interactions. Therefore, the transition amplitudes in the reactions we 
are discussing will be invariant under parity transformations (up to a phase factor). 

For the case of two body scattering such as $\pi N \rightarrow MB$ (where $M$ is a pseudoscalar meson and $B$ is a baryon), parity conservation leads to
\begin{equation} \label{PInvariance1}
M_{-\lambda_{B};-\lambda_{N}}(\theta)=(-1)^{\lambda_{N}-\lambda_{B}}M_{\lambda_{B};\lambda_{N}}(\theta),
\end{equation}
and reduces the number of independent transition amplitudes from 4 to 2. 

For a three body scattering process such as $\pi N \rightarrow M_{1}M_{2}B$ (where both $M_{1}$ and $M_{2}$ are pseudoscalar mesons) the relationship is
\begin{equation} \label{PInvariance2}
M_{-\lambda_{B};-\lambda_{N}}(\theta;\theta^{*},\phi^{*})=(-1)^{\lambda_{N}-\lambda_{B}}M_{\lambda_{B};\lambda_{N}}
(\theta;\theta^{*},2\pi-\phi^{*}).
\end{equation}
This relates two amplitudes at different kinematic points. Therefore, for the case of a 3-body final state, parity conservation
cannot be used to reduce the number of independent amplitudes at a single kinematic point.
For $\gamma N \rightarrow M_{1}M_{2}B$ we have
\begin{equation} \label{PInvariance3}
M_{-\lambda_{B};-\lambda_{\gamma}-\lambda_{N}}(\theta;\theta^{*},\phi^{*})=(-1)^{\lambda_{\gamma}-\lambda_{N}+\lambda_{B}}
M_{\lambda_{B};\lambda_{\gamma}\lambda_{N}}
(\theta;\theta^{*},2\pi-\phi^{*}).
\end{equation}
For the general case of two-body scattering with arbitrary spin, $A \ B \rightarrow a \ b$, the relations are
\begin{equation} \label{TransitionMatrixParityRelations}
 M_{-\lambda_{a} -\lambda_{b}; -\lambda_{A} -\lambda_{B}}(\theta) = \frac{\eta_{a} \eta_{b}}{\eta_{A} \eta_{B}}
 (-1)^{(\lambda_{b} - \lambda_{a}) - (\lambda_{B} - \lambda_{A})} M_{\lambda_{a} \lambda_{b}; \lambda_{A} \lambda_{B}}(\theta),
\end{equation}
where the $\lambda$'s and $\eta$'s are the helicities and intrinsic parities of the particles, respectively. 

Since we are assuming in this thesis that the photoproduction reaction occurs through a quasi-two-body state, we can apply this last
equation to the reactions $\gamma N \rightarrow VB$, $\gamma N \rightarrow M_{1} B_{3/2}$ and $\gamma N \rightarrow M_{2} B_{3/2}$,
\begin{equation}
 \begin{split}
 M_{-\lambda_{V} -\lambda_{B};  -\lambda_{\gamma} -\lambda_{N}}(\theta) & = 
 (-1)^{(\lambda_{V} - \lambda_{B}) - (\lambda_{\gamma} - \lambda_{N})} 
 M_{\lambda_{V}\lambda_{B};  \lambda_{\gamma} \lambda_{N}}(\theta), \\
 M_{ 0 -\lambda_{B_{3/2}};  -\lambda_{\gamma} -\lambda_{N}}(\theta) & = 
 (-1)^{- \lambda_{B_{3/2}} - (\lambda_{\gamma} - \lambda_{N})} 
 M_{0 \lambda_{B_{3/2}},  \lambda_{\gamma} \lambda_{N}}(\theta).
 \end{split}
\end{equation}

For the $V$ meson, these relations can be used with eqn. (\ref{DensityMatrixSigmaMatrixElementsExampleV}) in order to give

\begin{equation} \label{DensityMatrixParityV}
 \begin{split}
  & \rho^{0}_{-\lambda_{V}, -\lambda_{V}'}(V) = (-1)^{\lambda_{V} + \lambda_{V}'} \rho^{0}_{\lambda_{V}, \lambda_{V}'}(V), \\
  & \rho^{i (\gamma)}_{-\lambda_{V}, -\lambda_{V}'}(V) = \zeta^{(\gamma) i}(V) (-1)^{\lambda_{V} + \lambda_{V}'}
       \rho^{i (\gamma)}_{\lambda_{V}, \lambda_{V}'}(V), \\
  & \rho^{i (N)}_{-\lambda_{V}, -\lambda_{V}'}(V) = \zeta^{(N) i}(V) (-1)^{\lambda_{V} + \lambda_{V}'} 
       \rho^{i (N)}_{\lambda_{V}, \lambda_{V}'}(V), \\
  & \rho^{ij (\gamma N)}_{-\lambda_{V}, -\lambda_{V}'}(V) = \zeta^{(\gamma) i}(V) \zeta^{(N) j}(V) 
       (-1)^{\lambda_{V} + \lambda_{V}'} \rho^{ij (\gamma N)}_{\lambda_{V}, \lambda_{V}'}(V), \\
  & \zeta^{(\gamma) i}(V) = \begin{cases}
                       \hphantom{-} 1 & i = y, z, \\
                      -1 & i = x,
                     \end{cases} \\
  & \zeta^{(N) i}(V) = \begin{cases}
                       \hphantom{-} 1 & i = x, \\
                      -1 & i = y, z. 
                     \end{cases}                   
 \end{split}
\end{equation}
The values for the $\zeta^{i (\gamma)}(V)$'s and $\zeta^{i (N)}(V)$'s are due to the values of the intrinsic parities of the particles
involved in the interaction and from the fact that $(-1)^{2 \lambda_{\gamma}} = 1$, $(-1)^{2 \lambda_{N}} = -1$, and $(-1)^{2 \lambda_{B}} = -1$
for all $\lambda_{\gamma}$ (integer), $\lambda_{N}$ (half-integer), and $\lambda_{B}$ (half-integer). 
For the $B_{3/2}$ baryon, applying these parity relations to eqn. (\ref{DensityMatrixSigmaMatrixElementsExampleB'}) gives
\begin{align} \label{DensityMatrixParityB'}
  & \rho^{0}_{-\lambda_{B_{3/2}}, -\lambda_{B_{3/2}}'}(B_{3/2}) = (-1)^{\lambda_{B_{3/2}} + \lambda_{B_{3/2}}'} 
       \rho^{0}_{\lambda_{B_{3/2}}, \lambda_{B_{3/2}}'}(B_{3/2}),\nonumber \\
  & \rho^{i (\gamma)}_{-\lambda_{B_{3/2}}, -\lambda_{B_{3/2}}'}(B_{3/2}) = \zeta^{(\gamma) i}(B_{3/2}) (-1)^{\lambda_{B_{3/2}} + \lambda_{B_{3/2}}'}
       \rho^{i (\gamma)}_{\lambda_{B_{3/2}}, \lambda_{B_{3/2}}'}(B_{3/2}), \nonumber \\
  & \rho^{i (N)}_{-\lambda_{B_{3/2}}, -\lambda_{B_{3/2}}'}(B_{3/2}) = \zeta^{(N) i}(B_{3/2}) (-1)^{\lambda_{B_{3/2}} + \lambda_{B_{3/2}}'} 
       \rho^{i (N)}_{\lambda_{B_{3/2}}, \lambda_{B_{3/2}}'}(B_{3/2}), \\
  & \rho^{ij (\gamma N)}_{-\lambda_{B_{3/2}}, -\lambda_{B_{3/2}}'}(B_{3/2}) = \zeta^{(\gamma) i}(B_{3/2}) \zeta^{(N) j}(B_{3/2}) 
       (-1)^{\lambda_{B_{3/2}} + \lambda_{B_{3/2}}'} \rho^{ij (\gamma N)}_{\lambda_{B_{3/2}}, \lambda_{B_{3/2}}'}(B_{3/2}), \nonumber \\
  & \zeta^{(\gamma) i}(B_{3/2}) = \begin{cases}
                       \hphantom{-} 1 & i = x, \nonumber \\
                      -1 & i = y, z,
                     \end{cases} \nonumber \\
  & \zeta^{(N) i}(B_{3/2}) = \begin{cases}
                       \hphantom{-} 1 & i = y, z \nonumber \\
                      -1 & i = x. 
                     \end{cases}
\end{align}
Note that for the case of the $B_{3/2}$ baryon, the values of the
$\zeta^{i}$'s are different from those of the $V$ meson by a factor of $-1$. This is because the derivation does not involve the 
quantity $(-1)^{2 \lambda_{B}}$ since the hadron in the quasi-two-body state that accompanies the baryon $B_{3/2}$ is spinless. We therefore conclude that, aside from the factor of $(-1)^{\lambda_{V} + \lambda_{V}'}$ in the case of the $V$ 
meson and a factor of $(-1)^{\lambda_{B_{3/2}} + \lambda_{B_{3/2}}'}$ in the case of the $B_{3/2}$ baryon, 8 of the 16 matrices will get an
extra factor of $-1$ in these parity relations. It is this extra factor of $-1$ that will have a very specific consequence in using these
relations to simplify the expressions for the $W^{i}$'s in eqns. (\ref{WisRho}) and (\ref{WisDelta}).

While eqns. (\ref{DensityMatrixParityV}) and (\ref{DensityMatrixParityB'}) were derived for the case of a decaying vector meson and a decaying spin-3/2 baryon, respectively, the values of the $\zeta$ factors in eqn. (\ref{DensityMatrixParityV}) will be the same for any meson of any spin and intrinsic parity, while those in eqn. (\ref{DensityMatrixParityB'}) will be the same for any baryon of any spin and intrinsic parity. The reason it is independent of intrinsic parity is because when the parity relations in eqn. (\ref{TransitionMatrixParityRelations}) are applied in eqn. (\ref{InitialToFinalDensityMatrix}), each of the two transition matrices give you one factor of the same intrinsic parity. Since the only possible values are $1$ and $-1$, the overall factor will always be $1$ regardless of intrinsic parity. The $\zeta$ factor also depends on whether the decaying resonance is a baryon or meson, but not on its spin. For a quasi-two-body state consisting of a meson and a baryon of arbitrary spins $MB$, if the meson decays the derivation gives you factors of $(-1)^{2\lambda_{B}}$. But since $\lambda_{B}$ is a multiple of 1/2, this factor will always be $-1$ regardless of the spin. If it is the baryon that decays, the derivation instead gives you a factors of $(-1)^{2\lambda_{M}}$. But since $\lambda_{M}$ is a multiple of 1, this factor will always be 1 regardless of the spin. 

Using eqn. (\ref{STDefinition1}) along with the parity relations in eqns. (\ref{DensityMatrixParityV}) and (\ref{DensityMatrixParityB'}) leads to
\begin{equation} \label{STParity}
t^{i}_{LM} = \zeta^{i}(-1)^{L+M+2S}t^{i}_{L-M},
\end{equation}
for the statistical tensors.

The constraint of parity invariance in the production reaction will reduce the total number of independent parameters that describe the 16
matrices in eqns. (\ref{VMatrixDecomposition}) and (\ref{BMatrixDecomposition}), which means the expressions for the $W^{i}$'s will
simplify. Take for example the expression for $W^{y (\gamma)}(\theta^{*}, \phi^{*};V)$ in eqn. (\ref{W2DistributionV}). By using the relations in eqn. (\ref{DensityMatrixParityV}) and substituting into these equations, we see that all of the terms that are proportional to sine
functions of $\phi^{*}$ vanish. Since $\phi^{*}$ then only appears inside of cosine functions, $W^{y (\gamma)}(\theta^{*}, \phi^{*};V)$ is an even function of $\phi^{*}$,
\begin{align} \label{W2DistributionRhoSimplified}
W^{y (\gamma)}(\theta^{*},\phi^{*};V) = &
     \rho^{y (\gamma)}_{11}(V)\sin^{2}\theta^{*}
     +\rho^{y (\gamma)}_{00}(V) \cos^{2}\theta^{*} \\
     & -\sqrt{2}\Re \rho^{y (\gamma)}_{10}(V) \sin2\theta^{*}\cos\phi^{*} \\
      & -\Re \rho^{y (\gamma)}_{1-1}(V) \sin^{2}\theta^{*} \cos2\phi^{*} \nonumber
\end{align}
(the parity relations imply that $\rho^{y (\gamma)}_{1-1}(V)$ is purely real). We can also see that, for the case of 
$W^{x (\gamma)}(\theta^{*}, \phi^{*})$, it is the terms proportional to cosine functions of $\phi^{*}$ that vanish (as well as the
first two terms). Since the $\phi^{*}$ then only appears inside of sine functions, $W^{x (\gamma)}(\theta^{*}, \phi^{*};V)$ is an odd function of $\phi^{*}$,
\begin{align} \label{W1DistributionRhoSimplified}
W^{x (\gamma)}(\theta^{*},\phi^{*};V) = &
     \sqrt{2}\Im \rho^{x (\gamma)}_{10}(V) \sin2\theta^{*}\sin\phi^{*} \\
      & + \Im \rho^{x (\gamma)}_{1-1}(V) \sin^{2}\theta^{*}\sin2\phi^{*} \nonumber
\end{align}
(the parity relations imply that $\rho^{x (\gamma)}_{1-1}(V)$ is purely imaginary). We get the same results, as we should, if we express the
$W^{i}$'s in terms of statistical tensors instead. By using the parity and hermiticity relations of the tensors, eqns. (\ref{STParity}) and 
(\ref{STHermiticity}), in eqn. (\ref{W2TDistributionV}), we get the following simplified expression,
\begin{equation} \label{W2TDistributionRhoSimplified}
\begin{split}
 W^{y(\gamma)}(\theta^{*},\phi^{*};V) & = \frac{1}{\sqrt{3}}t^{y(\gamma)}_{00}
                      -\frac{1}{2\sqrt{6}}t^{y(\gamma)}_{20}(1 + 3 \cos 2 \theta^{*}) \\
       \mt +\Re[t^{y(\gamma)}_{21}] \sin 2 \theta^{*} \cos \phi^{*} \\
       \mt -\Re[t^{y(\gamma)}_{22}] \sin^{2}\theta^{*} \cos 2 \phi^{*},
\end{split}
\end{equation}
(parity invariance and hermiticity implies that $t^{y (\gamma)}_{21}$ and $t^{y (\gamma)}_{22}$ are purely real). In the same way, 
$W^{x(\gamma)}$ in terms of the statistical tensors is simplified when parity is conserved,
\begin{equation} \label{W1TDistributionRhoSimplified}
\begin{split}
W^{x(\gamma)}(\theta^{*},\phi^{*};V) & = -\Im[t^{x(\gamma)}_{21}] \sin 2 \theta^{*} \sin \phi^{*} \\
       \mt + \Im[t^{x(\gamma)}_{22}] \sin^{2} \theta \sin 2\phi^{*},  
\end{split}
\end{equation}
(parity invariance and hermiticity implies that $t^{x (\gamma)}_{21}$ and $t^{x (\gamma)}_{22}$ are purely imaginary, while
$t^{x (\gamma)}_{00}$ and $t^{x (\gamma)}_{20}$ are equal to zero).

For the case of the $B_{3/2}$ baryon, it is 
$W^{y (\gamma)}(\theta^{*}, \phi^{*}; B_{3/2})$ that is an odd function of $\phi^{*}$, while it is 
$W^{x (\gamma)}(\theta^{*}, \phi^{*}; B_{3/2})$ that is an even function of $\phi^{*}$,
\begin{equation} \label{W2DistributionDeltaSimplified}
\begin{split}
W^{y (\gamma)}(\theta^{*},\phi^{*};B_{3/2}) = &
     \sqrt{2}\Im \rho^{y (\gamma)}_{10}(B_{3/2}) \sin2\theta^{*}\sin\phi^{*} \\
      & + \Im \rho^{y (\gamma)}_{1-1}(B_{3/2}) \sin^{2}\theta^{*}\sin2\phi^{*},
 \end{split}
\end{equation}

\begin{equation} \label{W1DistributionDeltaSimplified}
\begin{split}
W^{x (\gamma)}(\theta^{*},\phi^{*};B_{3/2}) = &
     \rho^{x (\gamma)}_{11}(B_{3/2})\sin^{2}\theta^{*}
     +\rho^{x (\gamma)}_{00}(B_{3/2}) \cos^{2}\theta^{*} \\
     & -\sqrt{2}\Re \rho^{x (\gamma)}_{10}(B_{3/2}) \sin2\theta^{*}\cos\phi^{*} \\
      & -\Re \rho^{x (\gamma)}_{1-1}(B_{3/2}) \sin^{2}\theta \cos2\phi^{*}.
 \end{split}
\end{equation}
In terms of statistical tensors (again, by using the parity and hermiticity relations, eqns. (\ref{STParity}) and (\ref{STHermiticity})),
\begin{equation} \label{W2TDistributionDeltaSimplified}
\begin{split}
 W^{y (\gamma)}(\theta^{*},\phi^{*};B_{3/2}) & = -\sqrt{\frac{3}{2}}\Im[t^{y (\gamma)}_{21}(B_{3/2})] \sin 2\theta^{*} \sin \phi^{*} \\
       \mt +\frac{1}{2}\sqrt{\frac{3}{2}}\Im[t^{y (\gamma)}_{22}(B_{3/2})] \sin^{2}\theta^{*} \sin 2 \phi^{*},
\end{split}
\end{equation}
and
\begin{equation} \label{W1TDistributionDeltaSimplified}
 \begin{split}
 W^{x (\gamma)}(\theta^{*},\phi^{*};B_{3/2}) & = t^{x (\gamma)}_{00}(B_{3/2})
                      -\frac{1}{4}t^{x (\gamma)}_{20}(B_{3/2})(1 + 3 \cos 2 \theta^{*}) \\
       \mt +\sqrt{\frac{3}{2}}\Re[t^{x (\gamma)}_{21}(B_{3/2})] \sin 2 \theta^{*} \cos \phi^{*} \\
       \mt -\frac{1}{2}\sqrt{\frac{3}{2}}\Re[t^{x (\gamma)}_{22}(B_{3/2})] \sin^{2}\theta^{*} \cos 2 \phi^{*}.
\end{split}
\end{equation}
We will later show that it is in general true that once the parity constraints on the production reaction are applied, each of the $W^{i}$'s
and, therefore, the observables, will be either even or odd functions of $\phi^{*}$.  

\chapter{Angular Distributions and Polarization Observables When More Than One Quasi-Two-Body State Channel Contributes} \label{MoreThanOneQuasiTwoBodyState}
\section{Meson Interference}
To examine the case when more than one quasi-two-body state contributes to the photoproduction reaction, we will first examine the example of three specific channels,
\begin{equation} \label{ThreeMesonChannels}
\begin{split}
\gamma N \rightarrow S B \rightarrow M_{1} M_{2} B, \\
\gamma N \rightarrow V B \rightarrow M_{1} M_{2} B, \\
\gamma N \rightarrow T B \rightarrow M_{1} M_{2} B,
\end{split}
\end{equation}
where $S$, $V$ and $T$ are scalar, vector, and tensor (spin-$2$) mesons.

The main difference from the single-channel case is that in the multi-channel case the $\gamma N$ state transitions into a linear
superposition of the $SB$, $VB$, and $TB$ states,

\begin{equation} \label{LinearSuperposition}
    \ket{\Psi(SB, VB, TB)} = \ket{\Psi(SB)} + \ket{\Psi(VB)} + \ket{\Psi(TB)}.
\end{equation}
In matrix form, this is
\begin{equation} \label{LinearSuperpositionMatrix}
\ket{\Psi(B S, B V, B T)}= \begin{pmatrix}
                                \psi_{0 \frac{1}{2}}(SB) \\
                                \psi_{0 \minus \frac{1}{2}}(SB) \\
                                \psi_{1 \frac{1}{2}}(VB) \\
                                \psi_{1 -\frac{1}{2}}(VB) \\
                                \psi_{0 \frac{1}{2}}(VB) \\
                                \vdots \\
                                \psi_{\minus 1 \minus \frac{1}{2}}(VB) \\
                                \psi_{2 \frac{1}{2}}(TB) \\
                                \psi_{2 -\frac{1}{2}}(TB) \\
                                \psi_{1 \frac{1}{2}}(TB) \\
                                \vdots \\
                                \psi_{\minus 2 \minus \frac{1}{2}}(TB)
                               \end{pmatrix},
\end{equation}
where $\psi_{\lambda_{S} \lambda_{B}}(SB)$ is the probability amplitude for the state in the ensemble to be found in a
$SB$ state with respective spin projections  $\lambda_{S}$ and $\lambda_{B}$, and similarly for
$\psi_{\lambda_{V} \lambda_{B}}(VB)$ and $\psi_{\lambda_{T} \lambda_{B}(TB)}$. The first $2$ entries in this vector
are the amplitudes corresponding to the $B S$ state, the next $6$ correspond to the $B V$ state, and the last $10$
correspond to the $B T$ state, making its Hilbert space $18$-dimensional. Its spin density matrix will therefore be 
represented by a $18 \times 4$ matrix.

\begin{equation} \label{TransitionMatrixLinearSuperpositionMeson}
   \hat{M} = \begin{pmatrix} 
   \scriptscriptstyle M_{0\frac{1}{2};1\frac{1}{2}}(SB;\gamma N) &  \scriptscriptstyle M_{0\frac{1}{2};1-\frac{1}{2}}(SB;\gamma N)
    &  \scriptscriptstyle M_{0\frac{1}{2};-1\frac{1}{2}}(SB;\gamma N) &  \scriptscriptstyle M_{0\frac{1}{2};-1-\frac{1}{2}}(SB;\gamma N) \\
     \scriptscriptstyle M_{0-\frac{1}{2};1\frac{1}{2}}(SB;\gamma N) &  \scriptscriptstyle M_{0-\frac{1}{2};1-\frac{1}{2}}(SB;\gamma N)
    &  \scriptscriptstyle M_{0-\frac{1}{2};-1\frac{1}{2}}(SB;\gamma N) &  \scriptscriptstyle M_{0-\frac{1}{2};-1-\frac{1}{2}}(SB;\gamma N) \\
     \scriptscriptstyle M_{1\frac{1}{2};1\frac{1}{2}}(VB;\gamma N) &  \scriptscriptstyle M_{1\frac{1}{2};1-\frac{1}{2}}(VB;\gamma N)
    &  \scriptscriptstyle M_{1\frac{1}{2};-1\frac{1}{2}}(VB;\gamma N) &  \scriptscriptstyle M_{1\frac{1}{2};-1-\frac{1}{2}}(VB;\gamma N) \\
     \scriptscriptstyle M_{1-\frac{1}{2};1\frac{1}{2}}(VB;\gamma N) &   \scriptscriptstyle M_{1-\frac{1}{2};1-\frac{1}{2}}(VB;\gamma N)
    &   \scriptscriptstyle M_{1-\frac{1}{2};-1\frac{1}{2}}(VB;\gamma N) &  \scriptscriptstyle M_{1-\frac{1}{2};-1-\frac{1}{2}}(VB;\gamma N) \\
    \multicolumn{4}{c}{\vdots}\\
     \scriptscriptstyle M_{-1-\frac{1}{2};1\frac{1}{2}}(VB;\gamma N) &   \scriptscriptstyle M_{-1-\frac{1}{2};1-\frac{1}{2}}(VB;\gamma N)
    &   \scriptscriptstyle M_{-1-\frac{1}{2};-1\frac{1}{2}}(VB;\gamma N) &  \scriptscriptstyle M_{-1-\frac{1}{2};-1-\frac{1}{2}}(VB;\gamma N) \\
     \scriptscriptstyle M_{2\frac{1}{2};1\frac{1}{2}}(TB;\gamma N) &   \scriptscriptstyle M_{2\frac{1}{2};1-\frac{1}{2}}(TB;\gamma N)
    &   \scriptscriptstyle M_{2\frac{1}{2};-1\frac{1}{2}}(TB;\gamma N) &  \scriptscriptstyle M_{2\frac{1}{2};-1-\frac{1}{2}}(TB;\gamma N) \\
     \scriptscriptstyle M_{2-\frac{1}{2};1\frac{1}{2}}(TB;\gamma N) &  \scriptscriptstyle M_{2-\frac{1}{2};1-\frac{1}{2}}(TB;\gamma N)
    &  \scriptscriptstyle M_{2-\frac{1}{2};-1\frac{1}{2}}(TB;\gamma N) &  \scriptscriptstyle M_{2-\frac{1}{2};-1-\frac{1}{2}}(TB;\gamma N) \\
    \multicolumn{4}{c}{\vdots} \\
     \scriptscriptstyle M_{-2-\frac{1}{2};1\frac{1}{2}}(TB;\gamma N) &  \scriptscriptstyle M_{-2-\frac{1}{2};1-\frac{1}{2}}(TB;\gamma N)
    &  \scriptscriptstyle M_{-2-\frac{1}{2};-1\frac{1}{2}}(TB;\gamma N) &  \scriptscriptstyle M_{-2-\frac{1}{2};-1-\frac{1}{2}}(TB;\gamma N)
    \end{pmatrix}.
\end{equation}
In the previous matrix, the first two rows describe the transition of the initial system into the $SB$ system (first two entries in state vector in eqn. (\ref{LinearSuperpositionMatrix})). The next six rows describe the transition into the $VB$ system (next six entries in the state vector), and the last ten rows describe the transition into the $TB$ system (last ten entries in the state vector). Since this state vector is $18\times1$, its density matrix will be $18\times18$. 

As with the single-channel case, the initial and final density matrices are related by eqn. (\ref{InitialToFinalDensityMatrix}). But this time, since the indices in the transition matrix in eqn. (\ref{TransitionMatrixLinearSuperpositionMeson}) refer to different channels, we will express the resulting spin density matrix in terms of submatrices, 
\begin{equation} \label{BlockMatrix}
 \hat{\rho}(S,\ V,\ T)= \begin{pmatrix}
                 \hat{\rho}(S) & \hat{\xi}(S,V) & \hat{\eta}(S,T) \\
                 \hat{\xi}^{\dag}(V,S) & \hat{\rho}(V) & \hat{\chi}(V,T) \\
                 \hat{\eta}^{\dag}(T,S) & \hat{\chi}^{\dag}(T,V) & \hat{\rho}(T)
                \end{pmatrix}.
\end{equation}
The diagonal matrices are the ones that are obtained when the reaction is single-channel and no interference takes place. We will refer to the off-diagonal submatrices as interference matrices. We can also write it as
\begin{equation} \label{SumOfSubMatrices}
    \hat{\rho}(S, V, T) = \hat{\rho}(S) + \hat{\rho}(V) + \hat{\rho}(T) + \big[ \hat{\xi}(S, V) + \hat{\eta}(S, T) +\hat{\chi}(V, T) + \text{h.c.} \big],
\end{equation}
where the matrix elements of the interference matrices are
\begin{equation} \label{InterferenceMatrices}
    \begin{split}
        \xi(S,V)_{0\lambda_{V}} \equiv &                 \sum_{\substack{\lambda_{\gamma}\lambda_{N} \\ \lambda_{\gamma}'\lambda_{N}' \\
        \lambda_{B}}}M_{0\lambda_{B};\lambda_{\gamma}\lambda_{N}}\rho(\gamma N)_{\lambda_{\gamma}\lambda_{N};\lambda_{\gamma}'\lambda_{N}'}M^{*}_{\lambda_{V}\lambda_{B};\lambda_{\gamma}'\lambda_{N}'}, \\
        \eta(S,T)_{0\lambda_{T}} \equiv &                 \sum_{\substack{\lambda_{\gamma}\lambda_{N} \\ \lambda_{\gamma}'\lambda_{N}' \\
        \lambda_{B}}}M_{0\lambda_{B};\lambda_{\gamma}\lambda_{N}}\rho(\gamma N)_{\lambda_{\gamma}\lambda_{N};\lambda_{\gamma}'\lambda_{N}'}M^{*}_{\lambda_{T}\lambda_{B};\lambda_{\gamma}'\lambda_{N}'}, \\
        \chi(V,T)_{\lambda_{V}\lambda_{T}} \equiv &                 \sum_{\substack{\lambda_{\gamma}\lambda_{N} \\ \lambda_{\gamma}'\lambda_{N}' \\
        \lambda_{B}}}M_{\lambda_{V}\lambda_{B};\lambda_{\gamma}\lambda_{N}}\rho(\gamma N)_{\lambda_{\gamma}\lambda_{N};\lambda_{\gamma}'\lambda_{N}'}M^{*}_{\lambda_{T}\lambda_{B};\lambda_{\gamma}'\lambda_{N}'},
    \end{split}
\end{equation}
(note how there is a sum over $\lambda_{B}$, indicating that the partial trace over the recoil baryon has been taken). These interference matrices therefore have ``mixed'' indices: their ``row'' and ``column'' indices refer to a different meson. Since these mesons have different spin, these matrices are not square: $\hat{\xi}$ is $1\times3$, $\hat{\eta}$ is $1\times5$, and $\hat{\chi}$ is $3\times5$.  

We now proceed to find the angular distribution for the decay of the meson into $M_{1} M_{2}$,
\begin{equation} \label{DecayLinearSuperpositionMeson}
W(\theta^{*},\phi^{*}; S, V, B) = \hat{\cal{D}}^{\dagger}(\theta^{*}, \phi^{*})\hat{\rho}(S, V, T)\hat{\cal{D}}(\theta^{*}, \phi^{*}).
\end{equation}
This time, the form of the transition matrix $\hat{\cal{D}}^{\dagger}$ is
\begin{equation} \label{WignerDLinearSuperpositionMeson}
 \hat{\cal{D}}=\begin{pmatrix}
                D^{0}_{00}(\theta^{*},\phi^{*}) \\ D^{1}_{01}(\theta^{*},\phi^{*}) \\ D^{1}_{00}(\theta^{*},\phi^{*})
                \\ D^{1}_{0-1}(\theta^{*},\phi^{*}) \\ D^{2}_{02}(\theta^{*},\phi^{*}) \\ D^{2}_{01}(\theta^{*},\phi^{*})
                \\ \vdots \\ D^{2}_{0-2}(\theta^{*},\phi^{*})
               \end{pmatrix}.
\end{equation}
The distribution can be written as
\begin{align}
W(\theta^*,\phi^*;S,V,T)&=W^{(S)}(\theta^*,\phi^*)+W^{(V)}(\theta^*,\phi^*)+W^{(T)}(\theta^*,\phi^*)+W^{(SV)}(\theta^*,\phi^*)\nonumber\\
\mt
+W^{(ST)}(\theta^*,\phi^*)+W^{(TV)}(\theta^*,\phi^*).
\end{align}
In this expression $W^{(S)}$ is the contribution of a scalar intermediate meson, $W^{(V)}$ arise from a vector meson, and $W^{(T)}$ arises from a meson of spin 2. $W^{(SV)}$ arises from the interference between a scalar and vector meson, $W^{(ST)}$ comes from the interference between a scalar and a tensor, and $W^{(TV)}$ arises from interference between a vector and tensor meson. The explicit forms are

\begin{equation} \label{FirstEquation}
    W^{(S)}(\theta^*,\phi^*) = \rho_{00}(S),
\end{equation}

\begin{align}
        W^{(V)}(\theta^*,\phi^*) & = \frac{\rho_{00}(V)}{2}+\frac{\rho_{11}(V)+\rho_{-1-1}(V)}{4} + \bigg(\frac{\rho_{00}(V)}{2} -\frac{\rho_{11}(V)+\rho_{-1-1}(V)}{4}\bigg) \cos (2 \theta^* ) \nonumber \\
        \mt + \frac{1}{2}\bigg(\Im[\rho_{1-1}(V)]\sin (2 \phi^* )-\Re[\rho_{1-1}(V)]\cos(2\phi^*)\bigg)\bigg(1-\cos(2\theta^*)\bigg) \\
        \mt +\frac{\sin(2\theta^*)}{\sqrt{2}}\bigg(\Re[\rho_{0-1}(V)-\rho_{10}(V)]\cos (\phi^* )
    + \Im[\rho_{10}(V)-\rho_{0-1}(V)]\sin (\phi^*) \bigg),\nonumber
\end{align}

\begin{align}
    W^{(T)}(\theta^*,\phi^*) & = \frac{11+12\cos(2\theta^*)+9\cos(4\theta^*)}{32} \rho_{00}(T)+\bigg(1-\cos(4\theta^*)\bigg)\bigg(\frac{3}{16} \big(\rho_{11}(T)+\rho_{-1-1}(T)\big)\nonumber\\
    \mt - \frac{3}{8} \Big(\Re[\rho_{1-1}(T)]\cos(2\phi^*)-\Im[\rho_{1-1}(T)]\sin(2\phi^*)\Big)\bigg)\nonumber\\
   \mt+\frac{3}{64}\bigg(\rho_{22}(T)+\rho_{-2-2}(T)\bigg)\bigg(3-4\cos(2\theta^*)+\cos(4\theta^*)\bigg) \nonumber \\
   \mt+\frac{1}{16}
   \sqrt{\frac{3}{2}} \bigg(\Re[\rho_{0-2}(T)+\rho_{20}(T)]\cos(2\phi^*)\nonumber\\
   \mt -\Im[\rho_{0-2}(T)+\rho_{20}(T)]\sin(2\phi^*)\bigg)\bigg(3\cos(4\theta^*) -4\cos(2\theta^*)+1\bigg)\nonumber \\
   \mt -\frac{3}{32} \bigg(\Re[\rho_{2-2}(T)]\cos (4 \phi^* )-\Im[\rho_{2-2}(T)]\sin (4 \phi^* )\bigg)\bigg(3-4\cos(2\theta^*)+\cos(4\theta^*)\bigg)\nonumber \\
   \mt - \frac{1}{8}\sqrt{\frac{3}{2}} \bigg(\Re[\rho_{0-1}(T)-\rho_{10}(T)]\cos(2\phi^*) \\
   \mt-\Im[\rho_{0-1}(T)-\rho_{10}(T)]\sin(2\phi^*)\bigg)\bigg(2\sin(2\theta^*)+3\sin(4\theta^*)\bigg)\nonumber\\
    \mt + \frac{3}{16} \bigg)\bigg(\Re[\rho_{2-1}(T)-\rho_{1-2}(T)]\cos(3\phi^*)-\Im[\rho_{2-1}(T)-\rho_{1-2}(T)]\sin(3\phi^*)\bigg)\nonumber\\
   \mt + \bigg(\Re[\rho_{-1-2}(T)-\rho_{21}(T)]\cos(\phi^*)-\Im[\rho_{-1-2}(T)-\rho_{21}(T)]\sin(\phi^*)\bigg)\nonumber\\
   \mt\times\bigg(2 \sin (2 \theta^*)-\sin(4\theta^*)\bigg)\nonumber,
\end{align}
\begin{align}
W^{(SV)}(\theta^*,\phi^*) & = 2
   \Re[\xi_{00}]
\cos (\theta^* ) \nonumber\\
\mt
+\sqrt{2}\sin(\theta^*)\bigg(\Re[\xi_{0-1}-\xi_{01}]\cos(\phi^* )-\Im[\xi_{01}+\xi_{0-1}]\sin (\phi^*)\bigg),
\end{align}
\begin{align}
    W^{(ST)}(\theta^*,\phi^*) & = \frac{1}{4}\Re[\eta_{00}]
\bigg(1+3\cos (2 \theta^* )\bigg) \nonumber \\
\mt
+\sqrt{\frac{3}{2}}\sin(2\theta^*)\bigg(\Re[\eta_{0-1}-\eta_{01}]\cos(\phi^*)- \Im[\eta_{01}+\eta_{0-1}]\sin(\phi^*)\bigg) \\
\mt
+\frac{1}{2} \sqrt{\frac{3}{2}}\bigg(1-\cos(2\theta^*)\bigg)\bigg(\Re[\eta_{02} +\eta_{0-2}]\cos(2\phi^*)+\Im[\eta_{02}-\eta_{0-2}]\sin (2 \phi^* )\bigg),\nonumber
\end{align}
and
\begin{align} \label{SecondEquation}
    W^{(VT)}(\theta^*,\phi^*) & = \frac{1}{4} \Re[\chi_{00}](5\cos(\theta^*)+3\cos(3\theta^*)) \nonumber \\
\mt +(\cos(\theta^*)-\cos(3\theta^*))\Bigg\{ \frac{\sqrt{3}}{4}\bigg(\Re[\chi_{11}+\chi_{-1-1}] \nonumber \\
\mt
+\Re[\chi_{-11}-\chi_{1-1}]\cos(2\phi^*)
-\Im[\chi_{-11}+\chi_{1-1}]\sin(2\phi^*)\bigg)\nonumber\\
\mt+\frac{1}{4}
   \sqrt{\frac{3}{2}}\bigg(\Re[\chi_{02}+\chi_{0-2}]\cos(2\phi^*)
-\Im[\chi_{02}-\chi_{0-2}]\sin(2\phi^*)\bigg)\Bigg\} \nonumber \\
\mt
+ \frac{1}{4\sqrt{2}}\bigg(3\sin(3\theta^*)-\sin(\theta^*)\bigg)\bigg(\Re[\chi_{10}-\chi_{-10}]\cos(\phi^*)) \\
\mt +\Im[\chi_{10} +\chi_{-10}]\sin(\phi^*)\bigg) \nonumber \\
\mt
+ \frac{1}{2}\sqrt{\frac{3}{2}}\bigg(\sin(\theta^*)+\sin(3\theta^*)\bigg)\bigg(\Re[\chi_{0-1}-\chi_{01}]\cos(\phi^*) \nonumber \\
\mt -\Im[\chi_{01}+\chi_{0-1}]\sin(\phi^*)\bigg)\nonumber \\
\mt
+\frac{\sqrt{3}}{8}\bigg(3\sin(\theta^*)-\sin(3\theta^*)\bigg)\bigg(\Re[\chi_{-12}-\chi_{1-2}]\cos(3\phi^*) \nonumber \nonumber \\
\mt +\Im[\chi_{-12}+\chi_{1-2}]\sin(3\phi^*)\nonumber \\
\mt
+\Re[\chi_{-1-2}-\chi_{12}]\cos(\phi^*)-\Im[\chi_{-1-2}+\chi_{12}]\sin(\phi^*)\bigg).\nonumber
\end{align}
\section{Baryon Interference} \label{Interference}
We will now examine the case for the decay of photoproduced baryons in the reactions $\gamma N 
\rightarrow M_{1}B_{3/2} \rightarrow M_{1}M_{2}B$ and $\gamma N 
\rightarrow M_{2}B_{3/2} \rightarrow M_{1}M_{2}B$ under the assumption that three different baryons in the intermediate quasi-two-body state 
contribute of spins $1/2$, $3/2$, and $5/2$.

Just like in the case of the angular distribution of decaying mesons, the SDME's of the intermediate quasi-two-body
state, which is a linear superposition of the three baryons that contribute, can be grouped in to different blocks:
\begin{equation} \label{eqn 49}
\hat{\rho}(S)= \begin{pmatrix}
                 \hat{\rho}(1/2) & \hat{\xi} & \hat{\eta} \\
                 \hat{\xi}^{\dag} & \hat{\rho}(3/2) & \hat{\chi} \\
                 \hat{\eta}^{\dag} & \hat{\chi}^{\dag} & \hat{\rho}(5/2)
                \end{pmatrix}.
\end{equation}
$\hat{\rho}(1/2)$, $\hat{\rho}(3/2)$, and $\hat{\rho}(5/2)$ are the spin density matrices that the spin-$1/2$, 
$3/2$, and $5/2$ baryons would have if they were contributing to the process by themselves. The indices of these three matrices 
only correspond to one hadron (the spin-$1/2$, $3/2$, and $5/2$ baryons). The rest of the blocks have mixed indices:
one index corresponds to one hadron and the other to a different hadron. The left index of the matrix $\hat{\xi}$ corresponds to the 
spin-1/2 baryon, while its right index corresponds to the spin-3/2 baryon. For the matrix $\hat{\eta}$, the left index 
corresponds to the spin-1/2 baryon and the right index to the spin-5/2 baryon. Finally, for the matrix $\hat{\chi}$
the left index corresponds to the spin-3/2 baryon and the right index corresponds to the spin-5/2 baryons.
The matrices with the daggers are the conjugate transpose of their respective matrix.

The distribution can be written as
\begin{align}
W(\theta^*,\phi^*;1/2,3/2,5/2)&=W^{(1/2)}(\theta^*,\phi^*)+W^{(3/2)}(\theta^*,\phi^*)+W^{(5/2)}(\theta^*,\phi^*)\nonumber\\
\mt
+W^{(1/2,3/2)}(\theta^*,\phi^*)+W^{(1/2,5/2)}(\theta^*,\phi^*)+W^{(3/2,5/2)}(\theta^*,\phi^*).
\end{align}
In this expression $W^{(1/2)}$ is the contribution of a spin-$1/2$ intermediate baryon, $W^{(3/2)}$ arises from a baryon with spin 3/2, and $W^{(5/2)}$ arises from a baryon with spin 5/2. $W^{(1/2,3/2)}$ arises from the interference between the baryons with spin 1/2 and 3/2, $W^{(1/2,5/2)}$ comes from the interference between baryons with spin 1/2 and spin 5/2, and $W^{(3/2,5/2)}$ arises from interference between baryons with spin 3/2 and spin 5/2. The explicit forms are
\begin{align} \label{ThirdEquation}
    W^{(1/2)}(\theta^*,\phi^*) = \rho_{11}(1/2)+\rho_{-1-1}(1/2),
\end{align}
\begin{align}
    W^{(3/2)}(\theta^*,\phi^*) & = \frac{1}{8}\Bigg\{ 3\bigg(1-\cos(2\theta^*)\bigg)\bigg(\rho_{33}(3/2)+\rho_{-3-3}(3/2)\bigg)\nonumber \\
\mt
+\bigg(5+3\cos(2\theta^*)\bigg)\bigg(\rho_{11}(3/2)+\rho_{-1-1}(3/2)\bigg)\Bigg\} \\
\mt
-\frac{\sqrt{3}}{4}\bigg(1-\cos(2\theta^*)\bigg)\bigg(\Re[\rho_{1-3}(3/2)+\rho_{3-1}(3/2)]\cos(2\phi^*) \nonumber \\
\mt -\Im[\rho_{1-3}(3/2)+\rho_{3-1}(3/2)]\sin(2\phi^*)\bigg)\nonumber \\
\mt
+\frac{\sqrt{3}}{2}\sin(2\theta^*)\bigg(\Re[\rho_{-1-3}(3/2)-\rho_{31}(3/2)]\cos(\phi^*) \nonumber \\
\mt -\Im[\rho_{-1-3}(3/2)-\rho_{31}(3/2)]\sin(\phi^*)\bigg),\nonumber
\end{align}
\begin{align}
    W^{(5/2)}(\theta^*,\phi^*) & = \frac{1}{32}\bigg(\rho_{11}(5/2)+\rho_{-1-1}(5/2)\bigg)\bigg(12+15\cos(2\theta^*)+5\cos(4\theta^*)\bigg)\nonumber\\
\mt+\frac{1}{64}\bigg(\rho_{33}(5/2)+\rho_{-3-3}(5/2)\bigg)\bigg(19-4\cos(2\theta^*)-15\cos(4\theta^*)\bigg)\nonumber\\
\mt+\frac{5}{64}\bigg(\rho_{55}(5/2)+\rho_{-5-5}(5/2)\bigg)\bigg(3-4\cos(2\theta^*)+\cos(4\theta^*)\bigg) \nonumber\\
\mt
+\frac{\sqrt{5}}{8}\bigg(2\sin(2\theta^*)-\sin(4\theta^*)\bigg)\nonumber\\
\mt
\times\Bigg\{ \Re[\rho_{-3-5}(5/2)-\rho_{53}(5/2)]\cos(\phi^*)-\Im[\rho_{-3-5}(5/2)-\rho_{53}(5/2)]\sin(\phi^*) \nonumber \\ 
\mt
+\frac{1}{\sqrt{2}}\bigg(\Re[\rho_{5-1}(5/2)-\rho_{1-5}(5/2)]\cos(3\phi^*)-\Im[\rho_{5-1}(5/2)-\rho_{1-5}(5/2)]\sin(3\phi^*)\bigg)\Bigg\} \\ 
\mt
+\frac{\sqrt{5}}{32}\bigg(3-4\cos(2\theta^*)+\cos(4\theta^*)\bigg)\bigg(\Re[\rho_{5-3}(5/2)+\rho_{3-5}(5/2)]\cos(4\phi^*)\nonumber\\
\mt-\Im[\rho_{5-3}(5/2)+\rho_{3-5}(5/2)]\sin(4\phi^*)\bigg) \nonumber \\ 
\mt
+\frac{1}{16}\sqrt{\frac{5}{2}}\bigg(-1+4\cos(2\theta^*)-3\cos(4\theta^*)\bigg)\bigg(\Re[\rho_{51}(5/2)+\rho_{-1-5}(5/2)]\cos(4\phi^*)\nonumber\\
\mt
-\Im[\rho_{51}(5/2)+\rho_{-1-5}(5/2)]\sin(4\phi^*)\bigg) \nonumber \\ 
\mt
+\frac{1}{8\sqrt{2}}\bigg(6\sin(2\theta^*)-5\sin(4\theta^*)\bigg)\bigg(\Re[\rho_{-1-3}(5/2)-\rho_{31}(5/2)]\cos(\phi^*)\nonumber\\
\mt
-\Im[\rho_{-1-3}(5/2)-\rho_{31}(5/2)]\sin(\phi^*)\bigg) \nonumber \\ 
\mt
+\frac{1}{16\sqrt{2}}\bigg(9-4\cos(2\theta^*)-5\cos(4\theta^*)\bigg)\bigg(\Re[\rho_{1-3}(5/2)-\rho_{3-1}(5/2)]\cos(\phi^*)\nonumber\\
\mt
-\Im[\rho_{1-3}(5/2)-\rho_{3-1}(5/2)]\sin(\phi^*)\bigg), \nonumber 
\end{align}

\begin{align}
    W^{(1/2,3/2)}(\theta^*,\phi^*) & = 2\Re[\xi_{11}+\xi_{-1-1}]\cos (\theta^* ) \nonumber\\
\mt
+\sin(\theta^*)\Bigg\{
\bigg(\Re[\xi_{1-1}-\xi_{-11}]+\sqrt{3}\Re[\xi_{-1-3}-\xi_{13}]\bigg)\cos (\phi^* ) \nonumber\\ 
\mt-\bigg(\Im[\xi_{1-1}+\xi_{-11}]
+\sqrt{3}\Im[\xi_{-1-3}+\xi_{13}]\bigg)\sin(\phi^*)\Bigg\},
\end{align}
\begin{align}
    W^{(1/2,5/2)}(\theta^*,\phi^*) & = \frac{1}{2}\Re[\eta_{11}+\eta_{-1-1}]\bigg(1+3\cos(2\theta^*)\bigg) \nonumber \\ 
\mt
+\frac{1}{2\sqrt{2}}\bigg(1-\cos(2\theta^*)\bigg)\Bigg\{ \bigg(\Re[\eta_{1-3}+\eta_{-13}]+
   \sqrt{5} \Re[\eta_{15}+\eta_{-1-5}]\bigg)\cos (2 \phi^* ) \nonumber \\ 
\mt
\bigg(\Im[\eta_{-13}-\eta_{1-3}]+
   \sqrt{5} \Im[\eta_{15}-\eta_{-1-5}]\bigg)\sin (2 \phi^* )\Bigg\}  \\ 
\mt
+\sin(2\theta^*)\Bigg\{ \bigg(\Re[\eta_{1-1}-\eta_{-11}]+\sqrt{2}\Re[\eta_{-1-3}-\eta_{13}]\bigg)\cos (\phi^* ) \nonumber\\
\mt
-\bigg(\Im[\eta_{1-1}+\eta_{-11}]+\sqrt{2}\Im[\eta_{-1-3}+\eta_{13}]\bigg)\sin (\phi^* )\Bigg\},\nonumber 
\end{align}
and
\begin{align} \label{FourthEquation}
W^{(3/2,5/2)}(\theta^*,\phi^*) & = \frac{1}{8\sqrt{2}}\bigg(3\sin(\theta^*)-\sin(3\theta^*)\bigg)\Bigg\{ \sqrt{15}\bigg(\Re[\chi_{-3-5}-\chi_{35}]\cos(\phi^*) \nonumber \\
\mt -\Im[\chi_{-3-5}+\chi_{35}]\sin(\phi^*)\bigg)\nonumber \\
\mt
+\bigg(\sqrt{5}\Re[\chi_{-15}-\chi_{1-5}]+\sqrt{3}\Re[\chi_{-33}-\chi_{3-3}]\bigg)\cos(3\phi^*)\nonumber \\
\mt
+\bigg(\sqrt{5}\Im[\chi_{-15}+\chi_{1-5}]+\sqrt{3}\Im[\chi_{-33}+\chi_{3-3}]\bigg)\sin(3\phi^*)\Bigg\} \nonumber \\
\mt +\frac{1}{8}\bigg(\cos(\theta^*)-\cos(3\theta^*)\bigg)\Bigg\{ 2\sqrt{6}\Re[\chi_{33}+\chi_{-3-3}] \nonumber \\
\mt
+\bigg(\sqrt{10}\Re[\chi_{15}+\chi_{-1-5}]-2\sqrt{3}\Re[\chi_{3-1} \nonumber \\
\mt +\chi_{-31}]+\sqrt{3}\Re[\chi_{-13}+\chi_{1-3}]\bigg)\cos(2\phi^*)\nonumber\\
\mt
+\bigg(\sqrt{10}\Im[\chi_{15}-\chi_{-1-5}]-2\sqrt{3}\Im[\chi_{3-1}-\chi_{-31}] \\
\mt +\sqrt{3}\Im[\chi_{-13}-\chi_{1-3}]\bigg)\sin(2\phi^*)\Bigg\} \nonumber \\
\mt
+\frac{1}{2}\bigg(3\cos(\theta^*)+\cos(3\theta^*)\bigg)\Re[\chi_{11}+\chi_{-1-1}]\nonumber\\
\mt
+\frac{1}{8}\bigg(5\sin(\theta^*)+\sin(3\theta^*)\bigg)\bigg(\Re[\chi_{1-1}-\chi_{-11}]\cos(\phi^*) \nonumber \\
\mt -\Im[\chi_{1-1}+\chi_{-11}]\sin(\phi^*)\bigg) \nonumber \\
\mt
+\frac{1}{8\sqrt{2}}\bigg(11\sin(\theta^*)+7\sin(3\theta^*)\bigg)\bigg(\Re[\chi_{-1-3}-\chi_{13}]\cos(\phi^*)\nonumber \\
\mt
-\Im[\chi_{-1-3}+\chi_{13}]\sin(\phi^*)\bigg)\nonumber\\
\mt
+\frac{\sqrt{3}}{8}\bigg(3\sin(3\theta^*)-7\sin(\theta^*)\bigg)\bigg(\Re[\chi_{-3-1}-\chi_{31}]\cos(\phi^*)\nonumber \\
\mt
+\Im[\chi_{-3-1}+\chi_{31}]\sin(\phi^*)\bigg). \nonumber
\end{align}

There is one important aspect to consider in the case of decaying intermediate state baryons. Unlike the case of a decaying meson, if $M_1$ and $M_2$ are the same meson, then an individual baryon resonance has two possible decay channels: $B_{3/2} \rightarrow M_{1} B$, and $B_{3/2} \rightarrow M_{2} B$. This means that the decay should be considered as an interference of two different states, each with its own individual spin density 
matrix. Therefore, the density matrix of a single decaying baryon will have this form,
\begin{equation}
 \hat{\rho}(B_{3/2}) = \begin{pmatrix}
                  & \hat{\rho}(B_{3/2}^{\text{B}}) & \hat{\xi}(B_{3/2}^{\text{B}},B_{3/2}^{\text{C}}) \\
                  & \hat{\xi}^{\dagger}(B_{3/2}^{\text{B}},B_{3/2}^{\text{C}}) & \hat{\rho}(B_{3/2}^{\text{C}}) 
                 \end{pmatrix}.
\end{equation}
The diagonal spin density submatrices describe the spin state of the baryon resonance in pathways B and C. The off-diagonal submatrices is the interference matrix between the same baryon in pathways B and C.  

\section{Mesons and Baryon Interference}
Only minor modifications are needed for the case of interference of intermediate quasi-two-body states where some of them contain decaying
mesons and some of them contain decaying baryons. To illustrate this, we will use as an example the case where two states interfere: one 
intermediate state has a decaying spin-1 meson $V$, and the other has a decaying spin-3/2 baryon $B_{3/2}$,
\begin{equation}
 \begin{split}
  \gamma N \rightarrow V B \rightarrow M_{1} M_{2} B, \\
  \gamma N \rightarrow M_{1} B_{3/2} \rightarrow M_{1} M_{2} B
 \end{split}
\end{equation}
(for simplicity, we will only consider pathways A and B). In the distribution, there will be a set of terms that contain the SDME's of the interference submatrix $\hat{\xi}$, 
\begin{equation}
 W(V, B_{3/2}) = \sum_{\lambda_{V} \lambda_{B_{3/2}} \lambda_{B}} D^{1 *}_{0 \lambda_{V}}(\theta^{*}_{\text{A}}, \phi^{*}_{\text{A}}) 
 \xi_{\lambda_{V} \lambda_{B}; 0 \lambda_{B_{3/2}}} D^{3/2}_{\lambda_{B_{3/2}} \lambda_{B}}(\theta^{*}_{\text{B}}, \phi^{*}_{\text{B}}).
\end{equation}
Note that the angles in the arguments of the two Wigner-$\mathcal{D}$ are different. This is due to the fact that the angles for the two 
decaying resonances are defined in terms of different coordinate systems. Expressions relating these different angles will be given in appendix \ref{ChangingCoordinateSystems}.

In the cases in which only mesons or only baryons interfered, we could take the partial trace over the spectator hadron to get the spin density matrix of the decaying hadron.
But in this case, the helicity indices of the interference matrix have different spectator hadrons in the ``row'' and ``column'' indices: the baryon $B$ and the meson $M_{1}$ (since $M_{1}$ is spinless, its helicity index is equal to $0$). Therefore, we cannot define a partial trace on this interference matrix since the $\lambda_{B}$ helicity index
appears on both this matrix and on one of the Wigner-$\mathcal{D}$ functions. In this case, one option is to write the expression for the distribution in terms of this matrix without taking the partial trace. Another is to decompose the matrix into a sum of spin density matrices for the $VB$ system into eigenstates of total spin, which can take on values of $1/2$ or $3/2$ in this case,
\begin{align} \label{TotalSpinDecompositionMatrixElements}
    \xi_{\lambda_{V}\lambda_{B}; 0 \lambda_{B_{3/2}}} & = \sum_{\widetilde{S}, \widetilde{M}} 
      C^{\widetilde{S} \widetilde{M}}_{1 \lambda_{V}; \frac{1}{2} \lambda_{B}} 
      \widetilde{\xi}\bigg( \widetilde{S}, \frac{3}{2} \bigg)_{\widetilde{M} \lambda_{B_{3/2}}} \nonumber \\
    & = C^{\frac{1}{2} \ (\lambda_{V} + \lambda_{B})}_{1 \lambda_{V}; \frac{1}{2} \lambda_{B}} 
           \widetilde{\xi}\bigg(\frac{1}{2}, \frac{3}{2}\bigg)_{ (\lambda_{V} + \lambda_{B}) \ \lambda_{B_{3/2}}} \\
           \mt + \ \ C^{\frac{3}{2} \  (\lambda_{V} + \lambda_{B})}_{1 \lambda_{V}; \frac{1}{2} \lambda_{B}} 
           \widetilde{\xi}\bigg(\frac{3}{2}, \frac{3}{2}\bigg)_{ (\lambda_{V} + \lambda_{B}) \ \lambda_{B_{3/2}}}.\nonumber
\end{align}
In the previous expression, $\hat{\widetilde{\xi}}\big(\frac{1}{2}, \frac{3}{2}\big)$ and $\hat{\widetilde{\xi}}\big(\frac{3}{2}, \frac{3}{2}\big)$ are interference spin density matrices between $B_{3/2}$ and the $VB$ system in its $S=1/2$ eigenstate, and between $B_{3/2}$ and the $VB$ system in its $S=3/2$ eigenstate. The left index on these density matrices labels the total spin projection of the $VB$ system (which is set to $\widetilde{M} = \lambda_{V} + \lambda_{B}$ from conservation of angular momentum), while the right index labels the spin projection of the $B_{3/2}$ baryon. The expression for the distribution in equation
will therefore contain terms of this form,
\begin{align}
 W & = \cdots + \sum_{\substack{\widetilde{M} \lambda_{B_{3/2}} \\ \lambda_{B} \lambda_{V}}}
      D^{* 1}_{\lambda_{V} 0}(\theta^{*}_{A},\phi^{*}_{A}) C^{1/2 \widetilde{M}}_{1 \lambda_{V};1/2 \lambda_{B}}
                      \widetilde{\xi}\bigg(\frac{1}{2}, \frac{3}{2}\bigg)_{\widetilde{M} \lambda_{B_{3/2}}}
                      D^{3/2}_{\lambda_{B_{3/2}} \lambda_{B}}(\theta^{*}_{B},\phi^{*}_{B}) \nonumber \\
   \mt \hphantom{\cdots} + \sum_{\substack{\widetilde{M} \lambda_{B_{3/2}} \\ \lambda_{B} \lambda_{V}}}
       D^{* 1}_{\lambda_{V} 0}(\theta^{*}_{A},\phi^{*}_{A}) C^{3/2 \widetilde{M}}_{1 \lambda_{V};1/2 \lambda_{B}}
       \widetilde{\xi}\bigg(\frac{3}{2}, \frac{3}{2}\bigg)_{\widetilde{M} \lambda_{B_{3/2}}}
       D^{3/2}_{\lambda_{B_{3/2}} \lambda_{B}}(\theta^{*}_{B},\phi^{*}_{B})  + \cdots \\
   & = \cdots + \sum_{\widetilde{M}, \lambda_{B_{3/2}}} \widetilde{D}^{1/2*}_{\widetilde{M} \lambda_{B}} (\theta^{*}_{A},\phi^{*}_{A})
       \widetilde{\xi}\bigg(\frac{1}{2}, \frac{3}{2}\bigg)_{\widetilde{M} \lambda_{B_{3/2}}} 
       D^{3/2}_{\lambda_{B_{3/2}} \lambda_{B}}(\theta^{*}_{B},\phi^{*}_{B}) \nonumber \\
   \mt + \sum_{\widetilde{M}, \lambda_{B_{3/2}}} \widetilde{D}^{3/2*}_{\widetilde{M} \lambda_{B}} (\theta^{*}_{A},\phi^{*}_{A})
       \widetilde{\xi}\bigg(\frac{3}{2}, \frac{3}{2}\bigg)_{\widetilde{M} \lambda_{B_{3/2}}} 
       D^{3/2}_{\lambda_{B_{3/2}} \lambda_{B}}(\theta^{*}_{B},\phi^{*}_{B}) + \cdots , \nonumber                      
\end{align}
where
\begin{equation} \label{GeneralizedD}
 \widetilde{D}^{\widetilde{S}*}_{\widetilde{M} \lambda_{B}} (\theta^{*},\phi^{*}) \equiv 
      \sum_{\lambda_{V}}
       D^{* 1}_{\lambda_{V} 0}(\theta^{*},\phi^{*}) C^{\widetilde{S} \widetilde{M}}_{1/2 \lambda_{B},1 \lambda_{V}} \quad \text{with}\ \widetilde{S} = 1/2, \ 3/2.
\end{equation}
The quantity defined in the previous equation can be considered a generalized Wigner-$\mathcal{D}$ function, and also a decay
amplitudes for the $\widetilde{S}=1/2$ and $\widetilde{S}=3/2$ component or the $VB$ system to decay into $B$. Using the following properties of the Wigner-$\mathcal{D}$
functions and of the Clebsch-Gordan coefficients,
\begin{align}
  D^{* S}_{-\lambda -\lambda'} & = (-1)^{(\lambda - \lambda')} D^{S}_{\lambda \lambda'}, \nonumber \\
  C^{c \gamma}_{a \alpha; b \beta} & = (-1)^{a-\alpha} \sqrt{\frac{2c+1}{2b+1}} C^{b \beta}_{c \gamma; a -\alpha}, \\
  \gamma & = \alpha + \beta,\nonumber
\end{align}
the following property can be shown,
\begin{equation}
\begin{split}
\widetilde{D}^{* \widetilde{S}}_{-\widetilde{M} -\lambda_{N}} & = (-1)^{\widetilde{S} - \widetilde{M} + S_{R} + \frac{1}{2} - \lambda_{N}} 
     \widetilde{D}^{\widetilde{S}}_{\widetilde{M} \lambda_{N}}, \\
& = (-1)^{\widetilde{S} - \widetilde{M} + \frac{3}{2} - \lambda_{N}} \widetilde{D}^{\widetilde{S}}_{\widetilde{M} \lambda_{N}}
     \ \text{(for $S_{R} = 1$)},
\end{split}
\end{equation}
where $S_{R}$ is the spin of the meson resonance that decays (a $\rho$ meson, in this example). This property will be used later in this article.

\chapter{General Expressions for the Angular Distributions and Non-Recoil Polarization Observables} \label{GeneralExpression}
We will now derive the general expression for the angular distribution $W$ when an arbitrary number of quasi-two-body states of arbitrary
spin contribute to the reaction, with interference effects included.

\section{General Expression of the Decay Distribution in Terms of the SDME's}
As we have seen, the distribution will be a sum of different terms that contain
different submatrices,
\begin{equation} \label{Distribution}
	\begin{split}
    W(\theta^{*}, \phi^{*}) & = \sum_{i} \Tr[\hat{D}^{S_{i} \dagger}(\theta^{*}, \phi^{*}) \hat{\rho}(S_{i}) \hat{D}^{S_{i}}(\theta^{*}, \phi^{*})] 
	+ \sum_{i > j} \bigg\{ \Tr[\hat{D}^{S_{i} \dagger}(\theta^{*}, \phi^{*}) \hat{\xi}(S_{i}, S_{j}) \hat{D}^{S_{j}}(\theta^{*}, \phi^{*})] \\ 
	\mt \quad + \Tr[\hat{D}^{S_{j} \dagger}(\theta^{*}, \phi^{*}) \hat{\xi}(S_{j}, S_{i}) \hat{D}^{S_{i}}(\theta^{*}, \phi^{*})] \bigg\}.
	\end{split}
\end{equation}
The indices $i$ and $j$ are used to label the spins of all of the contributing intermediate state hadrons contributing to the reaction.
The $\hat{\rho}(S_{i})$'s are the spin density matrices of the intermediate states. The $\hat{\xi}(S_{i}, S_{j})$ are the interference spin density matrices between the hadrons labeled by the indices $i$ and $j$. Note that the sum goes over $i>j$ terms so as to not overcount the interfering hadron pairs. The $\hat{\mathcal{D}}^{S_{i}}$ are the Wigner-$\mathcal{D}$ matrices.

By using the hermiticity condition of the interference submatrices,
\begin{equation} \label{HermiticityInterferenceMatrices}
  \hat{\xi}(S_{j}, S_{i}) = \hat{\xi}^{\dagger}(S_{i}, S_{j}),
\end{equation}
we can rewrite the second term inside of the brackets in the previous expression as
\begin{align} \label{HermticityInterferenceMatricesRewrite}
    \Tr[\hat{D}^{S_{j} \dagger} \hat{\xi}(S_{j}, S_{i}) \hat{D}^{S_{i}}] 
    = \Tr[\hat{D}^{S_{j} \dagger} \hat{\xi}^{\dagger}(S_{i}, S_{j}) \hat{D}^{S_{i} \dagger}], \\
    = \Big(\Tr[\hat{D}^{S_{i}} \hat{\xi}(S_{i}, S_{j}) \hat{D}^{S_{j} \dagger}]\Big)^{*}.\nonumber
\end{align}
Equation \ref{Distribution} can therefore be written as
\begin{align} \label{DistributionHermiticityRewrite}
    W & = \sum_{i} \Tr[\hat{D}^{S_{i} \dagger} \hat{\rho}(S_{i}) \hat{D}^{S_{i}}] 
      + \sum_{i > j} \bigg\{ \Tr[\hat{D}^{S_{i} \dagger}
	\hat{\xi}(S_{i}, S_{j}) \hat{D}^{S_{j}}] \nonumber \\ 
      \mt \qquad + \Big(\Tr[\hat{D}^{S_{i}} \hat{\xi}(S_{i}, S_{j})
	\hat{D}^{S_{j}}]\Big)^{*} \bigg\} \\
	& = \sum_{i} \Tr[\hat{D}^{S_{i} \dagger} \hat{\rho}(S_{i}) \hat{D}^{S_{i}}]  + \sum_{i > j} 2 \Re \Tr[\hat{D}^{S_{i} \dagger} \hat{\xi}(S_{i}, S_{j}) 
	\hat{D}^{S_{j}}].\nonumber
\end{align}
Written using index notation, the expression is
\begin{equation} \label{DistributionIndexNotation}
  W = \sum_{i} \sum_{\lambda, \lambda', \eta} D^{S_{i}*}_{\lambda, \eta} \rho_{\lambda, \lambda'}(S_{i}) D^{S_{i}}_{\lambda', \eta}
	+ \sum_{i>j} \sum_{\lambda, \lambda', \eta} 2 \Re\Big[D^{* S_{i}}_{\lambda, \eta} \xi_{\lambda, \lambda'}(S_{i}, S_{j}) 
        D^{S_{j}}_{\lambda', \eta}\Big].
\end{equation}

Since the diagonal matrices are hermitian,
not all of its elements are independent. We can therefore simplify the expression by writing it only in terms of the independent elements,
which are the diagonal and top diagonal ones ($\lambda = \lambda'$ and $\lambda' > \lambda$, respectively). The first term in the last 
equation can therefore be written as
\begin{equation} \label{TopDiagonalTerms}
  \begin{split}
    \sum_{i} \sum_{\lambda, \lambda', \eta} D^{* S_{i}}_{\lambda, \eta} \rho_{\lambda, \lambda'}(S_{i}) D^{S_{i}}_{\lambda', \eta}
    & = \sum_{\lambda, \eta} D^{* S_{i}}_{\lambda, \eta} \rho_{\lambda, \lambda}(S_{i}) D^{S_{i}}_{\lambda, \eta} \\
    \mt + \sum_{\lambda' > \lambda, \eta} 
	  \Big\{D^{* S_{i}}_{\lambda, \eta} \rho_{\lambda, \lambda'}(S_{i}) D^{S_{i}}_{\lambda', \eta}
	  + D^{* S_{i}}_{\lambda', \eta} \rho_{\lambda', \lambda}(S_{i}) D^{S_{i}}_{\lambda, \eta}\Big\}.
  \end{split}
\end{equation}
The first term contains the diagonal terms. Inside the brackets, the first term is a sum over top diagonal elements, and the second is a
sum over the bottom diagonal elements. Using the hermitian property of the diagonal matrix,
\begin{equation} \label{Hermiticity2}
 \begin{split}
  \hat{\rho}(S_{i}) & = \hat{\rho}^{\dagger}(S_{i}), \\
  \rho(S_{i})_{\lambda', \lambda} & = \rho^{*}(S_{i})_{\lambda, \lambda'},
 \end{split}
\end{equation}
the expression inside the brackets becomes
\begin{equation} \label{BracketSimplification}
  \begin{split}
    D^{* S_{i}}_{\lambda, \eta} \rho_{\lambda, \lambda'}(S_{i}) D^{S_{i}}_{\lambda', \eta}
	+ D^{* S_{i}}_{\lambda', \eta} \rho_{\lambda', \lambda}(S_{i}) D^{S_{i}}_{\lambda, \eta} 
      & = D^{* S_{i}}_{\lambda, \eta} \rho_{\lambda, \lambda'}(S_{i}) D^{S_{i}}_{\lambda', \eta}
	    + \Big(D^{* S_{i}}_{\lambda, \eta} \rho_{\lambda, \lambda'}(S_{i}) D^{S_{i}}_{\lambda', \eta}\Big)^{*} \\
      & = 2 \Re[D^{* S_{i}}_{\lambda, \eta} \rho_{\lambda, \lambda'}(S_{i}) D^{S_{i}}_{\lambda', \eta}].
  \end{split}
\end{equation}
Substituting this last equation into eqn. (\ref{DistributionIndexNotation}), we get
\begin{equation} \label{DistributionReal}
  \begin{split}
    W & = \sum_{i} \Bigg\{\sum_{\lambda, \eta} D^{* S_{i}}_{\lambda, \eta} \rho_{\lambda, \lambda}(S_{i}) 
	D^{S_{i}}_{\lambda, \eta} + \sum_{\substack{\lambda > \lambda' \\ \eta}} 2 \Re \Big[ D^{* S_{i}}_{\lambda, \eta}
	\rho_{\lambda, \lambda'}(S_{i}) D^{S_{i}}_{\lambda', \eta} \Big] \Bigg\} \\
      \mt + \sum_{i > j} \sum_{\lambda, \lambda', \eta} 2 \Re \Big[ D^{* S_{i}}_{\lambda, \eta} \xi_{\lambda, \lambda'}(S_{i}, S_{j}) 
	D^{S_{j}}_{\lambda', \eta} \Big].
 \end{split}
\end{equation}
The first term in the brackets is a sum over diagonal elements, which are real. 
Therefore, this expression makes explicit the fact that the distribution is a real quantity.  

The $\phi^{*}$ dependence of the distribution occurs only in the Wigner-$D$ functions and in a very simple way,
\begin{equation} \label{LittleD}
 D^{S}_{\lambda, \lambda'}(\theta^{*}, \phi^{*}) = e^{-i \lambda \phi^{*}} d^{S}_{\lambda, \lambda'}(\theta^{*}),
\end{equation}
where these Wigner $d$-functions are well known. Substituting eqn. (\ref{LittleD}) into (\ref{DistributionReal}) and using the following identity for
complex numbers,
\begin{equation} \label{ComplexNumbersIdentity1}
 \Re[z_{1} z_{2}] = \Re[z_{1}] \Re[z_{2}] - \Im[z_{1}] \Im[z_{2}],
\end{equation}
we get,
\begin{equation} \label{DistributionFinal}
 \begin{split}
 W(\theta^{*}, \phi^{*}) & = \sum_{i} \Bigg\{\sum_{\lambda} \widetilde{d}^{S_{i}}_{\lambda, \lambda}(\theta^{*}) \rho_{\lambda, \lambda}(S_{i}) 
      + \sum_{\lambda > \lambda'} \widetilde{d}^{S_{i}}_{\lambda, \lambda'}(\theta^{*}) 
      \Bigg( \Re[\rho_{\lambda, \lambda'}(S_{i})] \cos[(\lambda-\lambda')\phi^{*})] \\ 
      \mt - \Im[\rho_{\lambda, \lambda'}(S_{i})] \sin[(\lambda-\lambda')\phi^{*})] \Bigg) \Bigg\} \\
       \mt + \sum_{i > j} \sum_{\lambda, \lambda'} \widetilde{d}^{S_{i}, S_{j}}_{\lambda, \lambda'}(\theta^{*}) 
       \Bigg(\Re[\xi_{\lambda, \lambda'}(S_{i}, S_{j})] \cos[(\lambda-\lambda')\phi^{*})] \\ 
      \mt - \Im[\xi_{\lambda, \lambda'}(S_{i}, S_{j})] \sin[(\lambda-\lambda')\phi^{*})] \Bigg)
 \end{split}
\end{equation}
where
\begin{equation} \label{WignerLittleDSumDefinition}
 \begin{split}
  \widetilde{d}^{S_{i}}_{\lambda, \lambda'}(\theta^{*}) & \equiv 2 \sum_{\eta} 
       d^{S_{i}}_{\eta, \lambda}(\theta^{*}) d^{S_{i}}_{\eta, \lambda'}(\theta^{*}), \\
  \widetilde{d}^{S_{i}, S_{j}}_{\lambda, \lambda'}(\theta^{*}) & \equiv 2 \sum_{\eta} 
       d^{S_{i}}_{\eta, \lambda}(\theta^{*}) d^{S_{j}}_{\eta, \lambda'}(\theta^{*}).
 \end{split}
\end{equation}
Therefore, the entire dependence on $\theta^{*}$ is contained in the functions $\widetilde{d}^{S_{i}}_{\lambda, \lambda'}(\theta^{*})$ and
$\widetilde{d}^{S_{i}, S_{j}}_{\lambda, \lambda'}(\theta^{*})$, which are simply bi-linear combinations of the well known $d$-functions, while
the entire $\phi^{*}$ dependence is only contained in cosine and sine functions. Every term in the expression for $W$ will either be
proportional to a cosine or sine function of $\phi^{*}$, or be independent of it.

Eqn. (\ref{DistributionFinal}) is general for any photoproduction reaction with two mesons in the final state that has any number of
quasi-two-body intermediate states of arbitrary spin contributing to the process. If we apply it to the case where only one 
intermediate state contributes (i.e., the index $i$ runs over only one value and the term with the sum over both $i$ and $j$ vanishes) with a decaying vector meson V ($S_{i} = 1$), we recover eqn. (\ref{WDistributionV}). If the intermediate states contains instead a decaying spin-3/2 baryon $B_{3/2}$ ($S_{i} = 3/2$), we recover eqn. (\ref{WDistributionBThreeHalfs}). 

\section{General Expression of Observables in Terms of SDME's}
Since the decomposition of the spin density matrix into 16 matrices shown in eqn. (\ref{16Matrices}) is general for any final state whose initial state has two particles each with two possible spin projections, and since it can be shown to also apply to interefence matrices, the matrices in eqn. (\ref{DistributionFinal}) can therefore also be decomposed. This will give us an expression of the form shown in eqn. (\ref{WDecomposition}), with each $W^{i}$ given by
\begin{equation} \label{DistributionFinal16}
 \begin{split}
    W^{i}(\theta^{*}, \phi^{*}) & = \sum_{k} \Bigg\{\sum_{\lambda} \widetilde{d}^{S_{k}}_{\lambda, \lambda}(\theta^{*}) \rho^{i}_{\lambda, \lambda}(S_{k})
      + \sum_{\lambda > \lambda'} \widetilde{d}^{S_{k}}_{\lambda, \lambda'}(\theta^{*}) 
      \Bigg( \Re[\rho^{i}_{\lambda, \lambda'}(S_{k})] \cos[(\lambda-\lambda')\phi^{*})] \\ 
      \mt - \Im[\rho^{i}_{\lambda, \lambda'}(S_{k})] \sin[(\lambda-\lambda')\phi^{*})] \Bigg) \Bigg\} \\
      \mt \sum_{l > k} \sum_{\lambda, \lambda'} \widetilde{d}^{S_{k}, S_{l}}_{\lambda, \lambda'}(\theta^{*}) 
      \Bigg(\Re[\xi^{i}_{\lambda, \lambda'}(S_{k}, S_{l})] \cos[(\lambda-\lambda')\phi^{*})] \\ 
      \mt - \Im[\xi^{i}_{\lambda, \lambda'}(S_{k}, S_{l})] \sin[(\lambda-\lambda')\phi^{*})] \Bigg),
 \end{split}
\end{equation}
the matrix elements are given by
\begin{equation} \label{DensityMatrixSigmaMatrixElements16}
  \begin{split}
    \rho^{0}_{\lambda \lambda'}(S_{k})
      & =  \frac{1}{4}\sum_{\substack{\lambda_{\gamma}\lambda_{N} \\ \lambda_{H}}} 
                        M_{\lambda\lambda_{H}; \lambda_{N} \lambda_{\gamma}}
                        M^{*}_{\lambda'\lambda_{H}; \lambda_{N} \lambda_{\gamma}}, \\
	\rho^{i (\gamma)}_{\lambda \lambda'}(S_{k})
      & =\frac{1}{4} \sum_{\substack{\lambda_{N} \lambda_{H}\\
                        \lambda_{\gamma}' \lambda_{\gamma}}} 
                        M_{\lambda\lambda_{H}; \lambda_{N} \lambda_{\gamma}}
                        \sigma^{i}_{\lambda_{\gamma} \lambda'_{\gamma}}
                        M^{*}_{\lambda'\lambda_{H}; \lambda_{N} \lambda'_{\gamma}}, \\
    \rho^{i (N)}_{\lambda \lambda'}(S_{k})
      & =  \frac{1}{4}\sum_{\substack{\lambda_{N} \lambda'_{N} \\
                        \lambda_{\gamma}\lambda_{H}}} 
                        M_{\lambda\lambda_{H}; \lambda_{N} \lambda_{\gamma}}
                        \sigma^{i}_{\lambda_{N} \lambda'_{N}}
                        M^{*}_{\lambda'\lambda_{H}; \lambda'_{N} \lambda_{\gamma}}, \\    
    \rho^{ij (\gamma N)}_{\lambda \lambda'}(S_{k})
      & =\frac{1}{4} \sum_{\substack{\lambda_{\gamma} \lambda'_{\gamma} 						\\ \lambda_{N} \lambda'_{N}
                         \\ \lambda_{H}}} 
                        M_{\lambda\lambda_{H}; \lambda_{N} \lambda_{\gamma}}\sigma^{i}_{\lambda_{\gamma}\lambda'_{\gamma}}
                        \sigma^{j}_{\lambda_{N} \lambda'_{N}}
                        M^{*}_{\lambda'\lambda_{H}; \lambda'_{N} \lambda'_{\gamma}},
  \end{split}
\end{equation}
and
\begin{align} \label{DensityMatrixSigmaMatrixElements16Int}
    \xi^{0}_{\lambda \lambda'}(S_{k},S_{l})
      & =  \frac{1}{4}\sum_{\substack{\lambda_{\gamma}\lambda_{N} \\ \lambda_{H}}} 
                        M_{\lambda\lambda_{H}; \lambda_{N} \lambda_{\gamma}}
                        M^{*}_{\lambda'\lambda_{H}; \lambda_{N} \lambda_{\gamma}}, \nonumber \\ 
    \xi^{i (\gamma)}_{\lambda \lambda'}(S_{k},S_{l})
      & =\frac{1}{4} \sum_{\substack{\lambda_{N} \lambda_{H}\\
                        \lambda_{\gamma} \lambda'_{\gamma}}} 
                        M_{\lambda\lambda_{H}; \lambda_{N} \lambda_{\gamma}}
                        \sigma^{i}_{\lambda_{\gamma} \lambda'_{\gamma}}
                        M^{*}_{\lambda'\lambda_{H}; \lambda_{N} \lambda'_{\gamma}}, \\
	\xi^{i (N)}_{\lambda \lambda'}(S_{k},S_{l})
      & =  \frac{1}{4}\sum_{\substack{\lambda_{N} \lambda'_{N} \\
                        \lambda_{\gamma}\lambda_{H}}} 
                        M_{\lambda\lambda_{H}; \lambda_{N} \lambda_{\gamma}}
                        \sigma^{i}_{\lambda_{N} \lambda'_{N}}
                        M^{*}_{\lambda'\lambda_{H}; \lambda'_{N} \lambda_{\gamma}}, \nonumber \\
    \xi^{ij (\gamma N)}_{\lambda \lambda'}(S_{k},S_{l})
      & =\frac{1}{4} \sum_{\substack{\lambda_{\gamma} \lambda'_{\gamma} 						\\ \lambda_{N} \lambda'_{N} \\ \lambda_{H}}} 
                        M_{\lambda\lambda_{H}; \lambda_{N}' \lambda_{\gamma}'}
                        \sigma^{i}_{\lambda_{\gamma}'\lambda_{\gamma}}\sigma^{j}_{\lambda_{N}' \lambda_{N}}
                        M^{*}_{\lambda'\lambda_{H}; \lambda_{N} \lambda_{\gamma}},\nonumber 
\end{align}
and $\lambda_{H}$ is the helicity of the spectator hadron, the one that does not decay, either a meson or a baryon. In the case of the interference matrices $\hat{\xi}$, $\lambda$ and $\lambda'$ are the helicities of the two hadrons that are interfering, whose spins are $S_{k}$ and $S_{l}$. We can use the expressions for the Pauli matrices in terms of the Kronecker delta in eqn. (\ref{PauliMatrixKroneckerDelta}) to simplify the previous expressions. For simplicity, we will show the simplified expressions only for the interference matrices, with the understanding that the only difference between them and the non-interference matrices is that in the latter, the indices $\lambda$ and $\lambda'$ refer both to helicities of a single hadron. The simplified expression for the density matrix related to the unpolarized cross section is
\begin{equation} \label{DensityMatrixHelicityAmplitude}
    \xi^{0}_{\lambda \lambda'}(S_{k}, S_{l})
     =  \frac{1}{4}\sum_{\substack{\lambda_{\gamma} \lambda_{N}\\ \lambda_{H}}} 
                        M_{\lambda \lambda_{H}; \lambda_{\gamma} \lambda_{N}}
                        M^{*}_{\lambda' \lambda_{H}; \lambda_{\gamma} \lambda_{N}}.
\end{equation}
The ones for the density matrices related to the beam observables are
\begin{equation} \label{DensityMatrixHelicityAmplitudesGamma}
\begin{split}
 & \xi^{x (\gamma)}_{\lambda \lambda' }(S_{k}, S_{l})
     =  \frac{1}{4}\sum_{\substack{\lambda_{\gamma} \lambda_{N}\\ \lambda_{H}}} 
                        M_{\lambda \lambda_{H}; \lambda_{\gamma} \lambda_{N}}
                        M^{*}_{\lambda' \lambda_{H}; -\lambda_{\gamma} \lambda_{N}}, \\
 & \xi^{y (\gamma)}_{\lambda \lambda' }(S_{k}, S_{l})
     =-\frac{i}{4} \sum_{\substack{\lambda_{\gamma} \lambda_{N}\\ \lambda_{H}}} 
                        \lambda_{\gamma} M_{\lambda \lambda_{H}; \lambda_{N} \lambda_{\gamma}}
                        M^{*}_{\lambda' \lambda_{H}; -\lambda_{\gamma} \lambda_{N}}, \\
 & \xi^{z (\gamma)}_{\lambda \lambda'}(S_{k}, S_{l})
     = \frac{1}{4} \sum_{\substack{\lambda_{\gamma} \lambda_{N}\\ \lambda_{H}}} 
                        \lambda_{\gamma} M_{\lambda \lambda_{H}; \lambda_{\gamma} \lambda_{N}}
                        M^{*}_{\lambda' \lambda_{H}; \lambda_{\gamma} \lambda_{N}},
\end{split}
\end{equation}
and the ones related to the target observables are
\begin{align} \label{DensityMatrixHelicityAmplitudesNucleon}
 & \xi^{x (N)}_{\lambda \lambda'}(S_{k}, S_{l})
     =  \frac{1}{4}\sum_{\substack{\lambda_{\gamma} \lambda_{N}\\ \lambda_{H}}} 
                        M_{\lambda \lambda_{H}; \lambda_{\gamma} \lambda_{N}}
                        M^{*}_{\lambda' \lambda_{H}; \lambda_{\gamma} - \lambda_{N}}, \nonumber \\
 & \xi^{y (N)}_{\lambda \lambda' }(S_{k}, S_{l})
     =-\frac{i}{4} \sum_{\substack{\lambda_{\gamma} \lambda_{N}\\ \lambda_{H}}} 
                        2 \lambda_{N} M_{\lambda \lambda_{H}; \lambda_{N} \lambda_{\gamma}}
                        M^{*}_{\lambda' \lambda_{H}; \lambda_{\gamma} - \lambda_{N}}, \\
 & \xi^{z (N)}_{\lambda \lambda'}(S_{k}, S_{l})
     = \frac{1}{4} \sum_{\substack{\lambda_{\gamma} \lambda_{N}\\ \lambda_{H}}} 
                        2 \lambda_{N} M_{\lambda \lambda_{H}; \lambda_{\gamma} \lambda_{N}}
                        M^{*}_{\lambda' \lambda_{H}; \lambda_{\gamma} \lambda_{N}}.\nonumber
\end{align}
We will not show the expressions for the ones related to the beam/target double polarization observables, but it is easy to see from the pattern of the previous expressions what they would look like. The first pattern is that matrices with superscript of 2 and 3 will have an extra factor of $-\lambda_{\gamma}$ or $-2 \lambda_{N}$ in the summation, the former for those with $(\gamma)$ in the superscript and the latter for those with $(N)$ in the superscript. Additionally, those with a superscript of 2 will have an extra factor of $i$. The second pattern is that matrices with superscript 1 and 2 will have the indices $\lambda_{\gamma}$ or $\lambda_{N}$ with opposite signs in each of the helicity amplitudes, the former for those with $(\gamma)$ in the superscript and the latter for those with $(N)$ in the superscript. These patterns will be important in deriving parity relations among these matrix elements. These in turn will be used to show which of the polarization observables will be even and which will be odd in $\phi^{*}$.

To relate the $W^{i}$'s in eqn. (\ref{DistributionFinal16}) to the polarization observables we follow the same procedure used in section \ref{ObservablesQuasiTwoBodyState} for the case of vector meson decays. To summarize, the trace is taken on both sides of eqn. (\ref{DensityMatrixObservables}). The terms on the right-hand side proportional to the Pauli matrices, $\hat{\sigma}^{i}$, vanish, while the left-hand side is equal to $W$. This expression is then compared to eqn. (\ref{WDecomposition}) to show that each $W^{i}$ is proportional to a polarization observable. We reproduce the results here for convenience,
\begin{equation} \label{WObservables3}
 \begin{split}
  W^{0}(\theta^{*}, \phi^{*}) & = I_{0}(\theta^{*}, \phi^{*}), \\
  W^{i (N)}(\theta^{*}, \phi^{*}) & = I_{0} P^{i}(\theta^{*}, \phi^{*}), \\
  W^{x (\gamma)}(\theta^{*}, \phi^{*}) & = I_{0} \cos2\beta I^{c}(\theta^{*}, \phi^{*}), \\
  W^{y (\gamma)}(\theta^{*}, \phi^{*}) & = I_{0} \sin2\beta I^{s}(\theta^{*}, \phi^{*}), \\
  W^{z (\gamma)}(\theta^{*}, \phi^{*}) & = I_{0} I^{\odot}(\theta^{*}, \phi^{*}), \\
  W^{ix (\gamma N)}(\theta^{*}, \phi^{*}) & = I_{0} \cos2\beta P^{ic}(\theta^{*}, \phi^{*}), \\
  W^{iy (\gamma N)}(\theta^{*}, \phi^{*}) & = I_{0} \sin2\beta P^{is}(\theta^{*}, \phi^{*}), \\
  W^{iz (\gamma N)}(\theta^{*}, \phi^{*}) & = I_{0} P^{i\odot}(\theta^{*}, \phi^{*}).
 \end{split}
\end{equation}

\section{General Parity Relations of the SDME's}
If the reaction is parity conserving, the helicity amplitudes can be related by equation (\ref{TransitionMatrixParityRelations}). When this is applied to eqns. (\ref{DensityMatrixHelicityAmplitude}) to (\ref{DensityMatrixHelicityAmplitudesNucleon}), we get that the parity relations of the 16 matrices will have the form
\begin{equation} \label{DensityMatrixParityRelation}
    \begin{split}
    \xi_{-\lambda -\lambda'} & = \zeta (-1)^{\lambda + \lambda'}\xi_{\lambda \lambda'}, \\
    \zeta & \in \{ 1, -1 \}.
    \end{split}
\end{equation}
The value of $\zeta$ depends on three factors: its superscript (e.g., $x(\gamma), z(N), zy(\gamma N)$, etc.), the intrinsic parities of the two unstable hadrons that are interfering, and on whether the density matrix describes the interference between mesons or baryons. We will now derive the previous equation and the conditions that determine the value of $\zeta$.  

When the parity relations are applied to eqns. (\ref{DensityMatrixHelicityAmplitude}) to (\ref{DensityMatrixHelicityAmplitudesNucleon}), four factors will appear, each equal to either $1$ or $-1$. Two of them are related to the properties of the hadrons involved, while the other two are related to the superscript of the 16 matrices. 

The first factor has to do with the intrinsic parities of the hadrons involved in the reaction. When the parity relations are applied to eqns. (\ref{DensityMatrixHelicityAmplitude}) to (\ref{DensityMatrixHelicityAmplitudesNucleon}), this factor appears:
\begin{equation} \label{IntrinsicParityVanish}
    \xi_{-\lambda -\lambda'} \propto \eta\eta'(\eta_{H})^{2}(\eta_{\gamma}\eta_{N})^{-2}\xi_{\lambda \lambda'} = \eta\eta'\xi_{\lambda \lambda'},
\end{equation}
where $\eta$ and $\eta'$ are the intrinsic parities of the two unstable resonances that are interfering, and $H$ labels the spectator hadron. The factors that are raised to the power of two are equal to one because all $\eta = \pm 1$. For the non-interference matrices $\eta = \eta'$, so that the factor $\eta \eta'$ in the previous equation will equal $1$. This last equation tells us that, when the parity relations of the helicity amplitudes are applied in eqns. (\ref{DensityMatrixHelicityAmplitude}) to (\ref{DensityMatrixHelicityAmplitudesNucleon}), the overall expression may gain a factor of $-1$ depending on the intrinsic parities of the unstable hadrons: If they have opposite parities, the expression will gain a factor of $-1$.

The second factor depends on whether the spectator hadron is a meson or a baryon. This is because of the factor
\begin{equation} \label{DensityMatrixDecayFactor}
    \xi_{-\lambda -\lambda'} \propto (-1)^{(\lambda + \lambda' - 2\lambda_{H})}\xi_{\lambda \lambda'} = \pm (-1)^{(\lambda + \lambda')}\xi_{\lambda \lambda'}
\end{equation}
where $\lambda_{H}$ is the helicity of the spectator hadron. If this hadron is a baryon, the expression gets a minus sign because $2\lambda_{H}$ is twice a half-integer, which is always an odd number. If it is a meson, the expression does not get the minus sign because $2\lambda_{H}$ is twice an integer, which is always an even number. Since the unstable resonance will be a meson (baryon) when the spectator hadron is a baryon (meson), we can equivalently say that the density matrix describing mesons will get a factor of $-1$, while those describing baryons will not.

The third factor depends on which helicity indices on the right-hand side of eqns. (\ref{DensityMatrixHelicityAmplitude}) to (\ref{DensityMatrixHelicityAmplitudesNucleon}) have minus signs. For example, all of the spin density matrices that have superscripts of $0$ or $z$ will have the factor (the symbol ``$\ |\ $'' represents the word ``or'')
\begin{equation} \label{DensityMatrix03Factor}
    \xi^{0|z}_{-\lambda -\lambda'} \propto (-1)^{2(\lambda_{\gamma}-\lambda_{N})}\xi^{0|z}_{\lambda \lambda'} = -\xi^{0|z}_{\lambda \lambda'},
\end{equation}
those with superscripts of $x$ or $y$ with $(\gamma)$ will have
\begin{equation} \label{DensityMatrix12FactorGamma}
    \xi^{x|y (\gamma)}_{-\lambda -\lambda'} \propto (-1)^{-2\lambda_{N}}\xi^{x|y (\gamma)}_{\lambda \lambda'} = -\xi^{x|y (\gamma)}_{\lambda \lambda'},
\end{equation}
and those with superscripts of $x$ or $y$ with $(N)$ will have
\begin{equation} \label{DensityMatrix12FactorN}
    \xi^{x|y (N)}_{-\lambda -\lambda'} \propto (-1)^{2\lambda_{\gamma}}\xi^{x|y (N)}_{\lambda \lambda'} = \xi^{x|y (N)}_{\lambda \lambda'}.
\end{equation}
Eqn. (\ref{DensityMatrix03Factor}) gets a $-1$ because $2(\lambda_{\gamma}-\lambda_{N})$ is twice the difference of a integer with a half-integer, which is always an odd integer. Eqn. (\ref{DensityMatrix12FactorGamma}) gets a $-1$ because $-2\lambda_{N}$ is twice a half-integer, which is always an odd integer. Eqn. (\ref{DensityMatrix12FactorN}) gets a $1$ because $2\lambda_{\gamma}$ is twice an integer, which is always an even integer. For the density matrices that have $(\gamma N)$ in their superscript, both factors that appear in the associated matrices with superscript $(\gamma)$ and $(N)$ appear. For example,
\begin{equation} \label{DensityMatrixFactorExample}
    \xi^{xz(\gamma N)}_{-\lambda -\lambda'} \propto - \xi^{xz(\gamma N)}_{\lambda \lambda'}
\end{equation}
because $\xi^{x (\gamma)}$ would get a factor of $-1$ while $\xi^{z (N)}$ would get a factor of $1$.

The fourth factor is related to whether the expression on the right-hand side of eqns. (\ref{DensityMatrixHelicityAmplitude}) to (\ref{DensityMatrixHelicityAmplitudesNucleon}) have a factor of $\lambda_{\gamma}$ or $\lambda_{N}$ in the summation. This follows from this identity, which is valid for any arbitrary function of an index $f(\lambda)$,
\begin{equation} \label{DummyIndex}
\sum_{\lambda}\lambda f(-\lambda) = -\sum_{\lambda}\lambda f(\lambda).
\end{equation}
The transformation $\lambda \rightarrow -\lambda$ can be done because $\lambda$ is a dummy index. Therefore, all matrices with superscript $2$ or $3$ will get a factor of $-1$. If the matrix has two indices in its superscript, you multiply the factors that would have appeared for the two associated matrices that have those indices. For example, the matrix $\xi^{xz(\gamma N)}$ would gets a factor of $-1$ because $\xi^{x(\gamma)}$ would get a factor of $1$ while $\xi^{z(N)}$ would get a factor of $-1$. 

When all of the previous properties are taken into account, we find that the parity relations of the 16 matrices are
\begin{align} \label{DensityMatrixParity}
  & \xi^{0}_{-\lambda, -\lambda'} = \bar{\eta} (-1)^{2 S^{*} +1}\zeta^{0}(-1)^{\lambda + \lambda'} \xi^{0}_{\lambda, \lambda'}, \nonumber \\
  & \xi^{i (\gamma)}_{-\lambda, -\lambda'} = \bar{\eta} (-1)^{2 S^{*} +1}\zeta^{(\gamma) i} (-1)^{\lambda + \lambda'}
       \xi^{i (\gamma)}_{\lambda, \lambda'}, \\
  & \xi^{i (N)}_{-\lambda, -\lambda'} = \bar{\eta} (-1)^{2 S^{*} +1}\zeta^{(N) i} (-1)^{\lambda + \lambda'} 
       \xi^{i (N)}_{\lambda, \lambda'}, \nonumber \\
  & \xi^{ij (\gamma N)}_{-\lambda, -\lambda'} = \bar{\eta} (-1)^{2 S^{*} +1}\zeta^{(\gamma) i} \zeta^{(N) j} 
       (-1)^{\lambda + \lambda'} \xi^{ij (\gamma N)}_{\lambda, \lambda'}, \nonumber \\
    & \nonumber \\
    & \qquad \qquad \qquad \bar{\eta} \equiv \eta\eta',\nonumber
\end{align}
where $\eta\eta'$ are the intrinsic parities of the two unstable hadrons that interfere. Therefore, $\bar{\eta}$ will be called the intrinsic parity factor. For non-interference spin density matrices, this factor will always equals $1$. $S^{*}$ is the spin of either one of the interfering hadrons. Since they will both have either integer or half-integer spin, it does not matter which one is used. Since the factor $(-1)^{2 S^{*} + 1}$ depends on whether the unstable hadron is a fermion or a boson, we will call this the statistics factor. The remaining factor depends only on the type of matrix, and are given by
\begin{equation} \label{SDMEFactor}
    \begin{split}
    \zeta^{0} & = -1, \\
   \zeta^{i (\gamma)} & = \begin{cases}
                       \hphantom{-} 1, & \text{if } i = y, z \\
                      -1, & \text{if } i = x
                     \end{cases}, \\
  \zeta^{i (N)} & = \begin{cases}
                       \hphantom{-} 1, & \text{if } i = x, z \\
                      -1, & \text{if } i = y. 
                     \end{cases}.              
 \end{split}
\end{equation}
We will therefore call this the SDME factor. As an example, suppose we want to find the parity relations for the non-interference spin density matrix $\rho^{z (\gamma)}_{\lambda \lambda'}(S = 2)$, for the decay of a spin-2 pseudoscalar meson. In this case, $\bar{\eta} = 1$ because it is a non-interference matrix, $(-1)^{2 S^{*} + 1} = -1$ because $S^{*} = 2$, and $\zeta^{z (\gamma)} = 1$ from eqn. (\ref{SDMEFactor}). The parity relation will therefore be
\begin{equation} \label{ParityRelationExample}
    \rho^{z (\gamma)}_{-\lambda - \lambda'}(S = 2) = - (-1)^{\lambda + \lambda'} \rho^{z (\gamma)}_{-\lambda - \lambda'}(S = 2).
\end{equation}

\section{Parity Considerations in General Observables in Terms of the SDME's}
We will now show why when the reaction is parity conserving only terms that are proportional to either cosine or sine functions will
appear in the expression. From the definitions of the $\widetilde{d}^{S_{i}}$ and $\widetilde{d}^{S_{i},S_{j}}$ in eqn. (\ref{WignerLittleDSumDefinition}) it can be shown that
\begin{equation} \label{dTildeProperty2}
	\begin{split}
		\widetilde{d}^{S_{i}}_{\lambda \lambda'} & = \widetilde{d}^{S_{i}}_{\lambda' \lambda} \\
		\widetilde{d}^{S_{i}, S_{j}}_{\lambda \lambda'} & = \widetilde{d}^{S_{i}, S_{j}}_{\lambda' \lambda}.
	\end{split}
\end{equation}
Also, the property
\begin{equation} \label{LittleDProperty}
  d^{S_{i}}_{-\eta -\lambda} = (-1)^{\eta - \lambda} d^{S_{i}}_{\eta \lambda}, 
\end{equation}
can be used to show that,
\begin{equation} \label{dTildeProperty}
  \begin{split}
    \widetilde{d}^{S_{i}}_{\lambda \lambda'} & = (-1)^{2S_{i}} (-1)^{\lambda + \lambda'} \widetilde{d}^{S_{i}}_{-\lambda -\lambda'}, \\
    \widetilde{d}^{S_{i}, S_{j}}_{\lambda \lambda'} & = (-1)^{2S_{i}} (-1)^{\lambda + \lambda'} 
      \widetilde{d}^{S_{i}, S_{j}}_{-\lambda -\lambda'}.
  \end{split}
\end{equation}
These last equations along with eqns. (\ref{DensityMatrixParity}), the hermiticity condition in
eqn. (\ref{Hermiticity}), the even and odd property of the 
trigonometric functions, and the following properties from the real and imaginary part functions,
\begin{equation} \label{ComplexNumbersIdentity2}
  \begin{split}
    \Re[z^{*}] & = \Re[z], \\
    \Im[z^{*}] & = - \Im[z], \\
    z & \in \mathbb{C},
  \end{split}
\end{equation}
can be used to shown that
\begin{equation} \label{ParityRelation1}
  \begin{split}
    \widetilde{d}^{S_{k}}_{\lambda \lambda'}(\theta^{*}) \Re[\rho^{i}_{\lambda \lambda'}(S_{k})] \cos[(\lambda - \lambda') \phi^{*}] & = 
      \zeta^{i} \widetilde{d}^{S_{k}}_{-\lambda' -\lambda}(\theta^{*}) \Re[\rho^{i}_{-\lambda' -\lambda}(S_{i})] \cos[(-\lambda' + \lambda) \phi^{*}], \\
    \widetilde{d}^{S_{k}}_{\lambda \lambda'}(\theta^{*}) \Im[\rho^{i}_{\lambda \lambda'}(S_{k})] \sin[(\lambda - \lambda') \phi^{*}] & = 
      -\zeta^{i} \widetilde{d}^{S_{i}}_{-\lambda' -\lambda}(\theta^{*}) \Im[\rho_{-\lambda' -\lambda}(S_{k})] \sin[(-\lambda' + \lambda) \phi^{*}].
  \end{split}
\end{equation}
Since $\zeta^{i} = \pm 1$, this means that applying the parity relations to each $W^{i}$ will give you either a plus or minus sign to each of the terms proportional to the real part function and the opposite sign to each of the terms proportional to the imaginary part function. The terms that get the minus signs will always cancel with another term, and therefore either all terms will be proportional to the cosine function or all terms will be proportional to the sine function will vanish. For example, the expression for $W^{x (\gamma)}$ for a reaction with a vector meson $V$ intermediate state with corresponding spin density matrix $\hat{\rho}^{x (\gamma)}(V)$ will contain the following terms,
\begin{equation} \label{ParityCancelation1}
  \begin{split}
    W^{x (\gamma)} & = \cdots + \bigg(\widetilde{d}^{1}_{1 0} \Re\Big[\rho^{x (\gamma)}_{1 0}(V)\Big]
	+ \widetilde{d}^{1}_{0 -1} \Re\Big[\rho^{x (\gamma)}_{0 -1}(V)\Big]\bigg)
	\cos(\phi^{*}) \\
	\mt - \bigg(\widetilde{d}^{1}_{1 0} \Im\Big[\rho^{x (\gamma)}_{1 0}(V)\Big] 
	+ \widetilde{d}^{1}_{0 -1} \Im\Big[\rho^{x (\gamma)}_{0 -1}(V)\Big]\bigg)
	\sin(\phi^{*}) + \cdots, \\
      & = \cdots + \bigg(\widetilde{d}^{1}_{1 0} \Re\Big[\rho^{x (\gamma)}_{1 0}(V)\Big] + \widetilde{d}^{1}_{1 0} \Re\Big[\rho^{x (\gamma)}_{1 0}(V)\Big]\bigg)
	\cos(\phi^{*}) \\
	\mt - \bigg(\widetilde{d}^{1}_{1 0} \Im\Big[\rho^{x (\gamma)}_{1 0}(V)\Big] - \widetilde{d}^{1}_{1 0} \Im\Big[\rho^{x (\gamma)}_{1 0}(V)\Big]\bigg)
	\sin(\phi^{*}) + \cdots, \\
      & = \cdots + 2 \widetilde{d}^{1}_{1 0} \Re\Big[\rho^{x (\gamma)}_{1 0}(V)\Big] \cos(\phi^{*}) + \cdots.
  \end{split}
\end{equation}
For the terms that contain the interference matrices, the property that the cosine function is even while the sine function is odd can be used to show this identity,
\begin{equation} \label{ParityRelation2}
  \begin{split}
    \widetilde{d}^{S_{i}, S_{j}}_{\lambda \lambda'}(\theta^{*}) \Re[\xi_{\lambda \lambda'}(S_{i}, S_{j})] \cos[(\lambda - \lambda') \phi^{*}] & = 
      \pm \widetilde{d}^{S_{i}, S_{j}}_{-\lambda -\lambda'}(\theta^{*}) \Re[\xi_{-\lambda - \lambda'}(S_{i}, S_{j})]
      \cos[(-\lambda + \lambda') \phi^{*}], \\
    \widetilde{d}^{S_{i}, S_{j}}_{\lambda \lambda'}(\theta^{*}) \Im[\xi_{\lambda \lambda'}(S_{i}, S_{j})] \sin[(\lambda - \lambda') \phi^{*}] & = 
      \mp \widetilde{d}^{S_{i}, S_{j}}_{-\lambda -\lambda'}(\theta^{*}) \Im[\xi_{-\lambda -\lambda'}(S_{i}, S_{j})]
      \sin[(-\lambda + \lambda') \phi^{*}],
  \end{split}
\end{equation}
which imply that for the terms proportional to the interference matrices, once again either all the terms proportional to the cosine 
functions vanish or all the terms proportional to the sine functions vanish. For example, the expression for $W^{x (\gamma)}$ for a 
reaction with a vector meson $V$ resonance interfering with a spin-2 meson $T$ resonance with corresponding spin density
matrix $\hat{\xi}^{x (\gamma)}(V, T)$ will have the following terms,
\begin{equation} \label{ParityCancelation2}
  \begin{split}
    W^{x (\gamma)} & = \cdots + \bigg(\widetilde{d}^{1, 2}_{1 -2} \Re\Big[\xi^{x (\gamma)}_{1 -2}(V ,T)\Big]
	+ \widetilde{d}^{1, 2}_{-1 2} \Re\Big[\xi^{x (\gamma)}_{-1 2}(V ,T)\Big]\bigg)
	\cos(3 \phi^{*}) \\
	\mt - \bigg(\widetilde{d}^{1, 2}_{1 -2} \Im\Big[\xi^{x (\gamma)}_{1 -2}(V ,T)\Big] 
	- \widetilde{d}^{1, 2}_{-1 2} \Im\Big[\xi^{x (\gamma)}_{-1 2}(V ,T)\Big]\bigg)
	\sin(3 \phi^{*}) + \cdots, \\
      & = \cdots + \bigg(\widetilde{d}^{1, 2}_{1 -2} \Re\Big[\xi^{x (\gamma)}_{1 -2}(V ,T)\Big]
	+ \widetilde{d}^{1, 2}_{1 -2} \Re\Big[\xi^{x (\gamma)}_{1 -2}(V ,T)\Big]\bigg)
	\cos(3 \phi^{*}) \\
	\mt - \bigg(\widetilde{d}^{1, 2}_{1 -2} \Im\Big[\xi^{x (\gamma)}_{1 -2}(V ,T)\Big] 
	- \widetilde{d}^{1, 2}_{1 -2} \Im\Big[\xi^{x (\gamma)}_{1 -2}(V ,T)\Big]\bigg)
	\sin(3 \phi^{*}) + \cdots, \\
      & = \cdots + 2 \widetilde{d}^{1, 2}_{1 -2} \Re\Big[\xi^{x (\gamma)}_{1 -2}(\rho)\Big] 
	\cos(3 \phi^{*}) + \cdots.
  \end{split}
\end{equation}

\section{General Expression of the Decay Distribution in Terms of Statistical Tensors}
We can also write the most general expression for the distribution, eqn. (\ref{DistributionFinal}), in terms of the statistical tensors.
First, we need to note that just like a spin density matrix can be decomposed into a sum of polarization operators as shown in
eqn. (\ref{TExpansion}), the same can be done for the interference spin density matrix between two hadrons of spins $S_{i}$ and $S_{j}$,
\begin{equation} \label{TExpansionInterference}
  \hat{\xi}(S_{i}, S_{j}) = \sum_{L = |S_{i} - S_{j}|}^{L = S_{i} + S_{j}} \sum_{M = -L}^{L} \tau_{LM}(S_{i}, S_{j})
    \hat{T}^{\dagger}_{LM}(S_{j}, S_{i}).
\end{equation}
The matrix elements of the interference polarization tensors, $\hat{T}_{LM}(S_{i}, S_{j})$, are now defined as
\begin{equation} \label{PolarizationTensorsInterference}
  \Big[T_{LM}(S_{i}, S_{j})\Big]_{m, m'} \equiv (-1)^{S_{j} - m'} C^{LM}_{S_{i}, m; S_{j}, -m'},
\end{equation}
and the $\tau_{LM}(S_{i}, S_{j})$'s are the interference statistical tensors between two particles of spin $S_{i}$ and $S_{j}$,
\begin{equation} \label{StatisticalTensorsInterference}
  \begin{split}
    \tau_{LM}(S_{i}, S_{j}) & = \Big<\hat{\xi}(S_{i}, S_{j}) \hat{T}_{LM}(S_{j}, S_{i})\Big> = 
      \Tr[\hat{\xi}(S_{i}, S_{j}) \hat{T}_{LM}(S_{j}, S_{i})] \\
    & = \sum_{m = -S_{i}}^{S_{i}} \sum_{m' = -S_{j}}^{S_{j}} (-1)^{S_{i} - m} \xi_{m, m'}(S_{i}, S_{j}) C^{LM}_{S_{j}, m'; S_{i}, -m}. 
  \end{split}
\end{equation}
The inverse relation of the previous equation is
\begin{equation} \label{StatisticalTensorsInterferenceInverseRelation}
        \xi_{mm'}(S_{i},\ S_{j}) = \sum_{L = - |S_{i}-S_{j}|}^{S_{i} + S_{j}} \sum_{M = -L}^{M = L} \tau_{LM}(S_{i},\ S_{j}) (-1)^{S_{j} - m} C^{LM}_{S_{j}m';S_{i}-m}.
\end{equation}

Substituting the expressions for spin density matrices in terms of statistical tensors, eqns. (\ref{TExpansion}) and 
(\ref{TExpansionInterference}), into equation (\ref{Distribution}), we get
\begin{equation} \label{DistributionStatisticalTensorsDerivation3}
  \begin{split}
    W(\theta^{*}, \phi^{*}) & = \sum_{i} \sum_{L = 0}^{2 S_{i}} \sum_{M = -L}^{L} t_{LM}(S_{i}) \Tr[\hat{\mathcal{D}}^{\dagger S_{i}} 
      \hat{T}^{\dagger}_{LM}(S_{i}) \hat{\mathcal{D}}^{S_{i}}] \\
      \mt + \sum_{i > j} \sum_{L = |S_{i} - S_{j}|}^{S_{i} + S_{j}} \sum_{M = -L}^{L}
	\Big\{\tau_{LM}(S_{i}, S_{j}) \Tr[\hat{\mathcal{D}}^{\dagger S_{i}} \hat{T}^{\dagger}_{LM}(S_{j}, S_{i})
	\hat{\mathcal{D}}^{S_{j}}] \\
      \mt + \tau_{LM}(S_{j}, S_{i}) \Tr[\hat{\mathcal{D}}^{\dagger S_{j}} \hat{T}^{\dagger}_{LM}(S_{i}, S_{j})
	\hat{\mathcal{D}}^{S_{i}}]\Big\}.
  \end{split}
\end{equation}
This expression can be further simplified by using certain properties of the polarization operators, $\hat{T}_{LM}(S_{i})$
and $\hat{T}_{LM}(S_{i}, S_{j})$, and of the statistical tensors, $t_{LM}(S_{i})$ and $\tau_{LM}(S_{i}, S_{j})$. By using these properties
of the Clebsch-Gordan coefficients, 
\begin{equation} \label{Clebsch-GordanProperties}
  \begin{split}
    C^{LM}_{S m; S' m'} & = (-1)^{S + S' -L} C^{LM}_{S' m'; S m}, \\
    C^{LM}_{S m; S' m'} & = (-1)^{S + S' -L} C^{L-M}_{S -m; S' -m'},
  \end{split}
\end{equation}
these properties of the elements of the polarization operators can be shown,
\begin{equation} \label{TProperties}
  \begin{split}
    [T_{LM}(S, S')]_{m m'} & = (-1)^{M + S - S'} [T_{L -M}(S', S)]_{m' m}, \\
    [T_{LM}(S, S')]_{m m'} & = (-1)^{L + S - S'} [T_{L -M}(S, S')]_{-m -m'}, \\
    [T_{LM}(S, S')]_{m m'} & = (-1)^{L + M} [T_{L M}(S', S)]_{-m' -m}.
  \end{split}
\end{equation}
They can also be used to show a generalized hermiticity property of the interference statistical tensors,
\begin{equation} \label{HermiticityInterferenceStatisticalTensors}
 \tau_{LM}(S, S') = (-1)^{S - S' + M} \tau^{*}_{L -M}(S', S).
\end{equation}
We can use these last two equations to rewrite the second term in the brackets in eqn. 
(\ref{DistributionStatisticalTensorsDerivation3}),
\begin{equation} \label{DistributionStatisticalTensorsBrackt}
  \begin{split}
    & \sum_{M = -L}^{L} \tau_{LM}(S_{j}, S_{i}) 
      \Tr[\hat{\mathcal{D}}^{\dagger S_{j}} \hat{T}^{\dagger}_{LM}(S_{i}, S_{j}) \hat{\mathcal{D}}^{S_{i}}] \\
    & \qquad \qquad = \sum_{M = -L}^{L} \Big(\tau_{LM}(S_{i}, S_{j}) 
      \Tr[\hat{\mathcal{D}}^{\dagger S_{i}} \hat{T}^{\dagger}_{LM}(S_{j}, S_{i}) \hat{\mathcal{D}}^{S_{j}}]\Big)^{*},
  \end{split}
\end{equation}
where we have used the fact that, since the index $M$ is a dummy index, when all of them in the expression are replaced by $-M$
the summation remains the same. Eqn. (\ref{DistributionStatisticalTensorsDerivation3}) therefore becomes
\begin{equation} \label{DistributionStatisticalTensorsDerivation4}
  \begin{split}
    W(\theta^{*}, \phi^{*}) & = \sum_{i} \sum_{L = 0}^{2 S_{i}} \sum_{M = -L}^{L} t_{LM}(S_{i}) \Tr[\hat{\mathcal{D}}^{\dagger S_{i}} 
      \hat{T}^{\dagger}_{LM}(S_{i}) \hat{\mathcal{D}}^{S_{i}}] \\
      \mt + \sum_{i > j} \sum_{L = |S_{i} - S_{j}|}^{S_{i} + S_{j}} \sum_{M = -L}^{L}
	2 \Re\Big\{\tau_{LM}(S_{i}, S_{j}) \Tr[\hat{\mathcal{D}}^{\dagger S_{i}} \hat{T}^{\dagger}_{LM}(S_{j}, S_{i})
	\hat{\mathcal{D}}^{S_{j}}]\Big\}.
  \end{split}
\end{equation}

We will now show how the two traces found in the previous equation can be further simplified. We will show it for the trace in the second
term, since the same derivation applies to the one in the first term but with $S_{i} = S_{j}$. In index notation, and for recoil baryon $B$
having spin $s$, that trace is
\begin{equation} \label{TraceIndexNotation}
  \Tr[\hat{\mathcal{D}}^{\dagger S_{i}} \hat{T}^{\dagger}_{LM}(S_{j}, S_{i})
      \hat{\mathcal{D}}^{S_{j}}]
    = \sum_{m = -S_{i}}^{S_{i}} \sum_{m' = -S_{j}}^{S_{j}} \sum_{n = -s}^{s} (-1)^{S_{i} - m}
      \mathcal{D}^{* S_{i}}_{m, n} C^{LM}_{S_{j}, m'; S_{i}, -m} \mathcal{D}^{S_{j}}_{m', n}.
\end{equation}
Certain properties of the Wigner-$\mathcal{D}$ functions and the Clebsch-Gordan coefficients will be used. First, 
\begin{equation} \label{WignerDProperty}
  \mathcal{D}^{S}_{m, m'} = (-1)^{m'-m} \mathcal{D}^{* S}_{-m, -m'}.
\end{equation}
Second is the Clebsch-Gordan series for the Wigner-$\mathcal{D}$ functions,
\begin{equation} \label{Clebsch-GordanSeries}
  \mathcal{D}^{S_{1}}_{m_{1}, n_{1}} \mathcal{D}^{S_{2}}_{m_{2}, n_{2}} = \sum_{J = |S_{1} - S_{2}|}^{J = S_{1} + S_{2}}
    \sum_{M, N = - J}^{J} C^{J M}_{S_{1}, m_{1}; S_{2}, m_{2}} \mathcal{D}^{J}_{M, N} C^{J N}_{S_{1}, n_{1}; S_{2}, n_{2}}.
\end{equation}
Third is the orthogonality condition of the Clebsch-Gordan coefficients,
\begin{equation} \label{Clebsch-GordanOrthogonality}
  \sum_{m_{1} = -j_{1}}^{j_{1}} \sum_{m_{2} = -j_{2}}^{j_{2}} C^{J M}_{j_{1}, m_{1}; j_{2}, m_{2}} C^{J' M'}_{j_{1}, m_{1}; j_{2}, m_{2}}
    = \delta_{JJ'} \delta_{M M'}.
\end{equation}
Finally, a relation between the Wigner-$\mathcal{D}$ functions and the spherical harmonics,
\begin{equation} \label{WignerDToSPhericalHarmonics}
  \mathcal{D}^{L}_{M, 0}(\theta, \phi) = \sqrt{\frac{4 \pi}{2 L + 1}} (-1)^M Y_{L,-M}(\theta, \phi).
\end{equation}
Using these last four equations, we can rewrite eqn. (\ref{DistributionStatisticalTensorsDerivation4}) in the simplified form
\begin{equation} \label{DistributionStatisticalTensorsSphericalHarmonics}
  \begin{split}
    & W(\theta^{*}, \phi^{*}; s) = \sum_{i} \sum_{L = 0}^{2 S_{i}} \sum_{M = -L}^{L} \kappa^{L}(S_{i}; s) (-1)^{M} t_{LM}(S_{i})
      Y_{L, -M}(\theta^{*}, \phi^{*}) \\
    & \qquad + \sum_{i > j} \sum_{L = |S_{i} - S_{j}|}^{S_{i} + S_{j}} \sum_{M = -L}^{L}
	2 \kappa^{L}(S_{i}, S_{j}; s) \Re\Big\{(-1)^{M} \tau_{LM}(S_{i}, S_{j}) Y_{L, -M}(\theta^{*}, \phi^{*})\Big\},
  \end{split}
\end{equation}
where
\begin{equation} \label{KappaDefinition}
  \begin{split}
    \kappa^{L}(S_{i}; s) & \equiv \sqrt{\frac{4 \pi}{2 L + 1}} \Tr[\hat{T}^{\dagger}_{L0}(S_{i})] \\
    & = \sqrt{\frac{4 \pi}{2 L + 1}} \sum_{m = -s}^{s} (-1)^{S_{i} - m} C^{L0}_{S_{i}, m; S_{i}, -m},
    \\ 
    \kappa^{L}(S_{i}, S_{j}; s) & \equiv \sqrt{\frac{4 \pi}{2 L + 1}} \Tr[\hat{T}^{\dagger}_{L0}(S_{j},\ S_{i})] \\
    & = \sqrt{\frac{4 \pi}{2 L + 1}} \sum_{m = -s}^{s} (-1)^{S_{i} - m} C^{L0}_{S_{j}, m; S_{i}, -m}.
  \end{split}
\end{equation}
This last equation shows that, when expressing the distribution $W$ in terms of the statistical tensors, it acquires a very simple form.
First, every single contributing hadron of spin $S_{i}$ will contribute a sum over the well known spherical harmonics for 
$0 \le L \le 2 S_{i}$, multiplied by the statistical tensors of the corresponding values for $L$ and $M$. The entire angular dependence 
for the decay of the hadron, $\theta^{*}$ and $\phi^{*}$, is contained in them. The prefactor for each term, $\kappa^{L}(S_{i}; s)$, 
depends only on the spin of the decaying hadron $S_{i}$, and the spin of the daughter hadron $s$. As seen in their definition in eqn. 
(\ref{KappaDefinition}), they are easily calculated for any combination of $S_{i}$ and $s$, since they are a linear combination of the well
known Clebsch-Gordan coefficients. Second, every pair of contributing hadrons with spins $S_{i}$ and $S_{j}$ will also contribute a linear
combination of spherical harmonics, this time for $|S_{i} - S_{j}| \le L \le S_{i} + S_{j}$. The prefactor, $\kappa^{L}(S_{i}, S_{j}; s)$,
is now a function of the spins of the interfering hadrons and the spin of the daughter hadron, but has a similar simple definition form as 
the previous one.

Note that this expression makes explicit the fact that distribution is a rotational scalar, since the contractions 
$\sum_{M} (-1)^M t_{LM} Y_{L -M}$ and $\sum_{M} (-1)^M \tau_{LM} Y_{L -M}$ are rotational invariants,
\begin{align} \label{RotationalInvarianceFirst}
    \sum_{M}(-1)^{M}t_{LM}Y_{L-M} \ \longrightarrow & \hphantom{=} \sum_{M,M',M''}(-1)^{M}D^{L}_{MM'}D^{L}_{-M-M''}t_{LM'}Y_{L-M''}\nonumber  \\
    & = \sum_{M,M',M''}(-1)^{M + M'' - M} D^{L}_{MM'}D^{L*}_{MM''} t_{LM'}Y_{L-M''} \\
    & = \sum_{M',M''}(-1)^{M''} \delta_{M'M''}
    t_{LM'}Y_{L-M''} \nonumber \\
    \hphantom{\sum_{M}(-1)^{M}t_{LM}Y_{L-M} \ } \longrightarrow & \hphantom{=} \sum_{\substack{M \\ \hphantom{M',M''}}}(-1)^{M}t_{LM}Y_{L-M}.\nonumber
\end{align}

One property should be noted about the coefficients $\kappa^{L}(S_{i}; s)$. By using the properties of the Clebsch-Gordan coefficients in
eqn. (\ref{Clebsch-GordanProperties}), and using the fact that $(-1)^{2 (S_{i} - m) = 1}$ for all $S_{i}$ and $m$ (since $S_{i}$ and $m$ are
both integer or half-integer, $2 (S_{i} - m)$ is twice an integer number, which is always an even number), we can rewrite
$\kappa^{L}(S_{i}; s)$ as
\begin{equation} \label{KappaPropertyDerivation}
  \begin{split}
    & \sum_{m = -s}^{s} (-1)^{S_{i} - m} C^{L0}_{S_{i}, m; S_{i}, -m} 
      = \sum_{m = -s}^{s} (-1)^{S_{i} - m + m - m + 2 S_{i} - L} C^{L0}_{S_{i}, -m; S_{i}, m} \\
    &  \qquad = \sum_{m = -s}^{s} (-1)^{S_{i} + m} (-1)^{2 (S_{i} -m)} (-1)^{-L} C^{L0}_{S_{i}, -m; S_{i}, m} \\
    &  \qquad = (-1)^{L} \sum_{m = -s}^{s} (-1)^{S_{i} + m} C^{L0}_{S_{i}, -m; S_{i}, m}.
   \end{split}
\end{equation}
In the last line, we used the fact that $(-1)^{-L} = (-1)^{L}$ since $L$ is an integer. Using the last equation, we can see that
$\kappa^{L}(S_{i}; s)$ has the property
\begin{equation} \label{KappaProperty}
  \kappa^{L}(S_{i}; s) = (-1)^{L} \kappa^{L}(S_{i}; s) \Rightarrow \kappa^{L}(S_{i}; s) = 0 \qquad \text{for} \ L \ \text{odd}. 
\end{equation}
Therefore, in the first term in eqn. (\ref{DistributionStatisticalTensorsSphericalHarmonics}), only the terms with even $L$ appear. Applying the same procedure to $\kappa^{L}(S_{i}, S_{j}; s)$, it can be shown that for the interference tensors $\tau_{LM}$, the $L$ odd terms vanish only if $S_{i} = S_{j}$.

This shows that not all tensors will appear on the expressions for the angular distribution and polarization observables. This may seem confusing at first, because this would seem to imply that the general expressions for the decay distribution in terms of the SDME's has more independent parameters than those in terms of the statistical tensors. However, upon closer inspection, we notice that when expressed in terms of the SDME's, the coefficients of each of the trigonometric functions in the sum are actually \textit{specific} linear combination of the SDME's. For example, in the case of vector meson decay, its density matrix has a total of 9 parameters. When expressed in terms of the statistical tensors in eqn. (\ref{SphericalHarmonics1}), only 6 independent parameters appear. But note that when expressed in terms of the SDME's as shown in eqn. (\ref{WDistributionRho}), there are only 6 coefficients because 3 of them are linear combinations of two SDME's. This is another reason that shows how using statistical tensors to represent the spin state of a mixed quantum state is more natural than with spin density matrices. 

The last step needed to simplify the general expression in eqn. (\ref{DistributionStatisticalTensorsSphericalHarmonics}) is to use the hermiticity condition of the non-interference tensors shown in eqn. (\ref{STHermiticity}), that shows that these tensors are not all independent. We first split the sum of the non-interference tensors, and then apply the hermiticity condition, and the Spherical Harmonic identity,
\begin{equation} \label{SphericalHarmonicIdentity}
    Y_{LM} = (-1)^{M} Y_{L-M}^{*},
\end{equation}
and the identity for complex numbers shown in eqn. (\ref{ComplexNumbersIdentity1}) to get
\begin{equation} \label{WSTHermiticitySimplification}
    \begin{split}
    & \sum_{M = -L}^{L} (-1)^{M} t_{LM}(S_{i})Y_{L,-M}(\theta^{*}, \phi^{*}) = t_{L0}(S_{i})Y_{L,0}(\theta^{*}, \phi^{*}) \\
    & \qquad \quad + \sum_{M > 0}^{L} \bigg\{ (-1)^{M} t_{LM}(S_{i})Y_{L,-M}(\theta^{*}, \phi^{*}) + (-1)^{-M} t_{L-M}(S_{i})Y_{L,M}(\theta^{*}, \phi^{*}) \bigg\} \\
    & \qquad = t_{L0}(S_{i})Y_{L,0}(\theta^{*}, \phi^{*}) + \sum_{M > 0}^{L} \bigg\{ (-1)^{M} t_{LM}(S_{i})Y_{L,-M}(\theta^{*}, \phi^{*}) \\
    & \qquad \quad + \Big[(-1)^{M} t_{LM}(S_{i})Y_{L,-M}(\theta^{*}, \phi^{*})\Big]^{*} \bigg\} \\
    & \qquad = t_{L0}(S_{i})Y_{L,0}(\theta^{*}, \phi^{*}) + 2\sum_{M > 0}^{L} \Re\Big[ (-1)^{M} t_{LM}(S_{i})Y_{L,-M}(\theta^{*}, \phi^{*}) \Big] \\
    & \qquad = t_{L0}(S_{i})Y_{L,0}(\theta^{*}, \phi^{*}) + 2\sum_{M > 0}^{L} (-1)^{M}\bigg(\Re\Big[t_{LM}(S_{i})\Big]\Re\Big[Y_{L-M}(\theta^{*}, \phi^{*})\Big] \\
    & \qquad \quad - \Im\Big[t_{LM}(S_{i})\Big]\Im\Big[Y_{L-M}(\theta^{*}, \phi^{*})\Big]\bigg).
    \end{split}
\end{equation}
Note that the hermiticity condition of the tensor implies that $t_{L0}$ is purely real. The previous expression is therefore manifestly purely real, as expected. The general expression therefore becomes
\begin{equation} \label{DistributionStatisticalTensorsSphericalHarmonics2}
  \begin{split}
    & W(\theta^{*}, \phi^{*}; s) = \sum_{i} \sum_{L = 0}^{2 S_{i}} \sum_{M \geq 0}^{L} 2 \kappa^{L}(S_{i}; s) (-1)^{M}\bigg(\Re\Big[t_{LM}(S_{i})\Big]\Re\Big[Y_{L-M}(\theta^{*}, \phi^{*})\Big] \\
    & \qquad \quad - \Im\Big[t_{LM}(S_{i})\Big]\Im\Big[Y_{L-M}(\theta^{*}, \phi^{*})\Big]\bigg) \\
    & \qquad + \sum_{i > j} \sum_{L = |S_{i} - S_{j}|}^{S_{i} + S_{j}} \sum_{M = -L}^{L}
	2 \kappa^{L}(S_{i}, S_{j}; s)(-1)^{M}\bigg(\Re\Big[\tau_{LM}(S_{i}, S_{j})\Big] \Re\Big[Y_{L, -M}(\theta^{*}, \phi^{*})\Big] \\
	& \qquad \quad - \Im\Big[\tau_{LM}(S_{i}, S_{j})\Big] \Im\Big[Y_{L, -M}(\theta^{*}, \phi^{*})\Big]\bigg),
  \end{split}
\end{equation}
where we have also used the complex numbers identity in eqn. (\ref{ComplexNumbersIdentity1}) to rewrite the expression in parenthesis in the second term in terms of real and imaginary parts of $\tau$. To simplify the notation, we included $t_{L0}$ inside the real and imaginary parts functions so that the sum can go over $M \geq 0$, but it must be remembered that $\Im[t_{L0}] = 0$. Note that in the second term of the previous equation the hermiticity relation for the interference tensor in eqn. (\ref{HermiticityInterferenceStatisticalTensors}) cannot be used to reduce the number of terms in the sum, because it relates $\tau(S_{i},S_{j})$ and $\tau(S_{j},S_{i})$, which are different tensors. Therefore, for the interference tensors the summation over $M$ goes from $-L$ to $L$. 
If we want the expression in terms of sine and cosine functions of $\phi^{*}$, we can write the spherical harmonics in terms of the associated Legendre polynomials,
\begin{equation} \label{SphericalHarmonicNorm}
    \begin{split}
    Y_{LM}(\theta^{*}, \phi^{*}) & = \sqrt{\frac{(2L+1)(L-M)!}{4\pi(L+M)!}} P^{M}_{L}(\cos\theta^{*}) e^{iM\phi^{*}} \\
    & = \sqrt{\frac{(2L+1)(L-M)!}{4\pi(L+M)!}} P^{M}_{L}(\cos\theta^{*})(\cos(M\phi^{*}) + i\sin(M\phi^{*})).
    \end{split}
\end{equation}
Note how the entire $\phi^{*}$ dependence is contained in the cosine and sine function. With this, we get for the general expression,
\begin{equation} \label{DistributionStatisticalTensorsSphericalHarmonics3}
  \begin{split}
    & W(\theta^{*}, \phi^{*}; s) = \sum_{i} \sum_{L = 0}^{2 S_{i}} \sum_{M \geq 0}^{L} \widetilde{\kappa}^{LM}(S_{i}; s) P^{M}_{L}(\cos\theta^{*})\bigg(\Re\Big[t_{LM}(S_{i})\Big]\cos(M\phi^{*}) \\
    & \qquad \quad + \Im\Big[t_{LM}(S_{i})\Big]\sin(M\phi^{*})\bigg) \\
    & \qquad + \sum_{i > j} \sum_{L = |S_{i} - S_{j}|}^{S_{i} + S_{j}} \sum_{M = -L}^{L}
	 \widetilde{\kappa}^{LM}(S_{i}, S_{j}; s)  P^{M}_{L}(\cos\theta^{*})\bigg(\Re\Big[\tau_{LM}(S_{i}, S_{j})\Big]\cos(M\phi^{*}) \\
	& \qquad \quad + \Im\Big[\tau_{LM}(S_{i}, S_{j})\Big]\sin(M\phi^{*})\bigg),
  \end{split}
\end{equation}
where
\begin{equation} \label{KappaTildeDefinition}
    \begin{split}
    \widetilde{\kappa}^{LM}(S_{i};s) & \equiv 2 \sqrt{\frac{(L-M)!}{(L+M)!}} \sum_{m = -s}^{s}(-1)^{S_{i}-m}C^{L0}_{S_{j}m;S_{i}-m}, \\
    \widetilde{\kappa}^{LM}(S_{i},S_{j};s) & \equiv 2 \sqrt{\frac{(L-M)!}{(L+M)!}} \sum_{m = -s}^{s}(-1)^{S_{i}-m}C^{L0}_{S_{j}m;S_{i}-m},
    \end{split}
\end{equation}
and we eliminated the factor of $(-1)^{M}$ by using the identity
\begin{equation} \label{SphericalHarmonicNormProperty}
     P^{-M}_{L}(\cos\theta^{*}) = (-1)^{M} P^{M}_{L}(\cos\theta^{*}).
\end{equation}

\section{General Expressions of Observables in Terms of Statistical Tensors}
The general expression for the 16 $W^{i}$'s shown in eqn. (\ref{DistributionFinal16}) can also be expressed in terms of the statistical tensors. The decomposition of the statistical tensors $t$ in terms or 16 other tensors shown in eqn. (\ref{Tensor16Tensors}) also apply to the interference tensors $\tau$. The relationship between the 16 tensors to the 16 matrices has the same form as eqn. (\ref{StatisticalTensorsInterferenceInverseRelation}), but with a superscript to label which member of the 16 matrices or tensors it refers to,
\begin{equation} \label{StatisticalTensorsInterference2}
  \begin{split}
    \tau^{i}_{LM}(S_{i}, S_{j}) & = \Big<\hat{\xi^{i}}(S_{i}, S_{j}) \hat{T}_{LM}(S_{j}, S_{i})\Big> = 
      \Tr[\hat{\xi^{i}}(S_{i}, S_{j}) \hat{T}_{LM}(S_{j}, S_{i})] \\
    & = \sum_{m = -S_{i}}^{S_{i}} \sum_{m' = -S_{j}}^{S_{j}} (-1)^{S_{i} - m} \xi^{i}_{m, m'}(S_{i}, S_{j}) C^{LM}_{S_{j}, m'; S_{i}, -m}. 
  \end{split}
\end{equation}
The inverse relation is
\begin{equation} \label{StatisticalTensorsInterferenceInverseRelation2}
    \begin{split}
        \xi^{i}_{mm'}(S_{i},\ S_{j}) & = \sum_{L = - |S_{i}-S_{j}|}^{S_{i} + S_{j}} \sum_{M = -L}^{M = L} \tau^{i}_{LM}(S_{i},\ S_{j}) (-1)^{S_{i} - m} C^{LM}_{S_{j}m';S_{i}-m}.
    \end{split}
\end{equation}
By substituting this last expression into equation eqn. (\ref{DistributionFinal16}) we get the most general expression for the $W^{i}$'s in terms or the statistical tensors, which has the same form as the most general expression of the decay angular distribution in eqn. (\ref{DistributionStatisticalTensorsSphericalHarmonics3}) but with the appropriate superscript,
\begin{equation} \label{DistributionStatisticalTensorsSphericalHarmonics316}
  \begin{split}
    & W^{i}(\theta^{*}, \phi^{*}; s) = \sum_{i} \sum_{L = 0}^{2 S_{i}} \sum_{M \geq 0}^{L} 2 \widetilde{\kappa}^{L}(S_{i}; s)P^{M}_{L}(\cos\theta^{*})\bigg(\Re\Big[t^{i}_{LM}(S_{i})\Big]\cos(M\phi^{*}) \\
    & \qquad \quad + \Im\Big[t^{i}_{LM}(S_{i})\Big]\sin(M\phi^{*})\bigg) \\
    & \qquad + \sum_{i > j} \sum_{L = |S_{i} - S_{j}|}^{S_{i} + S_{j}} \sum_{M = -L}^{L}
	2 \widetilde{\kappa}^{L}(S_{i}, S_{j}; s) P^{M}_{L}(\cos\theta^{*})\bigg(\Re\Big[\tau^{i}_{LM}(S_{i}, S_{j})\Big]\cos(M\phi^{*}) \\
	& \qquad \quad + \Im\Big[\tau^{i}_{LM}(S_{i}, S_{j})\Big]\sin(M\phi^{*})\bigg).
  \end{split}
\end{equation}
Remember that, as shown in eqn. (\ref{WObservables3}), the $W^{i}$ are proportional to polarization observables. 

\section{General Parity Relations of the Statistical Tensors}
To derive the parity relations of the statistical tensors, we need to use the parity relations of the SDME's shown in eqn. (\ref{DensityMatrixParity}), the relationship between the statistical tensors and the SDME's shown in eqn. (\ref{StatisticalTensorsInterference2}), and the properties of the Clebsch-Gordan coefficients shown in eqn. (\ref{Clebsch-GordanProperties}). With these, the parity properties of the statistical tensors can be shown to be
\begin{equation} \label{StatisticalTensorsParityNotSimplified}
    \begin{split}
        t^{k}_{LM}(S_{i}) & = \bar{\eta} (-1)^{2 S_{i} + 1}\zeta^{k} (-1)^{L + M + 2 S_{i}} t^{k}_{L-M}(S_{i}), \\
        \tau^{k}_{LM}(S_{i}, \ S_{j}) & = \bar{\eta} (-1)^{2 S_{i} + 1}\zeta^{k} (-1)^{L + M + S_{i} + S_{j}} \tau^{k}_{L-M}(S_{i}, \ S_{j}).
    \end{split}
\end{equation}
This relations can be simplified by using the following properties,
\begin{equation} \label{ExponentProperties}
    \begin{split}
        & (-1)^{2S_{i}}(-1)^{2S_{i}} = (-1)^{4S_{i}} = 1, \\
        & (-1)^{2S_{i}}(-1)^{S_{i}+S_{j}} = (-1)^{2S_{i} + 1}(-1)^{-S_{i}-S_{j}} = (-1)^{S_{i}-S_{j}}. 
    \end{split}
\end{equation}
These properties follow because $S_{i}$ and $S_{j}$ are either both integer or both half-integer. Therefore, $4S_{i}$ is an even integer, $S_{i}+S_{j}$ is an integer, and $(-1)^{n} = (-1)^{-n}$ for any integer $n$. The parity relations of the statistical tensors are
\begin{equation} \label{StatisticalTensorsParity}
    \begin{split}
        t^{k}_{LM}(S_{i}) & = - \bar{\eta} \zeta^{k} (-1)^{L + M} t^{k}_{L-M}(S_{i}), \\
        \tau^{k}_{LM}(S_{i}, \ S_{j}) & = - \bar{\eta} \zeta^{k} (-1)^{L + M + S_{i} - S_{j}} \tau^{k}_{L-M}(S_{i}, \ S_{j}),
    \end{split}
\end{equation}
where the expressions for the $\zeta^{k}$ were given in eqn. (\ref{SDMEFactor}), and $\bar{\eta}$ is the multiplication of the two intrinsic parities of the interfering hadrons. 

\section{Parity Considerations in General Observables in Terms of Statistical Tensors}
Just as in the case of the SDME's, when the scattering reaction is parity conserving, the expressions for the observables can be further simplified. Specifically, each of the $W^{i}$'s shown in eqn. (\ref{DistributionStatisticalTensorsSphericalHarmonics316}) will be either an even or odd function of the variable $\phi^{*}$ (i.e., all the terms proportional to either $\cos(M\phi^{*})$ or $\sin(M\phi^{*})$ vanish).

In the case of the non-interference tensors, parity conservation in the scattering process implies that each of the 16 tensors will be either purely real or purely imaginary. All that needs to be done to show this is use the hermiticity condition shown in eqn. (\ref{STHermiticity}) along with the parity relation shown in eqn. (\ref{StatisticalTensorsParity}),
\begin{equation} \label{StatisticalTensorsHermiticityAndParity}
    \begin{split}
    t^{k}_{LM}(S_{i}) & = (-1)^{M}t^{k *}_{L-M}(S_{i}) = -1(-1)^{L} \zeta^{k} t^{k *}_{LM}(S_{i}), \\
    t^{k}_{LM}(S_{i}) & = - \zeta^{k} t^{k *}_{LM}(S_{i})
    \end{split}
\end{equation}
The factor of $(-1)^{L}$ vanishes because only even terms appear in the sum for the non-interference tensors, as shown in eqn. (\ref{KappaProperty}). For the tensors of the type $t^{kl(\gamma N)}$, we instead have
\begin{equation} \label{StatisticalTensorsHermiticityAndParity2}
    t^{kl (\gamma N)}_{LM}(S_{i}) = - \zeta^{k}\zeta^{kl} t^{k (\gamma N) *}_{LM}(S_{i}).
\end{equation}

Since $\zeta^{k} = \pm 1$, the right-hand side of the second line of eqn. (\ref{StatisticalTensorsHermiticityAndParity}) will have a factor of either $1$ or $-1$. This factor depends only on the value of the superscript $k$, not on the spin or the statistics of the decaying hadron. From the general property of complex numbers shown in eqn. (\ref{ComplexNumbersIdentity2}), if this overall coefficient equals $1$ then $t^{k}(S_{i})$ is purely real, and if the coefficient is negative then it will be purely imaginary. The values of $\zeta^{k}$ shown in eqn. (\ref{SDMEFactor}) can be used to conclude that
\begin{equation} \label{StatisticalTensorsEvenOrOdd}
    \begin{split}
        \Re[t^{k}] & = 0, \text{ if } \zeta^{k} = 1 \rightarrow \text{ Observable odd in } \phi^{*}, \\
        \Im[t^{k}] & = 0, \text{ if } \zeta^{k} = -1\rightarrow \text{ Observable even in } \phi^{*}, \\
        & \\
        \Re[t^{kl}] & = 0, \text{ if } \zeta^{k}\zeta^{l} = 1\rightarrow \text{ Observable odd in } \phi^{*}, \\
        \Im[t^{kl}] & = 0, \text{ if } \zeta^{k}\zeta^{l} = -1\rightarrow \text{ Observable even in } \phi^{*}.
    \end{split}
\end{equation}
Since the cosine functions of $\phi^{*}$ are multiplied by real parts of tensors, while the sine functions are multiplied by imaginary parts, we conclude that the observable associated with $t^{k}$ or $t^{kl}$ will be even (odd) if $\zeta^{k}$ or $\zeta^{kl}$ equals $-1$($1$). Table \ref{table} summarizes the symmetry properties under $\phi^{*} \rightarrow -\phi^{*}$ for the unpolarized and single polarization observables.
\begin{table}
\centering
\caption{Symmetry of observables under $\phi^{*} \rightarrow -\phi^{*}$.}
{\renewcommand{\arraystretch}{1.3}
\begin{tabular}{| >{\centering} m{84pt} | >{\centering\arraybackslash} m{84pt} |} 
\hline
\multicolumn{2}{ | >{\raggedright} m{168pt} | }{Symmetry of observables under $\phi^{*} \rightarrow -\phi^{*}$} \\
\hline 
  Even & Odd \\
\hline
  $I^{0}$  & $I^{0}I^{s}$ \\
$I^{0}I^{c}$ & $I^{0}I^{\odot}$ \\
$I^{0}P_{y}$ & $I^{0}P_{x}$  \\
$I^{0}P_{y}^{c}$ & $I^{0}P_{z}$ \\
$I^{0}P_{x}^{s}$ & $I^{0}P_{x}^{c}$ \\
$I^{0}P_{z}^{s}$ & $I^{0}P_{z}^{c}$ \\
$I^{0}P_{x}^{\odot}$ & $I^{0}P_{y}^{s}$ \\
$I^{0}P_{z}^{\odot}$ & $I^{0}P_{y}^{\odot}$ \\
\hline
\end{tabular} }
\label{table}
\end{table}
We can derive similar results for the interference tensors $\tau$, but using a different procedure. To do so, we need to make use of some properties of $\kappa_{L}$, $P^{M}_{L}(\cos\theta^{*})$, $\cos(M\phi^{*})$, $\sin(M\phi^{*})$, and the parity relations for the interference tensors shown in the second line of eqn. (\ref{StatisticalTensorsParity}). First, in the general expression for the $W^{i}$'s shown in eqn. (\ref{DistributionStatisticalTensorsSphericalHarmonics316}), the summation over $M$ in the second term is split into two parts: a sum over the $M>0$ terms, and a sum over the $M<0$ terms,
\begin{align} \label{SplitSumRealParts}
    W^{i}(\theta^{*}, \ \phi^{*}) & = \cdots + \widetilde{\kappa}^{L}(S_{i}, S_{j}; s)P^{0}_{L}(\cos\theta^{*})\Re\Big[\tau^{i}_{L0}(S_{i}, S_{j})\Big] \nonumber \\
    & \hphantom{=} \quad + \widetilde{\kappa}^{L}(S_{i}, S_{j}; s)\sum_{M > 0}^{L}
	 \bigg(P^{M}_{L}(\cos\theta^{*})\Re\Big[\tau^{i}_{LM}(S_{i}, S_{j})\Big]\cos(M\phi^{*}) \\
	& \hphantom{=} \quad + P^{-M}_{L}(\cos\theta^{*})\Re\Big[\tau^{i}_{L-M}(S_{i}, S_{j})\Big]\cos(-M\phi^{*})\bigg) + \cdots.\nonumber
\end{align}
Four properties need to be used to simplify this last equation. First, using the properties of the Clebsch-Gordan coefficients shown in equation (\ref{Clebsch-GordanProperties}), it can be shown that
\begin{equation} \label{KappaProperty2}
\kappa_{L}(S_{i},\ S_{j}) = (-1)^{L + S_{i} - S_{j}}\kappa_{L}(S_{i},\ S_{j}).
\end{equation}
Second, the property of the associated Legendre polynomials shown in eqn. (\ref{SphericalHarmonicNormProperty}). Third,
\begin{equation} \label{CosineEven}
    \cos(-M\phi^{*}) = \cos(M\phi^{*}).
\end{equation}
And fourth, the parity property shown in the second line of eqn. (\ref{StatisticalTensorsParity}). Combining all these properties, eqn. (\ref{SplitSumRealParts}) becomes
\begin{align} \label{SplitSumRealParts2}
    W^{i}(\theta^{*}, \ \phi^{*}) & = \cdots + \widetilde{\kappa}^{L}(S_{i}, S_{j}; s)P^{0}_{L}(\cos\theta^{*})\Re\Big[\tau^{i}_{L0}(S_{i}, S_{j})\Big] \nonumber \\
    & \hphantom{=} \quad + \widetilde{\kappa}^{L}(S_{i}, S_{j}; s)\sum_{M > 0}^{L}
	 \bigg(P^{M}_{L}(\cos\theta^{*})\Re\Big[\tau^{i}_{LM}(S_{i}, S_{j})\Big]\cos(M\phi^{*}) \\
	& \hphantom{=} \quad -\eta \zeta^{i} P^{M}_{L}(\cos\theta^{*})\Re\Big[\tau^{i}_{LM}(S_{i}, S_{j})\Big]\cos(M\phi^{*})\bigg) + \cdots \nonumber
\end{align} 
Note the appearance of the factor $-\bar{\eta} \zeta^{i}$ in the second term inside of the parenthesis. This same procedure can be used with the terms proportional to the imaginary parts of the tensors, but since
\begin{equation} \label{SineOdd}
    \sin(-M\phi^{*}) = -\sin(-M\phi^{*}),
\end{equation}
we get instead
\begin{equation} \label{SplitSumImaginaryParts}
    \begin{split}
    W^{i}(\theta^{*}, \ \phi^{*}) & = \cdots + \widetilde{\kappa}^{L}(S_{i}, S_{j}; s)\sum_{M > 0}^{L}
	 \bigg(P^{M}_{L}(\cos\theta^{*})(\theta^{*})\Im\Big[\tau^{i}_{LM}(S_{i}, S_{j})\Big]\sin(M\phi^{*}) \\
	& \hphantom{=} \quad + \eta \zeta^{i} P^{M}_{L}(\cos\theta^{*})\Im\Big[\tau^{i}_{LM}(S_{i}, S_{j})\Big]\sin(M\phi^{*})\bigg) + \cdots.
	\end{split}
\end{equation}
The term with $\Im[\tau_{L0}]$ vanishes because $\sin(0) = 0$. Note how we get an extra minus sign, so that now the factor is $\eta \zeta^{i}$. The only possible values of this factor are $1$ and $-1$. If it is equal to $1$, the expression inside of the parenthesis in eqn. (\ref{SplitSumRealParts2}) vanishes and the observable will contain no terms proportional to cosines of $\phi^{*}$. If the factor instead equals $-1$, the expression inside of the parenthesis in eqn. (\ref{SplitSumImaginaryParts}) vanishes and the observable will contain no terms proportional to sines of $\phi^{*}$. This shows that once parity conservation is taken into account, the observables will be either even or odd in $\phi^{*}$. Also, for the term with $\Re[\tau^{i}_{L0}(S_{i},\ S_{j})]$, after applying all identities,
\begin{equation} \label{TauL0}
    \widetilde{\kappa}^{L}(S_{i}, S_{j})P^{M}_{L}(\cos\theta^{*}))\Re\Big[\tau^{i}_{L0}(S_{i}, S_{j})\Big] = -\widetilde{\eta} \zeta^{i} \kappa^{L}(S_{i}, S_{j})P^{M}_{L}(\cos\theta^{*})\Re\Big[\tau^{i}_{L0}(S_{i}, S_{j})\Big].
\end{equation}
From this we conclude that when the factor $\eta \zeta^{i} = 1$, this term vanishes (it does not necessarily imply that $\Re[\tau_{L0}] = 0$). As we've mentioned, when $\eta \zeta^{i} = 1$, all terms proportional to $\cos(M\phi^{*})$ vanish. We therefore conclude that the term with $\Re[\tau_{L0}]$ only appears when the observable is even in $\phi^{*}$.

The value of the factor $\eta \zeta^{i}$ only depends on two things. One is the intrinsic parity of the interacting hadrons. If their intrinsic parities are not the same, $\eta = -1$. It otherwise equals $1$. The other is which of the 16 tensors is related to the observables, since it determines the value of $\zeta^{i}$, which are shown in eqn. (\ref{SDMEFactor}). 

To summarize, once parity is taken into account, the expressions for the observables will either be
\begin{equation} \label{DistributionStatisticalTensorsSphericalHarmonicsEven}
  \begin{split}
    & W^{i}(\theta^{*}, \phi^{*}; s) = \sum_{i} \sum_{L = 0}^{2 S_{i}} \sum_{M \geq 0}^{L} 2 \widetilde{\kappa}^{L}(S_{i}; s)P^{M}_{L}(\cos\theta^{*})\Re\Big[t^{i}_{LM}(S_{i})\Big]\cos(M\phi^{*}) \\
    & \qquad + \sum_{i > j} \sum_{L = |S_{i} - S_{j}|}^{S_{i} + S_{j}} \sum_{M \geq 0}^{L}
	2 \widetilde{\kappa}^{L}(S_{i}, S_{j}; s) P^{M}_{L}(\cos\theta^{*})(-1)^{M}\Re\Big[\tau^{i}_{LM}(S_{i}, S_{j})\Big]\cos(M\phi^{*}),
  \end{split}
\end{equation}
or
\begin{equation} \label{DistributionStatisticalTensorsSphericalHarmonicsOdd}
  \begin{split}
    & W^{i}(\theta^{*}, \phi^{*}; s) = \sum_{i} \sum_{L = 0}^{2 S_{i}} \sum_{M \geq 0}^{L} 2 \widetilde{\kappa}^{L}(S_{i}; s) (-1)^{M}P^{M}_{L}(\cos\theta^{*})\Im\Big[t^{i}_{LM}(S_{i})\Big]\sin(M\phi^{*}) \\
    & \qquad + \sum_{i > j} \sum_{L = |S_{i} - S_{j}|}^{S_{i} + S_{j}} \sum_{M \geq 0}^{L}
	2 \widetilde{\kappa}^{L}(S_{i}, S_{j}; s) P^{M}_{L}(\cos\theta^{*})(-1)^{M}\Im\Big[\tau^{i}_{LM}(S_{i}, S_{j})\Big]\sin(M\phi^{*})
  \end{split}
\end{equation}
(we included the $M = 0$ term in the sum, even though it vanishes). Parity conservation therefore greatly reduces the number of independent statistical tensors, since all $M < 0$ are not independent.

For an interference matrix of two hadrons with opposite intrinsic parity, $\eta = -1$. However, we have to remember that the parity relations of transition matrices first shown in equation eqn. (\ref{TransitionMatrixParityRelations}) also applies to the transition matrix of the decay process (which, as shown, is proportional to a Wigner-$\mathcal{D}$ matrix). Therefore, under a parity transformation, both the transition matrix for the scattering process and the transition matrix for the decay process will produce a $-1$, the net result being a $1$. This shows that the value of $\eta$ ends up playing no role in determining whether an observable is even or odd in $\phi^{*}$. 

\chapter{General Expressions for the Recoil Observables}
\section{Derivation of General Expressions}
We will now derive expressions for the recoil baryon polarization observables for the case of decaying baryon resonances in $\gamma N \rightarrow M B^{*} \rightarrow MMB$. The polarization observables involving the recoil baryon in terms of the 16 expansion spin density matrices are given in eqns. (\ref{Observables4}), (\ref{Observables6}), (\ref{Observables7}), and (\ref{Observables8}) (for spin higher than 1/2, the appropriate spin operator, $S_{x}$, $S_{y}$ or $S_{z}$, is used instead of the Pauli matrices). These equations show that in order to calculate them, the 16 expansion matrices for the spin density matrix of the recoil baryon shown in eqn. (\ref{16Matrices}) must be multiplied by one of the three spin operators and then the trace must be taken. In this chapter, in order to refer to a generic polarization observable involving the recoil baryon, we will use the definition
\begin{equation} \label{GenericRecoilObservable}
	O_{ij} \equiv \Tr[\hat{\rho}^{i}(B)\hat{S}_{j}].
\end{equation}
The superscript $i$ is used to identify which of the 16 matrices in the expansion is being used, so it can take values such as $x (\gamma)$, $y (N)$, $z (\gamma N)$, etc. The index $j$ identifies which of the spin operators is used, so it takes values of $x$, $y$ or $z$.

We want expressions for the recoil observables in terms of the SDME's or statistical tensors of the decaying hadron, not those of the recoil baryon. We therefore need to use the equation that relates these two spin density matrices, which was shown in eqn. (\ref{InitialToFinalDensityMatrixThreeHalfsBaryon2}) for spin-3/2 resonances, but have the same form for higher-spin baryons, and it is also valid for any of the 16 matrices in the expansion,
\begin{equation} \label{BStartoB}
	\hat{\xi}^{i}(B) = \hat{D}^{\dagger}\hat{\xi}^{i}(B^{*})\hat{D}^{\dagger}
\end{equation}
(we have omitted explicitly showing that the spin density matrix on the left-had side of the equation depends on the meson $M$ since it is spinless) We take $\hat{\xi}$ to refer to either a non-interference matrix or an interference matrix. Two examples are
\begin{align} \label{RecoilObservableExample}
		O_{0y} \equiv P_{y'} & = \Tr[\hat{\rho}^{0}\hat{S}_{y}], \\
		O_{z(\gamma)x} \equiv P^{\odot}_{x'} & = \Tr[\hat{\rho}^{z(\gamma)}\hat{S}_{x}].\nonumber
\end{align}

We will derive the expressions in terms of the statistical tensors instead, since the final expressions will acquire a simpler form. If expressions in terms of SDME's are required, they are related to the statistical tensors by eqns. (\ref{STDefinition1}), (\ref{InverseRelation}), (\ref{StatisticalTensorsInterference}), (\ref{StatisticalTensorsInterferenceInverseRelation}). First, the spin density matrix $\hat{\xi}^{i}(B^{*})$ is expressed in terms of the statistical tensors using eqns. (\ref{InverseRelation}) and (\ref{StatisticalTensorsInterferenceInverseRelation}). Next, the two Wigner $\mathcal{D}$-matrices in eqn. (\ref{BStartoB}) are expressed in terms of only one using the Clebsch-Gordan series shown in equation (\ref{Clebsch-GordanSeries}). When the resulting expression is multiplied by a spin operator and then the trace is taken, we get
\begin{equation} \label{RecoilObservableStatisticalTensors}
  \begin{split}
    O_{ij}(\theta^{*}, \phi^{*}) & = \sum_{k} \sum_{L = 0}^{2 S_{k}}\  \sum_{M, N = -L}^{L} t^{i}_{LM}(S_{i}) D^{L}_{MN} \Tr[\hat{T}^{\dagger}_{NM}(S_{k}) \hat{S}_{j}] \\
      \mt + \sum_{k > l} \sum_{L = |S_{i} - S_{j}|}^{S_{k} + S_{l}} \sum_{M,N = -L}^{L}
	\Big\{\tau^{i}_{LM}(S_{k}, S_{l}) D^{L}_{MN} \Tr[\hat{T}^{\dagger}_{LN}(S_{l}, S_{k})
	\hat{S}_{j}] \\
      \mt + \tau^{i}_{L-M}(S_{l}, S_{k}) D^{L}_{-M-N} \Tr[\hat{T}^{\dagger}_{L-N}(S_{k}, S_{l})
	\hat{S}_{j}]\Big\}.
  \end{split}
\end{equation} 
The sum over $M$ in the second term in the expression inside of the curly brackets of the previous equation was written as $-M$ for later convenience (remember that $M$ is a dummy index). Since the polarization tensors $T_{LM}$ are spherical operators, writing the spin operator $S_{j}$ in terms of spherical operators will help in finding a more simplified version of the previous expression. The definitions of the spin operators in the spherical basis are
\begin{equation} \label{SphericalToCartesian}
	\begin{split}
		& \hat{S}_{+1} = -\frac{1}{\sqrt{2}}(\hat{S}_{x} 
			+ i\hat{S}_{y}), \\
		& \hat{S}_{-1} = \frac{1}{\sqrt{2}}(\hat{S}_{x} 
			- i\hat{S}_{y}), \\
		& \hat{S}_{0} = \hat{S}_{z}. 
	\end{split}
\end{equation}
The inverse relations are
\begin{equation} \label{CartesianToSpherical2}
	\begin{split}
		& \hat{S}_{x} = \frac{1}{\sqrt{2}}(\hat{S}_{-1} 
			- \hat{S}_{1}), \\
		& \hat{S}_{y} = \frac{i}{\sqrt{2}}(\hat{S}_{-1} 
			+ \hat{S}_{1}), \\
		& \hat{S}_{z} = \hat{S}_{0}. 
	\end{split}
\end{equation}
Note that since the spin operators are hermitian (their expectation values are real), eqn. (\ref{SphericalToCartesian}) implies that in the spherical basis the operators are also hermitian,
\begin{align} \label{HermiticitySphericalSpinOperators}
		& \hat{S}^{\dagger}_{+1} = \hat{S}_{+1} \nonumber \\
		& \hat{S}^{\dagger}_{-1} = \hat{S}_{-1} \\
		& \hat{S}^{\dagger}_{0} = \hat{S}_{0} \nonumber
\end{align}
In terms of polarization operators, the spin operators in the spherical basis are given by
\begin{equation} 
	\label{SphericalSpinOperatorAndPolarizationOperator} 
	\hat{S}_{M} = \sqrt{\frac{S(S+1)(2S+1)}{3}}\hat{T}_{1M}. 
\end{equation}
By using eqns. (\ref{TProperties}), (\ref{HermiticityInterferenceStatisticalTensors}) and (\ref{WignerDProperty}), it can be shown that
\begin{equation} \label{HermiticitySimplification}
	\begin{split}
		t^{i}_{L-M}(S_{k}) D^{L}_{-M-N} \Tr[\hat{T}^{\dagger}_{L-N}(S_{k}) \hat{S}_{j}] & = \Big(t^{i}_{LM}(S_{k}) D^{L}_{MN} \Tr[\hat{T}^{\dagger}_{LN}(S_{k}) \hat{S}_{j}]\Big)^{*}, \\
		\tau^{i}_{L-M}(S_{l}, S_{k}) D^{L}_{-M-N} \Tr[\hat{T}^{\dagger}_{L-N}(S_{l}, S_{k})\hat{S}_{j}] & = \Big(\tau^{i}_{LM}(S_{k}, S_{l}) D^{L}_{MN} \Tr[\hat{T}^{\dagger}_{LN}(S_{l}, S_{k})\hat{S}_{j}]\Big)^{*}.
	\end{split}
\end{equation}
These two equations can be used to simplify eqn. (\ref{RecoilObservableStatisticalTensors}),
\begin{equation} \label{RecoilObservableStatisticalTensorsSimplified}
  \begin{split}
    O_{ij}(\theta^{*}, \phi^{*}) & = \sum_{k} \sum_{L = 0}^{2 S_{k}}\  \sum_{M, N \geq 0}^{L} 2 \Re\Big\{t^{i}_{LM}(S_{i}) D^{L}_{MN} \Tr[\hat{T}^{\dagger}_{LN}(S_{k}) \hat{S}_{j}]\Big\} \\
      \mt + \sum_{k > l} \sum_{L = |S_{k} - S_{l}|}^{S_{k} + S_{l}}\  \sum_{M,N = -L}^{L} 2 \Re
	\Big\{\tau^{i}_{LM}(S_{k}, S_{l}) D^{L}_{MN} \Tr[\hat{T}^{\dagger}_{LN}(S_{l}, S_{k})
	\hat{S}_{j}]\Big\},
  \end{split}
\end{equation}
which shows explicitly that the observable is real. In the previous equation, the sum over $M$ and $N$ in the first term goes only over values greater or equal to $0$ because the first line of eqn. (\ref{HermiticitySimplification}) used to simplify eqn. (\ref{RecoilObservableStatisticalTensors}) shows that the $M, N < 0$ terms are not independent from the $M,N > 0$ terms.

To further simplify the previous equation, we first note that the real part function contain three complex factors as arguments: $t^{i}_{LM}$ (or $\tau^{i}_{LM}$), $D^{L}_{MN}$ and $\hat{S}_{j}$. We use the real part function identity shown in eqn. (\ref{ComplexNumbersIdentity1}) to rewrite the real part function as
\begin{equation} \label{RecoilObservableSimplification1}
	\begin{split}
	& \sum_{m,m'= -s}^{s}\Re\Big\{\tau^{i}_{LM} D^{L}_{MN} 
		(\hat{S}_{j})_{mm'}\Big\}
		\big(\hat{T}^{\dagger}_{LN}\big)_{m'm} 
		= \\ & \qquad \qquad \sum_{m,m'= -s}^{s}\Big\{\Re\Big[\tau^{i}_{LM} D^{L}_{MN}\Big]
		\Re\Big[(\hat{S}_{j})_{mm'}\Big] \\
		& \qquad \qquad \qquad - \Im\Big[\tau^{i}_{LM} D^{L}_{MN}\Big]
		\Im\Big[(\hat{S}_{j})_{mm'}\Big]\Big\}\big(\hat{T}^{\dagger}_{LN}\big)_{m'm},
	\end{split}
\end{equation}
where we are now using index notation to express the matrix multiplications. Note that $\big(\hat{T}^{\dagger}_{LN}\big)_{m'm}$ is purely real. Next, we notice that only one of the two terms in the previous equation in non-vanishing because, from eqns. (\ref{CartesianToSpherical2}) and (\ref{SphericalSpinOperatorAndPolarizationOperator}), $(\hat{S}_{j})_{mm'}$ is either purely real or purely imaginary. By also using the complex numbers identity in eqn. (\ref{ComplexNumbersIdentity1}) along with 
\begin{equation} \label{ComplexNumbersIdentity3}
	\Im[z_{1}z_{2}] = \Re[z_{1}]\Im[z_{2}] + \Im[z_{1}]\Re[z_{2}],
\end{equation}
eqn. (\ref{RecoilObservableSimplification1}) simplifies to
\begin{equation} \label{RecoilObservablesSimplification2}
	\begin{split}
		& \sum_{m,m'= -s}^{s}\Re\Big\{\tau^{i}_{LM} D^{L}_{MN} 
		(\hat{S}_{j})_{mm'}\Big\}
		\big(\hat{T}^{\dagger}_{LN}\big)_{m'm} 
		= \\ & \qquad \sum_{m,m' = -s}^{s}d^{L}_{MN}(\theta^{*})\big(\hat{T}^{\dagger}_{LN}\big)_{mm'}\Re\Big[\big(\hat{S}_{j}\big)_{m'm}\Big]\bigg(\Re\Big[\tau_{LM}\Big]\cos(M\phi^{*}) + \Im\Big[\tau_{LM}\Big]\sin(M\phi^{*})\bigg), \\
		& \\
		& \text{for } j = x,\ z,
	\end{split}
\end{equation}
or
\begin{equation} \label{RecoilObservablesSimplification3}
	\begin{split}
		& \sum_{m,m'= -s}^{s}\Re\Big\{\tau^{i}_{LM} D^{L}_{MN} 
		(\hat{S}_{j})_{mm'}\Big\}
		\big(\hat{T}^{\dagger}_{LN}\big)_{m'm} 
		= \\ & \qquad \sum_{m,m' = -s}^{s}d^{L}_{MN}(\theta^{*})\big(\hat{T}^{\dagger}_{LN}\big)_{mm'}\Im\Big[\big(\hat{S}_{j}\big)_{m'm}\Big]\bigg(\Im\Big[\tau_{LM}\Big]\cos(M\phi^{*}) - \Re\Big[\tau_{LM}\Big]\sin(M\phi^{*})\bigg), \\
		& \\
		& \text{for } j = y.
	\end{split}
\end{equation}
To further simplify, we will write the spin operator $\big(\hat{S}_{j}\big)_{m'm}$ in terms of polarization operators using eqn. (\ref{SphericalSpinOperatorAndPolarizationOperator}). Since each of the three spin operators have a different form when written in terms of the polarization operators, we will continue the derivation using $\big(\hat{S}_{y}\big)_{m'm}$, but the other two are derived in a similar way. We get, 
\begin{align} \label{RecoilObservablesSimplification4}
		& \sum_{N = -L}^{L}\sum_{m,m' = -s}^{s}d^{L}_{MN}(\theta^{*})\big[\hat{T}^{\dagger}_{LN}(S_{l},S_{k})\big]_{mm'}\Im\Big\{\big[\hat{S}_{j}\big]_{m'm}\Big\} = \nonumber \\
		& \qquad \qquad \sqrt{\frac{s(s+1)(2s+1)}{6}}\sum_{N = -L}^{L}\sum_{m,m' = -s}^{s}d^{L}_{MN}(\theta^{*}) \nonumber \\ 
		& \qquad \qquad \big[\hat{T}^{\dagger}_{LN}(S_{l},S_{k})\big]_{mm'}\bigg(\big[\hat{T}_{1-1}(s)\big]_{m'm} + \big[\hat{T}_{11}(s)\big]_{m'm}\bigg) \nonumber \\
		& \qquad = \sqrt{\frac{s(s+1)(2s+1)}{6}}\sum_{N = -L}^{L}\sum_{m,m' = -s}^{s}d^{L}_{MN}(\theta^{*})\\ 
		& \qquad \qquad (-1)^{S_{k} - m}C^{LN}_{S_{l}m';S_{k}-m}\bigg((-1)^{s-m}C^{1-1}_{sm';s-m} + (-1)^{s-m}C^{11}_{sm';s-m}\bigg) \nonumber \\
		& \qquad = \sqrt{\frac{s(s+1)(2s+1)}{6}}\sum_{m,m' = -s}^{s}\bigg(d^{L}_{M-1}(\theta^{*})(-1)^{S_{k} - s}C^{L-1}_{S_{l}m';S_{k}-m}C^{1-1}_{sm';s-m} \nonumber \\
		& \qquad \qquad + d^{L}_{M1}(\theta^{*})(-1)^{S_{k}-s}C^{L1}_{S_{l}m';S_{k}-m}C^{11}_{sm';s-m}\bigg), \nonumber
\end{align}
where in the second line we used the definitions of the polarization operators shown in eqns. (\ref{TDefinition1}) and (\ref{PolarizationTensorsInterference}), and in the last time only a $N = -1$ and $N = 1$ term survive because $N = m' - m = -1$ in one term and $N = m' - m = 1$ in another from the properties of the Clebsch-Gordan coefficients. Next, we use the properties shown in the second line of eqn. (\ref{Clebsch-GordanProperties}) in order to factor out the two Clebsch-Gordan coefficients in the two terms inside of the parenthesis of the last equation,
\begin{equation} \label{RecoilObservablesSimplification5}
	\begin{split}
	& \sum_{m,m' = -s}^{s}(-1)^{S_{k} - s}C^{L-1}_{S_{l}m';S_{k}-m}C^{1-1}_{sm';s-m} = \\
		& \qquad (-1)^{S_{l} + S_{k} + 2s - L  + 1} \sum_{m,m' = -s}^{s}(-1)^{S_{k} - s}C^{L1}_{S_{l}m';S_{k}-m}C^{11}_{sm';s-m},
	\end{split}
\end{equation}
where on the right-hand side of the previous equation we relabeled the dummy indices: $-m' \rightarrow m'$ and $m \rightarrow -m$. We will soon prove that, for the non-interference tensors, only the odd $L$ terms will appear in the final answer. Therefore, $(-1)^{-L+1}$ equals $1$ for the non-interference terms. Also, $s$ is half-integer because the decaying resonance is a baryon. This means that $(-1)^{2s}$ equals $-1$ in the previous equation. Additionally, note how when $S_k$ is equal to $S_l$, which includes the non-interference tensors, the factor $(-1)^{S_k+S_l+1}$ is equal to $1$ because $S_k+S_l$ equals an odd number, since $S_k$ and $S_l$ are half-integer.This gives us
\begin{equation} \label{RecoilObservablesSimplification6}
	\begin{split}
		& \sum_{N = -L}^{L}\sum_{m,m' = -s}^{s}d^{L}_{MN}(\theta^{*})\big[\hat{T}^{\dagger}_{LN}(S_{l},S_{k})\big]_{mm'}\Im\Big\{\big[\hat{S}_{j}\big]_{m'm}\Big\} = \\	
		& \qquad \sqrt{\frac{s(s+1)(2s+1)}{6}}\sum_{m,m' = -s}^{s}(-1)^{S_{k}-s}C^{L1}_{S_{l}m';S_{k}-m}C^{11}_{sm';s-m} \\
		& \qquad \qquad \times \bigg((-1)^{S_{k} + S_{l} - L}d^{L}_{M-1}(\theta^{*}) + d^{L}_{M1}(\theta^{*})\bigg).
	\end{split}
\end{equation}
With these simplifications we are ready to write the final forms of the recoil observables. These are
\begin{align} \label{RecoilObservablesX}
		O_{ix}(\theta^{*},\phi^{*}) & = \sum_{k}\sum_{L = 0}^{S_{k}}\sum_{M = -L}^{L} \kappa^{L}(S_{k};s)\bigg(d^{L}_{M-1}(\theta^{*}) - d^{L}_{M1}(\theta^{*})\bigg) \nonumber \\
		& \qquad \times \bigg(\Re\Big[t^{i}_{LM}(S_{k})\Big]\cos(M\phi^{*}) + \Im\Big[t^{i}_{LM}(S_{k})\Big]\sin(M\phi^{*})\bigg) \\
		\mt + \sum_{k > l} \sum_{L = |S_{k}-S_{l}|}^{S_{k} + S_{l}}\sum_{M = -L}^{L} \kappa^{L}(S_{k},S_{l};s)\bigg(\chi^L(S_{k},S_{l})d^{L}_{M-1}(\theta^{*}) - d^{L}_{M1}(\theta^{*})\bigg) \nonumber \\
		& \qquad \times \bigg(\Re\Big[\tau^{i}_{LM}(S_{k},S_{l})\Big]\cos(M\phi^{*}) + \Im\Big[\tau^{i}_{LM}(S_{k},S_{l})\Big]\sin(M\phi^{*})\bigg), \nonumber
\end{align}

\begin{align} \label{RecoilObservablesY}
		O_{iy}(\theta^{*},\phi^{*}) & = \sum_{k}\sum_{L = 0}^{S_{k}}\sum_{M = -L}^{L} \kappa^{L}(S_{k};s)\bigg(d^{L}_{M-1}(\theta^{*}) + d^{L}_{M1}(\theta^{*})\bigg) \nonumber \\
		& \qquad \times \bigg(\Im\Big[t^{i}_{LM}(S_{k})\Big]\cos(M\phi^{*}) - \Re\Big[t^{i}_{LM}(S_{k})\Big]\sin(M\phi^{*})\bigg) \\
		\mt + \sum_{k > l} \sum_{L = |S_{k}-S_{l}|}^{S_{k} + S_{l}}\sum_{M = -L}^{L} \kappa^{L}(S_{k},S_{l};s)\bigg(\chi^L(S_{k},S_{l})d^{L}_{M-1}(\theta^{*}) + d^{L}_{M1}(\theta^{*})\bigg) \nonumber \\
		& \qquad \times \bigg(\Im\Big[\tau^{i}_{LM}(S_{k},S_{l})\Big]\cos(M\phi^{*}) - \Re\Big[\tau^{i}_{LM}(S_{k},S_{l})\Big]\sin(M\phi^{*})\bigg), \nonumber
\end{align}
and
\begin{align} \label{RecoilObservablesZ}
		O_{iz}(\theta^{*},\phi^{*}) & = \sum_{k}\sum_{L = 0}^{S_{k}}\sum_{M = -L}^{L} \kappa^{L0}(S_{k};s)d^{L}_{M0}(\cos\theta^{*}) \nonumber \\
		& \qquad \times \bigg(\Re\Big[t^{i}_{LM}(S_{k})\Big]\cos(M\phi^{*}) + \Im\Big[t^{i}_{LM}(S_{k})\Big]\sin(M\phi^{*})\bigg) \\
		\mt + \sum_{k > l} \sum_{L = |S_{k}-S_{l}|}^{S_{k} + S_{l}}\sum_{M = -L}^{L} \kappa^{L0}(S_{k},S_{l};s) d^{L}_{M0}(\cos\theta^{*}) \nonumber \\
		& \qquad \times \bigg(\Re\Big[\tau^{i}_{LM}(S_{k},S_{l})\Big]\cos(M\phi^{*}) + \Im\Big[\tau^{i}_{LM}(S_{k},S_{l})\Big]\sin(M\phi^{*})\bigg), \nonumber
\end{align}
where we have defined
\begin{align} \label{KappaRecoilDefinitioin}
	& \kappa^{L1}(S_{k},S_{l};s) \equiv \sqrt{\frac{2s(s+1)(2s+1)}{3}}\sum_{m,m' = -s}^{s}(-1)^{S_{k}-s}C^{L1}_{S_{l}m';S_{k}-m}C^{11}_{sm';s-m}, \nonumber \\
	& \kappa^{L0}(S_{k},S_{l};s) \equiv \sqrt{\frac{2s(s+1)(2s+1)}{3}}\sum_{m,m' = -s}^{s}(-1)^{S_{k}-s}C^{L0}_{S_{l}m';S_{k}-m}C^{10}_{sm';s-m}, \\
	& \qquad \qquad \qquad \qquad \qquad \kappa^{L1}(S_{k};s) \equiv \kappa^{L1}(S_{k},S_{k};s), \nonumber \\
	& \qquad \qquad \qquad \qquad \qquad \kappa^{L0}(S_{k};s) \equiv \kappa^{L0}(S_{k},S_{k};s), \nonumber
\end{align}
and
\begin{equation} \label{ChiPhaseDefinition}
	\chi^L(S_{k},S_{l}) \equiv (-1)^{S_{k} + S_{l} - L} = \pm 1.
\end{equation}
The phase factor $\chi^L(S_k,S_l)$ shows that, depending on the value of $L$, the factors with the two Wigner d-functions may have a relative minus sign or not. Each of the three general expressions for the observables contains a sum over non-interference tensors and one over interference tensors. Each of these two summations is made up of three factors. One of them is the factor $\kappa^{L1}$ or $\kappa^{L0}$, which has no angular dependence and can be easily found from Clebsch-Gordan coefficients. The second is is a sum of two Wigner d-functions in the case of $O_{ix}$ and $O_{iy}$ or one Wigner-d function in the case of $O_{iz}$, which contain the dependence on $\theta^{*}$. The third factor contains the $\phi^{*}$ dependence, and is made up of a sum over cosines of $\phi^{*}$ and a sum over sines of $\phi^{*}$. In $O_{ix}$ and $O_{iz}$, the terms with cosine functions are proportional to the real parts of the statistical tensors and the terms with sine functions are proportional to the imaginary parts of the tensors. In $O_{iy}$, the terms with cosine functions are proportional to the imaginary parts of the statistical tensors, while the terms with sine functions are proportional to the real parts of the statistical tensors. 

By using the properties of the Clebsch-Gordan coefficients in the second line of eqn. (\ref{Clebsch-GordanProperties}), it can be shown that
\begin{equation} \label{KappaPropertyRecoil}
	\begin{split}
		\kappa^{L1}(S;s) & = (-1)^{L+1}\kappa^{L1}(S;s) \\
		\kappa^{L0}(S;s) & = (-1)^{L+1}\kappa^{L0}(S;s).
	\end{split}
\end{equation}
This implies that for the non-interference statistical tensors, only the terms with odd $L$ are non-vanishing. In comparison, in the polarization observables that don't involve the recoil baryon, it is the even $L$ terms that are non-vanishing for the non-interference statistical tensors.

As an example, if we apply eqn. (\ref{RecoilObservablesY}) to the observable $P^\odot_{y'}$ for the decay of a spin-3/2 baryon, we get
\begin{align}
	P^\odot_{y'} = & \frac{1}{40} \Bigg\{15\Big[1-\cos(2\theta )\Big]\Big[\cos (3 \phi ) \Im[t^{z(\gamma)}_{33}]-\sin (3 \phi ) \Re[t^{z(\gamma)}_{33}]\Big] \nonumber \\
\mt
+\sqrt{15}
    \Big[5 \cos (2 \theta )+3\Big]\Big[\cos (\phi ) \Im[t^{z(\gamma)}_{31}]-\sin (\phi ) \Re[t^{z(\gamma)}_{31}]\Big]\\
\mt
+10 \sqrt{6} \sin (2 \theta ) \Big[\cos (2 \phi )
    \Im[t^{z(\gamma)}_{32}]-\sin (2 \phi ) \Re[t^{z(\gamma)}_{32})\Big] \nonumber \\
\mt
+8 \sqrt{10} \Big[\sin (\phi ) \Re[t^{z(\gamma)}_{11}]-\cos (\phi ) \Im[t^{z(\gamma)}_{11}]\Big]\Bigg\} \nonumber
\end{align}

These expressions are valid for the case of a decaying baryon resonance, but not for the case of a decaying meson resonance, $\gamma N \rightarrow M^*B \rightarrow MMB$. The reason is because expressions for distributions and observables have been given in terms of the SDME's and statistical tensors of the decaying meson subsystem, and as shown in eqn. (\ref{PartialTraceVectorMeson}), these are found from the partial trace of the density matrix of entire $M^*B$ system, which requires a sum over the recoil baryon $B$'s helicity indices. When this partial trace is taken, we are unable to multiply by the spin operator of the recoil baryon, as is required by eqn. (\ref{GenericRecoilObservable}). When the partial trace is taken, we obtain the spin density matrix of the meson subsystem, which contains no information on the spin state of the recoil baryon. But the recoil observable requires a measurement of the recoil baryon's polarization, which means the expression should reference the spin state of this baryon. Therefore, the recoil observables can't be expressed in terms of the spin density matrix of the meson resonance.

Unlike the density matrix of the beam-target system, which can be written in terms of a tensor product of two spin density matrix as shown in eqn. (\ref{TensorProductDensityMatrix}), the spin density matrix of the $M^*B$ system cannot. It can be seen from the righ-hand side of eqn. (\ref{InitialToFinalDensityMatrixVBIndices}) for the spin density matrix of the quasi-two-body state that it cannot be factored into two separate spin density matrices. This indicates that when this state is produced in the reaction, the two subsystems are correlated. Therefore, the measuring process of the recoil baryon's polarization cannot be described independently from the decay of the meson resonance. The recoil observables in this case would necessarily have to be expressed in terms of the SDME's or statistical tensors of the full $M^*B$ system.

\section{Parity Considerations in the General Expressions for the Recoil Observables}
For parity conserving reactions the recoil polarization observables, just like the other polarization observables, will be either even or odd in $\phi^{*}$. For the non-interference tensors, we use  (\ref{STHermiticity}), (\ref{StatisticalTensorsParity}) and find
\begin{equation} \label{Non-InterferenceTensorParity}
	\begin{split}
    t^{i}_{LM}(S_{k}) & = (-1)^{M}t^{i *}_{L-M}(S_{k}) = -1(-1)^{L} \zeta^{i} t^{i *}_{LM}(S_{k}), \\
    t^{i}_{LM}(S_{k}) & = \zeta^{i} t^{i *}_{LM}(S_{k}).
    \end{split} 
\end{equation}
As was mentioned in this chapter, $(-1)^{L} = -1$ for the non-interference matrices because only the odd $L$ terms are non-vanishing. Compare this to all other non-recoil polarization observables, where it is the even $L$ terms that are non-vanishing. For the tensors of type $t^{ij(\gamma N)}_{LM}$, we have instead
\begin{equation} \label{StatisticalTensorsHermiticityAndParity3}
    t^{ij (\gamma N)}_{LM}(S_{k}) = \zeta^{i}\zeta^{j} t^{ij (\gamma N) *}_{LM}(S_{k}).
\end{equation}
Eqn. (\ref{SDMEFactor}) therefore implies
\begin{equation} \label{StatisticalTensorsEvenOrOdd2}
    \begin{split}
        \Re[t^{k}] & = 0, \text{ if } \zeta^{k} = - 1 \rightarrow \text{ Observable odd in } \phi^{*} \text{ for } O_{ix} \text{ and } O_{iz}, \text{ even in } \phi^{*} \text{ in } O_{iy}, \\
        \Im[t^{k}] & = 0, \text{ if } \zeta^{k} = \hphantom{-}1\rightarrow \text{ Observable odd in } \phi^{*} \text{ for } O_{ix} \text{ and } O_{iz}, \text{ even in } \phi^{*} \text{ in } O_{iy}, \\
        & \\
        \Re[t^{kl}] & = 0, \text{ if } \zeta^{k}\zeta^{l} = -1\rightarrow \text{ Observable odd in } \phi^{*} \text{ for } O_{ix} \text{ and } O_{iz}, \text{ even in } \phi^{*} \text{ in } O_{iy},, \\
        \Im[t^{kl}] & = 0, \text{ if } \zeta^{k}\zeta^{l} = \hphantom{-}1\rightarrow \text{ Observable odd in } \phi^{*} \text{ for } O_{ix} \text{ and } O_{iz}, \text{ even in } \phi^{*} \text{ in } O_{iy},.
    \end{split}
\end{equation}

Deriving these symmetries for the interference statistical tensors is only slightly more complicated. The parity properties can be used to  show that in the sum over $M$ for the interference statistical tensors, the terms with $M < 0$ can be related to those with $M > 0$. From the properties of the Clebsh-Gordan coefficients in the second line of eqn. (\ref{Clebsch-GordanProperties}), it can be shown that
\begin{equation} \label{KappaPropertyRecoil2}
	\begin{split}
		\kappa^{L1}(S_{k},S_{l};s) & = (-1)^{S_{k} + S_{l} + L}\kappa^{L1}(S_{k},S_{l};s) \\
		\kappa^{L0}(S_{k},S_{l};s) & = (-1)^{S_{k} + S_{l} + L}\kappa^{L0}(S_{k},S_{l};s).
	\end{split}.
\end{equation}
Using this property along with the definition of $\chi^{L}$ shown in eqn. (\ref{ChiPhaseDefinition}), the property of the Wigner d-functions,
\begin{equation}
	d^{L}_{-M-N} = (-1)^{M-N}d^{L}_{MN},
\end{equation}
the parity properties of the statistical tensors shown in eqn. (\ref{StatisticalTensorsParity}) (we ignore the factor of $\bar{\eta}$ because another parity factor comes from the decay amplitude) and the even or odd property of the cosine and sine functions shown in eqns. (\ref{CosineEven}) and (\ref{SineOdd}), it can be shown that in the expression for $O_{ix}$ we get
\begin{equation} \label{ParityRelationsInRecoilObservables1}
	\begin{split}
		& \kappa^{L}(S_{k},S_{l};s)\bigg(\chi^{L}(S_{k},S_{l};s)d^{L}_{-M-1}(\theta^{*}) - d^{L}_{-M1}(\theta^{*})\bigg) \\
		& \qquad \times \bigg(\Re\Big[\tau^{i}_{L-M}(S_{k},S_{l})\Big]\cos(-M\phi^{*}) + \Im\Big[\tau^{i}_{L-M}(S_{k},S_{l})\Big]\sin(-M\phi^{*})\bigg) = \\
		& \qquad \zeta^{i} \kappa^{L}(S_{k},S_{l};s)\bigg(\chi^{L}(S_{k},S_{l};s)d^{L}_{M-1}(\theta^{*}) - d^{L}_{M1}(\theta^{*})\bigg) \\
		& \qquad \times \bigg(\Re\Big[\tau^{i}_{LM}(S_{k},S_{l})\Big]\cos(M\phi^{*}) - \Im\Big[\tau^{i}_{LM}(S_{k},S_{l})\Big]\sin(M\phi^{*})\bigg),
	\end{split}
\end{equation}
in the expression for $O_{iy}$ we get
\begin{align} \label{ParityRelationsInRecoilObservables2}
		& \kappa^{L}(S_{k},S_{l};s)\bigg(\chi^{L}(S_{k},S_{l};s)d^{L}_{-M-1}(\theta^{*}) + d^{L}_{-M1}(\theta^{*})\bigg) \nonumber \\
		& \qquad \times \bigg(\Im\Big[\tau^{i}_{L-M}(S_{k},S_{l})\Big]\cos(-M\phi^{*}) - \Re\Big[\tau^{i}_{L-M}(S_{k},S_{l})\Big]\sin(-M\phi^{*})\bigg) = \\
		& \qquad - \zeta^{i} \kappa^{L}(S_{k},S_{l};s)\bigg(\chi^{L}(S_{k},S_{l};s)d^{L}_{M-1}(\theta^{*}) + d^{L}_{M1}(\theta^{*})\bigg) \nonumber \\
		& \qquad \times \bigg(-\Im\Big[\tau^{i}_{LM}(S_{k},S_{l})\Big]\cos(M\phi^{*}) - \Re\Big[\tau^{i}_{LM}(S_{k},S_{l})\Big]\sin(M\phi^{*})\bigg),\nonumber
\end{align}
and in the expression for $O_{iz}$ we get
\begin{equation} \label{ParityRelationsInRecoilObservables3}
	\begin{split}
		& \kappa^{L0}(S_{k},S_{l};s) d^{L}_{-M0}(\theta^{*}) \\
			& \qquad \times \bigg(\Re\Big[\tau^{i}_{L-M}(S_{k},S_{l})\Big]\cos(-M\phi^{*}) + \Im\Big[\tau^{i}_{L-M}(S_{k},S_{l})\Big]\sin(-M\phi^{*})\bigg) = \\
			& \qquad \zeta^{i}\kappa^{L0}(S_{k},S_{l};s) d^{L}_{M0}(\theta^{*}) \\
			& \qquad \times \bigg(\Re\Big[\tau^{i}_{LM}(S_{k},S_{l})\Big]\cos(M\phi^{*}) - \Im\Big[\tau^{i}_{LM}(S_{k},S_{l})\Big]\sin(M\phi^{*})\bigg).
	\end{split}
\end{equation}
In other words two things happen. First, in all three observables the terms proportional to sine functions of $\phi^{*}$ get a minus sign. And second, in $O_{ix}$ and $O_{iz}$ the expression gets an overall factor of $\zeta^{i}$ while in $O_{iy}$ the expression gets an overall factor of $-\zeta^{i}$ (for tensors of type $\tau^{ij(\gamma N)}_{LM}$, we instead get factors of $\zeta^{i}\zeta^{j}$ or $-\zeta^{i}\zeta^{j}$). This allows us to reach the same conclusions we reached for the non-interference tensors in eqn. (\ref{StatisticalTensorsEvenOrOdd2}). By using the values of $\zeta^{i}$ shown in eqn. (\ref{SDMEFactor}), we can see for example that (using our definition of $O_{0j} \equiv P_{j'}$)
\begin{equation} \label{EvenOrOddRecoilExamples}
	\begin{split}
		P_{x'}, P_{z'}: & \text{ odd in } \phi^{*}, \\
		P_{y'}: & \text{ even in } \phi^{*}.
	\end{split}
\end{equation}
The symmetry properties of the recoil polarization observables are shown in tables \ref{table2}, \ref{table3} and \ref{table4}.

\begin{table}
\centering
\caption{Symmetry of recoil observables under $\phi^{*} \rightarrow -\phi^{*}$ when the polarization is measured along the $x'$-axis.}
{\renewcommand{\arraystretch}{1.3}
\begin{tabular}{| >{\centering} m{84pt} | >{\centering\arraybackslash} m{84pt} |} 
\hline
\multicolumn{2}{ | >{\raggedright} m{168pt} | }{Symmetry of observables under $\phi^{*} \rightarrow -\phi^{*}$: Measurement along $x'$} \\
\hline 
  Even & Odd \\
\hline
$I^{0}P_{x'}^s$ &  $I^{0}P_{x'}$ \\ 
$I^{0}P_{x'}^{\odot}$ &  $I^{0}P_{x'}^c$ \\ 
$I^{0}\mathcal{O}_{xx'}$ & $I^{0}\mathcal{O}^{yx'}$ \\
$I^{0}\mathcal{O}_{zx'}$ & $I^{0}\mathcal{O}_{yx'}^c$ \\
$I^{0}\mathcal{O}_{xx'}^c$ & $I^{0}\mathcal{O}_{xx'}^s$ \\
$I^{0}\mathcal{O}_{zx'}^c$ & $I^{0}\mathcal{O}_{zx'}^s$ \\
$I^{0}\mathcal{O}_{yx'}^s$ &  $I^{0}\mathcal{O}_{xx'}^\odot$ \\
$I^{0}\mathcal{O}_{yx'}^\odot$  &  $I^{0}\mathcal{O}_{zx'}^\odot$ \\
\hline
\end{tabular} }
\label{table2}
\end{table}

\begin{table}
\centering
\caption{Symmetry of recoil observables under $\phi^{*} \rightarrow -\phi^{*}$ when the polarization is measured along the $y'$-axis.}
{\renewcommand{\arraystretch}{1.3}
\begin{tabular}{| >{\centering} m{84pt} | >{\centering\arraybackslash} m{84pt} |} 
\hline
\multicolumn{2}{ | >{\raggedright} m{168pt} | }{Symmetry of observables under $\phi^{*} \rightarrow -\phi^{*}$: Measurement along $y'$} \\
\hline 
  Even & Odd \\
\hline
$I_0 P_{y'}$ & $I^{0}P_{y'}^s$ \\
$I^{0}P_{y'}^c$ &  $I^{0}P_{y'}^\odot$ \\ 
$I_0 \mathcal{O}_{yy'}$ &  $I_0 \mathcal{O}_{xy'}$ \\ 
$I_0 \mathcal{O}_{yy'}^c$ & $I_0 \mathcal{O}_{zy'}$ \\
$I_0 \mathcal{O}_{xy'}^s$ & $I_0 \mathcal{O}_{xy'}^c$ \\
$I_0 \mathcal{O}_{zy'}^s$ &  $I_0 \mathcal{O}_{zy'}^c$ \\
$I_0 \mathcal{O}_{xy'}^\odot$ &  $I_0 \mathcal{O}_{yy'}^s$ \\
$I_0 \mathcal{O}_{zy'}^\odot$ &  $I_0 \mathcal{O}_{yy'}^\odot$  \\
\hline
\end{tabular} }
\label{table3}
\end{table}

\begin{table}
\centering
\caption{Symmetry of recoil observables under $\phi^{*} \rightarrow -\phi^{*}$ when the polarization is measured along the $z'$-axis.}
{\renewcommand{\arraystretch}{1.3}
\begin{tabular}{| >{\centering} m{84pt} | >{\centering\arraybackslash} m{84pt} |} 
\hline
\multicolumn{2}{ | >{\raggedright} m{168pt} | }{Symmetry of observables under $\phi^{*} \rightarrow -\phi^{*}$: Measurement along $z'$} \\
\hline 
  Even & Odd \\
\hline
$I_0 P_{z'}^s$ & $I_0 P_{z'}$ \\
$I_0 P_{z'}^\odot$ & $I_0 P_{z'}^c$  \\ 
$I_0 \mathcal{O}_{xz'}$ & $I_0 \mathcal{O}_{yz'}$ \\ 
$I_0 \mathcal{O}_{zz'}$ &  $I_0 \mathcal{O}_{yz'}^c$ \\
$I_0 \mathcal{O}_{xz'}^c$ & $I_0 \mathcal{O}_{xz'}^s$ \\
$I_0 \mathcal{O}_{zz'}^c$ & $I_0 \mathcal{O}_{zz'}^s$  \\
$I_0 \mathcal{O}_{yz'}^s$ &  $I_0 \mathcal{O}_{xz'}^\odot$ \\
$I_0 \mathcal{O}_{yz'}^\odot$ & $I_0 \mathcal{O}_{zz'}^\odot$ \\
\hline
\end{tabular} }
\label{table4}
\end{table}

Notice that the symmetries derived in this chapter show that recoil observables can be used to extract those SDME's or statistical tensors from fits that cannot be found from the non-recoil observables. For example, it was shown in chapter \ref{GeneralExpression} that the non-interference statistical tensors with odd $L$ cannot be found from fits to non-recoil observables because they do not appear in the expressions for those observables. However, in recoil observables, it is instead the odd $L$ terms that appear in the expressions, so these observables can be used to extract them. The same thing happens for the interference tensors. For example, if we wanted to extract the tensors $\Re[\tau^{x (N)}]$, $\Im[\tau^{x (N)}]$, $\Re[\tau^{y (N)}]$, $\Im[\tau^{y (N)}]$, $\Re[\tau^{z (N)}]$ and $\Im[\tau^{z (N)}]$ from the non-recoil observables, we would have to measure the observables $P_{x}$, $P_{y}$ and $P_{z}$. But since $P_{x}$ and $P_{z}$ are odd while $P_{y}$ is even, the tensors $\Re[\tau^{x (N)}]$, $\Im[\tau^{y (N)}]$ and $\Re[\tau^{z (N)}]$ do not appear in the expressions and therefore cannot be extracted. However, $\Re[\tau^{x (N)}]$ will appear in the expressions for $\mathcal{O}_{xx'}$, $\mathcal{O}_{xy'}$ and $\mathcal{O}_{xz'}$, $\Im[\tau^{y (N)}]$ will appear in the expressions for $\mathcal{O}_{yx'}$, $\mathcal{O}_{yy'}$ and $\mathcal{O}_{yz'}$, and $\Re[\tau^{z (N)}]$ will appear in the expressions for $\mathcal{O}_{zx'}$, $\mathcal{O}_{zy'}$, and $\mathcal{O}_{zz'}$.

\chapter{Summary}
For convenience, we will reproduce many of the important expressions derived in this chapter. In this thesis, we have derived general versions of expressions that have previously appeared in the literature for the decay distribution of a photoproduced vector meson, and for the decay distribution of a photoproduced spin-3/2 baryon, in terms of their SDME's,
\begin{equation} \label{WDistributionRho2} 
\begin{split}
W(\theta^{*},\phi^{*};V) & = \rho(V)_{00} \cos^{2}\theta^{*} + \frac{1}{2}(\rho(V)_{11}+\rho(V)_{-1-1}) \sin^{2}\theta^{*} \\
     \mt -\frac{1}{\sqrt{2}}\Re[\rho(V)_{10}-\rho(V)_{0-1}]\sin2\theta^{*}\cos\phi^{*} \\
     \mt +\frac{1}{\sqrt{2}}\Im[\rho(V)_{10}-\rho(V)_{0-1}]\sin2\theta^{*}\sin\phi^{*} \\
     \mt - \Re[\rho(V)_{1-1}]\sin^{2}\theta^{*} \cos2\phi^{*}
                    + \Im[\rho(V)_{1-1}]\sin^{2}\theta^{*} \sin2\phi^{*},                     
 \end{split}
\end{equation}
and
\begin{equation} \label{WDistributionDelta12}
\begin{split}
W(\theta^{*},\phi^{*};3/2) & = \frac{5}{8}(\rho_{11}(3/2) +\rho_{-1-1}(3/2))(1 + \frac{3}{5} \cos 2 \theta^{*}) \\
    \mt +\frac{3}{4}(\rho_{33}(3/2)+\rho_{-3-3}(3/2))\sin^{2}\theta^{*} \\
    \mt -\frac{\sqrt{3}}{2}\Re[\rho_{31}(3/2)-\rho_{-1-3}(3/2)]\sin 2 \theta^{*} \cos \phi^{*} \\
    \mt +\frac{\sqrt{3}}{2}\Im[\rho_{31}(3/2)-\rho_{-1-3}(3/2)]\sin 2 \theta^{*} \sin \phi^{*} \\
    \mt -\frac{\sqrt{3}}{2}\Re[\rho_{3-1}(3/2)+\rho_{1-3}(3/2)]\sin^{2} \theta^{*} \cos 2 \phi^{*} \\
    \mt +\frac{\sqrt{3}}{2}\Im[\rho_{3-1}(3/2)+\rho_{1-3}(3/2)] \sin^{2} \theta^{*} \sin 2 \phi^{*}.
\end{split}
\end{equation}

The expression in terms of the spin density matrix was shown in eqn. (\ref{DistributionFinal}),
\begin{align} \label{DistributionFinal2}
 W(\theta^{*}, \phi^{*}) & = \sum_{i} \Bigg\{ \sum_{\lambda}\widetilde{d}^{S_{i}}_{\lambda, \lambda}(\theta^{*}) \rho_{\lambda, \lambda}(S_{i}) 
      + \sum_{\lambda > \lambda'} \widetilde{d}^{S_{i}}_{\lambda, \lambda'}(\theta^{*}) 
      \Bigg( \Re[\rho_{\lambda, \lambda'}(S_{i})] \cos[(\lambda-\lambda')\phi^{*})] \nonumber \\ 
      \mt - \Im[\rho_{\lambda, \lambda'}(S_{i})] \sin[(\lambda-\lambda')\phi^{*})] \Bigg) \Bigg\} \\
       \mt + \sum_{i > j} \sum_{\lambda, \lambda'} \widetilde{d}^{S_{i}, S_{j}}_{\lambda, \lambda'}(\theta^{*}) 
       \Bigg(\Re[\xi_{\lambda, \lambda'}(S_{i}, S_{j})] \cos[(\lambda-\lambda')\phi^{*})] \nonumber \\ 
      \mt - \Im[\xi_{\lambda, \lambda'}(S_{i}, S_{j})] \sin[(\lambda-\lambda')\phi^{*})] \Bigg) \nonumber
\end{align}
where
\begin{equation} \label{WignerLittleDSumDefinition2}
 \begin{split}
  \widetilde{d}^{S_{i}}_{\lambda, \lambda'}(\theta^{*}) & \equiv 2 \sum_{\eta} 
       d^{S_{i}}_{\eta, \lambda}(\theta^{*}) d^{S_{i}}_{\eta, \lambda'}(\theta^{*}), \\
  \widetilde{d}^{S_{i}, S_{j}}_{\lambda, \lambda'}(\theta^{*}) & \equiv 2 \sum_{\eta} 
       d^{S_{i}}_{\eta, \lambda}(\theta^{*}) d^{S_{j}}_{\eta, \lambda'}(\theta^{*}).
 \end{split}
\end{equation}

Expressions for the polarization observables were also derived. Since the matrices $\hat{\rho}$ and $\hat{\xi}$ can be decomposed into 16 matrices,
\begin{equation} \label{16Matrices2}
\begin{split}
\hat{\rho}(S_{k}) & = \hat{\rho}^{0}+\sum_{i=1}^{3}\Lambda^{i (N)} \hat{\rho}^{i (N)} 
                            + \sum_{i=1}^{3} \Lambda^{i (\gamma)} \hat{\rho}^{i (\gamma)}
                            + \sum_{i,j=1}^{3} \Lambda^{i (N)} \Lambda^{j (\gamma)} \hat{\rho}^{ij (N \gamma)}, \\
\hat{\xi}(S_{k}, S_{l}) & =  \hat{\xi}^{0}+\sum_{i=1}^{3}\Lambda^{i (N)} \hat{\xi}^{i (N)} 
                            + \sum_{i=1}^{3} \Lambda^{i (\gamma)} \hat{\xi}^{i (\gamma)}
                            + \sum_{i,j=1}^{3} \Lambda^{i (N)} \Lambda^{j (\gamma)} \hat{\xi}^{ij (N \gamma)},
\end{split}
\end{equation}
$W$ can also be decomposed into 16 terms,
\begin{equation} \label{WDecomposition2}
 \begin{split}
     W(\theta^{*},\phi^{*}) & = W^{0}(\theta^{*},\phi^{*}) 
          + \sum_{i} \Lambda^{i (\gamma)} W^{i (\gamma)}(\theta^{*},\phi^{*}) \\
		\mt + \sum_{i} \Lambda^{i (N)} W^{i (N)}(\theta^{*},\phi^{*})
          + \sum_{i,j} \Lambda^{i (\gamma)} \Lambda^{j (N)} W^{ij (\gamma N)}(\theta^{*},\phi^{*}).
 \end{split}
\end{equation}
The forms of the $W^{i}$'s are same as $W$ in eqn. (\ref{DistributionFinal2}) but with the density matrices substituted by the appropriate matrix in the decomposition in eqn. (\ref{16Matrices2}),
\begin{equation} \label{DistributionFinal162}
 \begin{split}
    W^{i}(\theta^{*}, \phi^{*}) & = \sum_{j} \Bigg\{\sum_{\lambda}\widetilde{d}^{S_{j}}_{\lambda, \lambda}(\theta^{*}) \rho^{i}_{\lambda, \lambda}(S_{j})
      + \sum_{\lambda > \lambda'} \widetilde{d}^{S_{j}}_{\lambda, \lambda'}(\theta^{*}) 
      \Bigg( \Re[\rho^{i}_{\lambda, \lambda'}(S_{j})] \cos[(\lambda-\lambda')\phi^{*})] \\ 
      \mt - \Im[\rho^{i}_{\lambda, \lambda'}(S_{j})] \sin[(\lambda-\lambda')\phi^{*})] \Bigg) \Bigg\} \\
      \mt \sum_{j > k} \sum_{\lambda, \lambda'} \widetilde{d}^{S_{j}, S_{k}}_{\lambda, \lambda'}(\theta^{*}) 
      \Bigg(\Re[\xi^{i}_{\lambda, \lambda'}(S_{j}, S_{k})] \cos[(\lambda-\lambda')\phi^{*})] \\ 
      \mt - \Im[\xi^{i}_{\lambda, \lambda'}(S_{j}, S_{k})] \sin[(\lambda-\lambda')\phi^{*})] \Bigg).
 \end{split}
\end{equation}
These $W^{i}$'s are equal to the 16 polarization observables involving only the beam and/or target,
\begin{align} \label{WObservables2}
  W^{0}(\theta^{*}, \phi^{*}) & = I_{0}(\theta^{*}, \phi^{*}), \nonumber \\
  W^{i (N)}(\theta^{*}, \phi^{*}) & = I_{0} P^{i}(\theta^{*}, \phi^{*}), \nonumber \\
  W^{x (\gamma)}(\theta^{*}, \phi^{*}) & = I_{0} \cos2\beta I^{c}(\theta^{*}, \phi^{*}), \nonumber \\
  W^{y (\gamma)}(\theta^{*}, \phi^{*}) & = I_{0} \sin2\beta I^{s}(\theta^{*}, \phi^{*}), \\
  W^{z (\gamma)}(\theta^{*}, \phi^{*}) & = I_{0} I^{\odot}(\theta^{*}, \phi^{*}), \nonumber \\
  W^{ix (\gamma N)}(\theta^{*}, \phi^{*}) & = I_{0} \cos2\beta P^{ic}(\theta^{*}, \phi^{*}), \nonumber \\
  W^{iy (\gamma N)}(\theta^{*}, \phi^{*}) & = I_{0} \sin2\beta P^{is}(\theta^{*}, \phi^{*}), \nonumber \\
  W^{iz (\gamma N)}(\theta^{*}, \phi^{*}) & = I_{0} P^{i\odot}(\theta^{*}, \phi^{*}). \nonumber
\end{align}
The SDME's of eqn. (\ref{DistributionFinal162}) are given by
\begin{equation} \label{DensityMatrixSigmaMatrixElements2}
  \begin{split}
    \rho^{0}_{\lambda \lambda'}(S_{k})
      & =  \frac{1}{4}\sum_{\substack{\lambda_{\gamma} \lambda_{N}\\ \lambda_{H}}} 
                        M_{\lambda\lambda_{H}; \lambda_{\gamma} \lambda_{N}}
                        M^{*}_{\lambda'\lambda_{H}; \lambda_{\gamma} \lambda_{N}}, \\ 
    \rho^{i (\gamma)}_{\lambda \lambda'}(S_{k})
      & =\frac{1}{4} \sum_{\substack{\lambda_{N} \lambda_{H}\\
                        \lambda_{\gamma} \lambda'_{\gamma}}} 
                        M_{\lambda\lambda_{H}; \lambda_{\gamma} \lambda_{N}}
                        \sigma^{i}_{\lambda_{\gamma} \lambda'_{\gamma}}
                        M^{*}_{\lambda'\lambda_{H}; \lambda'_{\gamma} \lambda_{N}}, \\
	\rho^{i (N)}_{\lambda \lambda'}(S_{k})
      & =  \frac{1}{4}\sum_{\substack{\lambda_{N} \lambda'_{N} \\
                        \lambda_{\gamma}\lambda_{H}}} 
                        M_{\lambda\lambda_{H};\lambda_{\gamma} \lambda_{N}}
                        \sigma^{i}_{\lambda_{N} \lambda'_{N}}
                        M^{*}_{\lambda'\lambda_{H}; \lambda_{\gamma} \lambda'_{N}}, \\
    \rho^{ij (\gamma N)}_{\lambda \lambda'}(S_{k})
      & =\frac{1}{4} \sum_{\substack{\lambda_{\gamma} 				  				\lambda'_{\gamma}  \\ \lambda_{N} \lambda'_{N}
                         \\ \lambda_{H}}} 
                        M_{\lambda\lambda_{H};\lambda_{\gamma} \lambda_{N}} \sigma^{i}_{\lambda_{\gamma} \lambda'_{\gamma}}\sigma^{j}_{\lambda_{N} \lambda'_{N}}
                        M^{*}_{\lambda'\lambda_{H}; \lambda'_{N} \lambda'_{\gamma}},
  \end{split}
\end{equation}
and
\begin{align} \label{DensityMatrixSigmaMatrixElements16Int2}
    \xi^{0}_{\lambda \lambda'}(S_{k},S_{l})
      & =  \frac{1}{4}\sum_{\substack{\lambda_{\gamma} \lambda_{N} \\ \lambda_{H}}} 
                        M_{\lambda\lambda_{H}; \lambda_{\gamma} \lambda_{N} }
                        M^{*}_{\lambda'\lambda_{H}; \lambda_{\gamma} \lambda_{N} }, \nonumber \\ 
    \xi^{i (\gamma)}_{\lambda \lambda'}(S_{k},S_{l})
      & =\frac{1}{4} \sum_{\substack{\lambda_{N} \lambda_{H}\\
                        \lambda_{\gamma} \lambda'_{\gamma}}} 
                        M_{\lambda\lambda_{H}; \lambda_{\gamma} \lambda_{N}}
                        \sigma^{i}_{\lambda_{\gamma} \lambda'_{\gamma}}
                        M^{*}_{\lambda'\lambda_{H}; \lambda'_{\gamma} \lambda_{N} }, \\
	\xi^{i (N)}_{\lambda \lambda'}(S_{k},S_{l})
      & =  \frac{1}{4}\sum_{\substack{\lambda_{N} \lambda'_{N} \\
                        \lambda_{\gamma}\lambda_{H}}} 
                        M_{\lambda\lambda_{H}; \lambda_{\gamma} \lambda_{N}}
                        \sigma^{i}_{\lambda_{N} \lambda'_{N}}
                        M^{*}_{\lambda'\lambda_{H}; \lambda_{\gamma} \lambda'_{N} }, \nonumber \\
    \xi^{ij (\gamma N)}_{\lambda \lambda'}(S_{k},S_{l})
      & =\frac{1}{4} \sum_{\substack{\lambda_{\gamma} 										\lambda'_{\gamma} \\
                        \lambda_{N} \lambda'_{N} \\ \lambda_{H}}} 
                        M_{\lambda\lambda_{H};\lambda_{\gamma} \lambda_{N}}
                        \sigma^{i}_{\lambda_{\gamma}\lambda'_{\gamma}}\sigma^{j}_{\lambda_{N} \lambda'_{N}}
                        M^{*}_{\lambda'\lambda_{H}; \lambda'_{\gamma}\lambda'_{N}}. \nonumber
\end{align}
where $\lambda_{H}$ is the helicity of the spectator hadron. This is how some of those expressions look like when explicit values of the matrix elements of the Pauli matrices are used,
\begin{align} \label{DensityMatrixHelicityAmplitude2}
    \xi^{0}_{\lambda \lambda' }(S_{k}, S_{l})
     =  \frac{1}{4}\sum_{\substack{\lambda_{\gamma} \lambda_{N}\\ \lambda_{H}}} 
                        M_{\lambda \lambda_{H}; \lambda_{\gamma} \lambda_{N}}
                        M^{*}_{\lambda' \lambda_{H}; \lambda_{\gamma} \lambda_{N}},
\end{align}

\begin{align} \label{DensityMatrixHelicityAmplitudesGamma2}
 & \xi^{x (\gamma)}_{\lambda \lambda' }(S_{k}, S_{l})
     =  \frac{1}{4}\sum_{\substack{\lambda_{\gamma} \lambda_{N}\\ \lambda_{H}}} 
                        M_{\lambda \lambda_{H}; \lambda_{\gamma} \lambda_{N}}
                        M^{*}_{\lambda' \lambda_{H}; -\lambda_{\gamma} \lambda_{N}}, \nonumber \\
 & \xi^{y (\gamma)}_{\lambda \lambda' }(S_{k}, S_{l})
     =-\frac{i}{4} \sum_{\substack{\lambda_{\gamma} \lambda_{N}\\ \lambda_{H}}} 
                        \lambda_{\gamma} M_{\lambda \lambda_{H}; \lambda_{\gamma}\lambda_{N} }
                        M^{*}_{\lambda' \lambda_{H}; -\lambda_{\gamma} \lambda_{N}}, \\
 & \xi^{z (\gamma)}_{\lambda \lambda'}(S_{k}, S_{l})
     = \frac{1}{4} \sum_{\substack{\lambda_{\gamma} \lambda_{N}\\ \lambda_{H}}} 
                        \lambda_{\gamma} M_{\lambda \lambda_{H}; \lambda_{\gamma} \lambda_{N}}
                        M^{*}_{\lambda' \lambda_{H}; \lambda_{\gamma} \lambda_{N}}. \nonumber
\end{align}

\begin{align} \label{DensityMatrixHelicityAmplitudesNucleon2}
 & \xi^{x (N)}_{\lambda \lambda' }(S_{k}, S_{l})
     =  \frac{1}{4}\sum_{\substack{\lambda_{\gamma} \lambda_{N}\\ \lambda_{H}}} 
                        M_{\lambda \lambda_{H}; \lambda_{\gamma} \lambda_{N}}
                        M^{*}_{\lambda' \lambda_{H}; \lambda_{\gamma} - \lambda_{N}}, \nonumber \\
 & \xi^{y (N)}_{\lambda \lambda' }(S_{k}, S_{l})
     =-\frac{i}{4} \sum_{\substack{\lambda_{\gamma} \lambda_{N}\\ \lambda_{H}}} 
                        \lambda_{N} M_{\lambda \lambda_{H}; \lambda_{\gamma}\lambda_{N}}
                        M^{*}_{\lambda' \lambda_{H}; \lambda_{\gamma} - \lambda_{N}}, \nonumber \\
 & \xi^{z (N)}_{\lambda \lambda'}(S_{k}, S_{l})
     = \frac{1}{4} \sum_{\substack{\lambda_{\gamma} \lambda_{N}\\ \lambda_{H}}} 
                        \lambda_{N} M_{\lambda \lambda_{H}; \lambda_{\gamma} \lambda_{N}}
                        M^{*}_{\lambda' \lambda_{H}; \lambda_{\gamma} \lambda_{N}}. \nonumber 
\end{align}

The parity relations of these 16 spin density matrices were also derived. This equation is also valid for the non-interference matrices,
\begin{equation} \label{DensityMatrixParityLast}
 \begin{split} 
  & \xi^{0}_{-\lambda, -\lambda'} = \bar{\eta} (-1)^{2 S^{*} +1}\zeta^{0}(-1)^{\lambda + \lambda'} \xi^{0}_{\lambda, \lambda'}, \\
  & \xi^{i (\gamma)}_{-\lambda, -\lambda'} = \bar{\eta} (-1)^{2 S^{*} +1}\zeta^{(\gamma) i} (-1)^{\lambda + \lambda'}
       \xi^{i (\gamma)}_{\lambda, \lambda'}, \\
  & \xi^{i (N)}_{-\lambda, -\lambda'} = \bar{\eta} (-1)^{2 S^{*} +1}\zeta^{(N) i} (-1)^{\lambda + \lambda'} 
       \xi^{i (N)}_{\lambda, \lambda'}, \\
  & \xi^{ij (\gamma N)}_{-\lambda, -\lambda'} = \bar{\eta} (-1)^{2 S^{*} +1}\zeta^{(\gamma) i} \zeta^{(N) j} 
       (-1)^{\lambda + \lambda'} \xi^{ij (\gamma N)}_{\lambda, \lambda'}, \\
    & \\
    & \qquad \qquad \qquad \bar{\eta} \equiv \eta\eta'.
    \end{split}
\end{equation}
$\eta$ and $\eta'$ are the intrinsic parities of the two interfering hadrons, $S^{*}$ is the spin of either of the interference hadrons (it makes no difference because both have either integer or half-integer spin), and the $\zeta$ factors are characteristic of each of the 16 matrices,
\begin{equation} \label{SDMEFactorLast}
    \begin{split}
    \zeta^{0} & = -1, \\
   \zeta^{i (\gamma)} & = \begin{cases}
                       \hphantom{-} 1, & \text{if } i = 2, 3 \\
                      -1, & \text{if } i = 1
                     \end{cases}, \\
  \zeta^{i (N)} & = \begin{cases}
                       \hphantom{-} 1, & \text{if } i = 1, 3 \\
                      -1, & \text{if } i = 2. 
                     \end{cases}.              
 \end{split}
\end{equation}
There is therefore a factor dependent on the intrinsic parities ($\bar{\eta}$), one dependent on the statistics of the hadrons ($(-1)^{2S^{*}+1})$, one dependent on the type of matrix ($\zeta^{i}$), and one dependent on the matrix element ($(-1)^{\lambda+\lambda'}$). These relationships reduce the number of independent matrix elements when the reaction is parity conserving. 

An equivalent way to represent the spin state of a system is to use statistical tensors instead of SDME's. The relationships between them are, using matrix notation,
\begin{equation} \label{TExpansionLast}
    \begin{split}
    \hat{\rho}(S) & = \sum_{L=0}^{2S}\sum_{M=-L}^{L} t_{LM}(S)\hat{T}^{\dagger}_{LM}(S), \\
  \hat{\xi}(S_{i}, S_{j}) & = \sum_{L = |S_{i} - S_{j}|}^{L = S_{i} + S_{j}} \sum_{M = -L}^{L} \tau_{LM}(S_{i}, S_{j})
    \hat{T}^{\dagger}_{LM}(S_{j}, S_{i}),
    \end{split}
\end{equation}
where the elements of the polarization operators are
\begin{equation} \label{TDefinitionLast}
    \begin{split}
  [T_{LM}(S)]_{m m'} & \equiv (-1)^{S-m'} C^{LM}_{S m; S - m'}, \\
  [T_{LM}(S_{i}, S_{j})]_{m, m'} & \equiv (-1)^{S_{j} - m'} C^{LM}_{S_{i}, m; S_{j}, -m'}.
  \end{split}
\end{equation}
Using index notation, the expression is
\begin{equation} \label{IndexNotationLast}
    \begin{split}
  \rho_{mm'}(S) & = \sum_{L=0}^{2S} \sum_{M=-L}^{L} (-1)^{S-m} t_{LM}(S) C^{LM}_{Sm';S-m}, \\
  \xi_{mm'}(S_{i},\ S_{j}) & = \sum_{L = - |S_{i}-S_{j}|}^{S_{i} + S_{j}} \sum_{M = -L}^{M = L} \tau_{LM}(S_{i},\ S_{j}) (-1)^{S_{i} - m} C^{LM}_{S_{j}m';S_{i}-m}.
    \end{split}
\end{equation}
The inverse relations are
\begin{equation} \label{STDefinitionLast}
    \begin{split}
 & t_{LM}(S)=\langle \hat{T}_{LM}(S) \rangle = \Tr[\hat{\rho}(S)\hat{T}_{LM}(S)] \\
 & \quad = \sum_{m,m'=-S}^{S}(-1)^{S-m}\rho_{mm'}(S)C^{LM}_{sm';s-m}, \\
 & \tau_{LM}(S_{i}, S_{j}) = \Big<\hat{\xi}(S_{i}, S_{j}) \hat{T}_{LM}(S_{j}, S_{i})\Big> = 
      \Tr[\hat{\xi}(S_{i}, S_{j}) \hat{T}_{LM}(S_{j}, S_{i})] \\
 & \quad = \sum_{m = -S_{i}}^{S_{i}} \sum_{m' = -S_{j}}^{S_{j}} (-1)^{S_{i} - m} \xi_{m, m'}(S_{i}, S_{j}) C^{LM}_{S_{j}, m'; S_{i}, -m}. 
    \end{split}
\end{equation}
The last two equations show that the SDME's and the statistical tensors are linear combinations of each other. It is therefore easy to change from using one representation to the other.

The expressions for the general decay distributions and observables were also derived in terms of the statistical tensors. The decay distribution is
\begin{equation} \label{DistributionStatisticalTensorsSphericalHarmonicsLast}
  \begin{split}
    & W(\theta^{*}, \phi^{*}; s) = \sum_{i} \sum_{L = 0}^{2 S_{i}} \sum_{M = -L}^{L} \kappa^{L}(S_{i}; s) (-1)^{M} t_{LM}(S_{i})
      Y_{L, -M}(\theta^{*}, \phi^{*}) \\
    & \qquad + \sum_{i > j} \sum_{L = |S_{i} - S_{j}|}^{S_{i} + S_{j}} \sum_{M = -L}^{L}
	2 \kappa^{L}(S_{i}, S_{j}; s) \Re\Big\{(-1)^{M} \tau_{LM}(S_{i}, S_{j}) Y_{L, -M}(\theta^{*}, \phi^{*})\Big\},
  \end{split}
\end{equation}
where
\begin{equation} \label{KappaDefinition2}
  \begin{split}
    \kappa^{L}(S_{i}; s) & \equiv \sqrt{\frac{4 \pi}{2 L + 1}} \sum_{m = -s}^{s} (-1)^{S_{i} - m} C^{L0}_{S_{i}, m; S_{i}, -m} \\
    & = \sqrt{\frac{4 \pi}{2 L + 1}} \sum_{m = -s}^{s} [\hat{T}_{L0}(S_{i})]_{mm}, \\
    \kappa^{L}(S_{i}, S_{j}; s) & \equiv \sqrt{\frac{4 \pi}{2 L + 1}} \sum_{m = -s}^{s} (-1)^{S_{i} - m} C^{L0}_{S_{j}, m; S_{i}, -m} \\
    & = \sqrt{\frac{4 \pi}{2 L + 1}} \sum_{m = -s}^{s} [\hat{T}_{L0}(S_{i},\ S_{j})]_{mm},.
  \end{split}
\end{equation}
We have also shown how for the non-interference statistical tensors only the terms with even $L$ appear in the distributions because
\begin{equation} \label{VanishingKappa}
    \kappa^{L}(S_{i};\ s) = 0 \quad \text{for } L \text{ odd}.
\end{equation}

If written in terms of cosine and sine functions of $\phi^{*}$, 
\begin{align} \label{DistributionStatisticalTensorsSphericalHarmonics3Last}
    & W(\theta^{*}, \phi^{*}; s) = \sum_{i} \sum_{L = 0}^{2 S_{i}} \sum_{M \geq 0}^{L} 2 \widetilde{\kappa}^{L}(S_{i}; s)P^{M}_{L}(\cos\theta^{*})\bigg(\Re\Big[t_{LM}(S_{i})\Big]\cos(M\phi^{*})  \nonumber \\ 
    & \qquad \quad + \Im\Big[t_{LM}(S_{i})\Big]\sin(M\phi^{*})\bigg) \\
    & \qquad + \sum_{i > j} \sum_{L = |S_{i} - S_{j}|}^{S_{i} + S_{j}} \sum_{M = -L}^{L}
 	2 \widetilde{\kappa}^{L}(S_{i}, S_{j}; s) P^{M}_{L}(\cos\theta^{*})\bigg(\Re\Big[\tau_{LM}(S_{i}, S_{j})\Big]\cos(M\phi^{*})\nonumber \\
	& \qquad \quad + \Im\Big[\tau_{LM}(S_{i}, S_{j})\Big]\sin(M\phi^{*})\bigg), \nonumber
\end{align}
where $P^{M}_{L}(\cos\theta^{*})$ are the associated Legendre polynomials and
\begin{align} \label{KappaTildeDefinitionFinal}
    \widetilde{\kappa}^{LM}(S_{i};s) & \equiv 2 \sqrt{\frac{(L-M)!}{(L+M)!}} \sum_{m = -s}^{s}(-1)^{S_{i}-m}C^{L0}_{S_{j}m;S_{i}-m}, \\
    \widetilde{\kappa}^{LM}(S_{i},S_{j};s) & \equiv 2 \sqrt{\frac{(L-M)!}{(L+M)!}} \sum_{m = -s}^{s}(-1)^{S_{i}-m}C^{L0}_{S_{j}m;S_{i}-m}. \nonumber 
\end{align}

Just like the spin density matrices, the statistical tensors can also be decomposed into 16 other tensors, 
\begin{equation} \label{Tensor16TensorsLast}
    \begin{split}
 t_{LM} & = t_{LM}^{0} + \sum_{j} \Lambda^{j (\gamma)} t_{LM}^{j (\gamma)} + \sum_{i} \Lambda^{j N} t_{LM}^{j N} 
      + \sum_{jk} \Lambda^{j (\gamma)} \Lambda^{k N} t_{LM}^{jk (\gamma N)}, \\
 \tau_{LM} & = \tau_{LM}^{0} + \sum_{j} \Lambda^{j (\gamma)} \tau_{LM}^{j (\gamma)} + \sum_{i} \Lambda^{j N} \tau_{LM}^{j N} 
      + \sum_{jk} \Lambda^{j (\gamma)} \Lambda^{k N} \tau_{LM}^{jk (\gamma N)}.
      \end{split}
\end{equation}
The observables that involve only the beam and/or target in terms of statistical tensors can also be written, and they have the same mathematical form as the decay distribution shown in eqn. (\ref{DistributionStatisticalTensorsSphericalHarmonics3Last}) but with the appropriate superscript, 
\begin{equation} \label{DistributionStatisticalTensorsSphericalHarmonics316Last}
  \begin{split}
    & W^{i}(\theta^{*}, \phi^{*}; s) = \sum_{i} \sum_{L = 0}^{2 S_{i}} \sum_{M \geq 0}^{L} 2 \widetilde{\kappa}^{L}(S_{i}; s) (-1)^{M}P^{M}_{L}(\cos\theta^{*})\bigg(\Re\Big[t^{i}_{LM}(S_{i})\Big]\cos(M\phi^{*}) \\
    & \qquad \quad + \Im\Big[t^{i}_{LM}(S_{i})\Big]\sin(M\phi^{*})\bigg) \\
    & \qquad + \sum_{i > j} \sum_{L = |S_{i} - S_{j}|}^{S_{i} + S_{j}} \sum_{M = -L}^{L}
	2 \widetilde{\kappa}^{L}(S_{i}, S_{j}; s) P^{M}_{L}(\cos\theta^{*})\bigg(\Re\Big[\tau^{i}_{LM}(S_{i}, S_{j})\Big]\cos(M\phi^{*}) \\
	& \qquad \quad + \Im\Big[\tau^{i}_{LM}(S_{i}, S_{j})\Big]\sin(M\phi^{*})\bigg).
  \end{split}
\end{equation}
where the relationship between the $\hat{\rho}^{i}$'s and $\hat{\xi}^{i}$'s to the $t^{i}_{LM}$'s and $\tau^{i}_{LM}$'s has the same form as those shown in eqn. (\ref{IndexNotationLast}),
\begin{equation} \label{IndexNotationLastIndex}
    \begin{split}
  \rho^{k}_{mm'}(S) & = \sum_{L=0}^{2S} \sum_{M=-L}^{L} (-1)^{S-m} t^{k}_{LM}(S) C^{LM}_{Sm';S-m}, \\
  \xi^{k}_{mm'}(S_{i},\ S_{j}) & = \sum_{L = - |S_{i}-S_{j}|}^{S_{i} + S_{j}} \sum_{M = -L}^{M = L} \tau^{k}_{LM}(S_{i},\ S_{j}) (-1)^{S_{i} - m} C^{LM}_{S_{j}m';S_{i}-m}.
    \end{split}
\end{equation}

The parity relations for the statistical tensors were also derived, 
\begin{equation} \label{StatisticalTensorsParityFinal}
    \begin{split}
        t^{k}_{LM}(S_{i}) & = - \eta \zeta^{k} (-1)^{L + M} t^{k}_{L-M}(S_{i}), \\
        \tau^{k}_{LM}(S_{i}, \ S_{j}) & = - \eta \zeta^{k} (-1)^{L + M + S_{i} - S_{j}} \tau^{k}_{L-M}(S_{i}, \ S_{j}).
    \end{split}
\end{equation}
These relationships reduce the number of independent statistical tensors when the reaction is parity conserving. These can also be used to prove which observables are even or odd in $\phi^{*}$ when the reaction is parity conserving. Table \ref{table} shows which non-recoil observables are even and which ones are odd in the variable $\phi^{*}$.

Finally, we also derived general expression for the observables that involve the recoil baryon for the case of a decaying baryon resonance,
\begin{equation} \label{RecoilObservablesX2}
	\begin{split}
		O_{ix}(\theta^{*},\phi^{*}) & = \sum_{k}\sum_{L = 0}^{S_{k}}\sum_{M = -L}^{L} \kappa^{L}(S_{k};s)\bigg(d^{L}_{M-1}(\theta^{*}) - d^{L}_{M1}(\theta^{*})\bigg) \\
		& \qquad \times \bigg(\Re\Big[t^{i}_{LM}(S_{k})\Big]\cos(M\phi^{*}) + \Im\Big[t^{i}_{LM}(S_{k})\Big]\sin(M\phi^{*})\bigg) \\
		\mt + \sum_{k > l} \sum_{L = |S_{k}-S_{l}|}^{S_{k} + S_{l}}\sum_{M = -L}^{L} \kappa^{L}(S_{k},S_{l};s)\bigg(\chi^L(S_{k},S_{l})d^{L}_{M-1}(\theta^{*}) - d^{L}_{M1}(\theta^{*})\bigg) \\
		& \qquad \times \bigg(\Re\Big[\tau^{i}_{LM}(S_{k},S_{l})\Big]\cos(M\phi^{*}) + \Im\Big[\tau^{i}_{LM}(S_{k},S_{l})\Big]\sin(M\phi^{*})\bigg),
	\end{split}
\end{equation}

\begin{equation} \label{RecoilObservablesY2}
	\begin{split}
		O_{iy}(\theta^{*},\phi^{*}) & = \sum_{k}\sum_{L = 0}^{S_{k}}\sum_{M = -L}^{L} \kappa^{L}(S_{k};s)\bigg(d^{L}_{M-1}(\theta^{*}) + d^{L}_{M1}(\theta^{*})\bigg) \\
		& \qquad \times \bigg(\Im\Big[t^{i}_{LM}(S_{k})\Big]\cos(M\phi^{*}) - \Re\Big[t^{i}_{LM}(S_{k})\Big]\sin(M\phi^{*})\bigg) \\
		\mt + \sum_{k > l} \sum_{L = |S_{k}-S_{l}|}^{S_{k} + S_{l}}\sum_{M = -L}^{L} \kappa^{L}(S_{k},S_{l};s)\bigg(\chi^L(S_{k},S_{l})d^{L}_{M-1}(\theta^{*}) + d^{L}_{M1}(\theta^{*})\bigg) \\
		& \qquad \times \bigg(\Im\Big[\tau^{i}_{LM}(S_{k},S_{l})\Big]\cos(M\phi^{*}) - \Re\Big[\tau^{i}_{LM}(S_{k},S_{l})\Big]\sin(M\phi^{*})\bigg),
	\end{split}
\end{equation}
and
\begin{equation} \label{RecoilObservablesZ2}
	\begin{split}
		O_{iz}(\theta^{*},\phi^{*}) & = \sum_{k}\sum_{L = 0}^{S_{k}}\sum_{M = -L}^{L} \kappa^{L0}(S_{k};s)d^{L}_{M0}(\cos\theta^{*}) \\
		& \qquad \times \bigg(\Re\Big[t^{i}_{LM}(S_{k})\Big]\cos(M\phi^{*}) + \Im\Big[t^{i}_{LM}(S_{k})\Big]\sin(M\phi^{*})\bigg) \\
		\mt + \sum_{k > l} \sum_{L = |S_{k}-S_{l}|}^{S_{k} + S_{l}}\sum_{M = -L}^{L} \kappa^{L0}(S_{k},S_{l};s) d^{L}_{M0}(\cos\theta^{*}) \\
		& \qquad \times \bigg(\Re\Big[\tau^{i}_{LM}(S_{k},S_{l})\Big]\cos(M\phi^{*}) + \Im\Big[\tau^{i}_{LM}(S_{k},S_{l})\Big]\sin(M\phi^{*})\bigg),
	\end{split}
\end{equation}
where we have defined
\begin{align} \label{KappaRecoilDefinitioin2}
	& \kappa^{L1}(S_{k},S_{l};s) \equiv \sqrt{\frac{2s(s+1)(2s+1)}{3}}\sum_{m,m' = -s}^{s}(-1)^{S_{k}-s}C^{L1}_{S_{l}m';S_{k}-m}C^{11}_{sm';s-m}, \nonumber \\
	& \kappa^{L0}(S_{k},S_{l};s) \equiv \sqrt{\frac{2s(s+1)(2s+1)}{3}}\sum_{m,m' = -s}^{s}(-1)^{S_{k}-s}C^{L0}_{S_{l}m';S_{k}-m}C^{10}_{sm';s-m}, \\
	& \qquad \qquad \qquad \qquad \qquad \kappa^{L1}(S_{k};s) \equiv \kappa^{L1}(S_{k},S_{k};s), \nonumber \\
	& \qquad \qquad \qquad \qquad \qquad \kappa^{L0}(S_{k};s) \equiv \kappa^{L0}(S_{k},S_{k};s), \nonumber
\end{align}
and
\begin{equation} \label{ChiPhaseDefinition2}
	\chi^L(S_{k},S_{l}) \equiv (-1)^{S_{k} + S_{l} - L} = \pm 1.
\end{equation}
The second subscript in $O_{ij}$ refers to the axis along which the recoil baryon's polarization is measured. The first one refers to the superscript of the statistical tensor in the expression (e.g., $x(\gamma), z(N), zy(\gamma N)$, etc.). The observable $P^\odot_{x'}$, for example, is equal to $O_{zx}$. The parity relations in eqn. (\ref{StatisticalTensorsParityFinal}) can also be used to show which ones of these observables are even or odd in $\phi^*$ when the reaction is parity conserving. Tables \ref{table2}, \ref{table3} and \ref{table4} shows which recoil observables are even and which ones are odd in the variable $\phi^{*}$. These expressions are not valid for the case of a decaying meson resonance, since those cannot be expressed in terms of the statistical tensors of said meson. Rather, they would necessarily have to be expressed in terms of statistical tensors describing the entire meson-baryon quasi-two-body state.  

\chapter{Conclusions}
As stated at the beginning of this thesis, this work was initiated due to the current interest in photoproduction reactions in hadron spectroscopy. Because of the importance of polarization measurements in these experiments, we consider that it is important to understand as much as possible the relationship between polarization observables and the resonances that contribute to these reactions.

To this end, we have investigated the role that spin plays in two pseudoscalar (and scalar) photoproduction, $\gamma N \rightarrow M_{1}M_{2}B$, where $M_{1}$ and $M_{2}$ are mesons and $B$ is a baryon. But more specifically, for reaction channels involving the photoproduction of a quasi-two-body state, such as $\gamma N \rightarrow M^{*} B \rightarrow M_{1}M_{2}B$, $\gamma N \rightarrow M_{1} B^{*} \rightarrow M_{1}M_{2}B$ and $\gamma N \rightarrow M_{2} B^{*} \rightarrow M_{1}M_{2}B$, where $M^{*}$ and $B^{*}$ is an unstable meson and baryon, respectively, that undergo a two-body decay. 

The interest was in examining the connection between the spin state of the unstable hadron, $M^{*}$ or $B^{*}$, and polarization measurements, and also its connection with the decay angular distributions of the decaying hadron in a model-independent way. Another way of stating this goal is as follows: How much information can we gain about the spin state of the unstable hadron from experimental observables without having to rely on a specific model?

Note how we have used the generic labels $M_{1}$, $M_{2}$ and $B$ in order to emphasize that our conclusions are independent of the type of hadrons involved in the reaction. Therefore, our conclusions can also be applied to reactions such as
\begin{equation} 
    \begin{split}
        \gamma N & \rightarrow \pi \pi N, \\
        \gamma N & \rightarrow \pi \eta N, \\
        \gamma N & \rightarrow \eta \eta N, \\
        \gamma N & \rightarrow K \bar{K} N, \\
        \gamma N & \rightarrow \pi K Y, \\
        \gamma N & \rightarrow K K \Xi,
    \end{split}
\end{equation}
just to name a few. The only property of relevance in this research is the spin of the hadrons involved in the reaction.

This led us to deriving general expressions for the decay distributions of the contributing resonances and of the polarization observables in terms of quantities that describe the spin of the resonances: either SDME's or statistical tensors. Eqns. (\ref{WDistributionRho2}) and (\ref{WDistributionDelta12}), which have been previously derived in the literature, are special cases of the general expressions. These two equations had been used to aid in the interpretation of the data gathered from experiments, but can only be applied when vector mesons and spin-3/2 baryons contribute to the reaction. By contrast, the general expression derived in this work can be applied to contributions for hadrons of any spin, and it also takes into account the interference effects among the different reaction channels. As mentioned in the introduction, this is relevant because in the energy region where the ``missing'' resonances are expected to be found there are many broad and overlapping resonances, so it is useful to have an expression that can be applied to more general situations.

The general expression for the decay distribution of the resonance in terms of SDME's is shown in eqn. (\ref{DistributionFinal2}). It has a very simple structure: it consists of a linear sum of the real and imaginary parts of all of the independent spin density matrix elements (the hermiticity condition shown in eqn. (\ref{Hermiticity}) reduces the number of elements that are independent, which is the reason one of the sums only has the terms with $\lambda > \lambda'$). The terms with the real parts are proportional to cosine functions of $\phi^{*}$, while the imaginary parts are proportional to sine functions of this angle, of the form $\cos[n\phi^{*}]$ and $\sin[n\phi^{*}]$, where $n$ is an integer. The value of $n$ is completely determined by the SDME it has as its factor: it does not depend on the spin of the decaying hadron, but it is the difference between the two helicity indices of the matrix element, $n = \lambda - \lambda'$. Each of the terms will also be proportional to a trigonometric function of $\theta^{*}$, which is also a function of the spin and of the decaying hadron (or hadrons, for the interference matrix elements). The explicit form of these functions, $\widetilde{d}^{S_{i}}_{\lambda \lambda'}(\theta^{*})$ for the diagonal submatrices and $\widetilde{d}^{S_{i}S_{j}}_{\lambda \lambda'}(\theta^{*})$ for the off-diagonal interference submatrices, shown in eqn. (\ref{WignerLittleDSumDefinition2}), are bilinear combinations of Wigner $d$-functions. As such, while their explicit forms may be long and complicated, they are related to well known functions in an easy way. 

With this simple general expression, the decay distribution in terms of the SDME's can be generated for any number of contributing hadrons, all of arbitrary spin, and including the effects of interference. It can even be shown that the two equations from the literature, eqn. (\ref{WDistributionRho2}) and eqn. (\ref{WDistributionDelta12}), can be obtained from this general equation by including only one hadron with $S_{i} = 1$ or $S_{i} = 3/2$.

One reason for which eqns. (\ref{WDistributionRho2}) and (\ref{WDistributionDelta12}) are useful is because they can be used to extract the values of the SDME's from measurements of the decay distributions and fitting the data with the SDME's as fit parameters. Therefore, the derived general expression can be used in the same way but it can be applied in situations with an arbitrary number of contributing resonances of arbitrary spins.

Since the $\theta^{*}$ dependence of the distributions is a function of the spins of the hadrons that contribute, the fitting procedure can also be used to determine the values of spin that are not involved in the reaction: after doing the fit procedure, if the coefficient in a particular value of one of the $\widetilde{d}^{S_{i}}$ is equal to zero, this indicates that hadrons of spin $S_{i}$ are not contributing to the reaction.

The general expression for the decay distribution using statistical tensors is shown in eqn. (\ref{DistributionStatisticalTensorsSphericalHarmonicsLast}) in terms of spherical harmonics and in eqn. (\ref{DistributionStatisticalTensorsSphericalHarmonics3Last}) in terms of the associated Legendre polynomials. These equations show that the expression for the decay distribution acquires an even simpler form when written in terms of the tensors. One of the reasons is that the $\theta^{*}$ and $\phi^{*}$ dependence is entirely contained in the spherical harmonics, which are very well known functions. Another is that, unlike in the expression in terms of the SDME's, this expression is manifestly rotationally invariant because contractions between a contravariant and a covariant tensor, such as $t_{LM}$ and $(-1)^{M}Y_{L-M}$, are rotationally invariant. In fact, the constraint of rotational invariance is enough to conclude that each term in the distribution has to be proportional to this factor. When the distribution is instead expressed in terms of the Legendre polynomials, the $\theta^{*}$ dependence in contained in them, which are also well known functions. In this expression the $\phi^{*}$ dependence is entirely contained in cosine and sine functions, $\cos[M\phi^{*}]$ and $\sin[M\phi^{*}]$. Since the prefactor in their arguments is $M$, it is easy to know which trigonometric functions will be accompanying each of the tensors in the sum. The values of the prefactors, $\kappa^{L}$ and $\widetilde{\kappa}^{LM}$, are proportional to a factor very similar to the trace of the polarization operator $\hat{T}_{L0}$ (it will in general be a sum of only some of its diagonal elements). 

Another reason for its simplicity is that each of the cosine and sine functions of $\phi^{*}$ are multiplied by a single tensor. In contrast, when the distribution is expressed in terms of SDME's, each of the cosine and sine functions of $\phi^{*}$ are multiplied by a linear combination of them. This means that if these expressions for the distributions were to be used to perform fits to a sum of trigonometric functions of the form $\cos[n\phi^{*}]$ and $\sin[n\phi^{*}]$, with $n$ being an integer, the fit parameters will end up being sums of SDME's. See for example the decay distribution for vector mesons in eqn. (\ref{WDistributionRho2}) and (\ref{WDistributionDelta12}). Therefore, when the distribution is expressed in terms of the statistical tensors, it becomes manifest the fact that the number of independent parameters describing the spin of the resonance (the SDME's or the statistical tensors) that can be extracted from such fits is actually less than their total number. This is seen from eqn. (\ref{VanishingKappa}), which shows that for the non-interference tensors only those with even $L$ appear in the distribution. 

Expressions for the polarization observables were also derived and are shown in eqns. (\ref{DistributionFinal162}) and  (\ref{DistributionStatisticalTensorsSphericalHarmonics3Last}), where the $W^{i}$'s relation to the observables are shown in eqn. (\ref{WObservables2}). They have the same mathematical form as the decay distributions shown in eqns. (\ref{WDistributionRho2}) and (\ref{WDistributionDelta12}) but with the SDME's or statistical tensors substituted with the appropriate one of the 16 expansion SDME's and tensors in eqns. (\ref{16Matrices2}) and (\ref{Tensor16TensorsLast}). Note that $W^{i}$'s arise from the decomposition of the decay distribution $W$, as shown in eqn. (\ref{WDecomposition2}).

The 16 SDME's or statistical tensors in these expressions for the observables can be extracted by measuring the polarization observables and performing fits with these expressions with the SDME's and tensors as fit parameters. And since these expressions are general, they can be used for any resonance that contributes with any spin, and also for situations in which more than one resonance contributes to the reaction. 

This is of great importance because in order to accomplish a complete experiment, the helicity amplitudes of the process must be found. But these amplitudes are related to the SMDE's by eqns. (\ref{DensityMatrixSigmaMatrixElements2}) and (\ref{DensityMatrixSigmaMatrixElements16Int2}) (or to the statistical tensors, since they are related to these SDME's by eqns. (\ref{IndexNotationLast}) and (\ref{STDefinitionLast})). Once the helicity amplitudes are known, they can be used to perform partial wave analyses, which are of great help in the search of resonances. As has been mentioned, one of the reasons why these searches are important is because there is an interest in finding the ``missing'' resonances that have been predicted by quark models but not found. The general expressions derived in this thesis could therefore be of help in these searches.

For parity conserving reactions, we found the general parity properties of the 16 SDME's and statistical tensors in eqns. (\ref{DensityMatrixParityLast}), (\ref{SDMEFactorLast}), and (\ref{StatisticalTensorsParityFinal}). These parity relations can be used to find whether a polarization observable is even or on in the variable $\phi^{*}$. It turns out that when these parity relations are applied to the expressions for the observables, the factor $\zeta^{i}$ is the only one that does not vanish. Therefore, whether an observable is even or odd in $\phi^{*}$ is only dependent on the type of observable, not on the properties of the hadrons involved in the reaction. Being even in $\phi^{*}$ means that every term in the expression for the observables will be proportional to a cosine function of $\phi^{*}$, while being odd means the terms will be proportional to sine function of $\phi^{*}$. A summary of which observables are even or odd in $\phi^*$ can be found in table \ref{table}.

For the case of a decaying baryon resonance, the general expressions for polarization observables that involve a measurement of the recoil baryon's polarization where also derived and are shown in eqns. (\ref{RecoilObservablesX2}), (\ref{RecoilObservablesY2}), and (\ref{RecoilObservablesZ2}). It was shown that, for the non-interference tensors, only terms with odd $L$ show up in the expressions. By contrast, in the expressions for the observables that only involve the beam and/or target, only terms with even $L$ show up in the sum for the non-interference tensors. This shows how not all of the independent elements that describe the spin of the resonance (in this case, the statistical tensors) can be extracted using only polarized beams and targets. Experiments in which the polarization of the recoil baryon can be measured must also be performed in order to achieve a complete experiment.

Future work could aim to derive the general expressions for the recoil observables for the case of a decaying meson resonance. These would necessarily have to be expressed in terms of the statistical tensors of the full meson-baryon quasi-two-body state, rather than the statistical tensors of the meson subsystem.

It is important to note that for each resonance that contributes to these expressions, the angles used to describe its decays will be different because they can decay into three different particle pairs: $M_{1}M_{2}$, $M_{2}B$, and $M_{1}B$. This means that there are three different sets of $\theta^{*}$ and $\phi^{*}$ that will appear in the general expressions, even though only one of the sets is independent. Appendix \ref{ChangingCoordinateSystems} describes how to relate these different sets of angles, shown in eqns. (\ref{eqn 54}), (\ref{eqn 56}) and (\ref{eqn 57}). However, they are not related in a simple manner. If the general expressions are used for doing fits, instead of rewriting the non-independent sets of angles in terms of the dependent ones, all three sets could be measured in experiments and used in the general expressions to perform the fits. Future research could be devoted to finding a more compact way of expressing these relations between the different sets of angles, perhaps in terms of well known functions such as Wigner $\mathcal{D}$-functions or spherical harmonics.

It is important to remember the limitations of these general expression. They are only valid for reactions that only have contributions from quasi-two-body states. Also, as shown from eqns. (\ref{FirstEquation}) to (\ref{SecondEquation}) and from (\ref{ThirdEquation}) to (\ref{FourthEquation}), which are specific cases of the general expressions in which three resonances contribute to the reaction, the expressions can become very long, especially when the spin of the resonances is high and when many of them contribute. However, we emphasize that the general expressions are very simple, and shows that, while expressions for specific cases are long, it is straightforward to generate them from the general expressions.

Another thing to consider is the fact that the only property of the decaying resonance that plays a role in the form of it's distribution as a function of its SDME's or statistical tensors is its total spin. Therefore, for cases where you have more than one resonance with the same spin, the factors of the trigonometric functions in the expressions will be the sum of the SDME's or statistical tensors of all of the resonances of the same spin. As such, the general expressions along with measurements of the angles $\theta^{*}$ and $\phi^{*}$ will not be able to show on their own if more than one resonance of a particular spin is contributing to the reaction. 

\appendix
\chapter{Changing Phase Space Coordinates} \label{ChangingCoordinateSystems}
The expression for the angular distribution $\gamma N \rightarrow M_{1}M_{2}N$ is given in terms of Wigner $\mathcal{D}$-functions. The arguments to these functions are the decay angles of the products of the unstable resonance. However, as noted in section \ref{Kinematics}, each of the tree pathways A, B, and C, defined in eqns. (\ref{PathwayA}), (\ref{PathwayB}) and (\ref{PathwayC}), are described in terms of different angles, each defined in different reference frames. When the expression of the distribution involves more than one decaying resonance and interference terms, each of the arguments of the Wigner $\mathcal{D}$-functions will be different. It would be convenient to be able to relate these three different set of angles to each other, so that the expression for the decay distribution can be written in terms of only one set of angles. Refer to section \ref{Kinematics} for the definition of the kinematic variables. 

Since we are assuming an unstable resonance will decay into two of the hadrons in the three-body final state, its four-momentum will have three possible values,
\begin{align} \label{eqn 51}
  q_{12} & = q_{1} + q_{2} \\
  q_{31} & = p' + q_{1} \\
  q_{23} & = q_{2} + p' ,
\end{align}
where these four-momenta are defined in the overall center of mass frame. Decaying mesons will always have four-momentum equal to $q_{12}$,
while decaying baryons can have either four-momentum $q_{31}$ or $q_{23}$. As has been shown, W will be a function of Wigner $\mathcal{D}$-functions, 
and these contain the angular dependence on the decay angles. Since there are three different channels, each Wigner $\mathcal{D}$-function that appears
in the expression will be a function of one of three sets of angles: $\{\theta^{*}_{12}, \phi^{*}_{12}\}$, 
$\{\theta^{*}_{31}, \phi^{*}_{31}\}$,
$\{\theta^{*}_{23}, \phi^{*}_{23}\}$, which are the decay angles defined in the rest frame of the decaying resonance. These angles and the coordinate axes 
of these frames are defined analogously to the primed coordinate system defined in eqn. (\ref{eqn 9}),
\begin{align} \label{eqn 52}
 & \bm{\hat{z}}_{ij} = \frac{\bm{q}_{ij}}{|\bm{q}_{ij}|},\ \bm{\hat{y}}_{ij} = \frac{\bm{k} \times \bm{q}_{ij}}{|\bm{k} \times \bm{q}_{ij}|},\ 
      \bm{\hat{x}}_{ij} = \bm{\hat{y}}_{ij} \times \bm{\hat{z}}_{ij}, \nonumber \\
 & \cos(\theta^{*}_{ij})=\bm{\hat{\pi}}_{ij} \cdot \bm{\hat{z}}_{ij}, \ \cos(\phi^{*}_{ij})=\frac{\bm{\hat{y}}_{ij} \cdot (\bm{\hat{z}}_{ij} 
      \times \bm{\hat{\pi}}_{ij})}{|\bm{\hat{z}}_{ij} \times \bm{\hat{\pi}}_{ij}|}, \ \sin(\phi^{*}_{ij})=-\frac{\bm{\hat{x}}_{ij} \cdot
      (\bm{\hat{z}}_{ij} \times \bm{\hat{\pi}}_{ij})}{|\bm{\hat{z}}_{ij} \times \bm{\hat{\pi}}_{ij}|}, \ 
      \bm{\hat{\pi}}_{ij} = \frac{\bm{q^{*}_{i(ij)}}}{|\bm{q^{*}_{i(ij)}}|}, \\
 & i,j \in \{1, 2, 3\}, \nonumber 
\end{align}
where i and j label the three possible rest frames, $\bm{q}_{ij}$ is the spatial part of the four-momenta defined in eqn. (\ref{eqn 51}),
and $\bm{q^{*}_{i(ij)}}$ is the spatial part of the four-vector of the $i$-th final state particle in the center of mass frame of the $i$-th
and $j$-th final state particle, i.e., the frame in which $q^{*}_{ij}=0$ (The numbers in parentheses in $q^{*}_{i(ij)}$ are used to label the
coordinate frame).
The primed coordinate system that we defined in eqn. (\ref{eqn 9}) is the case when
$i=1$ and $j=2$. 
We will assume the set $\{\theta^{*}_{12}, \phi^{*}_{12}\}$ is the one chosen to describe the scattering reaction so we will derive expressions for the other two sets in terms of this one. 

The first step is to find the four-momenta of the two final state mesons $q_{1}$ and $q_{2}$ in the overall center of mass frame from their 
four-momenta in their center of mass frame $q^{*}_{1}$ and $q^{*}_{2}$. Eqns. (\ref{eqn 10}) and (\ref{eqn 11}) show the dependence of these 
two four-vectors on the phase space coordinates $s_{M_{1} M_{2}}$, $\theta^{*}_{12}$ and $\phi^{*}_{12}$. We find the four-momenta in the
overall center of mass frame by applying this boost and rotation,
\begin{equation} \label{eqn 53}
\begin{split}
  q_{i}^{\mu} & = \tensor{\Lambda}{^\mu _\nu} q_{i}^{* \nu}, \\
  \hat{\Lambda} & = \begin{pmatrix}
                     1 & 0 & 0 & 0 \\[+4pt]   
                     0 & \cos(\theta) & 0 & \sin(\theta) \\[+4pt]
                     0 & 0 & 1 & 0 \\[+4pt]
                     0 & -\sin(\theta) & 0 & \cos(\theta)
                    \end{pmatrix}
                    \begin{pmatrix}
                     \frac{E_{q_{12}}}{\sqrt{s_{M_{1} M_{2}}}} \hphantom{11} & 0 \hphantom{111}
                          & 0 \hphantom{11} & \frac{|\bm{q_{12}}|}{\sqrt{s_{M_{1} M_{2}}}} \\
                     0 \hphantom{11} & 1 \hphantom{111} 
                          & 0 \hphantom{11} & 0 \\
                     0 \hphantom{11} & 0 \hphantom{111}
                          & 1 \hphantom{11} & 0 \\
                     \frac{|\bm{q_{12}}|}{\sqrt{s_{M_{1} M_{2}}}} \hphantom{11} & 0 \hphantom{111} 
                          & 0 \hphantom{11} & \frac{E_{q_{12}}}{\sqrt{s_{M_{1} M_{2}}}}
                    \end{pmatrix},
\end{split}
\end{equation}
where $\theta$ is the scattering angle and $E_{q_{12}}$, $|\bm{q_{12}}|$ and $s_{M_{1},M_{2}}$ can be found in eqn. (\ref{eqn 6}). 
The expression for $q_{1}$ and $q_{2}$ is therefore

\begin{align} \label{eqn 54}
  q_{i} & = \begin{pmatrix}
             \frac{1}{\sqrt{s_{M_{1},M_{2}}}}(E_{q_{12}} E^{*}_{q_{i}} + \eta_{i} |\bm{q_{12}}| |\bm{q^{*}_{1}}| \cos(\theta^{*}_{12})) \\[+5pt]
             \eta_{i} |\bm{q^{*}_{1}}| \cos(\theta) \sin(\theta^{*}_{12}) \cos(\phi^{*}_{12}) 
                  + \frac{\sin(\theta)}{\sqrt{s_{M_{1},M_{2}}}}(|\bm{q_{12}}| E^{*}_{q_{i}} 
                  + \eta_{i} E_{q_{12}} |\bm{q^{*}_{1}}|) \\[+5pt]
             \eta_{i} |\bm{q^{*}_{1}}| \sin(\theta^{*}_{12}) \sin(\phi^{*}_{12}) \\[+5pt]
             -\eta_{i} |\bm{q^{*}_{1}}| \sin(\theta) \sin(\theta^{*}_{12}) \cos(\phi^{*}_{12}) 
             + \frac{\cos(\theta)}{\sqrt{s_{M_{1},M_{2}}}}(|\bm{q_{12}}|E^{*}_{q_{i}}
             + \eta_{i} E_{q_{12}}|\bm{q^{*}_{1}}|)
            \end{pmatrix}, \\
  \eta_{i} & = \begin{cases}
                1, & i=1 \\
                -1, & i=2
               \end{cases}, \nonumber
\end{align}
where $E^{*}_{q_{i}}$ and $|\bm{q_{1}}|$ are defined in eqn. (\ref{eqn 11}). This leads us to the expression for the unit vector $\bm{\hat{n}}$
normal to the plane that contains the three-momenta of the three reaction products,
\begin{equation} \label{eqn 55}
 \frac{\bm{q_{2} \times \bm{q_{1}}}}{|\bm{q_{2} \times \bm{q_{1}}}|} = \bm{\hat{n}} = \begin{pmatrix}
                                                                                    -\cos(\theta) \sin(\phi^{*}_{12}) \\
                                                                                    \cos(\phi^{*}_{12}) \\
                                                                                    \sin(\theta) \sin(\phi^{*}_{12})
                                                                                   \end{pmatrix},
\end{equation}
where $\bm{q_{2}}$ and $\bm{q_{1}}$ are the spatial parts of the four-vectors $q_{2}$ and $q_{1}$. 

Eqn. (\ref{eqn 54}), along with eqn. (\ref{eqn 5}), gives the expressions for the four-momenta of the final state particles, $q_{1}$,
$q_{2}$, and $p'$, as functions of the 5 independent phase space variables and the masses of the particles involved, $m_{B}$, $m_{M_{1}}$,
and $m_{M_{2}}$. These four-vectors will be used to find
the other set of angles that we are interested in, either $\{\theta^{*}_{31}, \phi^{*}_{31}\}$ or $\{\theta^{*}_{23}, \phi^{*}_{23}\}$. In order
to accomplish this, we must find the expression for the four-vector $q^{*}_{3(31)}$ if we are interested in $\{\theta^{*}_{31}, \phi^{*}_{31}\}$
or for the four-vector $q^{*}_{2(23)}$ if we are interested in $\{\theta^{*}_{23}, \phi^{*}_{23}\}$. For generality, will refer to the
four-vector of interest as $q^{*}_{i(ij)}$ for $(i,j) \in \{(3,1),(2,3)\}$, in what follows.

To find $q^{*}_{i(ij)}$, we first rotate the $\{\bm{\hat{x}}, \bm{\hat{y}}, \bm{\hat{z}}\}$ system into the 
$\{\bm{\hat{x}}_{ij}, \bm{\hat{y}}_{ij}, \bm{\hat{z}}_{ij}\}$ system by applying a rotation around the $\bm{\hat{z}}$ axis followed by a 
rotation around the $\bm{\hat{y}}$. We then apply a boost in the $\bm{\hat{z}_{ij}}$ direction to reach the frame in which $\bm{q^{*}_{ij}}=0$. 
$q^{*}_{i(ij)}$ is therefore related to $q_{i}$ by
\begin{equation} \label{eqn 56}
\begin{split}
  q_{i(ij)}^{* \mu} & = \tensor{\Lambda}{^\mu _\nu} q_{i}^{\nu}, \\
  \hat{\Lambda} & = \begin{pmatrix}
                     \frac{P^{0}_{ij}}{(q_{ij})^{2}} \hphantom{11} & 0 \hphantom{111}
                          & 0 \hphantom{11} & \frac{|\bm{q_{ij}}|}{(q_{ij})^{2}} \\
                     0 \hphantom{11} & 1 \hphantom{111} 
                          & 0 \hphantom{11} & 0 \\
                     0 \hphantom{11} & 0 \hphantom{111}
                          & 1 \hphantom{11} & 0 \\
                     \frac{|\bm{q_{ij}}|}{(q_{ij})^{2}} \hphantom{11} & 0 \hphantom{111} 
                          & 0 \hphantom{11} & \frac{P^{0}_{ij}}{(q_{ij})^{2}}
                    \end{pmatrix}
                    \begin{pmatrix}
                     1 & 0 & 0 & 0 \\[+4pt]   
                     0 & \cos(\theta_{ij}) & 0 & \sin(\theta_{ij}) \\[+4pt]
                     0 & 0 & 1 & 0 \\[+4pt]
                     0 & -\sin(\theta_{ij}) & 0 & \cos(\theta_{ij})
                    \end{pmatrix} \\
		\mt	\times	\begin{pmatrix}
                     1 & 0 & 0 & 0 \\[+4pt]   
                     0 & \cos(\phi_{ij}) & \sin(\phi_{ij}) & 0 \\[+4pt]
                     0 & -\sin(\phi_{ij}) & \cos(\phi_{ij}) & 0 \\[+4pt]
                     0 & 0 & 0 & 1
                    \end{pmatrix},
\end{split} 
\end{equation} 
where $(q_{ij})^{2}=q_{\mu}q^{\mu}$ and $\theta_{ij}$ and $\phi_{ij}$ are the polar and azimuthal angles of $\bm{q_{ij}}$. $\theta_{ij}$
and $\phi_{ij}$ are given by
\begin{equation} \label{eqn 57}
 \begin{split}
  \tan(\theta^{*}_{ij})& =\frac{\sqrt{(q^{1}_{ij})^{2}+(q^{2}_{ij})^{2}}}{q^{3}_{ij}}, \\
  \tan(\phi^{*}_{ij})& =\frac{q^{2}_{ij}}{q^{1}_{ij}},
 \end{split}
\end{equation}
where the expressions for $q^{1}_{ij}$, $q^{2}_{ij}$, and $q^{3}_{ij}$ can be found from eqns. (\ref{eqn 5}), (\ref{eqn 6}), (\ref{eqn 11}),
(\ref{eqn 51}) and (\ref{eqn 54}) and the correct solution for $\phi_{ij}$ is chosen based on the signs of $q^{1}_{ij}$ and $q^{2}_{ij}$. After 
applying this boost to find $q_{i(ij)}^{*}$ as a function of the phase space coordinates, the polar and azimuthal angles of $q_{i(ij)}^{*}$ are found in the 
standard way,
\begin{equation} \label{eqn 58}
 \begin{split}
  \tan(\theta^{*}_{ij})& =\frac{\sqrt{(q^{*1}_{i(ij)})^{2}+(q^{*2}_{i(ij)})^{2}}}{q^{*3}_{i(ij)}}, \\
  \tan(\phi^{*}_{ij})& =\frac{q^{*2}_{i(ij)}}{q^{*1}_{i(ij)}},
 \end{split}
\end{equation}
where, again, the correct solution for $\phi^{*}_{ij}$ is chosen based on the signs of $q^{*1}_{i(ij)}$ and $q^{*2}_{i(ij)}$. 

\bibliography{References}

\begin{biosketch}
The author was born and raised in the town of Villalba, Puerto Rico. He obtained a bachelor's degree in Chemical Engineering from the University of Puerto Rico, Mayag\"{u}ez Campus. His deep fascination with the field of physics that started even before he finished his high school education led him to pursue a doctoral degree in the field after obtaining his bachelor's degree. 
\end{biosketch}

\end{document}